\documentclass[english]{report}
\usepackage{mfirstuc}
\usepackage{xfor}
\usepackage{datatool}
\usepackage{substr}
\usepackage[acronym]{glossaries}

\makeglossaries
\usepackage[utf8]{inputenc}
\usepackage[english]{babel}
\usepackage[T1]{fontenc}
\usepackage{longtable}
\usepackage{tabularx}
\usepackage{multirow}
\usepackage{listliketab}
\usepackage{listings}
\usepackage{color}
\usepackage{xcolor}
\usepackage{amssymb} 
\usepackage{fancyhdr}
\usepackage{rotating}
\usepackage{caption}
\usepackage{subcaption}

\usepackage{mathptmx}
\usepackage{fontenc}
\usepackage{textcomp}
\usepackage[squaren]{SIunits}
\usepackage{graphicx}
\usepackage{url}

\usepackage{geometry}
\usepackage[absolute]{textpos}
\usepackage{makeidx}
\usepackage{colortbl}
\usepackage{pdflscape}
\usepackage{pdfpages}
\usepackage{tabularx}
\usepackage{lmodern}
\usepackage{array}
\usepackage{float}
\PassOptionsToPackage{hyphens}{url}
\usepackage[plainpages=false]{hyperref}
\usepackage{wallpaper} 
\usepackage[numbers,square]{natbib} 
\usepackage{ragged2e}  
\usepackage{booktabs}  
\usepackage{varwidth}
\usepackage{paralist}
\usepackage{sectsty}
\usepackage{enumitem}
\usepackage[ruled,vlined]{algorithm2e} 
\usepackage{tikz}
\usepackage{diagbox} 
\usepackage{lscape}
\usepackage[hyphenbreaks]{breakurl}
\usepackage[hyphens]{url}
\allsectionsfont{\rmfamily}

\newcolumntype{Y}{>{\RaggedRight\arraybackslash}X}

\makeindex

\textblockorigin{10mm}{10mm} 

\definecolor{gray}{rgb}{0.95,0.95,0.95}
\definecolor{darkgray}{rgb}{0.4,0.4,0.4}
\definecolor{lgray}{RGB}{250,250,250}
\definecolor{lgreen}{RGB}{63,127,95}
\definecolor{lred}{RGB}{127,0,85}
\definecolor{lblue}{RGB}{42,0,255}

\definecolor{orange}{rgb}{1,0.5,0}

\colorlet{punct}{red!60!black}
\definecolor{background}{HTML}{EEEEEE}
\definecolor{delim}{RGB}{20,105,176}
\colorlet{numb}{magenta!60!black}
\lstdefinelanguage{json}{
	basicstyle=\normalfont\ttfamily,
	numbers=left,
	numberstyle=\scriptsize,
	stepnumber=1,
	numbersep=6pt,
	showstringspaces=false,
	breaklines=true,
	frame=lines,
	backgroundcolor=\color{gray}, 
	string=[s]{"}{"},
	comment=[l]{:\ "},
	morecomment=[l]{:"},
	literate=
	*{0}{{{\color{numb}0}}}{1}
	{1}{{{\color{numb}1}}}{1}
	{2}{{{\color{numb}2}}}{1}
	{3}{{{\color{numb}3}}}{1}
	{4}{{{\color{numb}4}}}{1}
	{5}{{{\color{numb}5}}}{1}
	{6}{{{\color{numb}6}}}{1}
	{7}{{{\color{numb}7}}}{1}
	{8}{{{\color{numb}8}}}{1}
	{9}{{{\color{numb}9}}}{1}
}

\newcommand*\circled[1]{\tikz[baseline=(char.base)]{
		\node[shape=circle,draw,inner sep=1.5pt] (char) {#1};}}

\newglossaryentry{SMALI}
{
	name={Smali},
	text={smali},
	description={a Android specific programming language for the DEX 
	files}
}

\newglossaryentry{Tracker}
{
	name={Tracker},
	text={tracker},
	plural={trackers},
	description={A piece of software use to collect user usage data. Often 
	used in advertisement}
}

\newglossaryentry{Standard App}
{
	name={Standard App},
	text={standard app},
	plural={Standard Apps},
	description={An Android app without special system permissions that can be 
	installed over an app market}
}

\newglossaryentry{System App}
{
	name={System App},
	text={system app},
	plural={System Apps},
	description={An Android app with special system permissions that is 
	pre-installed in the Android firmware}
}

\newglossaryentry{Firmware archive}
{
	name={Firmware archive},
	text={firmware archive},
	plural={Firmware archives},
	description={A compressed file that contains all parts of an Android firmware 
	images like the kernel or the system partition}
}

\newglossaryentry{Root of Trust}
{
	name={Root of Trust},
	text={root of trust},
	description={A cryptographic key that starts a chain of trust}
}

\newglossaryentry{ssdeep}
{
	name={Ssdeep},
	text={ssdeep},
	description={A fuzzy hashing technique}
}

\newglossaryentry{TLSH}
{
	name={TLSH},
	text={TLSH},
	description={A fuzzy hashing technique created by Trend Micro}
}

\newglossaryentry{Yaffs2}
{
	name={Yaffs2},
	text={Yaffs2},
	description={Yet Another Flash File System is a log-structured file-system for 
	flash devices}
}

\newglossaryentry{RRO}
{
	name={Runtime Resource Overlays},
	text={Runtime Resource Overlays},
	description={A package that can change the resources of a target package}
}

\newglossaryentry{fsVerity}
{
	name={Fs-verity},
	text={fs-verity},
	description={A integrity protection system for read+write partitions}
}

\newglossaryentry{SparseImage}
{
	name={Sparse image},
	text={sparse image},
	description={A partition format for Android}
}

\newacronym{aosp}{AOSP}{Android Open Source Project}
\newacronym{rom}{ROM}{Read-Only Memory}
\newacronym{fbe}{FBE}{File-based encryption}
\newacronym{odm}{ODM}{Original Design Manufacturer}
\newacronym{oem}{OEM}{Original Equipment Manufacturer}
\newacronym{soc}{SoC}{System-on-Chip}
\newacronym{ota}{OTA}{over-the-Air}
\newacronym{ubifs}{UBIFS}{Unsorted Block Image File System}
\newacronym{avb}{AVB}{Android Verified Boot}
\newacronym{dbi}{DBI}{Dynamic Binary Analysis}
\newacronym{bsp}{BSP}{board-support packages}
\newacronym{hal}{HAL}{Hardware Abstraction Layer}
\newacronym{sku}{SKU}{Stock Keeping Units}
\newacronym{rro}{RRO}{Runtime Resource Overlays}
\newacronym{gki}{GKI}{Generic Kernel Image}
\newacronym{fec}{FEC}{Forward-Error-Correction}
\newacronym{pm}{PM}{Package Manager}
\newacronym{dac}{DAC}{Discretionary access control}
\newacronym{mac}{MAC}{Mandatory access control}
\newacronym{aid}{AID}{Android ID}
\newacronym{pim}{PIM}{Personal Information Manager}
\begin{document}
	\bibliographystyle{IEEEtran}
	\pagenumbering{roman}
	\begin{titlepage}

\ThisTileWallPaper{\paperwidth}{\paperheight}{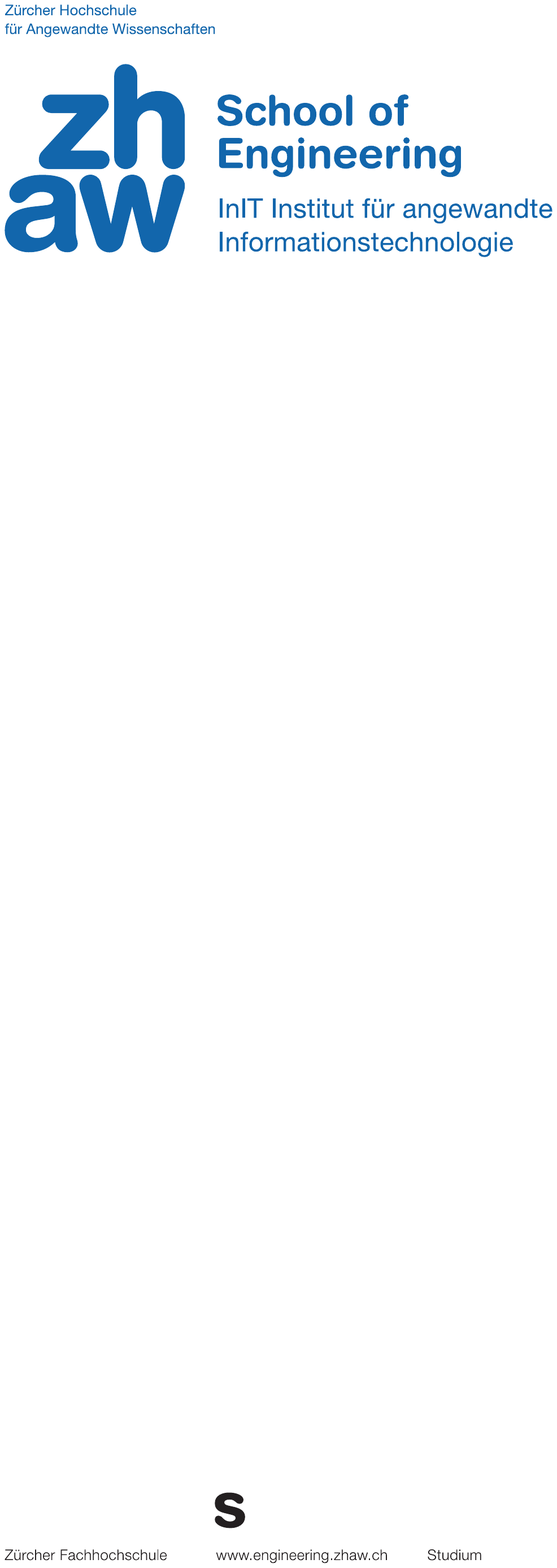}
{\fontfamily{lmss}\selectfont
\begin{minipage}[b]{0.117\textwidth}
\hskip 0.05cm
\end{minipage}
\begin{minipage}[b]{0.91\textwidth}
\begin{tiny}.\end{tiny}\vskip 2.8cm
	{\huge
	
	\textbf{\underline{Master of Science in Engineering}}\\
	
	FirmwareDroid: \\Security Analysis of the Android Firmware Eco-System\
	\vskip 0.5cm}
	
	\begin{minipage}[b]{0.27\textwidth}
	\hrule\vskip 0.5cm
		\textbf{Author}\\
	\end{minipage}
	\begin{minipage}[b]{0.03\textwidth}
	\hskip 0.5cm
	\end{minipage}
	\begin{minipage}[b]{0.7\textwidth}
	\hrule\vskip 0.5cm
		Thomas Sutter\\
	\end{minipage}
	
	\begin{minipage}[b]{0.27\textwidth}
	\hrule\vskip 0.5cm
		\textbf{Supervisor}\\
		\\
	\end{minipage}
	\begin{minipage}[b]{0.03\textwidth}
	\hskip 0.5cm
	\end{minipage}
	\begin{minipage}[b]{0.7\textwidth}
	\hrule\vskip 0.5cm
		Prof. Dr. Bernhard Tellenbach\\
		\\
	\end{minipage}	
	
	\begin{minipage}[b]{0.27\textwidth}
	\hrule\vskip 0.5cm
		\textbf{Date}
	\end{minipage}
	\begin{minipage}[b]{0.03\textwidth}
	\hskip 0.5cm
	\end{minipage}
	\begin{minipage}[b]{0.7\textwidth}
	\hrule\vskip 0.5cm
		31.01.2021
	\end{minipage}

\end{minipage}
\vskip 0.5cm
}
\end{titlepage}

	\newpage
	\thispagestyle{plain} 
	\mbox{}
	\chapter*{DECLARATION OF ORIGINALITY}

\section*{Declaration concerning the independent drafting of a project at the School of  
	Engineering}

By submitting this Master’s thesis, the undersigned student confirms that this thesis is his/her own
work and was written without the help of a third party.

\noindent
\\
The student declares that all sources in the text (including Internet pages) and appendices have been
correctly disclosed. This means that there has been no plagiarism, i.e. no sections of the Master’s
thesis have been partially or wholly taken from other texts and represented as the student’s own work
or included without being correctly referenced.

\noindent
\\
Any misconduct will be dealt with according to paragraphs 39 and 40 of the General Academic
Regulations for Bachelor’s and Master’s Degree courses at the Zurich University of Applied Sciences
(Rahmenprüfungsordnung ZHAW (RPO)) and subject to the provisions for disciplinary action
stipulated in the University regulations.
\vspace{25mm}

\noindent
\textbf{City, Date:} \hspace{0.05cm} \makebox[1.5in]{\hrulefill} \hspace{0.5cm}
\textbf{Signature:} \hspace{0.05cm} \makebox[1.5in]{\hrulefill}

	\newpage
	\thispagestyle{plain} 
	\mbox{}
	\chapter*{Abstract}

The \gls{aosp} is probably the most used and customized operating system for 
smartphones and IoT devices worldwide. Its market share and high adaptability 
makes Android an interesting operating system for many developers. Nowadays, 
we use Android firmware in smartphones, TVs, smartwatches, cars, and other 
devices by various vendors and manufacturers. 

The sheer amount of customized Android firmware and devices makes it hard for 
security analysts to detect potentially harmful applications. Another fact is that 
many vendors include apps from 3rd party developers. Such bloatware usually has 
more privileges than standard apps and cannot be removed by the user without 
rooting the device. In recent years several cases were reported where 3rd party 
developers could include malicious apps into the Android built chain. Media reports 
claim that pre-installed malware like Chamois and Triade we able to infect several 
million devices. Such cases demonstrate the need for better strategies for 
analyzing Android firmware.

In our study, we analyze the Android firmware eco-system in various ways. We 
collected a dataset with several thousand Android firmware archives and show that 
several terabytes of firmware data are waiting on the web to be analyzed. We 
develop a web service called FirmwareDroid for analyzing Android firmware 
archives and pre-installed apps and create a dataset of firmware samples. 
Focusing on Android apps, we automated the process of extracting and scanning 
pre-installed apps with state of the art open-source tools. 

We demonstrate on real data that pre-installed apps are, in fact, a 
a threat to Android's users, and we can detect several hundred malware samples 
using scanners like VirusTotal, AndroGuard, and APKiD. With state of the art 
tools, we could scan more than 900'000 apps during our research and give unique 
insights into Android custom ROMs. Moreover, we enhanced FirmwareDroid with 
fuzzy hashing algorithms and used them to detect similarities between binaries.

We show that automated security testing of Android firmware is feasible and that 
we can detect pre-installed malware with open-source analysis tools. We explain 
the challenges of dynamic analysis of Android pre-installed apps and discuss the 
limits of fuzzy hashing algorithms for similarity detection.
	\newpage
	\thispagestyle{plain} 
	\mbox{}
	\thispagestyle{empty}
	
	\begingroup
	\let\cleardoublepage\relax
	\let\clearpage\relax
	\renewcommand{\glsnamefont}[1]{\textbf{#1}}
	\setlength\LTleft{0pt}
	\setlength\LTright{0pt}
	\setlength\glsdescwidth{0.8\hsize}
	\printglossary
	\newpage
	\printglossary[title=Acronyms,type=\acronymtype, style=long]
	\setcounter{section}{0}
	\newpage
	\tableofcontents
	\endgroup
	\pagenumbering{arabic}
\setcounter{page}{1}
\chapter{Introduction} \label{Introduction}

In this thesis, we introduce a new security framework for Android, FirmwareDroid. 
Our study analyzes pre-installed apps with static analysis tools and gives 
unique insights into Android's firmware eco-system.

In recent years, several incidents happened where attackers could infiltrate 
different vendors' supply chain and include malicious apps to Android firmware, 
exposing several millions of users to potential security and privacy risks. Malware 
families like Cosiloon \cite{Cosiloon}, Chamois \cite{Chamois} and Triada 
\cite{Triada} are just some examples. Attackers seem to focus on pre-installed 
apps specifically, and it is unknown how much malware is still present in Android 
firmware today.

In recent years several open-source tools for static \cite{APKiD, QARK, 
Androwarn, AndroGuard, Exodus} and dynamic \cite{AndroPyTool, Armandroid, 
FridaSourceCode} analysis of Android apps were published. To our knowledge, 
there are no open-source tools that allow automated analysis of Android firmware 
and the containing Android pre-installed apps. Therefore we developed 
FirmwareDroid-- a tool to automated Android firmware analysis. FirmwareDroid 
allows to extract files like apps from Android firmware and does an automated 
static analysis of the extracted apps. 

Since every year a new Android version is released, and firmware updates are 
more frequently rolled-out, the need for automated analysis tools grows. We use 
Android as an operating system on smartphones; other devices like TVs and cars 
use customized AOSP builds as firmware. The demand for automated security 
assessment is high enough for large IT companies like Google, Facebook, and 
Amazon to invest in security certifications like the ioXt \cite{ioxtAlliance}.

Most studies in Android App Security have focused on analyzing apps from the 
Google Playstore or other markets. Some studies examine Android pre-installed 
apps, and firmware \cite{DroidRay, AnAnalysisofPreinstalled, firmscope-2020}, but 
to our knowledge, none of these studies released an open-source tool for 
analyzing pre-installed apps. With FirmwareDroid, we aim to provide an 
open-source framework to analyze large datasets of Android pre-installed apps in a 
reasonable amount of time. Our goal is to provide one tool that can extract apps 
from all major Android versions and is extendable for further research.

Furthermore, we use FirmwareDroid to conduct a large scale study on 5'931 
Android firmware archives and over 900'000 apps. We explain how we collected 
firmware archives from various vendors and will give some unique insight into the 
Android firmware eco-system. 

At the beginning of this thesis, we conduct a literature search in Section 
\ref{Introduction:LiteraturSearch} and discuss related work in the following Section 
\ref{Introduction:RelatedWork}. In Chapter \ref{Fundamentals} we explain our 
methodology to collect Android firmware and some of the fundamental concepts of 
Android's security system. At the beginning of Chapter \ref{Analysis} we describe 
which static analysis tools we selected for our study. After that, we progress with 
analyzing the data we gathered using these static analysis tools in 
Sections \ref{Analysis:BuildPropAnalysis} to 
\ref{Analysis:VirusTotalQarkAnalysis}. 
To further explore Android pre-installed apps we integrated \gls{TLSH} into 
FirmwareDroid and evaluate its performance in Section 
\ref{Analysis:FuzzyHashing}. In Chapter 
\ref{Implementation} we describe our software architecture and give an overview of 
FirmwareDroid's REST API, and in Chapter \ref{Results} we summarize our 
findings and 
discuss them. We end the thesis in Chapter \ref{FutureWork} by discussing future 
work and our conclusion of this project.

\section{Project Goals, Conditions and Scope}\label{Introduction:Projectgoals}
We conduct a study on Android Firmware and pre-installed app analysis and define 
the following research questions:

\begin{itemize}
	\item Can we automate the extraction of Android pre-installed apps from 
	existing android firmware archives?
	
	\item Can we detect android malware within the standard Android apps like 
	calculator, calendar, browsers etc.?
	
	\item Can we automate the static and dynamic analysis of Android firmware and 
	apps with the existing open-source tools?
\end{itemize}

Additional to the defined research questions we define a hypothesis to test:
\begin{itemize}
	\item 0.1\% of the Android apps in our corpus are malicious.
\end{itemize}

We define technical goals for the implementation of the FirmwareDroid framework:
\begin{itemize}
	\item Automate the process of scanning several thousand Android app's 
	with state of the art static and dynamic analysis tools.
	
	\item Use fuzzy hashing techniques to detect similarities between Android 
	apps.
\end{itemize}

Due to the time limitation of this master thesis, we have to focus on specific parts 
of the Android firmware and define some elements to be out of scope for this 
project:
\begin{itemize}
	\item It is not part of this work to analyze all aspects of android firmware. The 
	main focus is on analyzing pre-installed android apps (.apk).
	
	\item The main focus of this study lies on newer (>= 8) Android versions. Other 
	versions can be mentioned but are not in the focus of this research.
	
	\item We focus on analyzing the system partition of the Android firmware. Other 
	parts of the firmware like for example, the kernel, is out of scope for this 
	study.
	
	\item It is not part of this study to analyse malware in detail or to provide 
	exploits.
\end{itemize}

\section{Literature Search} \label{Introduction:LiteraturSearch}
In this section, we describe our methodology for our literature search, and in 
Section \ref{Introduction:RelatedWork} we discuss related work. We use the tool 
Publish or Parish \cite{PublishorParish} to conduct a keyword-based literature 
search on schoolar.google.com. Table \ref{tab:Search_Terms} shows all the 
keywords used for our investigation. We search every time for a main keyword 
together with one of the search terms. For every combination of keyword and 
search term, we save the top 20 search results. In total, this results in a list of 980 
publications. We filter patents, HTML, or any other kind of reports that are not 
scientific publications and not available as pdf files. As the next step, we removed 
all older publications than ten years and published before 01.01.2011. Moreover, 
we filter all papers that are not in English. 
As the final step, we read from the remaining 902 papers the title, abstract, and, if 
necessary, the results section to determine if the publication is relevant for our 
work or related to our study. We show the results of this literature search in 
Appendix \ref{Appendix:LiteratureSearchResults}.

\begin{table}[H]
	\resizebox{\textwidth}{!}{
	\begin{tabular}{|l|l|}
		\hline
		\rowcolor[HTML]{EFEFEF} 
		\textbf{Main Keyword} & \textbf{Search terms} \\ \hline
		Fuzzy Hashing & \begin{tabular}[c]{@{}l@{}}tools, ssdeep, TLSH, TLSH 
		clustering, lempel-ziv jaccard, sdhash, \\ performance, clustering\end{tabular} 
		\\ 
		\hline
		Android & \begin{tabular}[c]{@{}l@{}}emulator detection, root detection, 
		obfuscator, deobfuscator, \\ packer, unpacker, code analysis, app fuzzing, \\ 
		malware analysis, permissions, privacy, \\ verified boot, verified boot 2.0, 
		avb, 
		dm-verity\end{tabular} \\ \hline
		Android firmware & \begin{tabular}[c]{@{}l@{}}analysis,  dataset,  samples, 
		extraction, \\ kernel fuzzing, kernel extraction, app extraction\end{tabular} \\ 
		\hline
		Android pre-installed apps & \begin{tabular}[c]{@{}l@{}}bloatware analysis, 
		analysis, static analysis, dynamic analysis, \\ malware\end{tabular} \\ \hline
		Android library & detection, fingerprinting \\ \hline
		Android static analysis & \begin{tabular}[c]{@{}l@{}}tools, reverse 
		engineering, 
		permission extraction, \\ manifest parsing\end{tabular} \\ \hline
		Android dynamic analysis & \begin{tabular}[c]{@{}l@{}}firmware, pre-installed 
		app, traffic analysis, \\ docker, emulator, frida, api monitoring, file access 
		monitoring\end{tabular} \\ \hline
	\end{tabular}
}
	\caption{Overview of search terms used for our literature search in Appendix 
	\ref{Appendix:LiteratureSearchResults}.}
	\label{tab:Search_Terms}
\end{table}

\section{Related Work} \label{Introduction:RelatedWork}
Our literature search in the Section \ref{Introduction:LiteraturSearch} 
shows that several studies exist with similar aims when it comes to analyzing 
Android firmware or pre-installed app. In this section, we review literature related to 
Android pre-installed app analysis and other related topics.

\textbf{Android firmware analysis:}
One of the first publications analyzing the pre-installed app eco-system is from 
2014 by Zheng et al. \cite{DroidRay}. Min Zheng et al. studied in \cite{DroidRay} 
250 Android firmware archives for vulnerabilities and found that 19 
firmware samples contained
malware. Moreover, Zheng et al. found out that 1'947 apps had vulnerable 
signatures. A study from Julien Gamba et al. \cite{AnAnalysisofPreinstalled} 
scanned pre-installed apps of over 200 vendors for vulnerabilities and analyze the 
traffic of the apps with the 
Lumen\footnote{\url{https://androidobservatory.com/datasets/}/} dataset. 
Their work identified 36 
potentially privacy-intrusive behaviors of pre-installed apps. Furthermore, Julien 
Gamba et al. identified the top 15 advertising and tracking services and identified 
the most requested permissions. Another recent study from Mohamed Elsabagh et 
al. analyzed in \cite{firmscope-2020} 2'017 firmware archives and extracted 
331'342 apps. Their work claims to find over 850 new vulnerabilities for 
pre-installed apps using a novel static taint analysis tool. 

In another direction goes the Uraniborg \cite{uraniborg-scoring-2020} project. It is a 
collaboration of the University of Cambridge and the Johannes Kepler University 
and aims to create a security risk scoring system for Android firmware. Similar to 
\cite{ioxtAlliance} it's aim is to create a verification and certification process for 
Android firmware to increase the security of Android firmware overall.

Grant Hernandez et al. show in \cite{BicMac} that it is possible to develop a tool 
for testing Android's permission policies on static Android firmware. Their approach 
can test DAC and MAC policies without actually running the firmware on a real 
device. They demonstrate the capabilities of their framework by detecting novel 
privilege escalation attacks on Android firmware. 

\textbf{Fuzzy hashing:} 
\label{Introduction:FuzzyHashingAndSimiliarityDetection}
Like cryptographic hash functions, fuzzy hash algorithms create a hash of any 
given data like strings or files. In contrast to cryptographic hash functions like 
MD5 or SHA1, we use fuzzy hashes to detect similarities between binaries and 
not exact matches. Therefore, small changes in the data do not significantly 
impact the fuzzy hash's digest like it would have with a cryptographic hash. 
Several implementations of fuzzy hash algorithms exist. For example, TLSH 
\cite{TLSH},  Ssdeep \cite{SSDEEP}, Sdhash \cite{sdhash}, Lempel-Ziv Jaccard 
distance \cite{LempelZiv}. All of these hashes have different approaches to 
detecting similarities. 

One of the challenges when using fuzzy hashes is how to compare the hashes 
efficiently and at scale. In a naive approach, we can compute all fuzzy hashes of 
a 
dataset and then compare them against each other. The result will be a matrix of 
$n * n$ rows and columns with an exponential growth of $\mathcal{O}(n^{2})$ for 
the computation of all values. More 
computational efficient approaches use a Locality-sensitive hashing where similar 
hashes lay near each other in the vector space or other optimizations like filtering 
the search space to reduce the number of needed comparisons.

One use case of fuzzy hashes is to identify malware families. For example, Nitin 
Naik et al. use in \cite{ClusteringCMeans} fuzzy hashes combined with the 
fuzzy c-means clustering to detect ransomware families. A different approach in 
clustering malware families was proposed by Parvez Faruki et al. in 
\cite{AndroSimilar}. With SDHash, their prototype can cluster known malware 
families with an accuracy of 76\%. Another more recent study from Muqeet Ali, 
Josiah Hagen, and Jonathan Oliver 
\cite{ali2020scalable} shows that a multi-stage clustering with k-means and HAC-T 
is applicable for clustering TLSH digests on the scale.

\textbf{Library detection:} 3rd party libraries are commonly used on Android apps. 
Several researchers propose techniques to detect these 3rd party libraries for 
vulnerability detection. However, in some cases, obfuscation makes detection of 
such libraries a challenging task.

Michael Backes et al. describe in \cite{ReliableThirdPartyLibrary} their approach to 
detect vulnerable libraries for apps in the Google Play Store. Their results suggest 
that apps have slow library patching circles and that 296 apps have severe misuse 
of cryptographic APIs.

Ziang Ma et al. claim in \cite{LibRadar} that their approach based on the frequency 
of different Android API calls is in most cases resilient to obfuscation. Their 
system collects features by pre-processing 1'027'584 apps from the Google Play 
Store and using these features to identify similar libraries. As a result, they have 
created a list of 29,279 potential libraries for Google Play apps.

Another approach to overcome the problem of obfuscated library code is shown by 
Menghao Li et al. in \cite{LibD}. Their solution is based on feature hashing, and the 
usage of internal code dependencies. In their work, they 
were able to detect 60'729 different libraries. A different solution is proposed by 
Yuan Zhang et al. in \cite{libpecker}. Their system, called LibPecker, adopts 
signature matching of application and library classes to find similarities with high 
precisions and recall. LibPecker outperforms other state-of-the-art tools with 91\% 
recall and 98.1\% precision.

\textbf{Vulnerability analysis:} There is a vast number of tools and papers that 
attempt to automate the process of vulnerability analysis. For example, Francisco 
Palma et al. developed a system to detect app vulnerabilities for app stores in 
\cite{AutomatedSecurityTesting}. Or Amr Amin et al. describes in 
\cite{AndroShield} a web-service to automate the process of static- and dynamic 
app analysis.

\subsection{Conference Talks}
There is also a large number of conference talks providing insights into 
many aspects of the Android eco-system. We found the talks listed in Table 
\ref{tab:conferences} to be very insightful for anyone wanting to do a 
deep-dive into the topic of Android security and the analysis of Android 
apps. However, these lectures are not relevant for this work. They do not 
have a direct relation to the contributions.

\begin{table}[H]
	\centering
	\resizebox{\textwidth}{!}{%
		\begin{tabular}{|l|l|}
			\hline
			\rowcolor[HTML]{EFEFEF} 
			\textbf{Conference} & \textbf{Author - Title} \\ \hline
			RSAConference 2020 & \begin{tabular}[c]{@{}l@{}}Łukasz Siewierski - 
			Challenges in Android Supply \\ Chain Analysis 
			\cite{ChallengesSupplyChain}\end{tabular} \\ \hline
			IEEE Symposium on Security and Privacy 2020 & 
			\begin{tabular}[c]{@{}l@{}}Julien Gamba et. al. - An Analysis of 
			Pre-installed \\ Android Software 
			\cite{ConferenceAnAnalysisOfPreInstalled}\end{tabular} \\ \hline
			BlackHat 2020 & \begin{tabular}[c]{@{}l@{}}Maddie Stone - Securing the 
			System: A Deep Dive into Reversing \\ Android Pre-Installed Apps 
			\cite{SecuringTheSystem}\end{tabular} \\ \hline
			RSAConference 2018 & \begin{tabular}[c]{@{}l@{}}Giovanni Vigna - How 
			Automated Vulnerability Analysis Discovered \\ Hundreds of Android 
			0-days \cite{HowToAutomate}\end{tabular} \\ \hline
			BlackHat EU 2015 & \begin{tabular}[c]{@{}l@{}}Yu-Cheng Lin - Androbugs 
			Framework: An Android \\ Application Security Vilnerability Scanner 
			\cite{AndroBugs}\end{tabular} \\ \hline
			Black Hat 2013 & Kevin McNamee - How to Build a SpyPhone 
			\cite{SpyPhone} \\ \hline
			DFRWS EU 2019 & \begin{tabular}[c]{@{}l@{}}Lorenz Liebler and Harald 
			Baier - Toward Exact and Inexact\\ Approximate Matching Of Executable 
			Binaries \cite{TowardExact}\end{tabular} \\ \hline
		\end{tabular}%
	}
	\caption{List of topic related conference talks.}
	\label{tab:conferences}
\end{table}

	\chapter{Fundamentals} \label{Fundamentals} 

This chapter explains how we created our dataset and how to extract 
pre-installed apps from Android firmware. One of the first challenges we 
have to overcome in analyzing pre-installed apps is to collect firmware 
archives, and that we need to know where the pre-installed apps within the 
firmware are stored. 

Moreover we introduce and discusses (Android) concepts and knowledge 
which is essential to understand the challenges and solutions presented in 
the reminder of this thesis.

\section{Android Firmware Structure} 
\label{Fundamentals:AndroidFirmwareStructure}
A \gls{Firmware archive} is a compressed file that contains all parts of the Android 
firmware. An Android firmware has several partitions or image files 
\cite{DevImages}. Depending on the Android version, these partition files have 
other formats and are named differently \cite{DevParitions}. Their content may as 
well vary depending on the smartphone vendor. For example, the content 
and basic structure of an Android 10 firmware image might look as follows: 

\begin{itemize}
	\item \textbf{bootloader.img:} As the name suggests, it contains the 
	smartphone's bootloader. The bootloader on Android assures that only genuine 
	partitions are loaded at start \cite{DevBootload}. The bootloader starts the 
	recovery partition or the kernel of the OS.
	
	\item \textbf{boot.img:} This image contains the kernel and ramdisk image. 
	Android uses the ramdisk image to start the initial process \textit{initd}, and it 
	includes a file-system with the \textit{/init} directory. 
	
	\item \textbf{system.img:} The system image contains Android's main 
	framework with generic code meant for sharing on multiple devices. This 
	partition is usually mounted read-only and contains the main file system. The 
	file-system structure is similar to the one shown in Figure 
	\ref{fig:08_OdmPartition}, but 
	instead of the odm folder, it uses the system folder. In the folders \textit{/app} 
	and \textit{/priv-app} contain most of the pre-installed apps. 
	
	\item \textbf{vendor.img:} The vendor partition is used for including 
	proprietary binaries into Android's file-system. It allows to overlay 
	existing files or to add new ones \cite{DevVendorOverlays}. Its purpose is 
	that vendors can integrate customizations for their smartphones without 
	having to modify the system image.
	
	\item \textbf{oem.img and odm.img:} Similar to the vendor image an \gls{odm} 
	or \gls{oem} can include customizations over the oem.img or odm.img 
	partitions. These images are optional and represent an extension to the vendor 
	images. They can be used to create \gls{rro} for android apps 
	\cite{DevRuntimeResourceOverlays} or to add any customizations for 
	board-specific components. Figure \ref{fig:08_OdmPartition} shows an example 
	odm image file-system. From Figure \ref{fig:08_OdmPartition} we can see that 
	an \acrshort{odm} can include any new file like for example: native libraries, 
	selinux policies, system apps or kernel modules.
	
	\begin{figure}[H]
		\centering
		\includegraphics[width=1\linewidth]{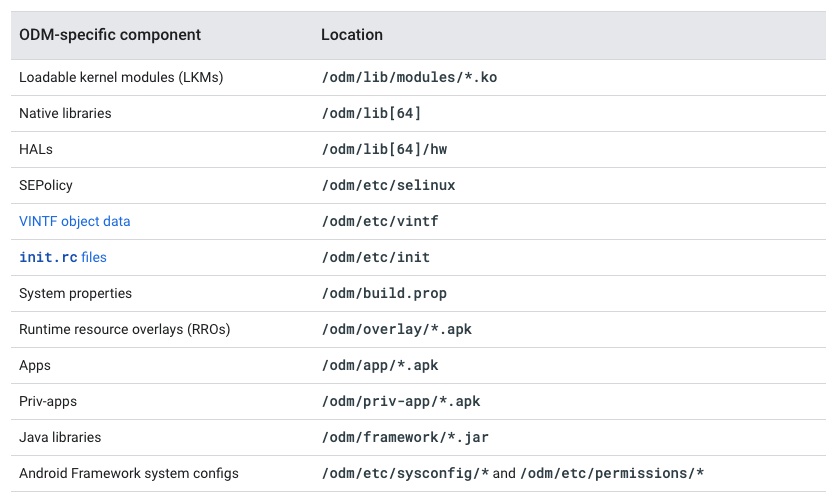}
		\caption{Overview of ODM-specific components. Image source 
		\cite{DevODMPartition}}
		\label{fig:08_OdmPartition}
	\end{figure}
		
	\item \textbf{userdata.img:} Another possibility to include customized data into 
	the operating system is over the user-data image. This partition is optional and 
	can consist of user-installed apps and data \cite{DevODMPartition}.
	
	\item \textbf{radio.img:} Usually, this image is a proprietary binary that includes 
	the base-radio firmware.
\end{itemize}
Note: More information about partition files can be found on the official Google 
developer website. See \cite{DevParitions, 
DevImages, DevODMPartition} fore more information.

The images mentioned above are just an example for Android 10. Other Android 
versions have different partitions. For example, in Android 11, changes to the boot 
image were made \cite{DevAndroid11Release}. Android 11 introduced so-called 
\Glspl{gki} \cite{DevVendorBootPartitions} to separate generic kernel code from 
\Glspl{soc} code. However, we do not focus on the Android kernel and its boot 
images for this project. Our goal is to analyze pre-installed apps, and therefore our 
main focus is on analyzing the system image. Other images 
with pre-installed apps like the optional OEM image are interesting for our analysis 
but out of scope.

\section{Collecting Firmware Samples} 
\label{Fundamentals:CollectingFirmwareSamples}
Most smartphone vendors do not publish download links for their firmware. At the 
time of writing, Google is one of the only vendors with its firmware available for 
free. 
From their official webpage, \cite{DevGoogleFirmware} we can download all 
releases for their devices, allowing us to get a dataset of around 1TB of firmware 
samples.

To check whether other vendors offer the same service, we searched 
through their official webpages. We show the result of this search in Table 
\ref{tab:VendorLinks}. To our knowledge, it seems that except for Google and 
Sony, all other vendors do not offer the possibility to download their firmware 
images for free. Therefore, to have a wider variety of samples, we used alternative 
firmware sources from unofficial webpages.

\begin{table}[H]
	\centering
	\resizebox{0.8\textwidth}{!}{
	\begin{tabular}{|l|l|l|}
		\hline
		\textbf{Vendor} & \textbf{Link}                                          & 
		\textbf{Firmware} \\ \hline
		Alcatel         & \url{https://www.alcatelmobile.com/}                   & 
		None              \\ \hline
		Google          & \url{https://developers.google.com/android/images}     & Free 
		complete     \\ \hline
		Hitec Mobile    & \url{https://hitecmobile.com.sg/}                      & 
		None              \\ \hline
		Huawei          & \url{https://www.huawei.com/}                          & 
		None              \\ \hline
		Intex           & \url{https://www.intex.in/}                            & None              
		\\ \hline
		Karbonn         & \url{https://www.karbonnmobiles.com/}                  & 
		None              \\ \hline
		LG              & \url{https://www.lg.com}                               & None              
		\\ \hline
		Maximus         & \url{http://www.maximus-mobile.com/}                   & 
		None              \\ \hline
		Micromax        & \url{https://www.micromaxinfo.com/}                    & 
		None              \\ \hline
		Mobicel         & \url{https://mobicel.co.za/smart-phones/}              & 
		None              \\ \hline
		Motorola        & \url{https://www.motorola.com/}                        & 
		None              \\ \hline
		Nokia           & \url{https://www.nokia.com/}                           & 
		None              \\ \hline
		Oppo            & \url{https://www.oppo.com/in/}                         & 
		None              \\ \hline
		Panasonic       & \url{https://mobile.panasonic.com/in/}                 & 
		None              \\ \hline
		Samsung         & \url{https://www.samsung.com/}                         & 
		None              \\ \hline
		Sony            & \url{https://developer.sony.com/develop/open-devices/} & 
		Partially         \\ \hline
		Symphony        & \url{https://www.symphony-mobile.com/}                 & 
		None              \\ \hline
		Tecno           & \url{https://www.tecno-mobile.com/home/}               & 
		None              \\ \hline
		Vivo            & \url{https://www.vivo.com/in/}                         & 
		None              \\ \hline
		Walton          & \url{https://waltonbd.com/smart-phone}                 & 
		None              \\ \hline
		Xiamoi          & \url{https://www.mi.com/in/}                           & 
		None              \\ \hline
		ZTE             & \url{https://www.ztedevices.com/de/}                   & 
		None              \\ \hline
	\end{tabular}
	}
	\caption{Overview of Android OS vendors and the checked links.}
	\label{tab:VendorLinks}
\end{table}

When we search the web for Android firmware, we can find many websites offering 
firmware files for more or less any known vendor on the market. Some websites 
provide their Android firmware for free, and some even have built commercial 
service around custom firmware. We visited these websites and found out that 
their firmware is, in many cases, accessible for everyone. Most sites use file 
hosters like Google Drive, Mega.nz, Uploaded.net, MediaFire.com, and 
AndroidFileHost.com to store their firmware. With this information in mind, we 
decided to test if we could use the webpages in Table \ref{tab:CustomROMSites} 
as an additional source of Android firmware for our study. To download the 
firmware, we developed a web crawler with puppeteer \cite{Puppeteer} and a 
headless browser that collected all download links from the webpage in Table 
\ref{tab:CustomROMSites}.

\begin{table}[H]
	\centering
	\resizebox{0.6\textwidth}{!}{
\begin{tabular}{|c|l|c|}
	\hline
	\rowcolor[HTML]{EFEFEF} 
	\multicolumn{1}{|l|}{\cellcolor[HTML]{EFEFEF}\textit{\textbf{\#}}} & 
	\textbf{Custom ROM Website} & \textit{\textbf{Estimated Firmware Data}} \\ 
	\hline
	1 & \url{https://firmwarefeeds.com/} & \textit{~4.17TiB} \\ \hline
	2 & \url{https://firmwarefile.com} & \textit{~4.18TiB} \\ \hline
	3 & \url{https://vnrom.net/} & \textit{~0.72TiB} \\ \hline
	4 & \url{https://firmwarex.net/} & \textit{~2.64TiB} \\ \hline
	5 & \url{http://huawei-firmware.com/} & \textit{Not Crawled} \\ \hline
	6 & \url{https://androidfilehost.com/} & \textit{Not Crawled} \\ \hline
	7 & \url{https://mirrors.lolinet.com/firmware/} & \textit{Not Crawled} \\ \hline
	8 & \url{https://firmware.center/firmware/} & \textit{~2.84TiB} \\ \hline
	9 & \url{https://fire-firmware.com/} & \textit{Not Crawled} \\ \hline
	10 & \url{https://droidfilehost.com/} & \textit{Not Crawled} \\ \hline
	11 & \url{https://www.ytechb.com/} & \textit{~17.2Gib} \\ \hline
	12 & \url{https://stockromfiles.com/} & ~4.52GiB \\ \hline
	13 & \url{https://www.androidinfotech.com/} & ~8.24GiB \\ \hline
	14 & \url{https://www.xda-developers.com/} & \textit{Not Crawled} \\ \hline
	15 & \url{https://androidfre.com/} & \textit{Not Crawled} \\ \hline
	16 & \url{https://samsungfirmware.net/} & ~7.51GiB \\ \hline
	17 & \url{https://www.samsungsfour.com/} & \textit{Not Crawled} \\ \hline
	18 & \url{https://www.androidsage.com/} & ~1.39TiB \\ \hline
	19 & \url{https://addrom.com/} & ~13.73TiB \\ \hline
	20 & \url{https://www.getdroidtips.com/} & ~7.04TiB \\ \hline
	21 & \url{https://www.teamandroid.com/} & \textit{~0.67TiB} \\ \hline
	22 & \url{https://www.entertainmentbox.com/} & \textit{Not Crawled} \\ \hline
	23 & \url{https://androidtvbox.eu/} & ~0.15TiB \\ \hline
	24 & \url{https://androidpctv.com/} & \textit{~0.12TiB} \\ \hline
	25 & \url{https://www.cubot.net/} & \textit{Not Crawled} \\ \hline
\end{tabular}
	}
\caption{List of websites offering Android firmware. Note: the crawled number of 
	bytes is without duplicates links.}
	\label{tab:CustomROMSites}
\end{table}

To test our approach we started with two web pages, number one and two in Table 
\ref{tab:CustomROMSites}. These two pages were selected, because they offer 
firmware for more than 50 vendors and therefore seemed to have a reasonable 
variety of Android versions. 
Against our expectation our crawler was able to collect unique links of around 
7'000 files with a storage need of around 10TB only using these two webpages. As 
shown in Table \ref{tab:CustomROMSites} we further crawled through other 
webpages and found several thousand potential links with Android firmware. We 
then filtered the list of links in JDownloader2 \cite{JDownloader2} to remove 
duplicates, files with small size (<=100MB) and files which filename indicates that 
they are not Android firmware like for example, drivers or hacking tools. Since our 
storage capacity is limited we decided to download only the firmware archives 
from firmwarefeeds.com and firmwarefile.com as a starting point for our study. 
Using firmware from these unofficial sources has some drawbacks we would like 
to mention at this point. 

First, since we use unofficial firmware samples from untrusted sources, we cannot 
assume that it is the vendor's original content. We have to consider that a third 
party could have manipulated all downloaded firmware. The only exception is for 
the Google firmware sample since we download them from the official source.

Second, another drawback is that we cannot filter the firmware by version before 
we have downloaded them. We cannot distinguish which Android version the file is 
if the filename does not contain the version name or number. In some cases, it is 
possible to filter older Android versions by their filename, but the files do not have 
a naming schema in most cases. Therefore, it is likely that we will download many 
older Android versions, which are not interesting for our study.

Nevertheless, after collecting around 10TB of firmware samples from official and 
unofficial sources, we have to solve two problems to continue our analysis:
\begin{itemize}
	\item How do we access the data within the firmware images?
	
	\item How do we identify the version of the firmware?
\end{itemize}
We will discuss how we solved these problems in the next two Sections.

\section{Extracting pre-installed Apps}  
\label{Fundamentals:ExtractingPreInstalledApps}

\begin{figure}[H]
	\centering
	\includegraphics[width=1\linewidth]{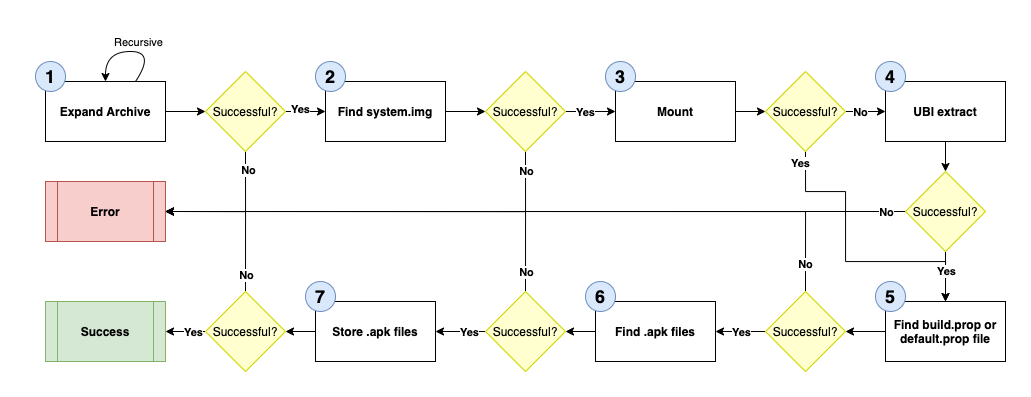}
	\caption{Overview of the extraction process.}
	\label{fig:01_MassImport}
\end{figure}

To access the firmware data, we have to understand where the data of interest is 
stored and then access it. As explained in Section 
\ref{Fundamentals:AndroidFirmwareStructure} pre-installed apps are stored in the 
system.img image. If we want to extract a file from the system.img or any other 
image, we need to mount or extract it. Since Android is customizable, vendors use 
different file formats, naming schemas, and compression algorithms for their 
images. These facts make automation more challenging since we have to deal 
with those vendor-specific customizations if we want to extract the data from their 
firmware.

We have designed a process to export android apps from the system.img. We 
show an overview of our approach in Figure \ref{fig:01_MassImport}. To extract the 
Android apps, we implemented a process with seven steps:

\begin{enumerate}
	\item \textbf{Decompress archives:} Android vendors compress firmware 
	archives with various compression formats. If we want to access the data, it is 
	necessary to expand these archives. Following an incomplete list of archive 
	formats we have detected in our dataset:
	\begin{itemize}
		\item .zip
		\item .tar and tar.md5
		\item .pac
		\item .bin
		\item .dat
		\item .lz4
		\item .nb0
	\end{itemize}
	In some cases, vendors pack their firmware files more than once, and 
	consequently, it is necessary to expand the nested archives as well. We can 
	recursively expand the nested archives to get the data. A problem that can 
	occur 
	is that the libraries are encrypted and need a password to extract. In these 
	cases, 
	we cannot expand the archive without the password or key. Another problem is 
	the 
	use of unknown compression formats. In this case, we cannot extract the 
	archive 
	without adding the support for the additional format. The list of formats above 
	shows which formats we currently support in FirmwareDroid.
	
	\item \textbf{Find system.img:} As soon as we have decompressed all nested 
	archives on the file system, we search for the system.img file. One of the main 
	problems when we search for this file, is that there is no strict naming 
	convention. Consequently, vendors name the system.img file in many ways, for 
	example, system.rfs, system.bin, system.ext4.img, system-sign.img, 
	sign-system.img, system.img\_sparsechunk.0, system\_6.img and so on. To 
	solve this problem, we use a priority list of regex patterns to search for the 
	correct file. This approach is not 100\% error resistance but shows good enough 
	results to import several thousand firmware samples. With the regex priority list, 
	we search for exact matches, then for small changes, and at the end for larger 
	changes in the naming. If we cannot find the file, the process stops, and the 
	firmware cannot be import.
	
	\item \textbf{Mount:} In case the system.img file is found, we can attempt to 
	mount the file with the Linux \textit{mount} command. The system.img can be a 
	\gls{Yaffs2} or a \gls{SparseImage} \cite{DevImages}. In case it is a sparse 
	image we 
	can use 
	simg2img \cite{Simg2img} to reformat the file into an ext4 or f2fs partition and 
	mount the file. This approach works for many firmware files, but it does not 
	work for all samples since some vendors use custom ext4 headers, encrypted 
	firmware files, or other formats that we cannot mount without further processing. 
	To mount as many samples as possible, we tested several pre-processing 
	strategies and came up with the following one:
	\begin{enumerate}[label={\arabic*.}]
		\item Simg2img and mount (Std. / fuse)
		\item mount
		\item fuse and fuseext2 \cite{fuseext2}
		\item Resize and mount (Std.)
		\item Repair, resize and mount (Std.)
	\end{enumerate}
	
	Testing showed that using the Linux kernel extension fuse allows mounting 
	some ext4 partitions with customized headers. It seems that fuse on Debian 
	ignores some customer headers, where mount often refuses to perform the 
	mount operation. Therefore we use fuseext2 as a supplement tool for 
	performing mount operations.
	
	Another problem is when vendors split the system.img files into chunks. 
	Simg2img can process some chunk formats and generate one complete file 
	from the chunks. Nevertheless, these files often could not be mounted in our 
	process due to invalid sparse formats. Therefore, the current FirmwareDroid 
	version cannot mount chunked firmware files. Moreover, we have 
	password-protected zip files, and we cannot unpack them without the correct 
	password. In the current version of FirmwareDroid we ignore these files.
	
	\item \textbf{UBI extract:} Some vendors use the \gls{ubifs} alternative to ext4. 
	To extract the data from such files, we use an open-source python code 
	\cite{UbiReader} that allows us to extract such UBI partitions directly to the 
	file-system without mounting.
	
	\item \textbf{Find build.prop or default.prop:} After it is possible to mount the 
	system partition, we can search through the file-system. Like other studies 
	\cite{firmscope-2020, AnAnalysisofPreinstalled} one of the files we extract is 
	the \textit{build.prop} file. It contains a key-value store with the system 
	properties. We can use it to extract the meta-data of the firmware. Figure 
	\ref{lst:buildProp} shows an example build.prop file. For example, we can find 
	some data like the Android version, brand, manufacturer, language, or build 
	version in the build.prop file, and we can parse the file to store it in a database.
	
	\item \textbf{Find .apk files:} Similar to the build.prop file, we can extract apks 
	from the mounted or extracted file-system. We search for filenames with the 
	.apk extension and store their location.
	
	\item \textbf{Store .apk files:} All the found apks are then copied from the 
	system.img partition to the host file-system for further processing. Copying the 
	apks has the advantage that we can use other analysis tools directly on the 
	files without the need to extract and mount the system.img on every access. 
	One drawback of this approach is that we need more disk storage for storing 
	the files. Together with copying the files, we create some meta-data of the apks 
	and store them in the database.
\end{enumerate}

We tested the process shown in Figure \ref{fig:01_MassImport} and could import 
5'931 firmware archives to our database successfully. However, with the current 
design, we were not able to import another 3'224 potential firmware files. We think 
this is due to customized, old, or unknown file formats, unique naming schemas, 
or files that aren't Android firmware.

\newpage
\section{Version Identification}

There are various ways to identify the version of an Andrioid firmware. For 
example, we can look for files that only exist on specific Android versions. Such a 
file can be, for example, the 'com.android.egg' app, which is different on every 
Android version and always has the name in the form of "Android Q Ester Egg" or 
any other system app which is unique to an Android version. Another way is to 
read the contents of the build.prop or default.prop file. As shown in Listing 
\ref{lst:buildProp} this file holds a key-value store with properties. The key-value 
pairs represent the system properties of the operating system. As shown on Line 
18 in Listing \ref{lst:buildProp} we can identify the version by the property 
\textit{ro.build.version.release}. Moreover, we can find other information like the 
phone manufacturer, the last security patch date, or the product model in the 
build.prop file.

\begin{lstlisting}[caption={Contens of an exmple build.prop file.}, 
label=lst:buildProp, language=make, basicstyle=\scriptsize] 
# PRODUCT_OEM_PROPERTIES\part{title}
import /oem/oem.prop ro.config.ringtone
import /oem/oem.prop ro.config.notification_sound
import /oem/oem.prop ro.config.alarm_alert
import /oem/oem.prop ro.config.wallpaper
import /oem/oem.prop ro.config.wallpaper_component
import /oem/oem.prop ro.oem.*
import /oem/oem.prop oem.*
# begin build properties
# autogenerated by buildinfo.sh
ro.build.id=MDA89D
ro.build.display.id=MDA89D
ro.build.version.incremental=2294819
ro.build.version.sdk=23
ro.build.version.preview_sdk=0
ro.build.version.codename=REL
ro.build.version.all_codenames=REL
ro.build.version.release=6.0
ro.build.version.security_patch=2015-10-01
ro.build.date=Wed Sep 30 00:50:26 UTC 2015
ro.build.date.utc=1443574226
ro.build.type=user
ro.build.user=android-build
ro.build.host=wpix10.hot.corp.google.com
ro.build.tags=release-keys
ro.build.flavor=angler-user
ro.product.model=Nexus 6P
ro.product.brand=google
ro.product.name=angler
ro.product.device=angler
ro.product.board=angler
# ro.product.cpu.abi and ro.product.cpu.abi2 are obsolete,
# use ro.product.cpu.abilist instead.
ro.product.cpu.abi=arm64-v8a
ro.product.cpu.abilist=arm64-v8a,armeabi-v7a,armeabi
ro.product.cpu.abilist32=armeabi-v7a,armeabi
ro.product.cpu.abilist64=arm64-v8a
ro.product.manufacturer=Huawei
ro.product.locale=en-US
ro.board.platform=msm8994
# ro.build.product is obsolete; use ro.product.device
ro.build.product=angler
# Do not try to parse description, fingerprint, or thumbprint
ro.build.description=angler-user 6.0 MDA89D 2294819 release-keys
ro.build.fingerprint=google/angler/angler:6.0/MDA89D/2294819:user/release-keys
ro.build.characteristics=nosdcard
\end{lstlisting}

A drawback of using the build.prop file is that customized Android ROMs may 
modify the file's values and show us false information. Therefore, we cannot 
completely trust the found information in the build.prop file for custom ROMs, and 
when using it for the Android version detection, we have to keep in mind that we 
could detect a false version. Nevertheless, in FirmwareDroid we use the build.prop 
file to detect the firmware version.

\section{Verified Boot} \label{Fundamentals:VerifiedBoot}
Our dataset is likely to contain modified Android firmware because we downloaded 
the most samples from untrusted sources. To understand how we can distinguish 
between an original firmware file and a modified one, we need to know Android's 
integrity, and boot protection. We, therefore, explain in this section the main 
components of \gls{avb} 2. We will not go through all details of \acrshort{avb} 
because this would go beyond this project's scope. We will explain parts of the 
\acrshort{avb} also called Verified Boot 2.0 that Google introduced with Android 
Oreo. We do not discuss earlier versions of Android that use Verified Boot 1.0. 

\begin{figure}[H]
	\centering
	\includegraphics[width=1\linewidth]{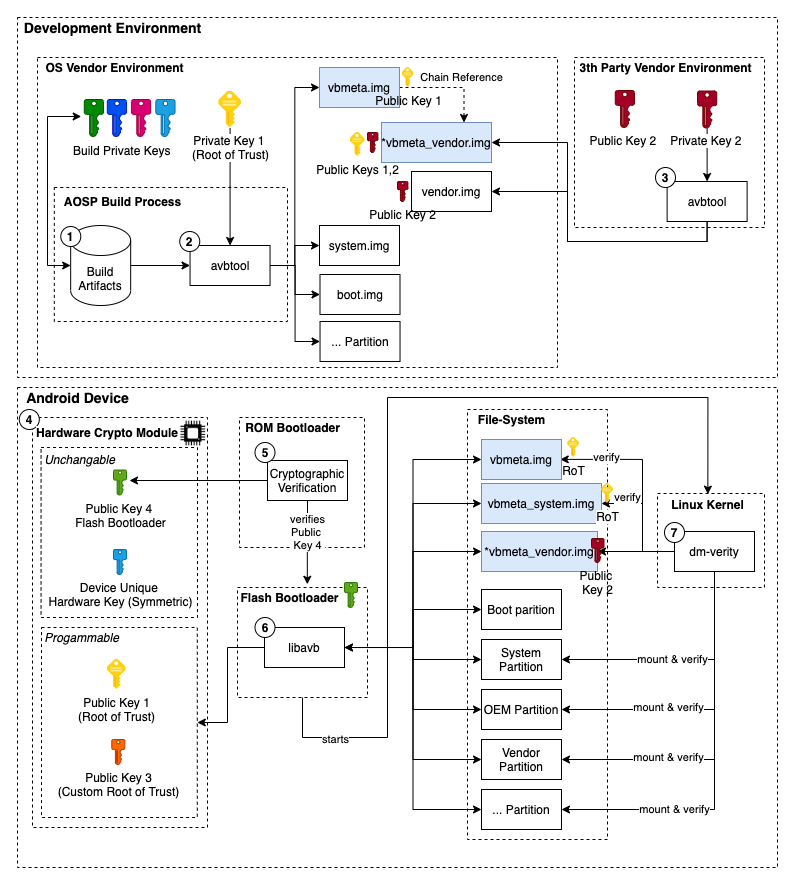}
	\caption{Android Verified Boot Architecture.}
	\label{fig:09_VerifiedBoot}
\end{figure}

First of all, the implementation of the Android Boot process depends on the actual 
phone. Smartphone manufacturers like for example, Samsung 
\cite{SamsungSecureBoot} or Huawei \cite{HuaweiSecureBoot}, have their 
implementations of Secure Boot with different designs depending on the CPU 
architecture. Nevertheless, we describe an example design to give the reader 
some insights into modern smartphones' security measurements. Figure 
\ref{fig:09_VerifiedBoot} illustrates an overview of the firmware signing process at 
the development and the verification process on the Android device. We will go 
through every step of the process, beginning at the Android firmware's build 
phases.

\textbf{\circled{1} AOSP Build:}  When we build the \gls{aosp} we need to provide 
at least four platform keys for signing different parts of the operating system. We 
can add as well additional keys for signing specific apps. The default configuration 
uses the following four keys:

\begin{itemize}
	\item Platform: A key used for the main packages of the Android framework. 
	We use this key for signing the pre-installed system apps and the Android core 
	framework package (framework-res.apk) if not explicitly configured different. 
	
	\item Shared: We sign with this keys packages of he home/contacts system. 
		
	\item Media: We sign with this key packages of the media/download system.
		
	\item Testkey: The key is used as default if no other key is defined.
\end{itemize}

If we do not provide custom private keys, then the AOSP default keys are used. 
We shouldn't use the default keys for a release builds because they are publicly 
available online \cite{aospMirrorGithub} and everyone could use them for signing. 
These keys are not directly necessary for Android Verified Boot, but we use them 
later for signing system apps. We will discuss in Section 
\ref{Fundamentals:AppSigning} how we use the platform keys. After building the 
AOSP, we have several build artifacts that we want to protect with Android Verified 
Boot.

\begin{table}[H]
	\resizebox{\textwidth}{!}{
		\begin{tabular}{|l|l|l|}
			\hline
			\rowcolor[HTML]{EFEFEF} 
			\textbf{Type} &
			\textbf{Variable Name} &
			\textbf{Code Comment} \\ \hline
			uint8\_t &
			magic{[}4{]}; &
			0: Four bytes equal to "AVB0" (AVB\_MAGIC). \\ \hline
			uint32\_t &
			required\_libavb\_version\_major; &
			4: The major version of libavb required for this header. \\ \hline
			uint32\_t &
			required\_libavb\_version\_minor; &
			8: The minor version of libavb required for this header. \\ \hline
			uint64\_t &
			authentication\_data\_block\_size; &
			12: The size of the signature block. \\ \hline
			uint64\_t &
			auxiliary\_data\_block\_size; &
			20: The size of the auxiliary data block \\ \hline
			uint32\_t &
			algorithm\_type; &
			28: The verification algorithm used, see |AvbAlgorithmType| enum. \\ \hline
			uint64\_t &
			hash\_offset; &
			32: Offset into the "Authentication data" block of hash data. \\ \hline
			uint64\_t &
			hash\_size; &
			40: Length of the hash data. \\ \hline
			uint64\_t &
			signature\_offset; &
			48: Offset into the "Authentication data" block of signature data. \\ \hline
			uint64\_t &
			signature\_size; &
			56: Length of the signature data. \\ \hline
			uint64\_t &
			public\_key\_offset; &
			64: Offset into the "Auxiliary data" block of public key data. \\ \hline
			uint64\_t &
			public\_key\_size; &
			72: Length of the public key data. \\ \hline
			uint64\_t &
			public\_key\_metadata\_offset; &
			80: Offset into the "Auxiliary data" block of public key metadata. \\ \hline
			uint64\_t &
			public\_key\_metadata\_size; &
			\begin{tabular}[c]{@{}l@{}}88: Length of the public key metadata. Must be 
				set to 
				zero if there\\ is no public key metadata.\end{tabular} \\ \hline
			uint64\_t &
			descriptors\_offset; &
			96: Offset into the "Auxiliary data" block of descriptor data. \\ \hline
			uint64\_t &
			descriptors\_size; &
			104: Length of descriptor data. \\ \hline
			uint64\_t &
			rollback\_index; &
			112: The rollback index which can be used to prevent rollback to older 
			versions. 
			\\ \hline
			uint32\_t &
			flags; &
			\begin{tabular}[c]{@{}l@{}}120: Flags from the AvbVBMetaImageFlags 
				enumeration. This \\ must be set to zero if the vbmeta image is not a 
				top-level 
				image.\end{tabular} \\ \hline
			uint32\_t &
			rollback\_index\_location; &
			\begin{tabular}[c]{@{}l@{}}124: The location of the rollback index defined in 
				this 
				header. \\ Only valid for the main vbmeta. For chained partitions, the 
				rollback \\ 
				index location must be specified in the AvbChainPartitionDescriptorand 
				\\ 
				this 
				value must be set to 0.\end{tabular} \\ \hline
			uint8\_t &
			release\_string{[}256{]}; &
			\begin{tabular}[c]{@{}l@{}}128: The release string from avbtool, e.g. 
				"avbtool 
				1.0.0" or "avbtool 1.0.0 \\ xyz\_board Git-234abde89". Is guaranteed to 
				be 
				NUL 
				terminated. \\ Applications must not make assumptions about how this 
				string is 
				formatted.\end{tabular} \\ \hline
			uint8\_t &
			reserved{[}80{]}; &
			176: Padding to ensure struct is size 256 bytes. This must be set to 
			zeroes. \\ 
			\hline
		\end{tabular}
	}
	\caption{AvbVBMetaImageHeader struct from \textit{avb\_meta\_image.h.}
		Table source: \cite{AndroidVerifiedBoot2Repo}}
	\label{tab:AvbVBMetaImageHeader}
\end{table}

\textbf{\circled{2} Partition Signing:} As mentioned in Section 
\ref{Fundamentals:AndroidFirmwareStructure} an Android firmware archive 
consists of several partition files. With Android Verified Boot we can protect these 
images with a cryptographic signature. We use the 
avbtool \cite{AndroidVerifiedBoot2Repo} as shown in Figure 
\ref{fig:09_VerifiedBoot} and generate a verification partition called 
\textit{vbmeta.img}. Before we can understand how we can us the vbmeta.img we 
need some background information about the data-structure it contains, called 
\textit{VBMeta struct}. 

We explain the data-structure shown in Table \ref{tab:AvbVBMetaImageHeader} 
which is used in the one of the official avbtool implementations 
\cite{AndroidVerifiedBoot2Repo}. 
Within the repository (see \cite{AndroidVerifiedBoot2Repo}), we can find a C 
implementation of avb in the libavb directory with C code comments in various 
header files. We know that a VBMeta struct has three parts from the comments 
and the code: header, authentication, and auxiliary. In version 1.2.0 of the avblib, 
the header has a fixed size of 256 bytes. Moreover, it includes the block sizes of 
the authentication and auxiliary parts. The authentication and auxiliary parts have 
as shown in Table \ref{tab:AvbVBMetaImageHeader} variable sizes. The 
authentication part contains a signature for verification, and the auxiliary part a list 
of descriptors and a public key. 

As shown on Lines 1-4 in in Listing \ref{lst:AvbDescriptor} a basic descriptor 
contains two variables: \textit{tag} and \textit{num\_bytes\_following}. All other 
descriptors extend this basic structure with additional variables. From the libavb 
implementation, we can see that there are five types of descriptors:

\begin{enumerate}
	\item \textit{Property descriptor:} As shown on Lines 6-7 in Listing 
	\ref{lst:AvbDescriptor} this descriptor is for general purposes and can contain 
	any meta-data in key-value form
	
	\item \textit{Hash-Tree descriptor:} A Merkle-tree is calculated over the partition 
	bytes and later verified by libavb and dm-verity to ensure the partition integrity. 
	The hash-tree descriptor holds the Merkle-tree information necessary to verify 
	the integrity of a partition as shown in Listing \ref{lst:AvbDescriptor} on Lines 
	13-28. In other words, this structure holds information about the root hash of the 
	Merkle-tree and the salt used for verification and other variables for \gls{fec}.
	
	\item \textit{Hash descriptor:} Like the hash-tree descriptor holds the hash 
	descriptor, the cryptographic data necessary for integrity verification. As shown 
	on Lines 31-40 in Listing \ref{lst:AvbDescriptor} the \textit{AvbHashDescriptor} 
	struct
	has variables for the hash algorithm, digest length and salt length. We use this 
	descriptor when we create a hash over the complete partition instead of a hash 
	tree. We think this is usually the case for the boot.img and the ramdisk.img 
	image files.
	
	\item \textit{Kernel cmdline descriptor:} We use this descriptor for passing 
	arguments to a kernel shell. For example, we can add here a flag that 
	deactivates dm-verity. The boot-loader then passes the flag to the kernel at 
	startup. Depending on the kernel build, we can use other descriptors.
	
	\item \textit{Chain partition descriptor:} The last descriptor type is for delegating 
	the authority over a specific partition. With this descriptor, we add a partition 
	that is signed and verified by another asymmetric key pair. In this case, we 
	either create a second vbmeta image (see \circled{3} in Figure 
	\ref{fig:09_VerifiedBoot}) for verifying the 
	signature, or we add a vbmeta footer to the existing partition. In both cases, we 
	save the public key of the delegated partition in the chain partition descriptor to 
	build a verification chain. A footer, as shown in Listing \ref{lst:AvbDescriptor} on 
	Lines 56-64 contains a complete VBMeta struct with header, authentication, 
	and auxiliary blocks. Using a footer is optional. As we can see from Line 50 in 
	Listing \ref{lst:AvbDescriptor} and from Table \ref{tab:AvbVBMetaImageHeader} 
	avb includes a rollback index for chained partitions and the vbmeta image. The 
	rollback index is used to prevent downgrade attacks and should be updated with 
	every new firmware version.
\end{enumerate}

\newpage
\begin{lstlisting}[caption={Overview of avb descriptor structs from 
\cite{AndroidVerifiedBoot2Repo}.}, 
label=lst:AvbDescriptor, language=c, basicstyle=\scriptsize] 
typedef struct AvbDescriptor {
	uint64_t tag;
	uint64_t num_bytes_following;
} AVB_ATTR_PACKED AvbDescriptor;

typedef struct AvbPropertyDescriptor {
	AvbDescriptor parent_descriptor;
	uint64_t key_num_bytes;
	uint64_t value_num_bytes;
} AVB_ATTR_PACKED AvbPropertyDescriptor;

typedef struct AvbHashtreeDescriptor {
	AvbDescriptor parent_descriptor;
	uint32_t dm_verity_version;
	uint64_t image_size;
	uint64_t tree_offset;
	uint64_t tree_size;
	uint32_t data_block_size;
	uint32_t hash_block_size;
	uint32_t fec_num_roots;
	uint64_t fec_offset;
	uint64_t fec_size;
	uint8_t hash_algorithm[32];
	uint32_t partition_name_len;
	uint32_t salt_len;
	uint32_t root_digest_len;
	uint32_t flags;
	uint8_t reserved[60];
} AVB_ATTR_PACKED AvbHashtreeDescriptor;

typedef struct AvbHashDescriptor {
	AvbDescriptor parent_descriptor;
	uint64_t image_size;
	uint8_t hash_algorithm[32];
	uint32_t partition_name_len;
	uint32_t salt_len;
	uint32_t digest_len;
	uint32_t flags;
	uint8_t reserved[60];
} AVB_ATTR_PACKED AvbHashDescriptor;

typedef struct AvbKernelCmdlineDescriptor {
	AvbDescriptor parent_descriptor;
	uint32_t flags;
	uint32_t kernel_cmdline_length;
} AVB_ATTR_PACKED AvbKernelCmdlineDescriptor;

typedef struct AvbChainPartitionDescriptor {
	AvbDescriptor parent_descriptor;
	uint32_t rollback_index_location;
	uint32_t partition_name_len;
	uint32_t public_key_len;
	uint8_t reserved[64];
} AVB_ATTR_PACKED AvbChainPartitionDescriptor;

typedef struct AvbFooter {
	uint8_t magic[AVB_FOOTER_MAGIC_LEN];
	uint32_t version_major;
	uint32_t version_minor;
	uint64_t original_image_size;
	uint64_t vbmeta_offset; // The offset of the |AvbVBMetaImageHeader| struct
	uint64_t vbmeta_size; // size of the vbmeta block (header + auth + aux blocks)
	uint8_t reserved[28];
} AVB_ATTR_PACKED AvbFooter;
\end{lstlisting}

We create descriptors for every partition of the firmware in the vbmeta.img. 
Depending on the partition, we use different descriptors. We use a hash descriptor 
for the boot partition because its file size isn't too large to get entirely loaded into 
the RAM and verified in a short time. For the system, vendor, and oem partitions, 
we use either hash-tree descriptors or chain partition descriptors because these 
partitions are usually too large to verify in time within the memory. After creating 
all descriptors and some parts of the header, avb creates the authentication block. 
This block contains a cryptographic signature over the complete vbmeta partition. 
As shown in Figure \ref{fig:09_VerifiedBoot} it is signed by the private key of the 
operating system vendor, and the vendor includes its public key into the VBMeta 
struct.

The VBMeta structure allows that the bootloader can verify the integrity of the boot 
image at startup. The verification process is done depending on the descriptor 
used. For a hash descriptor, the bootloader creates a digest over the complete 
partition and compares it to the digest in the vbmeta.img. For a hash-tree 
descriptor is the verification different. The official documentation 
\cite{AndroidVerifiedBoot2Repo} does not explain how avb verifies hash-trees 
descriptors, but we find the answer in the avb\_slot\_verify.c file. Looking at the 
avb\_slot\_verify and load\_and\_verify\_vbmeta methods, we can go through 
every step of the verification process step by step. However, we will not go 
through every line of code. Instead, we will follow the program flow to see where 
the descriptors are verified. 

We go through the code in the avb\_slot\_verify method of the 
avb\_slot\_verify.c file in \cite{AndroidVerifiedBoot2Repo}. We added a copy of the 
method's code to Appendix 
\ref{Appendix:avbOutput}. At Line 93 of the avb\_slot\_verify method the script 
executes a call to the load\_and\_verify\_vbmeta depending on the existence of 
the vbmeta partition or footer. If we follow the program flow and look at the 
load\_and\_verify\_vbmeta method, we can find the descriptor verification process 
in a switch case statement. Listing \ref{lst:AvbDescriptorValidation} shows parts of 
the switch cases statement with the verification of the hash-tree descriptor and the 
original code comments. The code comments state that avb only verifies a 
hash-tree descriptor when we use a persistent digest. 
Otherwise, the OS should conduct the checks and not avb. Even if the comments 
state that avb does no verification, we can see from the code that avb calls the 
method avb\_hashtree\_descriptor\_validate\_and\_byteswap. This method checks 
if the descriptor is in the right format and contains all necessary fields. At this 
point, avb makes no comparison between the hash-tree descriptor and the actual 
data on disk. To our knowledge, the official documentation does not explain why 
avb's implements the verification process like this. We can only assume that the 
hash descriptors' verification in combination with dm-verity is enough to ensure the 
partition's integrity. Another possibility is that the use of persistent digests is more 
common. As explained in \cite{AndroidVerifiedBoot2Repo} a persistent digest is 
not stored as a descriptor; instead, avb sets the hash length in the hash-tree 
descriptor to zero, and it uses a named persistent value. Vendors store named 
persistent values on tamper-evident storage in a key-value store. AVB can retrieve 
these values from the tamper-evident store by using a combination of the partition 
name and a prefix (avb.persistent\_digest.) as a key. Avb can then use the 
persistent digest for comparison.

In conclusion, if a vendor uses a persistent digest, they store the hash tree's root 
digest within a tamper-evident storage. AVB can access this storage and use its 
values for verification with the data found in the vbmeta.img. Moreover, we know 
that avb verifies hash descriptors by calculating a complete hash over the data 
loaded in memory. We can verify this by looking at the 
load\_and\_verify\_hash\_partition method. Consequently, avb can protect the 
kernel from being tampered under the assumption that a hash descriptor is used 
for the boot.img and correctly verified. Other partitions like the system.img, which 
uses a hash-tree descriptor, are verified by the kernel with dm-verity.

\newpage
\begin{lstlisting}[caption={Descriptor validation found in the 
load\_and\_verify\_vbmeta method. filename: avb\_slot\_verify.c }, 
label=lst:AvbDescriptorValidation, language=c, basicstyle=\scriptsize] 
...
	/* Now go through all descriptors and take the appropriate action:
	*
	* - hash descriptor: Load data from partition, calculate hash, and
	*   checks that it matches what's in the hash descriptor.
	*
	* - hashtree descriptor: Do nothing since verification happens
	*   on-the-fly from within the OS. (Unless the descriptor uses a
	*   persistent digest, in which case we need to find it).
	*
	* - chained partition descriptor: Load the footer, load the vbmeta
	*   image, verify vbmeta image (includes rollback checks, hash
	*   checks, bail on chained partitions).
	*/
  descriptors = avb_descriptor_get_all(vbmeta_buf, vbmeta_num_read, 
  &num_descriptors);
for (n = 0; n < num_descriptors; n++) {
	AvbDescriptor desc;
	
	if (!avb_descriptor_validate_and_byteswap(descriptors[n], &desc)) {
		avb_errorv(full_partition_name, ": Descriptor is invalid.\n", NULL);
		ret = AVB_SLOT_VERIFY_RESULT_ERROR_INVALID_METADATA;
		goto out;
	}
	switch (desc.tag) {
		case AVB_DESCRIPTOR_TAG_HASH: {
			AvbSlotVerifyResult sub_ret;
			sub_ret = load_and_verify_hash_partition{...};
			if (sub_ret != AVB_SLOT_VERIFY_RESULT_OK) {...}
		} 
		break;
		case AVB_DESCRIPTOR_TAG_CHAIN_PARTITION: {...}
		break;
		case AVB_DESCRIPTOR_TAG_KERNEL_CMDLINE: {...}
		case AVB_DESCRIPTOR_TAG_HASHTREE: {
			AvbHashtreeDescriptor hashtree_desc;
			if (!avb_hashtree_descriptor_validate_and_byteswap(
				(AvbHashtreeDescriptor*)descriptors[n], &hashtree_desc)) {
				avb_errorv(
				full_partition_name, ": Hashtree descriptor is invalid.\n", NULL);
				ret = AVB_SLOT_VERIFY_RESULT_ERROR_INVALID_METADATA;
				goto out;
			}
			/* We only need to continue when there is no digest in the descriptor.
			* This is because the only processing here is to find the digest and
			* make it available on the kernel command line.
			*/
			if (hashtree_desc.root_digest_len == 0) {...}
		} break;
		case AVB_DESCRIPTOR_TAG_PROPERTY:
		/* Do nothing. */
		break;
  }
}
...
\end{lstlisting}

For interested readers, we have added the python code for the verification process 
in Appendix \ref{Appendix:avbOutput}. The python code verification process is 
different from the C implementation. In python, a hash-tree is generated over the 
complete image file and compared to the data in the vbmeta.img. The C code 
example with all files can be found in the official avb repository 
\cite{AndroidVerifiedBoot2Repo}.

\textbf{\circled{4} Hardware-based crypto module:} So far, we have discussed how 
the partitions are created, signed, and verified in the development environment. 
We now proceed with how Android Verified Boot runs on the device. In the 
literature, the term secure element tends to be used to refer to hardened software 
and hardware environments with limited access. To avoid confusion, we use in our 
example a smartphone with a hardware-based crypto module as illustrated in 
Figure 
\ref{fig:09_VerifiedBoot} and do not discuss the boot process with software-based 
secure elements.

Smartphone manufacturers tend to implement crypto modules with unchangeable 
and programmable memory parts. In the immutable memory of our device is a 
symmetric key stored. The devices uses this key for encryption and decryption 
operations, and the key is unique for every device. The key itself is burnt into the 
device during manufacturing and, in theory, cannot directly be accessed or 
changed by any other component. Within the programmable part of the chip is the 
OS vendor's public key stored, the so-called \textit{Root of Trust}. 

\textbf{\circled{5} ROM bootloader:} This bootloader is in literature often referred to 
as the first stage or SoC bootloader. The ROM bootloader verifies the flash 
bootloader (or second stage bootloader) integrity. The ROM bootloader does this 
by verifying the flash bootloader's signature with a public key stored in the 
hardware crypto module.

\textbf{\circled{6} Libavb:} The flashable bootloader integrates a version of the 
libavb library and implements the interfaces to the tamper-evident storage if 
necessary. The libavb will then go through the verification process of the 
vbmeta.img and all the read-only partitions as mentioned above. Keep in mind that 
Android verified boot can only assure the integrity of read-only partitions. 
Read-write partitions like the user data partition are not integrity protected by AVB. 
In case the verification succeeds, the bootload loads the kernel from the boot.img. 
Some Android phone manufacturers allow their users to installed custom firmware. 
In this case, the user can sign the custom firmware with an own private key and 
store the public key as the new root of trust. The flash bootloader will then load the 
custom Root of Trust instead of the OS vendors Root of Trust for verification.

\textbf{\circled{7} Dm-verity:} The dm in dm-verity stands for device-mapper. Verity 
is a Linux kernel extension that allows integrity protection for read-only block 
devices using the kernels crypto API. Android has a custom implementation of the 
Linux kernel extension dm-verity \cite{KernelDMVerity, 
KernelDMVerityWikiChrome} and uses it for the verification of the partitions at 
mount and runtime.  As mentioned before, libavb does not validate the integrity of 
the tree-hash descriptors in the vbmeta.img to the actual data on disk. Therefore, 
AVB can only ensure the integrity of partitions with hash descriptors or persistent 
digests. However, dm-verify checks the hash-tree descriptors by comparing the 
bytes on disk to the hash-tree at runtime.

One of the problems with using hash-trees over block devices is that blocks can 
have bit-errors. In the case of a bit-error, the hash would differ, and the dm-verity 
verification would fail. A forward error correction code like Reed-Solomon is used 
\cite{StrictlyEnforcedVerifiedBoot} to solve this problem.

Another problem is that generating hash-trees over relatively large block devices 
can take a significant amount of time and system resources. To solve this 
problem, dm-verity only verifies a block when it's accessed. Dm-verity can even 
be configured only to verify a block once when it's loaded the first time into the 
memory \cite{DevDMVerity}.

More information about \acrshort{avb} and dm-verity can be found in the official 
repository \cite{AndroidVerifiedBoot2Repo} of avb, and on the Google developers 
webpage \cite{DevDMVerity, DevVerifiedBootFlow}.

\section{Modifying Firmware} \label{Fundamentals:CreatingCustomROM}
\begin{figure}[H]
	\centering
	\includegraphics[width=1\linewidth]{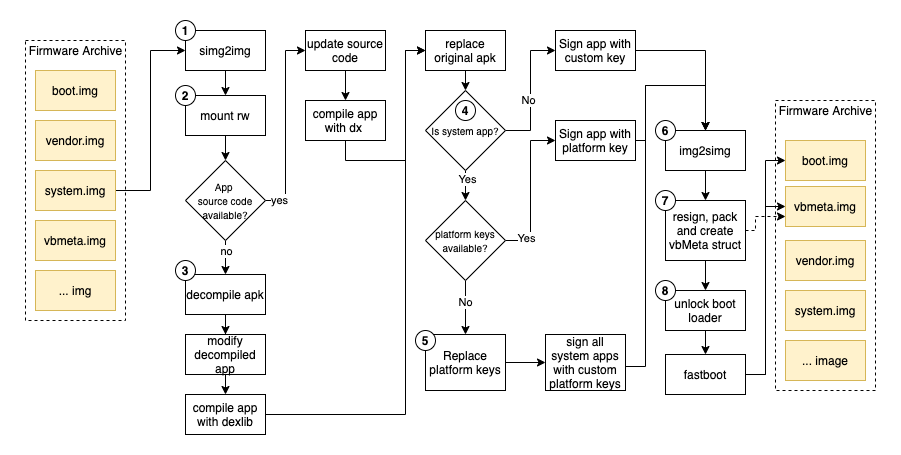}
	\caption{Process of modifying an existing Android firmware.}
	\label{fig:12_StockRomTampering}
\end{figure}

In Section \ref{Fundamentals:VerifiedBoot} we explained Android's integrity 
protection for partitions. This section discusses how we can modify existing 
Android firmware to create our own custom ROM. We have two ways to create a 
custom ROM--- building from source or modifying an existing build. We will not 
discuss how to build the AOSP from source since it is outside of this project's 
scope, and there are plenty of tutorials on the web on how to do it. However, we 
will explain how we can modify an existing firmware and customize some of the 
apps within the firmware. Figure \ref{fig:12_StockRomTampering} shows the 
process of modifying an existing firmware archive, and we will go through the 
numbered steps in detail.

\begin{enumerate}
	\item When we have a firmware archive, we need to extract the firmware 
	partitions first and then create an ext4 image from Android's sparse image 
	format. We can use the tool simg2img \cite{Simg2img} to convert a partition 
	file from sparse format into ext4 format.
	
	\item After we have converted the sparse image into an ext4 image, we can 
	mount it in read+write mode; this allows us to modify the partition. For example, 
	we want to change an existing app without having the source code of the app.
	
	\item If we do not have the source code, we need to decompile the apk; 
	otherwise, we could have compiled a new version from the app's source code. 
	One technique to modify android apps without the source code is to decompile 
	the app into \gls{SMALI} or java code, add modifications and then recompile the 
	modified source code to dex byte code. Such modifications can be done for 
	example with the jadx \cite{Jadx} or apktool \cite{Apktool} tools. When 
	modifying an app like this, an attacker has to resign the app. Otherwise, the 
	system will detect an incorrect app signature and not allow it to be 
	installed/updated.
	
	\item Depending on the kind of apk we are modifying, we have to use another 
	key to sign the apk. If we modify a \gls{Standard App} we can sign it with a 
	self-signed certificate. If we change a \gls{System App} we need to resign it 
	with 
	the 
	right platform key; otherwise, the app may not have the correct permissions. 
	We will discuss the app signing process in detail in Section 
	\ref{Fundamentals:AppSigning}
	
	\item In case we do not have access to the platform keys used for building the 
	firmware, we need to replace the existing ones. When we use custom platform 
	keys, we have to resign all the system apps. Otherwise, the verification of the 
	apks signature would fail.
	
	\item After we have finished modifying the partition, we can use img2simg to 
	convert our ext4 partition into a sparse image.  
	
	\item Since we have modified an existing firmware, we will need to resign all 
	partitions as described in Section \ref{Fundamentals:VerifiedBoot} with the 
	avbtool. After signing, we need to pack the partition back into the firmware 
	archive and then load it onto the device.
	
	\item To successfully flash our custom Android ROM to a device, we either 
	unlock the bootloader or sign the ROM with the device manufacturer's private 
	key. In our example, we modify an existing ROM, and we do not have the 
	manufacturer's private key. Therefore we need to unlock the bootloader and add 
	our public key to the device as the \gls{Root of Trust}. We can then use tools 
	like 
	fastboot to load the firmware to the device. If we have signed everything 
	correctly, the device should start with our custom firmware.
\end{enumerate}

As described above, the process of modifying an existing Android firmware only 
works if we have an unlocked bootloader. Otherwise, the Android Verified Boot 
would detect a signature mismatch and would stop the boot process. In the next 
section, we will go through the app signing process of system apps and explain 
how the system and standard apps are differentiated.

\section{APK signing and permissions} \label{Fundamentals:AppSigning}
In this section, we discuss Android's permission architecture for app processes. 
We start by explaining how standard and system apps are signed. We then move 
forward and explain how SELinux policies work together with the Android 
framework permissions to ensure app sandboxing. At the time of writing, the app 
signature schema has four versions \cite{DevSignatureSchemeV3, 
DevSignatureSchemeV2} and we will only discuss the newest version v4 
\cite{DevSignatureSchemeV4}. We will not explain all of the v4 signing schema 
features because this goes outside of our project scope, but we will show what 
benefit the v4 schema has compared to the v3 schema.

\begin{figure}[H]
	\centering
	\includegraphics[width=1\linewidth]{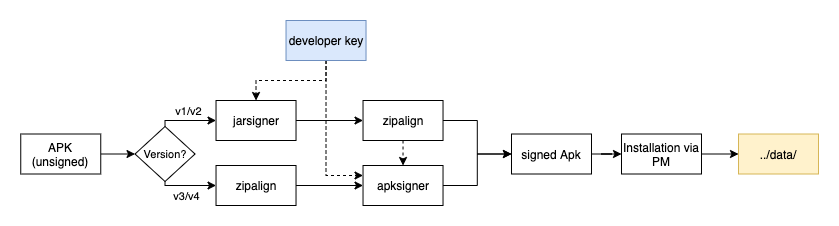}
	\caption{Overview of the standard app signing process.}
	\label{fig:13_AppSigning}
\end{figure}

When we develop an app in Android Studio, we will 
need to compile the Java or Kotlin code into an ART compatible dex byte code. If 
we want to upload our app to an app market like, for example, the Google Play 
Store, the market usually enforces that we have to sign our app. The signing will 
assure the app's integrity and allow us to use Android's app update mechanism. 
Android's package manager verifies the signature of every app at installation. 
Android apps do not use a public key infrastructure for app certificates. Without a 
PKI, it is possible to use self-signed certificates for signing Android apps. 
Depending on the signature schema used, we have to use other tools to generate 
a valid signature. To generate a v4 signature, we can sign an apk with the 
apksigner tool \cite{DevAPKsigner} as shown in Figure \ref{fig:13_AppSigning}. 
Before we can sign an apk, we need to use the zipalign tool \cite{DevZiplalign} to 
optimize the alignment of files within the apk. After signing the apk, we can upload 
it to an app store or directly install it on a device. Android's \gls{pm} will assure 
that the signature is valid and then install the apk into the \textit{/system/data/} 
folder.

In v3, the apksinger will append the signing data to the apk in a so called signing 
block. In v4, the signature schema uses parts of the v2 or v3 schema to generate 
a new separate signature file. The apk digest is first verified and then integrated 
into the v4 signature file. As stated in \cite{DevSignatureSchemeV4} the signature 
file contains a \gls{fsVerity} \cite{FSVerity, FSVerityPaper} compatible data 
structure. 
Fs-verity is similar to dm-verity and is used to generate a Merkle-tree over a 
read-only file on read+write partitions. It ensures a file's integrity by verifying the 
data and signature each time it is paged into the memory. According to the 
fs-verity documentation \cite{FSVerity, FSVerityPaper} 
These runtime checks ensure that the data can't be corrupted during runtime, even 
running on malicious firmware. The fs-verity approach helps to prevent an app 
from 
being tampered at runtime, and it is an additional protection on read+write 
partitions. As we have already discussed, the system and vendor partitions are 
integrity protected by dm-verity. However, dm-verity cannot detect the corruption 
of a file on read+write partitions like the user-data partition. In this case, fs-verity 
could ensure the integrity of specific files by making them read-only and verifying 
the signature at runtime. However, we will not further discuss how fs-verity is 
implemented in detail and discuss system apps' signing schema. Interested 
readers can find more information about fs-verity and the signature v4 schema in 
\cite{FSVerity, DevSignatureSchemeV4, FSVerityPaper}.

\begin{figure}[H]
	\centering
	\includegraphics[width=1\linewidth]{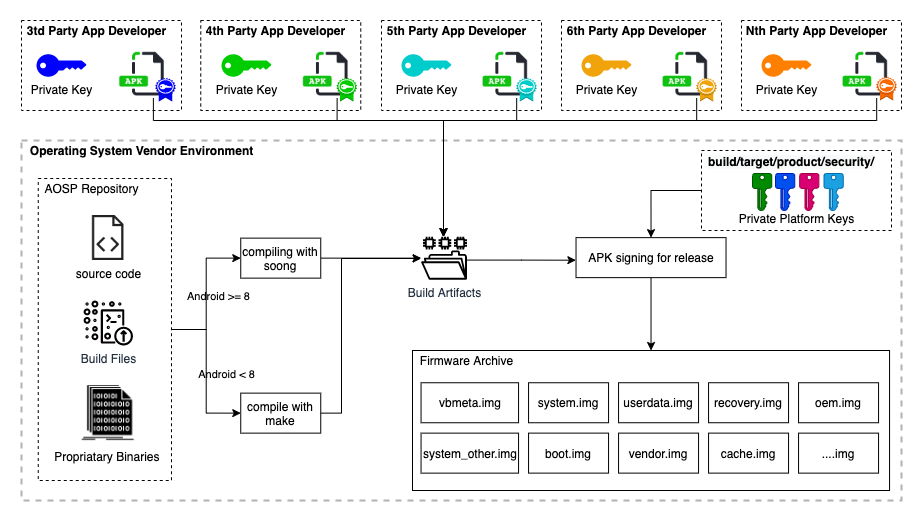}
	\caption{Example build process for system apps.}
	\label{fig:10_AOSPBuild}
\end{figure}

System apps use the same apk signature schemas as standard apps. One of the 
key questions to ask when we talk about system apps' security is how Android 
knows the difference between standard- and system apps? Answer: We sign 
system- and standard apps with different keys. App developers sign Android apps 
with their app keys and OS vendors sign system apps with their platform keys. As 
mentioned in Section \ref{Fundamentals:VerifiedBoot}, Android has at least four 
platform keys used for signing system apps. It is possible to add additional 
platform keys for signing specific apps. A reason for including more platform keys 
is when a 3rd party app vendor wants to be pre-installed on the system. In this 
case, the OS vendor has, to our knowledge, three ways to include it into the 
system. The first way is the 3rd party vendor signs it's app with their private key 
and then sends the signed apk to the OS vendor. The 
OS vendor will then add the apk to the firmware as illustrated in Figure 
\ref{fig:10_AOSPBuild}. Usually, this type of app is then placed within the 
firmware in the \textit{/system/app/} folder or it can be placed in other partitions 
like the \textit{/vendor/app/} and \textit{/product/app/} folders. Apps in these folders 
are called system apps and cannot be removed by the user because it is then 
stored on a read-only partition. However, as long as this app is not signed with a 
platform key, they inherit no special permissions. 

In case a developer wants to update its system app, then the new version of the 
app is installed on the writable user-data partition. The package manager will 
check the signature and version of the update and install the new apk into the 
\textit{/data} folder. The user can remove the update but not the original system 
app residing in the \textit{../app/} folder. If an OS vendor wants to remove or 
update the system app on one of the read-only partitions, it needs to roll-out an 
\Gls{ota} update. In such an update, the device is booted into recovery mode to 
load a new firmware version. These changes are then permanently stored on the 
phone.

Another way to integrate a system app is to sign it with a platform key and place it 
in the \textit{../priv-app/} folder, which is read-only. Apps signed by the platform 
key can request system permissions that grant access to protected or hidden 
APIs in the framework. Since Android Oreo, apps have to be allow-listed in a 
\textit{../etc/permissions/privapp-permissions.xml}\footnote{Filename is variable 
and can be changed.} 
file of the partition to grant or deny access to a specific system permission 
\cite{DevPrivilegedPermissionAllowlisting}. This permission check can be 
activated with the build property \textit{ro.control\_privapp\_permissions=enforce}. 
According to \cite{DevPrivilegedPermissionAllowlisting} the device will not boot if 
an app has permissions that are not allow-listed. When we want to understand how 
Android handles system app permissions, we need some information about 
Android's security architecture. 

Android uses a \gls{dac} \cite{DevDiscretionaryAccessControl} and a \gls{mac} 
\cite{DevSELinuxConceptsMAC} to manage access to specific API- and OS 
resources. The \acrshort{dac} uses users- (UID), groups-(GID), and access mode 
bits (RWX) to determine the access rights of a process. In Android, every app 
gets a custom UID/GID in the range of 10'000 to 19'999 at installation time. As 
with 
other Linux distributions, the UID/GID 0 is reserved for the root user and 
hardcoded into the kernel. The Android framework defines differently from other 
Linux distributions that all UIDs in range 1'000 to 9'999 are for system use as 
partly shown in Listing \ref{lst:FileSystemConfig}. Moreover, it defines the term 
\gls{aid} for identifying the human-readable username and UID/GID 
\cite{DevDiscretionaryAccessControl}. We can find a complete list of pre-defined 
system users and other defined ranges in the \textit{android\_filesystem\_config.h} 
file \cite{DevAndroidFilesystemConfig}. The listed aid are used at the built time to 
create the users and groups.

\newpage
\begin{lstlisting}[caption={Lines 48 to 55 from android\_filesystem\_config.h. 
Source \cite{DevAndroidFilesystemConfig}}, label=lst:FileSystemConfig, 
language=c, basicstyle=\scriptsize] 
...
#define AID_SYSTEM 1000 /* system server */
#define AID_RADIO 1001           /* telephony subsystem, RIL */
#define AID_BLUETOOTH 1002       /* bluetooth subsystem */
#define AID_GRAPHICS 1003        /* graphics devices */
#define AID_INPUT 1004           /* input devices */
#define AID_AUDIO 1005           /* audio devices */
#define AID_CAMERA 1006          /* camera devices */
...
\end{lstlisting}

The kernel maps the system services shown in Listing \ref{lst:FileSystemConfig} 
to the Android framework's permissions. The XML file \textit{platform.xml} can be 
found in the \textit{<partition>/etc/permissions} folder on every Android firmware. It 
contains a mapping of the GID to Android API permission for one partition. The 
Android framework has a set of pre-defined API permissions that change with 
every API version. We can find a complete list of all Android permissions on the 
Google developers webpage (see \cite{DevManifestPermission}). The mapping 
between Android's API permission and the GID assures that an app can access a 
specific resource if it has the API permission. When an app needs permission to a 
specific service, for example, Bluetooth, the app needs to request this permission 
during installation or runtime. Android's package manager service will then check 
that the app is allowed to have the permission and, if so, adds the app's UID to the 
group (GID) of the service. The kernel will then grant or denied access to the 
resources with DAC and MAC.

One problem with DAC is that if misconfigured, it allows setting world-writeable 
files. In the worst case, world-writable files can lead to privileged escalation 
attacks. SELinux is used to enhance Android's security architecture further and to 
prevent such misconfiguration. It enforces mandatory access control for every 
process, even for processes running under the root user. Per default, SELinux 
denies all access that is not explicitly granted by a policy. It hardens Android's 
sandbox model by ensuring that even when a malicious process runs under root, it 
cannot access all system resources unless allowed by the SELinux policies. In 
contrast to DAC, we cannot change SELinux policies during runtime. The 
SE-policies of AOSP is available online \cite{DevSELinuxPolicies}. OS vendors 
can customize their policies and extend the existing \acrshort{aosp} policies. It is 
a fact that the SELinux files are stored on different directories depending on the 
OS vendor's customizations. Per default the SELinux policy files are store in 
\textit{/system/etc/selinux}. The folder contains the context files with the policies 
that SELinux enforces. A list of key files can be found on the official Google 
developers page \cite{DevImplementingSELinux}. We will not further discuss how 
to create SELinux policies. Instead, we will explain how DAC and MAC are used in 
the framework to control app permissions. More information about the SELinux 
implementation can be found in \cite{SELinuxTrebleImplementation, 
TheSELinuxNotebook}.

Android stores the core files of the framework in the 
\textit{/system/framework} directory. It contains the framework-res.apk and the 
Java libraries. The framework-res.apk contains the graphical user interface's 
resources, and its AndroidManifest.xml contains the framework's initial permission 
definitions. In Android, every app can define its custom permissions in the 
AndroidManifest.xml. An example permission is shown in Listing 
\ref{lst:XMLPermissions} on Lines 2-7. When an app wants to access a service 
outside of the sandbox, like, for example, the internet, it needs to add the 
permission defined in the framework-res.apk in its own 
AndroidManifest.xml with a <uses-permission> XML tag. An example of the 
<uses-permission> tag is shown in Listing \ref{lst:XMLPermissions}. 

\begin{lstlisting}[caption={Example of a xml permission defintion and usage.}, 
label=lst:XMLPermissions, language=xml, basicstyle=\scriptsize] 
...
<permission 
	android:description="@string/permdesc_createNetworkSockets" 
	android:label="@string/permlab_createNetworkSockets" 
	android:name="android.permission.INTERNET" 
	android:protectionLevel="normal"
/>
..
<permission 		
	android:name="android.permission.REVOKE_RUNTIME_PERMISSIONS" 
	android:protectionLevel="installer|signature|verifier"
/>
...
<uses-permission 
	android:name="android.permission.INTERNET"
/>
...
\end{lstlisting}

\begin{figure}[H]
	\centering
	\includegraphics[width=0.8\linewidth]{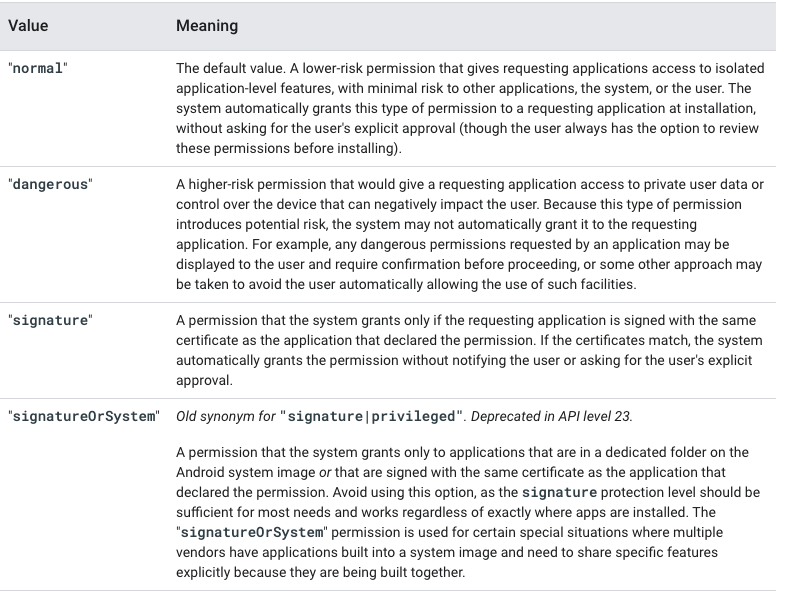}
	\caption{Overview of Androids access levels. Image source 
	\cite{DevManifestTagPermission}}
	\label{fig:15_Permission_Levels}
\end{figure}

Android defines a set of base permissions as shown in Figure 
\ref{fig:15_Permission_Levels}. Depending on the base 
permission and protection level flags the system follows other procedures when an 
app requests permissions. Moreover, the protection levels indicate the risk to the 
user. A base permission level can have zero or more flags, and they can extend 
the base permission with additional permissions as shown in Listing 
\ref{lst:XMLPermissions}. For example, the flag \textit{vendorPrivileged} with value 
8000 can grant additional permissions for vendor apps. Permission must have a 
unique name attribute. The package manager will not install an app that defines an 
already existing permission name.

As mentioned before, every app has its UID and runs in a sandboxed 
environment. However, apps can share data by using the same signing key and 
the same UID. Two apps can define in their AndroidManifest.xml the 
\textit{sharedUserId} attribute to use the same UID. Before Android10, system 
apps could set the sharedUserId to \textit{ “android.uid.system"} to use Android's 
pre-defined system user and run in the same context. With Android API 29, this 
function is deprecated because it leads to unpredictable behavior in the package 
manager \cite{DevManifest}. As described in Figure \ref{fig:15_Permission_Levels} 
the \textit{signature} base permission allows an app with the same certificate to 
use the permission. As already mentioned, pre-installed apps can be signed with 
the platform key and can therefore use the signature permissions of the 
framework-res.apk without notifying the user. 

In summary, we have discussed the basic structure of Android firmware images, 
and we gave some insights on the build process of Android firmware. This section 
explains how apk's are signed and how DAC, together with MAC, are used in 
Android to implement a permission architecture that allows process sandboxing 
and API protection. In terms of analyzing the Android firmware eco-system, we 
have seen that Android stores pre-installed system apps on read-only partitions 
and that \acrshort{avb} and dm-verity ensures integrity protection. In conclusion, 
we can say that Android's security framework is a complex structure, and we 
cannot go through all details of the framework within this work, but we have 
described in Section \ref{Fundamentals:CreatingCustomROM} 
how we can extract pre-installed apps from the system partition at scale. We will 
progress in Chapter \ref{Analysis} with analyzing the extracted pre-installed apps.

	\chapter{Dataset Analysis} \label{Analysis}

As described in Chapter \ref{Fundamentals} we were able to collect a dataset of 
5'931 firmware archives from various sources. From these firmware archives, we 
extracted 902'262 Android apps by only crawling the system partitions of the 
firmware archives. In Chapter\ref{Fundamentals} we explained how custom ROM's 
were built and which security techniques ensure partition integrity. Before we 
progress with analyzing the pre-installed apps in detail, we will present our 
analysis methodology:

\begin{enumerate}
	\item \textit{Create a dataset:} First, we collect the necessary data to conduct 
	our study. In our case, we collect Android firmware and pre-installed apps as 
	described in Section \ref{Fundamentals:CollectingFirmwareSamples}.
	
	\item \textit{Experiment:} As the next step of our experiment, we select tools to 
	extract security-relevant data from the Android apps. In detail, that means we 
	search for tools that can extract security-relevant data like certificates, 
	permissions, call graphs, and runtime information like network traffic or memory 
	states. We follow a similar approach to that used by other researchers 
	\cite{firmscope-2020, AnAnalysisofPreinstalled} and use mainly static analysis 
	tools for our experiment. We will evaluate available tools analysis in Section 
	\ref{Analysis:StateOfTheArtTools} and discuss the scan results in the following 
	sections. 
	
	\item \textit{Analyse Data:} After we have scanned the pre-installed apps, we 
	will analyze the collected data. We will see which data we can extract at scale 
	and what we can learn from the data in Sections 
	\ref{Analysis:BuildPropAnalysis} to \ref{Analysis:VirusTotalQarkAnalysis}.	
	
	\item \textit{Improve experiment:} Depending on the results, we will see if we 
	have to improve our experiment setup. Possible enhancements could be the 
	use of more tools or other analysis techniques. We will discuss potential 
	improvements of our approach in Chapter \ref{FutureWork}.
\end{enumerate}

In the next Chapter \ref{Implementation} we will then explain how we developed 
FirmwareDroid to automate the data analysis.

\section{Selecting State of the Art Tools} \label{Analysis:StateOfTheArtTools}
The open-source community has created many tools to analyze Android apps. It is 
outside of this project's scope to test every tool on the market in detail and 
integrate it into FirmwareDroid. However, we attempted to identify up to date static 
and dynamic analysis tools that we could include in our project, serving our 
interests. It's not the goal of this evaluation to find the best analysis tool or make 
any assumptions about a tool's quality. The purpose is to determine potential tools 
that we could use for our analysis, which would work within our environment. We, 
therefore, define the following criteria for our selection process:

\begin{enumerate}[label=\Alph*]
	\item \textbf{Purpose}: Does the tool's purpose fits our project goal?
	
	\item \textbf{Maintenance:} Is the tool still maintained, and is it likely to be 
	maintained in the future? (Yes/No)
	
	\item \textbf{Compatibility:} Is the integration of the tool into our environment 
	possible?
	This includes if the tools is compatible with docker and python 3.8. (Yes/No)
	
	\item \textbf{Scaling:} Is it possible to use parallel computation, and can it scan 
	our dataset in the given project time? (Yes/No/Unknown)
	
	\item \textbf{Dependencies and Requirements:} Can the tools requirements and 
	dependencies be fulfilled in our environment? (Yes/No)
	
	\item \textbf{Copyright:} Does the tool has an open-source license with free 
	usage for non-commercial use? 
	(Yes/No/Unknown)
\end{enumerate} 

\begin{table}[H]
	\centering
	\resizebox{\textwidth}{!}{
		
\begin{tabular}{|l|c|c|c|c|c|c|}
	\hline
	\rowcolor[HTML]{EFEFEF} 
	\textbf{\backslashbox[60mm]{Tool}{Criteria}} &
	\multicolumn{1}{l|}{\cellcolor[HTML]{EFEFEF}\textbf{Purpose}} &
	\multicolumn{1}{l|}{\cellcolor[HTML]{EFEFEF}\textbf{Maintenance}} &
	\multicolumn{1}{l|}{\cellcolor[HTML]{EFEFEF}\textbf{Compatibility}} &
	\multicolumn{1}{l|}{\cellcolor[HTML]{EFEFEF}\textbf{Scaling}} &
	\multicolumn{1}{l|}{\cellcolor[HTML]{EFEFEF}\textbf{Dependencies}} &
	\multicolumn{1}{l|}{\cellcolor[HTML]{EFEFEF}\textbf{Copyright}} \\ \hline
	Androbug \cite{AndroBugs}                          & Yes & No  & No      & Yes     & 
	No  & Yes     \\ \hline
	AndroGuard \cite{AndroGuard}                       & Yes & Yes & Yes     & Yes     & 
	Yes & Yes     \\ \hline
	Androwarn \cite{Androwarn}                         & Yes & Yes & Yes     & Yes     & 
	Yes & Yes     \\ \hline
	AndrODet \cite{AndrODet}                           & No  & Yes & Yes     & Yes     & 
	No  & Yes     \\ \hline
	AndroPyTool \cite{AndroPyTool}                     & Yes & No  & No      & Yes     & 
	Yes & Yes     \\ \hline
	apkanalyzer \cite{apkanalyzer}                     & Yes & Yes & No      & Yes     & 
	Yes & Unknown \\ \hline
	APKiD \cite{APKiD}                                 & Yes & Yes & Yes     & Yes     & 
	Yes & Yes     \\ \hline
	Armandroid \cite{Armandroid}                       & Yes & No  & No      & No      & 
	No  & Yes     \\ \hline
	ClassyShark \cite{ClassyShark}                     & Nes & Yes & Yes     & No      & 
	Yes & Yes     \\ \hline
	Droidbox \cite{droidbox}                           & Yes & No  & No      & Yes     & No  
	& Unknown \\ \hline
	DroidSafe \cite{perkins2016droidsafe}              & No  & No  & Unknown & 
	No      & Yes & Yes     \\ \hline
	Droidstatx \cite{droidstatx}                       & Yes & Yes & Yes     & Unknown & 
	Yes & Yes     \\ \hline
	Exodus \cite{Exodus}                               & Yes & Yes & Yes     & Yes     & 
	Yes & Yes     \\ \hline
	FlowDroid \cite{FlowDroid}                         & Yes & Yes & No      & No      & 
	Yes & Yes     \\ \hline
	Maldrolyzer \cite{maldrolyzer}                     & Yes & No  & No      & Unknown & 
	No  & Yes     \\ \hline
	Medusa \cite{Medusa}                               & Yes & Yes & Yes     & Yes     & 
	Yes & Unknown \\ \hline
	Mobile-Security-Framework \cite{MobSF}             & Yes & Yes & No      & 
	Yes     & No  & Yes     \\ \hline
	QARK \cite{QARK}                                   & Yes & Yes & Yes     & Unknown & 
	Yes & Yes     \\ \hline
	Quark-Engine \cite{QuarkEngine}                    & Yes & Yes & Yes     & Yes     
	& No  & Yes     \\ \hline
	RiskInDroid \cite{RiskInDroid}                     & Yes & Yes & Yes     & Yes     & 
	Yes & Yes     \\ \hline
	SmaliCFGs \cite{SmaliCFGs}                         & No  & No  & No      & No      & 
	No  & U       \\ \hline
	SUPER Android Analyzer \cite{SUPERAndroidAnalyzer} & Yes & Yes & Yes     
	& Yes     & Yes & Yes     \\ \hline
\end{tabular}

	}
	\caption{Results of the tools evaluation.}
	\label{tab:ToolEvaluation}
\end{table}
We searched the web for analysis tools and then evaluated them according to the 
defined criteria above. Table \ref{tab:ToolEvaluation} shows the result of our 
evaluation. With the info from Table \ref{tab:ToolEvaluation} we can select some of 
the tools for integration into FirmwareDroid. As we can see from Table 
\ref{tab:ToolEvaluation} some Android tools are no longer maintained and outdated. 
Other tools are regularly maintained but do not scale well in our infrastructure or 
seem to have compatibility problems. We, therefore, selected four tools that would 
meet our criteria for the static analysis. The static analysis tools used are 
AndroGuard (v. 3.3.5), Androwarn (v. 1.6.1), APKiD (v. 
2.2.1), and QARK (v. 4.0.0). As the next step, we will explain why we selected 
these tools and what features the tools offer.

\begin{itemize}
	\item \textbf{AndroGuard} \cite{AndroGuard}: AndroGuard is a static analysis 
	tool that has been used in another study as well \cite{AnAnalysisofPreinstalled, 
	AndroPyTool}. AndroGuard implements a command-line tool that can decompile 
	apks and extract meta-data from the manifest and the decompile smali code. 
	Its main API allows to extract of the following data:
	
	\begin{itemize}
		\item AndroidManifest.xml parsing: this allows to extract apk meta-data like 
		the version of the apk. Moreover, AndroGuard can extract the permissions, 
		activities, services, and receivers declared and used in the manifest.
		
		\item String extraction: AndoGuard can extract strings from the smali code. 
		AndoGuard uses specific upcodes in the smali code to detect strings.
		
		\item App certificates: The integrated certificate parser allows to extract the 
		signing certificate of the apk in DER format. The API offers as well the 
		possibility to export the certificate date in a human-readable form.
		
		\item Call Graph: AndroGuard has an integrated crossreference module. This 
		module allows the creation of call graphs for classes, methods, and fields.
	\end{itemize}

	We choose AndroGuard because it is well maintained, documented, and offers 
	essential features for our study. The extraction of the AndroidManifest.xml data 
	and the app certificates can help us understand how pre-installed apps are 
	configured and by whom.

	\item \textbf{Androwarn} \cite{Androwarn}: Androwarn is a malware scanner 
	based on AndroGuard. It can scan apps for potentially harmful behavior and can 
	generate a JSON report. One of Androwarn's main features is detecting 
	dynamic code loading techniques like JNI to load dex files or UNIX commands 
	execution. We choose Androwarn because it scales well and gives us insights 
	into commonly known malicious behavior and privacy issues.
	
	\item \textbf{APKiD} \cite{APKiD}: In 2016 Eduardo Novella Lorente et al. 
	released a tool called APKiD \cite{APKiD}. With APKiD, we can fingerprint 
	Android app compilers and packers by pattern matching. The tools collect 
	features from the classes.dex and Manifest.xml and match them to a compiler 
	or packer if possible. The APKiD developers use in \cite{APKiD} APKiD to 
	detect malicious apps by fingerprinting suspicious compilers. Their work shows 
	that APKiD is a useful tool to detect repackaged apps.
	
	\item \textbf{QARK} \cite{QARK}: QARK is a vulnerability scanner that 
	decompiles an app to find vulnerabilities. The QARK scanner can check for 
	vulnerabilities like leaked private keys, activated debug flags, or weak 
	cryptography usage. QARK is the only of our tool that uses Java, and from 
	other studies, it is not clear how well it scales. Nevertheless, we decided to 
	integrate QARK into FirmwareDroid since its vulnerability report has more 
	information than Androwarn.
\end{itemize}

Additional to the above-mentioned tools we use VirusTotal \cite{VirusTotal} for 
scanning. VirusTotal is well known in the security community, and its API allows to 
scan binaries against the most common anti-virus products on the market. We 
used all the scanners mentioned above to scan all the apks in our dataset, and we 
will discuss the results of these scans in the next sections. In Chapter 
\ref{Implementation} we will see that our software architecture allows us to 
integrate more tools in future versions of FirmwareDroid.

\section{Build Property Analysis} \label{Analysis:BuildPropAnalysis}
Before we can discuss our static analysis scan results, we give the reader an 
overview of the firmware data in this section. We have extracted from every 
firmware archive the build.prop or default.prop file to get the firmware version by 
using the \textit{ro.build.version.release} key. Table \ref{tab:FirmwareVersions} 
shows the result of this extraction. Moreover, we show the numbers of apps and 
package names by version.

\begin{table}[H]
	\centering
	\resizebox{\textwidth}{!}{%
		\begin{tabular}{|c|c|c|c|c|}
			\hline
			\rowcolor[HTML]{EFEFEF} 
			\textbf{\begin{tabular}[c]{@{}c@{}}Android version\\ 
			(ro.build.version.release)\end{tabular}} & \textbf{Firmware count} & 
			\textbf{App count} & \textbf{\begin{tabular}[c]{@{}c@{}}Unique \\ 
			packagename\\ count\end{tabular}} & 
			\textbf{\begin{tabular}[c]{@{}c@{}}Firmware size\\ on Disk\end{tabular}} \\ 
			\hline
			2 & 18 (0.30\%) & 1'876 (0.2\%) & 407 & 0.003TB \\ \hline
			3 & 1 (0.01\%) & 151 (0.02\%) & 151 & 0.397GB \\ \hline
			4 & 2'511 (42.31\%) & 317'250 (35.61\%) & 8'203 & 1.447TB \\ \hline
			5 & 1'000 (16.85\%) & 141'575 (15.69\%) & 4'806 & 0.950TB \\ \hline
			6 & 795 (13.98) & 114'406 (12.68\%) & 4'290 & 0.924TB \\ \hline
			7 & 685 (11.54\%) & 134'254 (14.88\%) & 3'968 & 1.119TB \\ \hline
			8 & 400 (6.74\%) & 72'976 (8.09\%) & 1'911 & 0.658TB \\ \hline
			9 & 234 (3.94\%) & 55'965 (6.02\%) & 1'375 & 0.568TB \\ \hline
			10 & 136 (2.91\%) & 15'078 (1.67\%) & 242 & 0.243TB \\ \hline
			Unknown & 154 (2.6\%) & 48'731 (5.40\%) & 1'922 & 0.502TB \\ \hline
			\rowcolor[HTML]{EFEFEF} 
			\textit{\textbf{Total}} & \textit{\textbf{5'934}} & \textit{\textbf{902'262}} & 
			\textit{\textbf{27'275}} & \textbf{6.4TB} \\ \hline
		\end{tabular}%
	}
	\caption{Number of firmware archives by version.}
	\label{tab:FirmwareVersions}
\end{table}

The numbers in Table \ref{tab:FirmwareVersions} show that our dataset contains 
mostly Android 4 (42.31\%) samples followed by Android 5 with 16.85\%. The 
dataset contains around 13.59\% of firmware archives from major (>=v8) Android 
versions. From 902'262 apps, we have a total number of 27'275 unique packages 
with 3'528 apps from major versions. These numbers show the variance of 
package 
names or, in other words, how many different app builds we have. Comparing the 
number of unique package names to the app total tells us that we have more apps 
with different builds than individual apps. We think this is not surprising because 
system apps on other firmware usually uses the same package name. For 
example, the settings app uses the "com.android.settings" 
package name over several Android versions. Another fact is that we have 154 
unknown firmware versions because we couldn't detect the ro.build.version.release 
in the build property file. These files are likely newer versions that use multiple 
build property files or have modified build properties.

In total, we have detected 862 brands and 796 manufacturers. The difference 
between the manufacturer property and the brand property is that a company can 
produce a phone, but the phone can have another companies brand. For example, 
HTC manufactured the first Pixel device, but the branding is by Google. The model 
property is the name of the phone series—for example, Pixel 2 or SM-A605G.

\begin{table}[H]
	\centering
	\resizebox{0.8\textwidth}{!}{%
		\begin{tabular}{|c|c|c|}
			\hline
			\rowcolor[HTML]{EFEFEF} 
			\textbf{\begin{tabular}[c]{@{}c@{}}Brands\\ (ro.product.brand)\end{tabular}} 
			& \textbf{\begin{tabular}[c]{@{}c@{}}Manufacturers\\ 
			(ro.product.manufacturer)\end{tabular}} & 
			\textbf{\begin{tabular}[c]{@{}c@{}}Model\\ (ro.product.model)\end{tabular}} 
			\\ \hline
			6 & 5 & 16 \\ \hline
			1 & 1 & 1 \\ \hline
			414 & 370 & 2004 \\ \hline
			193 & 186 & 760 \\ \hline
			134 & 120 & 543 \\ \hline
			81 & 79 & 407 \\ \hline
			21 & 24 & 186 \\ \hline
			10 & 9 & 101 \\ \hline
			- & - & - \\ \hline
			2 & 2 & 2 \\ \hline
			\rowcolor[HTML]{EFEFEF} 
			862 & 796 & 420 \\ \hline
		\end{tabular}%
	}
	\caption{Overview of unique values in brand, manufacturer, and model build 
	properties by Android version}
	\label{tab:FirmwareBrands}
\end{table}

The numbers in Table \ref{tab:FirmwareBrands} show that we have various brands, 
manufacturers, and phone models in our dataset. Note that we could not extract 
the Android 10 build properties for the model, brand, and manufacturer due to the 
mentioned problem of using referenced build property files on other partitions. 
Nevertheless, if we look at Table \ref{tab:FirmwareBrandsv89} we can see the 
distribution of brands for Android 8 and 9. We have mostly firmware branded by 
Google and Samsung. We show similar results in Table 
\ref{tab:FirmwareManufacturersv89} where Samsung and Google are on top of the 
most detected manufacturers.

\begin{minipage}[h]{\textwidth}
	\centering
	\begin{minipage}[b]{0.49\textwidth}
		\begin{table}[H]
			\begin{tabular}{|l|c|c|}
				\hline
				\rowcolor[HTML]{EFEFEF} 
				\multicolumn{1}{|c|}{\cellcolor[HTML]{EFEFEF}\textbf{\backslashbox[40mm]{Brand}{Version}}}
				& \textbf{8} & \textbf{9} \\ \hline
				Allview & 1 & 0 \\ \hline
				Alps & 1 & 2 \\ \hline
				Fih & 1 & 0 \\ \hline
				Google & 147 & 102 \\ \hline
				Helio & 1 & 0 \\ \hline
				Infinix & 11 & 21 \\ \hline
				Itel & 52 & 0 \\ \hline
				Kddi & 1 & 1 \\ \hline
				Lava & 19 & 1 \\ \hline
				Lmkj & 1 & 0 \\ \hline
				Micromax & 10 & 0 \\ \hline
				Motorola & 0 & 3 \\ \hline
				Nokia & 2 & 2 \\ \hline
				Samsung & 106 & 96 \\ \hline
				Symphony & 23 & 0 \\ \hline
				Tecno & 6 & 0 \\ \hline
				Verizon & 2 & 0 \\ \hline
				Vivo & 6 & 5 \\ \hline
				Vkworld & 2 & 0 \\ \hline
				Walton & 3 & 0 \\ \hline
				Wiko & 3 & 0 \\ \hline
				Zte & 2 & 1 \\ \hline
			\end{tabular}
			\caption{Count of detected brands \\ for Android 8 and 9. \\}
			\label{tab:FirmwareBrandsv89}
		\end{table}
	\end{minipage}
	\hfill
	\begin{minipage}[b]{0.49\textwidth}
		\begin{table}[H]
			\begin{tabular}{|l|c|c|}
				\hline
				\rowcolor[HTML]{EFEFEF} 
				\textbf{\backslashbox[40mm]{Manufacturer}{Version}} & 
				\textbf{8} & 
				\textbf{9} \\ \hline
				Allview & 1 & 0 \\ \hline
				Alps & 3 & 4 \\ \hline
				Asus & 2 & 0 \\ \hline
				Fih & 1 & 0 \\ \hline
				Google & 102 & 102 \\ \hline
				Helio & 1 & 0 \\ \hline
				Hmd Global & 0 & 2 \\ \hline
				Huawei & 22 & 0 \\ \hline
				Infinix Mobility Limited & 11 & 19 \\ \hline
				Itel & 52 & 0 \\ \hline
				Lava & 19 & 1 \\ \hline
				Leimin & 1 & 0 \\ \hline
				Lge & 21 & 0 \\ \hline
				Micromax & 10 & 0 \\ \hline
				Motorola & 0 & 3 \\ \hline
				Samsung & 109 & 97 \\ \hline
				Symphony & 23 & 0 \\ \hline
				Tecno Mobile Limited & 6 & 0 \\ \hline
				Vivo & 6 & 5 \\ \hline
				Vkworld & 2 & 0 \\ \hline
				Walton & 3 & 0 \\ \hline
				Wiko & 3 & 0 \\ \hline
				Zte & 2 & 1 \\ \hline
			\end{tabular}
			\caption{Count of detected manufacturers \\ for Android 8 and 9.}
			\label{tab:FirmwareManufacturersv89}
		\end{table}
	\end{minipage}
	\hfill
\end{minipage}

We extracted as well how many apps set security-relevant build properties to true 
(1) or false 
(0). Table \ref{tab:BuildPropsSecure} shows the result of this extraction in 
numbers. What we can see is that seven firmware archives deactivate SELinux 
(ro.secure), twelve deactivate ADB authentication (ro.adb.secure), and nine 
firmware archives activate debugging (ro.debuggable). We can assume that these 
firmware archives are insecure custom ROMs and not made for production. 
Moreover, the numbers in Table \ref{tab:BuildPropsSecure} show that we cannot 
extract all the build properties for every firmware archive using the build property 
file on the system partition. Since build.prop files can have import statements, it is 
possible to refer to other build.prop files. These files can lay on another partition, 
and we would need to extract all the data from all partitions to get a complete 
overview of build properties. However, for some build properties, we can extract 
the complete numbers.

\begin{table}[H]
	\centering
	\begin{tabular}{|l|c|c|}
		\hline
		\rowcolor[HTML]{EFEFEF} 
		\textbf{Build property key} & \textbf{1} & \textbf{0} \\ \hline
		\rowcolor[HTML]{FFFFFF} 
		ro.secure & 153 & 7 \\ \hline
		\rowcolor[HTML]{FFFFFF} 
		ro.adb.secure & 810 & 12 \\ \hline
		\rowcolor[HTML]{FFFFFF} 
		ro.debuggable & 9 & 153 \\ \hline
		\rowcolor[HTML]{FFFFFF} 
		ro.oem.unlock.supported & 177 & 0 \\ \hline
	\end{tabular}
	\caption{Detected numbers of security relevant build properties.}
	\label{tab:BuildPropsSecure}
\end{table}

The \textit{ro.build.tags} property gives us some insight on which keys the 
developers used to build the firmware. 3'675 firmware archives have the 
ro.build.tags set to \textit{"release-keys"}, 1'871 use \textit{"test-keys"}, 194 
use\textit{"dev-keys"} and the rest (194) has set a custom string or does not set 
the property at all in the build.props file. We conclude that we have at least 2'065 
(34.8\%) firmware builds in our dataset that shouldn't be used for productive builds 
due to possible insecure key usage.

In summary, we can say that our dataset contains a larger amount of old (< v8) 
Android versions. Having a lot of old Android firmware is one of the drawbacks of 
our crawling web approach. We cannot determine the Android version until we 
have downloaded the data. However, one of the benefits of this dataset is that we 
have a wide range of firmware samples and enough pre-installed apps to test our 
software. We will progress with discussing the results of the individual scanners in 
the next sections.

\section{AndroGuard Scan Result Analysis} \label{Analysis:AndroGuardAnalysis}
In this section, we will discuss how we can use AndroGuard reports to analyze our 
Firmware dataset. When we scan an app with AndroGuard, we can parse and 
export the data from the AndroidManifest.xml. Within the manifest, we find the 
component and permission decelerations. Using FirmwareDroid, we can scan 
every app in our dataset with AndroGuard and store all the extracted data from the 
AndroidManifest.xml in our database. We use the data to create firmware specific 
statistics about the permission usage of the pre-installed apps. 

In general, extracting permission data allows us to generate a various number of 
statistics. We implemented the functionality to aggregate the app data over any 
set of firmware archives. We can use this function to compare arbitrary sets of 
firmware to each other. For example, we can generate statistics for firmware 
samples of two OS vendors and compare them against each other. 

With AndroGuard, we cannot only extract manifest data. As we will see in 
Sections \ref{Analysis:Certificates} and \ref{Implementation:StringEnrichment} we 
can use AndroGuard to extract app certificates and strings. Moreover, AndroGuard 
can create a complete code call graph with class, method, and field references. 
However, due to the high performance necessary to create complete call graphs 
with AndroGuard we do not use this feature in the current version of 
FirmwareDroid. 

\subsection{Permissions} \label{Analysis:Permissions}
In this section, we discuss the permissions per Android version 
and do not compare them to each other because this could lead to wrong 
conclusions. Comparing arbitrary sets of firmware is not always meaningful. In 
some cases, it can give us misleading information if we compare groups of 
unequal size or different Android versions. As shown in 
Tables \ref{tab:FirmwareVersions} and Table \ref{tab:FirmwareBrands} we have an 
unbalanced number of android apps per firmware version, and the number of 
brands per Android versions differs as well. Our goal for this analysis is to show 
different statistics for Android 8, 9, and 10 to demonstrate how we can use the 
permission data extracted with 
AndroGuard and not to compare the versions to each other.

\subsubsection{Android 10: Permissions statistics}
\begin{figure}[H]
	\centering
	\includegraphics[width=\linewidth]{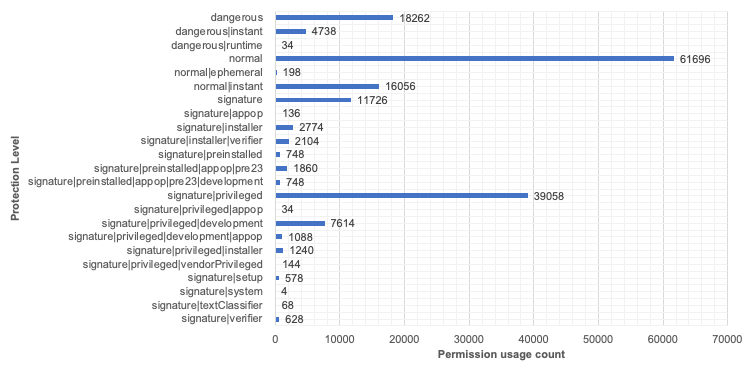}
	\caption{Android10: Permission usage by protection level.}
	\label{fig:21_permissions_v10_barh}
\end{figure}

\begin{figure}[H]
	\centering
	\includegraphics[width=0.4\linewidth]{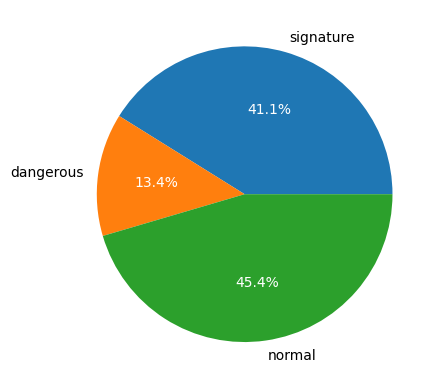}
	\caption{Android10: Total permission usage grouped by base permissions.}
	\label{fig:16_pie_v10_permission_by_level_grouped_count_plot}
\end{figure}

We start with examining the app permissions of our Android 10 firmware by 
showing some permission usage statistics. We extracted all the permissions and 
corresponding protection flags for every app and aggregate the data to see how 
common some protection levels are. Figure \ref{fig:21_permissions_v10_barh} 
illustrates the protection levels and their frequency. We can see that the base 
permission \textit{normal} is used the most times with a frequency of 61'696. The 
signature|privileged and dangerous permission are used the second and third most 
with 39'058 and 18'262 times. We can see from Figure 
\ref{fig:21_permissions_v10_barh} that we have some protection levels compared 
to others with a rather low count. For example, the signature|system or the 
dangerous|runtime times. Note: That to our knowledge, the official Google 
developers page for permissions \cite{DevManifestPermission} does not even list 
some of these protection levels. Without further investigation, we cannot 
determine their purpose. Anyways, we group the protection levels by their base 
permission to better illustrate the permission usages in Figure 
\ref{fig:16_pie_v10_permission_by_level_grouped_count_plot}. As shown in Figure 
\ref{fig:16_pie_v10_permission_by_level_grouped_count_plot} our pre-installed 
apps on Android 10 use 45.4\% normal (total: 77'950), 41.1\% signature (total: 
70'552) and 13.4\% dangerous (total: 23'034) permissions.

We determine as well how many apps declare custom permissions. Figure 
\ref{fig:16_pie_v10_has_permissions_requested_third_party_plot_dict} shows that 
44,9\% of our Android 10 apps request at least on third party permission. Figure 
\ref{fig:16_pie_v10_has_permissions_declared_plot_dict} illustrates the 
number of apps that define at least one custom permission. We see that 24\% 
of our Android 10 apps declare at least one custom permission. 

\begin{minipage}{\textwidth}
	\begin{minipage}[b]{0.1\linewidth}
		
	\end{minipage}
	\begin{minipage}[b]{0.4\linewidth}
		\begin{figure}[H]
			\includegraphics[width=\linewidth]{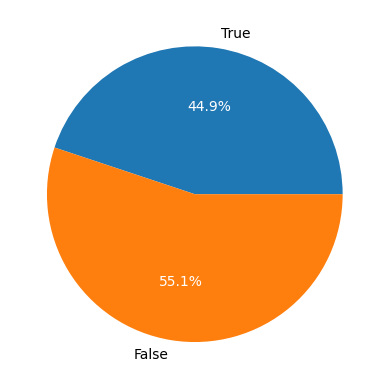}
			\caption{Android10: Percentage of apps that request at least one 
				3rd party permission. True = Requests at least one 3rd party 
				permission, False = requests no 3rd party permission.}
			\label{fig:16_pie_v10_has_permissions_requested_third_party_plot_dict}
		\end{figure}
	\end{minipage}
	\hfill
	\begin{minipage}[b]{0.4\linewidth}
		\begin{figure}[H]
			\includegraphics[width=\linewidth]{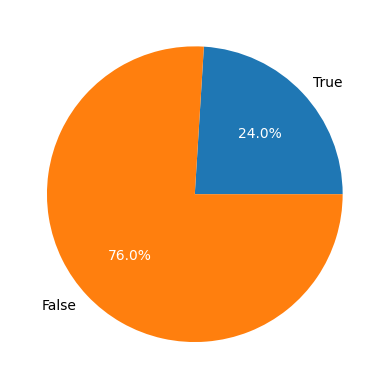}
			\caption{Android10: Percentage of apps that declare at least one 
				permission in their AndroidManifest.xml. True = Declares at least one 
				permission, False = declares 
				no permission.}
			\label{fig:16_pie_v10_has_permissions_declared_plot_dict}
		\end{figure}
	\end{minipage}
	\hfill
	\begin{minipage}[b]{0.1\linewidth}

	\end{minipage}
\end{minipage}

\subsubsection{Android 9: Permissions statistics}
\begin{figure}[H]
	\centering
	\includegraphics[width=\linewidth]{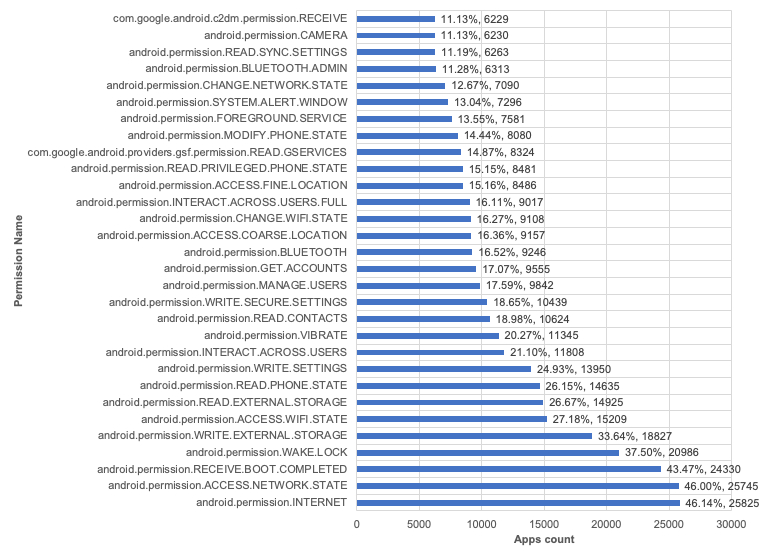}
	\caption{Android 9: Top 30 most used permissions.}
	\label{fig:23_top30_permissions_android9}
\end{figure}

We have extracted 2'806 different permissions and counted the frequency of their 
usages from our Android 9 dataset. Figure 
\ref{fig:23_top30_permissions_android9} shows the top 30 most used 
permissions for our Android 9 apps. Please note that we added in Appendix 
\ref{Appendix:AndroGuardStats} a graphic with the top 100 permissions. We can 
see from the top 30 that the internet and network state permission seems to be on 
top of the most used permissions. Moreover, we have several dangerous 
permissions in the top 30. To be more precise, eight dangerous, 13 normal, eight 
signature, and one permission that Google lists as "not used by third-parties 
applications"\footnote{android.permission.WRITE\_SECURE\_SETTINGS 
according to \cite{DevManifestPermission}}. Nearly every third app (33,64\%) in 
our Android 9 dataset has write access to the external storage, and 26.67\% have 
read access. 26.15\% apps use the read phone state permission, which gives 
access to the unique phone identifiers. Overall we conclude that the use of 
dangerous permissions for our Android 9 apps seems to be common.

\begin{figure}[H]
	\centering
	\includegraphics[width=\linewidth]{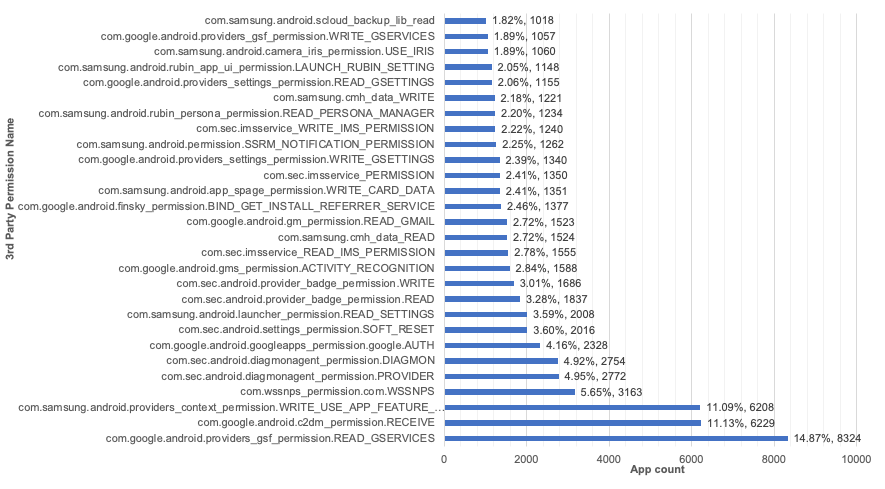}
	\caption{Android 9: Top 30 most used 3rd party permissions.}
	\label{fig:24_top30_thirdparty_android9}
\end{figure}

To further investigate the usage of the permissions we list in Figure 
\ref{fig:24_top30_thirdparty_android9} the top 30 most requested 3rd party 
permissions. On top of the most requested permissions is access to the Google 
Play service. Looking at the top 30, it seems that we have mainly custom Google 
and Samsung permissions. For most permission, we can only assume their 
purpose, and without manually analyzing each of the permissions, we cannot 
conclude if their usage is a risk for the user.

\subsubsection{Android 8: Permission statistics}

Another way to use the permission data is to aggregate the total number of 
permissions grouped by protection level to show the protection levels' distribution. 
Figure \ref{fig:25_dangerous_v8} shows a boxplot of the dangerous protection level 
without any additional protection flags. Thy y-axis represents the total number of 
permissions an app uses for the individual protection level. The illustration shows 
that we have several out-liners with unusually high usage of dangerous 
permissions. On top is one app that uses 25 dangerous permissions. Overall, 
most apps that use dangerous permissions do not use more than five dangerous 
permissions. The median lays at two permissions.

\begin{minipage}{\textwidth}
	
\begin{minipage}[b]{0.05\linewidth}
	
\end{minipage}
\hfill
	\begin{minipage}[b]{0.4\linewidth}
\begin{figure}[H]
	\centering
	\includegraphics[width=\linewidth]{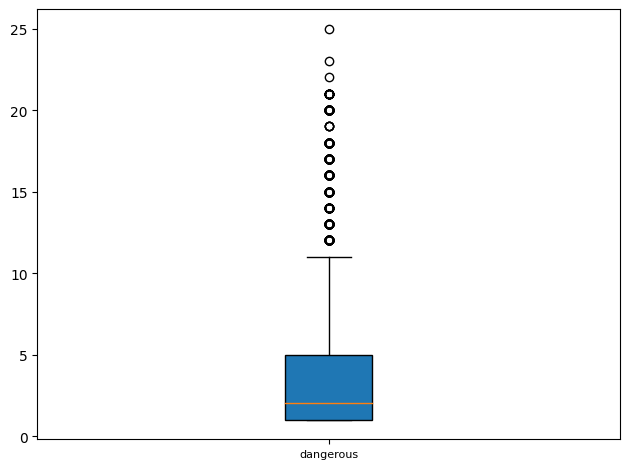}
	\caption{Android 8: Distribution of the dangerous protection level.}
	\label{fig:25_dangerous_v8}
\end{figure}
	\end{minipage}
	\hfill
	\centering
	\begin{minipage}[b]{0.4\linewidth}
\begin{figure}[H]
	\centering
	\includegraphics[width=\linewidth]{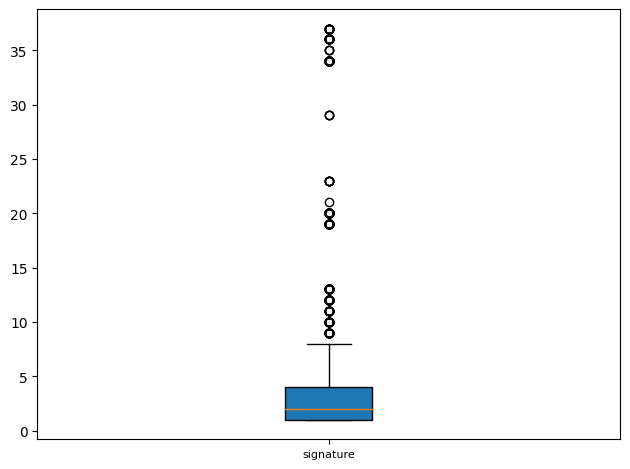}
	\caption{Android 8: Distribution of the signature protection level.}
	\label{fig:26_signature_v8}
\end{figure}
	\end{minipage}
	\hfill
	\begin{minipage}[b]{0.1\linewidth}
	\end{minipage}
\end{minipage}

\begin{figure}[H]
	\centering
	\includegraphics[width=0.5\linewidth]{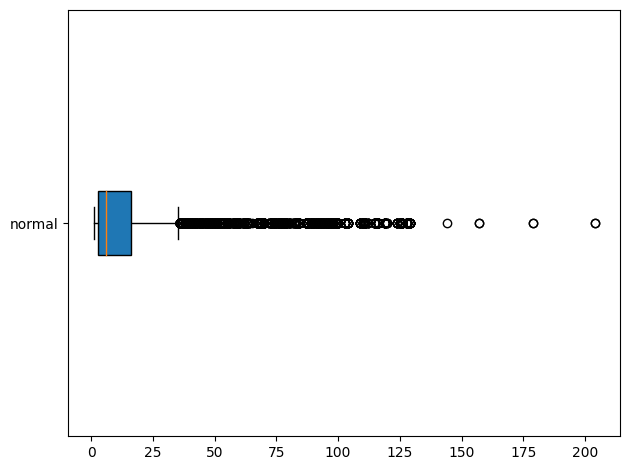}
	\caption{Android 8: Distribution of the normal protection level.}
	\label{fig:27_normal_v8}
\end{figure}

We illustrate in Figure \ref{fig:26_signature_v8} the signature protection level 
distribution without any additional flags. Similar to the dangerous protection level, 
we have out-liners with rather high usage of signature permissions. Overall, 
signature permissions' usage lies between one and four, with a median of two 
signature permissions.
If we look at Figure \ref{fig:27_normal_v8} we see the distribution of the normal 
permission level. Note that the scale of normal permissions compared to the 
dangerous and signature permissions is higher. The x-axis in Figure 
\ref{fig:27_normal_v8} shows the number of normal permissions used, and we 
have out-liners that use more than 100 normal permissions. Overall the Q1 and Q3 
percentiles show that most apps use between five to 20 normal permissions.

In summary, we have seen the percentages of apps that declare and 
use custom permissions on Android 10. For Android 9, we have shown which 
permissions 
pre-installed app developers use the most overall and which 3rd party permissions 
are more used than others. For Android 8, we showed the distribution of 
dangerous, signature, and normal permissions, and we show that we can identify 
unusual permission usages by looking at the out-liners in Figures 
\ref{fig:25_dangerous_v8}, \ref{fig:26_signature_v8} and \ref{fig:27_normal_v8}. 
Keep in mind that the shown numbers only represent the trend in our dataset. We 
cannot generalize these trends to all Android 8, 9, or 10 firmware. The depicted 
examples demonstrate that the collected data is useful to identify out-liners and 
trends. As mentioned before, we can generate such statistics for any arbitrary set 
of firmware with FirmwareDroid and compare the results.

\newpage
\subsection{Certificates} \label{Analysis:Certificates}
With AndroGuard, we extract as well the certificates of all pre-installed apps. Keep 
in mind that we cannot check Android app certificates against a PKI. Therefore we 
cannot verify that the certificate data provided is genuine. It's unlikely that all 
certificates are fake and not from an official vendor; however, some certificates 
contain counterfeit data.

As with the permissions, we can use the data to generate statistics per Android 
firmware and over arbitrary sets of apps. For example, if we are interested in a 
specific app, we can query FirmwareDroid's database for one app and search for 
all apps with the same signing certificate. Another possibility is to query how 
common a specific certificate is to determine how many apps are using the same 
certificate. Table \ref{tab:Certificates_Unique} shows how many certificates we 
have extracted and how many unique certificates we have for every Android 
version. Note that an app can have more than one certificate.

\begin{table}[H]
	\centering
	\begin{tabular}{|c|c|c|c|}
		\hline
		\rowcolor[HTML]{EFEFEF} 
		\textbf{\begin{tabular}[c]{@{}c@{}}Android \\ Version\end{tabular}} & 
		\multicolumn{1}{l|}{\cellcolor[HTML]{EFEFEF}\textbf{\begin{tabular}[c]{@{}l@{}}Total
		 number\\ of certificates\\ extracted\end{tabular}}} & 
		\textbf{\begin{tabular}[c]{@{}c@{}}Number of \\ unique\\ 
		certificates\end{tabular}} & \textbf{\begin{tabular}[c]{@{}c@{}}Ratio unique \\ 
		packagename to \\ unique certificate\end{tabular}} \\ \hline
		8 & 72'076 & 447 & 1'991/447 \\ \hline
		9 & 55'291 & 320 & 1'375/320 \\ \hline
		10 & 15'130 & 92 & 242/92 \\ \hline
	\end{tabular}
	\caption{Number of unique certificates in our database.}
	\label{tab:Certificates_Unique}
\end{table}

We can calculate the average of how many apps use the same certificates 
by dividing the unique number of package names with the unique number of 
certificates. Our Android 8 certificates are, on average, used by 4.45 apps, our 
Android 9 by 4.3 apps, and our Android 10 by 2.63 apps. These numbers show 
that, on average, the companies use their app certificates more than once. We 
give a further example of how often certificates are reused by searching for AOSP 
development keys. To be more precise, we search for how common the AOSP 
platform certificate (sha1: 27 19 6E 38 6B 87 5E 76 AD F7 00 E7 EA 84 E4 C6 EE 
E3 3D FA) is in our dataset. As a result, we can say that we have 108'983 
(12,07\%) apps that use the AOSP certificate. Note that the certificate is insecure 
since its private keys are publicly available \cite{GithubAOSPPlatformKeys} and 
everyone can sign apps with this certificate. To further explore the most common 
certificates, we analyze the top ten for Android 8, 9, 10. 

\begin{table}[H]
	\centering
	\resizebox{\textwidth}{!}{
	\begin{tabular}{|c|l|c|c|}
		\hline
		\rowcolor[HTML]{EFEFEF} 
		\textbf{\begin{tabular}[c]{@{}c@{}}App developer\\ and \\ example 
		packagename\end{tabular}} & 
		\multicolumn{1}{c|}{\cellcolor[HTML]{EFEFEF}\textbf{Certificate SHA1}} & 
		\textbf{\begin{tabular}[c]{@{}c@{}}Number of certificates\\ in 
		database\end{tabular}} & \textbf{\begin{tabular}[c]{@{}c@{}}Percantage of 
		total \\ apps in database \\ (902'262)\end{tabular}} \\ \hline
		\rowcolor[HTML]{EFEFEF} 
		\multicolumn{4}{|c|}{\cellcolor[HTML]{EFEFEF}\textit{Android 10}} \\ \hline
		\begin{tabular}[c]{@{}c@{}}Google Inc.\\ 
		com.quicinc.cne.CNEService\end{tabular} & \begin{tabular}[c]{@{}l@{}}E3 22 
		16 17 B1 D6  B5 10 DB 06 \\ 4E DB 38 15 02 BD FD 3B AC 47\end{tabular} 
		& 1316 & 0.15\% \\ \hline
		Google Inc. & \begin{tabular}[c]{@{}l@{}}38 91 8A 45 3D 07 19 93 54 F8 \\ B1 
		9A F0 5E C6 56 2C ED 57 88\end{tabular} & 1134 & 0.13\% \\ \hline
		\begin{tabular}[c]{@{}c@{}}Google Inc.\\ 
		com.android.sharedstoragebackup\end{tabular} & 
		\begin{tabular}[c]{@{}l@{}}38 D8 1C FB 64 BD 5D 78 BA \\ ED 39 3C 25 E7 
		E9 2B 91 60 20 0E\end{tabular} & 869 & 0.10\% \\ \hline
		\begin{tabular}[c]{@{}c@{}}Google Inc.\\ Android Easter Egg\end{tabular} & 
		\begin{tabular}[c]{@{}l@{}}69 25 F4 EE 29 7C 96 E8 30 5C \\ 59 EA 02 6B FA 
		74 A8 DC E1 91\end{tabular} & 869 & 0.10\% \\ \hline
		\begin{tabular}[c]{@{}c@{}}Google Inc.\\ com.android.vpndialogs\end{tabular} 
		& \begin{tabular}[c]{@{}l@{}}81 C2 B1 2C BA 14 66 FE CF \\ 45 29 BE 3F 08 
		14 B7 F5 0B 44 83\end{tabular} & 814 & 0.09\% \\ \hline
		\begin{tabular}[c]{@{}c@{}}Google Inc.\\ com.android.vpndialogs\end{tabular} 
		& \begin{tabular}[c]{@{}l@{}}81 90 2F C3 62 B6 A8 85 40 B2 \\ 7F 04 90 9C 
		2C 85 0A E8 46 85\end{tabular} & 814 & 0.09\% \\ \hline
		\begin{tabular}[c]{@{}c@{}}Google Inc.\\ 
		com.quicinc.cne.CNEService\end{tabular} & \begin{tabular}[c]{@{}l@{}}2A 5F 
		D0 80 D1 5E D5 C6 91 36 \\ B3 9A 9F 3E 64 C6 65 B9 91 A2\end{tabular} & 
		703 & 0.08\% \\ \hline
		\begin{tabular}[c]{@{}c@{}}Google Inc.\\ android\end{tabular} & 
		\begin{tabular}[c]{@{}l@{}}D4 D3 D7 EF 20 B8 59 8A 43 71 \\ 49 9C AD 13 01 
		37 AA 30 FF 3B\end{tabular} & 703 & 0.08\% \\ \hline
		\begin{tabular}[c]{@{}c@{}}Google Inc.\\ X Google enrollment\end{tabular} & 
		\begin{tabular}[c]{@{}l@{}}5F 3F 5A D9 2D 23 D6 40 2A 26 \\ 31 70 D8 7A 0D 
		8A 44 C5 90 AC\end{tabular} & 634 & 0.07\% \\ \hline
		\begin{tabular}[c]{@{}c@{}}Google Inc.\\ Default Print Service\end{tabular} & 
		\begin{tabular}[c]{@{}l@{}}70 62 71 04 12 02 F8 0C E2 AB\\ 09 DD 7C 22   95 
		9D 2D 92 DB 0C\end{tabular} & 634 & 0.07\% \\ \hline
		\rowcolor[HTML]{EFEFEF} 
		\multicolumn{4}{|c|}{\cellcolor[HTML]{EFEFEF}\textit{Android 9}} \\ \hline
		\begin{tabular}[c]{@{}c@{}}Samsung\\ 
		com.samsung.aasaservice\end{tabular} & \begin{tabular}[c]{@{}l@{}}9C A5 
		17 0F 38 19 19 DF E0 44 \\ 6F CD AB 18 B1 9A 14 3B 31 63\end{tabular} & 
		21272 & 2.36\% \\ \hline
		Google Inc. & \begin{tabular}[c]{@{}l@{}}38 91 8A 45 3D 07 19 93 54 F8 B1\\  
		9A F0 5E C6 56 2C ED 57 88\end{tabular} & 5577 & 0.62\% \\ \hline
		\begin{tabular}[c]{@{}c@{}}InfinixMobility\\ android\end{tabular} & 
		\begin{tabular}[c]{@{}l@{}}BF D5 79 80 68 4E B3 7C 5E 86 \\ DA 0D D0 8A 
		9D DF E4 81 87 41\end{tabular} & 1869 & 0.21\% \\ \hline
		\begin{tabular}[c]{@{}c@{}}Samsung (DMC)\\ 
		com.android.dreams.basic\end{tabular} & \begin{tabular}[c]{@{}l@{}}97 41 A0 
		F3 30 DC 2E 86 19 B7 \\ 6A 25 97 F3 08 C3 7D BE 30 A2\end{tabular} & 
		1686 & 0.19\% \\ \hline
		\begin{tabular}[c]{@{}c@{}}Samsung Electronics Co. Ltd.\\ 
		com.samsung.aasaservice\end{tabular} & \begin{tabular}[c]{@{}l@{}}29 C6 47 
		CB CC 9A 5F BD 6C 0C\\  96 1E 05 71 2B D1 53 52 A1 F5\end{tabular} & 
		1333 & 0.15\% \\ \hline
		\begin{tabular}[c]{@{}c@{}}Google Inc.\\ com.android.bluetooth\end{tabular} & 
		\begin{tabular}[c]{@{}l@{}}73 69 74 B3 71 23 FA 90 07 CF 05\\  CD C1 FB   
		43 D9 15 91 76 22\end{tabular} & 1116 & 0.12\% \\ \hline
		\begin{tabular}[c]{@{}c@{}}Google Inc.\\ com.android.bluetooth\end{tabular} & 
		\begin{tabular}[c]{@{}l@{}}B8 41 56 6D C2 B4 69 F3 11 14\\  BB 27 17 14   
		B5 DC A6 44 FD 80\end{tabular} & 1116 & 0.12\% \\ \hline
		Google, Inc & \begin{tabular}[c]{@{}l@{}}24 BB 24 C0 5E 47 E0 AE FA 68\\ 
		A5 8A 76 61 79 D9 B6 13 A6 00\end{tabular} & 926 & 0.10\% \\ \hline
		\begin{tabular}[c]{@{}c@{}}Google Inc.\\ android\end{tabular} & 
		\begin{tabular}[c]{@{}l@{}}D4 D3 D7 EF 20 B8 59 8A 43 71 \\ 49 9C AD 13 01 
		37 AA 30 FF 3B\end{tabular} & 919 & 0.10\% \\ \hline
		\begin{tabular}[c]{@{}c@{}}Google Inc.\\ 
		com.quicinc.cne.CNEService\end{tabular} & \begin{tabular}[c]{@{}l@{}}2A 5F 
		D0 80 D1 5E D5 C6 91 36 \\ B3 9A 9F 3E   64 C6 65 B9 91 A2\end{tabular} 
		& 919 & 0.10\% \\ \hline
		\rowcolor[HTML]{EFEFEF} 
		\multicolumn{4}{|c|}{\cellcolor[HTML]{EFEFEF}\textit{Android 8}} \\ \hline
		\begin{tabular}[c]{@{}c@{}}Samsung\\ 
		com.samsung.aasaservice\end{tabular} & \begin{tabular}[c]{@{}l@{}}9C A5 
		17 0F 38 19 19 DF E0 44 6F \\ CD AB 18 B1 9A 14 3B 31 63\end{tabular} & 
		19837 & 2.20\% \\ \hline
		Google Inc. & \begin{tabular}[c]{@{}l@{}}38 91 8A 45 3D 07 19 93 54 F8 \\ B1 
		9A F0 5E C6 56 2C ED 57 88\end{tabular} & 8731 & 0.97\% \\ \hline
		\begin{tabular}[c]{@{}c@{}}Transission\\ 
		com.mediatek.batterywarning\end{tabular} & \begin{tabular}[c]{@{}l@{}}BE 8C 
		B9 F9 5B CB 5B FB 04 \\ 04 50 34 E5 18 26 34 A2 FC A1 FA\end{tabular} & 
		2421 & 0.27\% \\ \hline
		\begin{tabular}[c]{@{}c@{}}Samsung\\ 
		com.samsung.aasaservice\end{tabular} & \begin{tabular}[c]{@{}l@{}}29 C6 47 
		CB CC 9A 5F BD 6C \\ 0C 96 1E 05 71 2B D1 53 52 A1 F5\end{tabular} & 
		1921 & 0.21\% \\ \hline
		Google, Inc & \begin{tabular}[c]{@{}l@{}}24 BB 24 C0 5E 47 E0 AE FA 68 \\ 
		A5 8A 76 61 79 D9 B6 13 A6 00\end{tabular} & 1852 & 0.21\% \\ \hline
		Samsung Corporation (DMC) & \begin{tabular}[c]{@{}l@{}}97 41 A0 F3 30 DC 
		2E 86 19 B7 \\ 6A 25 97 F3 08 C3 7D BE 30 A2\end{tabular} & 1720 & 
		0.19\% \\ \hline
		\begin{tabular}[c]{@{}c@{}}Zhantang\\ com.sprd.engineermode\end{tabular} & 
		\begin{tabular}[c]{@{}l@{}}F1 A5 4A 3F 02 4A 8D 1B 74 D1 \\ FF 1F 74 D3 
		BE 66 ED 79 31 2E\end{tabular} & 1478 & 0.16\% \\ \hline
		\begin{tabular}[c]{@{}c@{}}Google Inc.\\ com.android.egg\end{tabular} & 
		\begin{tabular}[c]{@{}l@{}}69 25 F4 EE 29 7C 96 E8 30 5C \\ 59 EA 02 6B FA 
		74 A8 DC E1 91\end{tabular} & 1407 & 0.16\% \\ \hline
		Tecno & \begin{tabular}[c]{@{}l@{}}48 19 D1 56 9B D0 64 AD 33 94 \\ 61 73 
		79 32 FB 76 1C 1B E1 77\end{tabular} & 1323 & 0.15\% \\ \hline
		\begin{tabular}[c]{@{}c@{}}Lava\\ com.adups.fota\end{tabular} & 
		\begin{tabular}[c]{@{}l@{}}FB 01 33 26 78 40 78 4C A3 00 \\ 9E 26 F2 69 F7 
		43 2C 09 19 EC\end{tabular} & 1184 & 0.13\% \\ \hline
	\end{tabular}
}
	\caption{Top ten most used certificates for Android 8, 9, and 10 in our database.}
	\label{tab:Certificates_Top5perVersion}
\end{table}

In Table \ref{tab:Certificates_Top5perVersion} we show the most common 
certificates for Android 8, 9, and 10 in our database. For Android 10, we have only 
Google certificates in the top ten. We think this is not surprising as we have for 
Android 10 mostly official stock firmware from Google. For Android 8 and 9, we 
have two certificates from Samsung at the top followed by Google certificates. We 
have as well other Samsung certificates in the top ten. If we compare the numbers 
of certificates for Samsung, we can see that there seems to be a large gap 
between the first certificate and the others in numbers. We compare the three 
Samsung certificates for Android 9 to see if we can manually spot an anomaly in 
the certificate data. Most of the data like subject or issuer seem to be the same, 
but the validity timestamps aren't. Two of the certificates have a "not-valid-before" 
field from 2011, and the other certificate (sha1: 29 C6 47 CB CC 9A 5F BD 6C 0C 
96 1E 05 71 2B D1 53 52 A1 F5) has it set to 2018. It seems odd that an Android 
9 device from Samsung would use a not-valid-before date from November 2018 
since Android 9 was released in August 2018. However, it could be possible that 
they 
created a new certificate at that time, so we have to investigate further to see if 
that certificate is part of a malicious campaign. So we check the first apk in our 
database that shows up when we search for the certificate 
(md5:ad5803db6c96b5494a4fd822cb7812ec, package 
name:com.Samsung.aasaservice) in our list with FirmwareDroid:

\begin{itemize}
	\item VirusTotal shows that three scanners detect the apk as malicious.
	
	\item APKiD does not detect a suspicious compiler. It detects the dx compiler.
	
	\item Qark lists 36 issues. Twenty-five issues are general infos, and eleven are 
	warnings. We list only the warnings that we think are relevant:
	
	\begin{itemize}
		\item This permission can be obtained by malicious apps installed prior to 
		this one, without the proper signature. Applicable to Android Devices prior to 
		L (Lollipop).
		
		\item Backups enabled: Potential for data theft via local attacks via adb 
		backup, if the device has USB debugging enabled (not common).
		
		\item The receiver com.adups.library.ac.NetworkReceiver is exported, but not 
		protected by any permissions. Failing to protect receiver tags could leave 
		them vulnerable to attack by malicious apps. The receiver tags should be 
		reviewed for vulnerabilities, such as injection and information leakage.
		 
		\item The activity com.adups.fota.GoogleOtaClient is exported, but not 
		protected by any permissions. Failing to protect activity tags could leave 
		them vulnerable to attack by malicious apps. The activity tags should be 
		reviewed for vulnerabilities, such as injection and information leakage.
		
		\item This results in AMS either resuming the earlier activity or loads it in a 
		task with same affinity or the activity is started as a new task. This may 
		result in Task Poisoning. 
	\end{itemize}
	
	\item Androwarn reports that at least two UNIX commands are executed and 
	that the app accessed some phone identifiers and can conduct phone calls.
	
	\item AndroGuard shows us that the application accesses four dangerous 
	permissions and eleven normal permissions. Looking at the normal 
	permissions, we think it is interesting that the app requests permission to install 
	and remove shortcuts. The app uses the following dangerous permissions:
	\begin{itemize}
		\item android\_permission\_GET\_ACCOUNTS
		
		\item android\_permission\_READ\_PHONE\_STATE
		
		\item android\_permission\_READ\_EXTERNAL\_STORAGE
		
		\item android\_permission\_WRITE\_EXTERNAL\_STORAGE
	\end{itemize}
\end{itemize}

The point that VirusTotal reports this apk as malicious undermines our assumption 
that the certificate is malicious. However, APKiD does not detect that the apk is 
repackaged, and it could be a coincidence. We check ten other apks of the same 
certificate with FirmwareDroid but could not detect any anomalies. Therefore, we 
decide two compare an apk with the same package name but signed with one of 
the other Samsung certificates (sha1: 9C A5 17 0F 38 19 19 DF E0 44 6F CD AB 
18 B1 9A 14 3B 31 63). Following the scanner results for the potential genuine 
com.samsung.aasaservice apk (md5:8ca66bf726e316c79590399ab00377ab):

\begin{itemize}
	\item VirusTotal does not detect any findings.
	
	\item APKiD shows that the Jack 3.x and dx compilers are used.
	
	\item Qark lists four issues-- two warnings and two infos. Following the 
	warnings reported by Qark:
	\begin{itemize}
		\item Backups enabled: Potential for data theft via local attacks via adb 
		backup, if the device has USB debugging enabled (not common). 
		
		\item This permission can be obtained by malicious apps installed prior to 
		this one, without the proper signature. Applicable to Android Devices prior to 
		L (Lollipop). 
	\end{itemize}
	
	\item Androwarn does not report any suspicious findings. 
	
	\item AndroGuard shows that the app uses one dangerous and one signature 
	permission as shown below. In total, only 6 permissions are used. 
	\begin{itemize}
		\item android\_permission\_REAL\_GET\_TASKS
		
		\item android\_permission\_WRITE\_EXTERNAL\_STORAGE
	\end{itemize}
\end{itemize}

Comparing these two apps shows significant differences in permission usages and 
findings by the individual scanners. A detailed analysis of the apps signed with the 
potential malicious Samsung certificate (sha1: 29 C6 47 CB CC 9A 5F BD 6C 0C 
96 1E 05 71 2B D1 53 52 A1 F5) is out of the scope of this project. However, even 
if it seems likely that we have found at least one malicious app singed with a 
suspicious-looking Samsung certificate, we cannot prove that the certificate is 
fake. Without a PKI, the only way to check the authenticity of this certificate is by 
asking Samsung if it is, in fact, one of their certificates. In case the certificate is 
malicious, we could assume that all the apps signed with this certificate are 
harmful.

In summary, analyzing our certificate data for anomalies can help detect 
potentially harmful applications, but without a PKI, we cannot prove a certificate's 
authenticity. If vendors would begin to publish their official app certificates, we 
could confirm every certificate's authenticity in our dataset and find more potential 
malicious certificates. A manual analysis of every certificate in our database is not 
feasible during this project, and as shown in Table 
\ref{tab:Certificates_Top5perVersion} we have more certificates that we could 
check. Nevertheless, we have demonstrated a possible use of the collected 
certificate data. We can use the data in various other ways, such as showing the 
relations between firmware and app vendors like other researchers 
\cite{AnAnalysisofPreinstalled} have done. We continue in the next section with 
our APKiD scan result analysis.

\newpage 
\section{APKiD Scan Result Analysis} \label{Analysis:APKiDAnalysis}
In this section, we will discuss the scanning result of the APKiD tool. We will 
provide for every APKiD feature a table with the raw scanning result and discuss 
some of the results in detail. 

\begin{table}[H]
	\centering
	\resizebox{0.8\textwidth}{!}{
		\begin{tabular}{|l|l|l|l|}
			\hline
			\rowcolor[HTML]{EFEFEF} 
			\textbf{Compiler} & \textbf{Decompiler} & \textbf{Obfuscators / Packers} & 
			\textbf{Deobfuscators / Unpackers} \\ \hline
			Jack \cite{DevJack} & backsmali \cite{SmaliBacksmali} & AAMO 
			\cite{AAMO} 
			& GDA(GJoy Dex Analysizer) \cite{GDA} \\ \hline
			Dx & Jad \cite{JD_JAD} & ADVobfuscator \cite{ADVobfuscator} & 
			Deoptfuscator \cite{Deoptfuscator} \\ \hline
			D8 &  & Alipay & JEB \cite{JEB} \\ \hline
			smali \cite{SmaliBacksmali} &  & Allatori \cite{Allatori} & apktool 
			\cite{Apktool} 
			\\ \hline
			Mercury &  & AntiSkid & jadx \cite{Jadx} \\ \hline
			C++Builder &  & ApkProtect \cite{ApkProtect} & deobf \cite{deobf} \\ \hline
			Intel C++ Compiler (icc) &  & AppSuit \cite{AppSuit} & Paranoid 
			Deobfuscator 
			\cite{ParanoidDeobfuscator} \\ \hline
			Visual C++ (cl) &  & Arxan GuardIT &  \\ \hline
			Iodine &  & ByteGuard &  \\ \hline
			B4X &  & DashO \cite{Dasho} &  \\ \hline
			&  & Dexguard \cite{DexGuard} &  \\ \hline
			&  & DexProtector \cite{DexProtector} &  \\ \hline
			&  & Firehash &  \\ \hline
			&  & Gemalto \cite{Gemalto} &  \\ \hline
			&  & Kiwi encrypter \cite{KiwiEncrypter} &  \\ \hline
			&  & Obfuscapk \cite{obfuscapk} &  \\ \hline
			&  & Obfuscator-LLVM \cite{Obfuscator-LLVM} &  \\ \hline
			&  & Safeengine \cite{Safeengine} &  \\ \hline
			&  & Shield4j &  \\ \hline
			&  & SnapProtect &  \\ \hline
			&  & Stringer \cite{Stringer} &  \\ \hline
			&  & PromonShield \cite{PromonShield} &  \\ \hline
			&  & Enigma \cite{ObEnigma} &  \\ \hline
			&  & Paranoid \cite{Paranoid} &  \\ \hline
			&  & Simple-obfs-android \cite{simpleObfs} &  \\ \hline
			&  & AndroidLibrary \cite{AndroidLibrary} &  \\ \hline
			&  & AESJniEncrypt \cite{AESJniEncrypt} &  \\ \hline
			&  & Obfuscapk \cite{obfuscapk} &  \\ \hline
			&  & AabResGuard \cite{AabResGuard} &  \\ \hline
			&  & rotacsufbo \cite{rotacsufbo} &  \\ \hline
			&  & Robfuscate \cite{Robfuscate} &  \\ \hline
			&  & AndrOpGAN \cite{AndrOpGAN} &  \\ \hline
			&  & AEON \cite{AEON} &  \\ \hline
			&  & ICFGO \cite{ICFGO} &  \\ \hline
		\end{tabular}
	}
	\caption{Incomplete list of compilers, decompilers, de-/obfuscators and 
		un-/packers for Android.}
	\label{tab:OverviewCompilersObf}
\end{table}

To fully understand the potential of APKiD, we need some more information about 
Android compilers and decompilers. In Table \ref{tab:OverviewCompilersObf} we 
collected a list of de-/compilers, de-/obfuscators and un-/packers. The list is not 
complete, but it shows some common tools used on Android. When we look at the 
first column \textit{"Compiler"}, we can see the entries Jack and DX. 
These two are the main toolchains used for building Android apps since there are 
the official tools supported by Google. Jack and DX were deprecated in 2017 
\cite{DevJack}, and the migration to the newer D8 toolchain is currently in progress 
\cite{DevBlogDXDepcrecation}.

With APKiD, we can scan an apk, and in some cases, it can detect which 
compiler was used based on pattern recognition with Yara \cite{Yara}. We can then 
use this data to detect if an attacker has used an unknown or suspicious compiler. 
For example, tools like Jadx \cite{Jadx} and the apktool \cite{Apktool} use smali 
and backsmali \cite{SmaliBacksmali} for de-/compiling. Such compilers have 
different patterns than the DX compiler since they use dexlib for compiling. We 
considered it unlikely that the original app developer would need to use a tool like 
Jadx to compile their source code. In this case, we think the app was likely 
re-packaged, or the developer uses a packer to protect the app.

Pre-installed apps are protected by \acrshort{avb} against re-packaging. As we 
have already discussed in Section \ref{Fundamentals:VerifiedBoot} Android 
Verified Boot prevents an attacker from corrupting a firmware partition after it is 
signed. Therefore system apps are protected by dm-verity and can't be re-package 
under the assumption that \acrshort{avb} is secure and active. However, to our 
knowledge, during the build process and before the OS vendor signs the 
pre-installed apps and partitions, no known protection mechanisms are in use. OS 
vendors like Google likely have their processes to verify a pre-installed app's 
integrity and authenticity before adding the app to a firmware build. Nevertheless, 
we can assume that some OS vendors do not have such security processes or 
less mature ones and that re-packaging of 3td party apps is possible before the 
OS vendor adds the app to the firmware.

Another reason for re-packaging pre-installed apps is for customization purposes. 
Since custom ROMs use their key material, it is possible for their developers to 
re-package a pre-installed app and resign it with their platform key. Taking this into 
consideration, we have scanned all the pre-installed in our dataset with APKiD (v. 
2.2.1). We were able to scan 897'570 out of 902'837 apps. We could not scan 
5'267 due to incompatibility and technical problems with APKiD. Table 
\ref{tab:APKIDCompilerAll} shows the result of APKiD's compiler detection.

\begin{table}[H]
	\centering
	\resizebox{\textwidth}{!}{
	\begin{tabular}{|l|c|c|c|c|c|c|c|c|c|c|c|}
		\hline
		\rowcolor[HTML]{EFEFEF} 
		\textbf{Compiler} & \cellcolor[HTML]{EFEFEF}\textbf{Unknown} & \textbf{2} & 
		\textbf{3} & \textbf{4} & \textbf{5} & \textbf{6} & \textbf{7} & \textbf{8} & 
		\textbf{9} & \textbf{10} & \cellcolor[HTML]{EFEFEF}\textbf{Total} \\ \hline
		dx & 100530 & 597 & 43 & 187553 & 55468 & 66722 & 160542 & 98872 & 
		91087 & 3946 & 765360 \\ \hline
		dx (possible dexmerge) & 4230 & 3 & 0 & 11633 & 3183 & 3902 & 12122 & 
		4923 & 4765 & 234 & 44995 \\ \hline
		Jack (unknown version) & 0 & 0 & 0 & 9 & 101 & 10535 & 966 & 2 & 2 & 0 & 
		11615 \\ \hline
		Jack 4.12 & 0 & 0 & 0 & 0 & 0 & 1 & 60 & 48 & 35 & 0 & 144 \\ \hline
		Jack 4.x & 604 & 0 & 0 & 0 & 0 & 8 & 517 & 22928 & 297 & 0 & 24354 \\ 
		\hline
		Jack 3.x & 9 & 0 & 0 & 0 & 0 & 52 & 39079 & 52 & 8 & 0 & 39200 \\ \hline
		dexlib 2.x & 89 & 0 & 0 & 1054 & 1272 & 1018 & 1156 & 125 & 255 & 0 & 
		4969 \\ \hline
		dexlib 1.x & 33 & 5 & 0 & 10283 & 354 & 386 & 1239 & 133 & 40 & 0 & 
		12473 \\ \hline
		Unknown & 24430 & 2 & 0 & 905 & 718 & 1822 & 1543 & 405 & 32943 & 
		11134 & 73902 \\ \hline
	\end{tabular}
}
	\caption{APKiD scan result for compilers.}
	\label{tab:APKIDCompilerAll}
\end{table}

If APKiD can't detect the compiler, it flags it either as unknown or does not show 
any report results. Moreover, APKiD can detect several compilers, obfuscators, or 
packers per apk because an apk can contain more than one dex, elf, or other 
binary files. In Table \ref{tab:APKIDCompilerAll} we can see that APKiD detected 
apps compiled with dexlib one and two. As mentioned before, attackers use the 
dexlib compiler for re-packaging, or developers use packers for protection. As a 
consequence, we assume that some of these samples may be malicious but not 
necessarily. For example, packers can use re-packaging techniques to secure and 
apk.

A detailed analysis of every potential re-packaged app is not feasible during this 
project. However, we pick one of the dexlib 2.x detected apks as an example to 
show how we can progress after finding it. We take the "SkinPackViVo.apk" (md5: 
5f73cf1c9018514954387aed3547f3d4) that was detected in Android 9 firmware 
archives.

First, we can search in our database for more information about the 
SkinPackViVo.apk. It seems that the SkinPackViVo.apk is only 379KB in size and 
stored in /system/app/SkinPackViVo. We search for another sample of the same 
apk, which was not compiled with dexlib. If we find one sample, we may have a 
version of the apk that we can use for comparison and not re-packaged. In the 
case of the SkinPackViVo.apk we do have a sample in our database that was 
compiled with dx (md5: 3e6334e22abc18a355aa5ab2f088fe16). So we can 
compare the certificates of these apk's to each other to see if there is any 
difference. In this case, we see that both certificates are identical and, therefore, 
from the same developer. We think this is an indicator that not a third party has 
tampered with the apk because otherwise, we would have two different certificate 
signatures. The only other possibility would be that an attack has managed to 
steal the original key material to sign the re-packaged apk.

As the second step, we scan the apk with VirusTotal. The VirusTotal online scan 
\cite{IkarusVirustotal} shows that one scanner (Ikarus) detects the file as 
malicious with the category \textit{PUA.AndroidOS.Cootek}. Searching the web for 
the apk and the malware category shows that we have found an app plugin 
developed by the Chinese company CooTek. Some online articles claim 
\cite{Malwarebytes, AndroidPolice} that CooTek apps have shown overaggressive 
advertising and were temporary removed from the Google Play Store. We were not 
expecting these results when looking for re-packaged apps, but it still shows that 
we can find suspicious apps with APKiD. We will progress with other APKiD 
results.

If we have a look at the obfuscator detection results of APKiD in Table 
\ref{tab:APKIDObfuscator} it seems that most pre-installed apps don't use an 
obfuscator, or APKiD can't detect it at least. We could detect 6'986 obfuscated 
files in 897'570 apps with APKiD. There is so far no study for APKiD that 
compares the detection rate for obfuscators to our knowledge. Other tools like 
\cite{AndroidObfuscationFingerprinter, AndroidCodeObfuscationChecker} may 
have better detection rates but comparing these tools is outside of our project 
scope. However, it seems unlikely that the official Android Studio obfuscators, 
ProGuard \cite{ProGuard} and R8, are not detected. We think it seems that 
APKiD's obfuscator and packer detection is still in development and that these are 
likely to be added in later versions. At the time of writing, several Github issues on 
the official APKiD repository \cite{APKiD} are open for adding new detection rules 
for obfuscators and packers. We can conclude that the numbers from Table 
\ref{tab:APKIDObfuscator} are not representative of all obfuscators that exist. Still, 
for DexGuard and Arxan, we can say that app developers use them on all Android 
versions in our dataset. 

\begin{table}[H]
		\centering
	\resizebox{\textwidth}{!}{
\begin{tabular}{|l|c|c|c|c|c|c|c|c|c|c|c|}
	\hline
	\rowcolor[HTML]{EFEFEF} 
	\textbf{Obfuscator} & \textbf{Unknown} & \textbf{2} & \textbf{3} & \textbf{4} & 
	\textbf{5} & \textbf{6} & \textbf{7} & \textbf{8} & \textbf{9} & \textbf{10} & 
	\textbf{Total} \\ \hline
	DexGuard & 128 & 0 & 0 & 309 & 602 & 785 & 544 & 156 & 425 & 68 & 3017 \\ 
	\hline
	Arxan & 23 & 0 & 0 & 32 & 22 & 168 & 570 & 245 & 158 & 34 & 1252 \\ \hline
	Arxan (multidex) & 0 & 0 & 0 & 25 & 0 & 0 & 0 & 0 & 0 & 0 & 25 \\ \hline
	Obfuscator-LLVM version 3.4 & 5 & 0 & 0 & 47 & 18 & 75 & 240 & 10 & 12 & 0 
	& 407 \\ \hline
	Obfuscator-LLVM version 3.5 & 1 & 0 & 0 & 21 & 8 & 5 & 44 & 75 & 0 & 0 & 
	154 \\ \hline
	Obfuscator-LLVM version 3.6.1 & 634 & 0 & 0 & 1 & 4 & 0 & 9 & 6 & 0 & 0 & 
	654 \\ \hline
	unreadable method names & 27 & 0 & 0 & 48 & 60 & 281 & 120 & 69 & 49 & 0 
	& 654 \\ \hline
	unreadable field names & 99 & 0 & 0 & 48 & 74 & 279 & 166 & 84 & 66 & 0 & 
	816 \\ \hline
	DexProtector & 0 & 0 & 0 & 0 & 0 & 0 & 1 & 0 & 0 & 0 & 1 \\ \hline
	ADVobfuscator & 0 & 0 & 0 & 0 & 0 & 0 & 2 & 0 & 0 & 0 & 2 \\ \hline
	Allatori demo & 0 & 0 & 0 & 3 & 0 & 0 & 0 & 0 & 0 & 0 & 3 \\ \hline
	Alipay & 1 & 0 & 0 & 0 & 0 & 0 & 0 & 0 & 0 & 0 & 1 \\ \hline
\end{tabular}
}
	\caption{APKiD obfuscators scan results for different Android versions.}
	\label{tab:APKIDObfuscator}
\end{table}

Another fact is that we detected the Obfuscator-LLVM in all versions except 
Android 10. We think it seems unlikely that an obfuscator is used in all versions 
before, and then app developers stopped using it. However, we have to keep in 
mind that not all versions of Obfuscator-LLVM are detected and that we have 634 
detected files from 154 unknown firmware versions. It could be that some of the 
unknown firmware are Android10 or even a higher Android version. Moreover, it 
could be just a coincidence since we have different samples of firmware for every 
version.

\begin{table}[H]
		\centering
	\resizebox{\textwidth}{!}{
	\begin{tabular}{|l|c|c|c|c|c|c|c|c|c|c|c|}
		\hline
		\rowcolor[HTML]{EFEFEF} 
		\textbf{Packer} & \textbf{Unknown} & \textbf{2} & \textbf{3} & \textbf{4} & 
		\textbf{5} & \textbf{6} & \textbf{7} & \textbf{8} & \textbf{9} & \textbf{10} & 
		\textbf{Total} \\ \hline
		APKProtect & 0 & 0 & 0 & 38 & 14 & 0 & 0 & 0 & 0 & 0 & 52 \\ \hline
		Jiagu & 0 & 0 & 0 & 14 & 7 & 11 & 4 & 0 & 0 & 0 & 36 \\ \hline
		UPX 3.91 (unmodified) & 0 & 0 & 0 & 34 & 26 & 2 & 0 & 0 & 0 & 0 & 62 \\ 
		\hline
		Unicom SDK Loader & 1 & 0 & 0 & 71 & 2 & 0 & 0 & 0 & 0 & 0 & 74 \\ \hline
		Ijiami & 0 & 0 & 0 & 11 & 3 & 0 & 3 & 2 & 0 & 0 & 19 \\ \hline
		Ijiami (UPX) & 0 & 0 & 0 & 14 & 3 & 0 & 0 & 0 & 0 & 0 & 17 \\ \hline
		SecNeo.A & 0 & 0 & 0 & 1 & 2 & 16 & 6 & 0 & 0 & 0 & 25 \\ \hline
		Mobile Tencent Protect & 0 & 0 & 0 & 0 & 4 & 10 & 4 & 0 & 1 & 0 & 19 \\ 
		\hline
		Bangcle & 0 & 0 & 0 & 40 & 7 & 0 & 0 & 0 & 0 & 0 & 47 \\ \hline
		newer-style Bangcle/SecNeo (UPX) & 0 & 0 & 0 & 74 & 14 & 0 & 0 & 0 & 0 
		& 0 & 88 \\ \hline
		SecNeo.B & 0 & 0 & 0 & 0 & 0 & 1 & 5 & 0 & 0 & 0 & 6 \\ \hline
		Bangcle/SecNeo (UPX) & 0 & 0 & 0 & 3 & 0 & 0 & 0 & 0 & 0 & 0 & 3 \\ \hline
		Bangcle (SecShell) & 0 & 0 & 0 & 0 & 0 & 0 & 3 & 0 & 0 & 0 & 3 \\ \hline
		UPX (unknown, modified) & 0 & 0 & 0 & 0 & 0 & 0 & 6 & 24 & 0 & 0 & 30 \\ 
		\hline
		sharelib UPX & 3 & 0 & 0 & 0 & 2 & 0 & 0 & 0 & 0 & 0 & 5 \\ \hline
		DexProtector & 0 & 0 & 0 & 0 & 0 & 1 & 0 & 0 & 0 & 0 & 1 \\ \hline
		Tencent's Legu & 0 & 0 & 0 & 0 & 0 & 0 & 0 & 0 & 1 & 0 & 1 \\ \hline
	\end{tabular}
}
	\caption{APKiD scan result for packers.}
	\label{tab:APKIDPacker}
\end{table}

When we look at the packer detection rates in Table \ref{tab:APKIDPacker} we can 
see that we have detected 487 packers in total. This number seems rather small 
compared to the number of apps we have scanned. However, as mentioned 
before, the APKiD's packer detection is still under development, and additional 
packers are 
likely to be added in future versions.

\begin{table}[H]
		\centering
	\resizebox{\textwidth}{!}{
	\begin{tabular}{|l|c|c|c|c|c|c|c|c|c|c|c|}
		\hline
		\rowcolor[HTML]{EFEFEF} 
		\textbf{Anti-VM} & \textbf{Unknown} & \textbf{2} & \textbf{3} & \textbf{4} & 
		\textbf{5} & \textbf{6} & \textbf{7} & \textbf{8} & \textbf{9} & \textbf{10} & 
		\textbf{Total} \\ \hline
		Build.FINGERPRINT check & 9656 & 71 & 7 & 24944 & 6912 & 17659 & 
		26963 & 15586 & 16438 & 2504 & 120740 \\ \hline
		possible Build.SERIAL check & 12679 & 13 & 2 & 17345 & 5596 & 13442 & 
		20012 & 10948 & 12421 & 510 & 92968 \\ \hline
		Build.MANUFACTURER check & 14085 & 82 & 6 & 39863 & 12081 & 25815 
		& 37256 & 19330 & 19174 & 2170 & 169862 \\ \hline
		Build.HARDWARE check & 3096 & 0 & 0 & 3390 & 1767 & 5473 & 6873 & 
		6042 & 6132 & 1300 & 34073 \\ \hline
		Build.MODEL check & 8186 & 23 & 3 & 15664 & 5165 & 11551 & 19975 & 
		9671 & 10090 & 630 & 80958 \\ \hline
		Build.PRODUCT check & 4251 & 15 & 0 & 13572 & 4298 & 9825 & 13133 & 
		6077 & 5902 & 450 & 57523 \\ \hline
		Build.TAGS check & 2586 & 1 & 0 & 3638 & 2109 & 5174 & 7319 & 3501 & 
		3249 & 510 & 28087 \\ \hline
		Build.BOARD check & 3812 & 29 & 2 & 19539 & 5010 & 9238 & 12104 & 
		6129 & 5826 & 842 & 62531 \\ \hline
		Build.USER check & 108 & 0 & 0 & 5 & 4 & 0 & 1 & 113 & 114 & 0 & 345 \\ 
		\hline
		Build.BRAND check & 235 & 2 & 0 & 58 & 35 & 98 & 883 & 244 & 223 & 34 
		& 1812 \\ \hline
		Build.DEVICE check & 267 & 2 & 0 & 236 & 31 & 81 & 903 & 306 & 312 & 
		34 & 2172 \\ \hline
		Build.ID check & 0 & 11 & 0 & 0 & 0 & 0 & 0 & 0 & 0 & 0 & 11 \\ \hline
		ro.hardware check & 1 & 0 & 0 & 1 & 2 & 0 & 3 & 10 & 5 & 0 & 22 \\ \hline
		ro.product.device check & 632 & 0 & 0 & 382 & 372 & 1168 & 2195 & 1037 
		& 616 & 34 & 6436 \\ \hline
		ro.build.type check & 2946 & 15 & 0 & 3852 & 911 & 917 & 5523 & 4020 & 
		3082 & 68 & 21334 \\ \hline
		ro.kernel.qemu check & 567 & 15 & 1 & 2597 & 534 & 719 & 1592 & 663 & 
		894 & 204 & 7786 \\ \hline
		possible ro.secure check & 84 & 1 & 0 & 1373 & 89 & 159 & 680 & 275 & 
		135 & 34 & 2830 \\ \hline
		subscriber ID check & 4588 & 21 & 1 & 11934 & 3506 & 5654 & 9763 & 6035 
		& 5692 & 712 & 47906 \\ \hline
		possible VM check & 1194 & 4 & 1 & 2381 & 862 & 1145 & 1531 & 551 & 
		900 & 34 & 8603 \\ \hline
		device ID check & 1663 & 13 & 0 & 6506 & 2232 & 2415 & 4615 & 2806 & 
		1671 & 68 & 21989 \\ \hline
		SIM operator check & 5689 & 25 & 1 & 16548 & 4746 & 9557 & 17445 & 
		8770 & 7032 & 510 & 70323 \\ \hline
		voice mail number check & 2 & 0 & 0 & 0 & 0 & 0 & 0 & 0 & 0 & 0 & 2 \\ 
		\hline
		network operator name check & 4363 & 31 & 2 & 18248 & 4897 & 9756 & 
		13203 & 6389 & 5974 & 434 & 63297 \\ \hline
		emulator file check & 407 & 0 & 0 & 479 & 171 & 303 & 735 & 341 & 219 & 
		0 & 2655 \\ \hline
		network interface name check & 4 & 0 & 0 & 36 & 6 & 6 & 22 & 3 & 0 & 0 & 
		77 \\ \hline
		/proc/cpuinfo check & 1 & 0 & 0 & 1 & 3 & 0 & 0 & 0 & 0 & 0 & 5 \\ \hline
	\end{tabular}
}
	\caption{APKiD scan result for Anti-VM checks.}
	\label{tab:APKiDAntiVM}
\end{table}

Two other features of APKiD is the detection of potential anti-vm and 
anti-disassembly techniques. To be more precise, APKiD attempts to detects if 
apps access some specific environment values. It is outside of this project's 
scope to discuss these techniques in detail, but we can see from Table 
\ref{tab:APKiDAntiVM} the results of the APKiD anti-vm scan. It seems that 
APKiD mainly scans for the access of build properties and specific files in the 
file-system.

We show the anti-disassembly results of the APKiD scan in Table 
\ref{tab:APKIDAntiDisasambly}. APKiD detected overall versions of the usage of 
illegal class names. We assume this is probably due to the use of packers or 
obfuscators.

\begin{table}[H]
		\centering
	\resizebox{\textwidth}{!}{
	\begin{tabular}{|l|c|c|c|c|c|c|c|c|c|c|c|}
		\hline
		\rowcolor[HTML]{EFEFEF} 
		\textbf{Anti-Disassembly} & \textbf{Unknown} & \textbf{2} & \textbf{3} & 
		\textbf{4} & \textbf{5} & \textbf{6} & \textbf{7} & \textbf{8} & \textbf{9} & 
		\textbf{10} & \textbf{Total} \\ \hline
		illegal class name & 4021 & 3 & 0 & 6404 & 2991 & 9518 & 13287 & 7570 & 
		8866 & 2272 & 54932 \\ \hline
		non-zero link size & 0 & 0 & 0 & 142 & 19 & 1 & 0 & 0 & 1 & 0 & 163 \\ \hline
		non-zero link offset & 0 & 0 & 0 & 142 & 19 & 1 & 0 & 0 & 1 & 0 & 163 \\ 
		\hline
	\end{tabular}
}
	\caption{APKiD scan result for Anti-Disassembly checks.}
	\label{tab:APKIDAntiDisasambly}
\end{table}

In summary, we have demonstrated that we can use APKiD's fingerprinting 
mechanism to scan large amounts of Android apps. We have shown in Tables 
\ref{tab:APKIDCompilerAll} to \ref{tab:APKIDAntiDisasambly} the scan results of 
all APKiD scans by version. Some of the results can be used to further investigate 
specific apps. Overall 
we assume that the obfuscator and packer detection rates are rather low 
compared to the number of unique packages. Future versions of APKiD will detect 
more 
obfuscators and 
packers on the same dataset. In conclusion, we think that the current dataset and 
state of APKiD are not sufficient to conclude how common a specific obfuscator 
or packer is for pre-installed apps. Nevertheless, APKiD's compiler fingerprinting is 
useful for detecting re-package or suspicious apps.

\section{Androwarn Report Analysis} \label{Analysis:AndrowarnAnalysis}
In this section, we discuss the scan results of the Androwarn tool. We scanned 
895'950 apps with Androwarn. We could not scan 6'887 apps due to compatibility 
problems with Androwarn. Compared to other tools lays Androwarn's focus more 
on finding privacy-related data than vulnerabilities. An Androwarn report can have 
several hundred messages depending on the scanned app. An overview of 
features can be found on Androwarn's official  \cite{Androwarn} Github repository. 

We can use the generated Androwarn reports in various ways. For example, we 
can compare two versions of the same app to see if we detected privacy-related 
changes. Such a change can be access to the location provider or the inclusion of 
a native library. Furthermore, it is thinkable to generate a report for every version 
of an app and create histories over time to track changes. However, for our 
Androwarn analysis, we will focus on showing examples reports for Android10 
apps. The idea of showing Android10 reports is to give the reader some insights 
into the raw data generated by Androwarn and how we can use it. We will start with 
Androwarn's privacy-related data and continue with UNIX command shell and 
native library detection. Table \ref{tab:AndrowarnPrivacy} shows the numbers of 
detected apps per privacy category. The numbers in Table 
\ref{tab:AndrowarnPrivacy} show just the app count and do not represent unique 
packages. We can see that Androwarn checks if the app accesses the location, 
audio, or telephone data. It is outside of this project's scope to determine how 
accurate Androwarn's detection rates are. 

\begin{table}[H]
				\centering
	\resizebox{\textwidth}{!}{
\begin{tabular}{|l|c|}
	\hline
	\rowcolor[HTML]{EFEFEF} 
	\textbf{Androwarn Category} & 
	\multicolumn{1}{l|}{\cellcolor[HTML]{EFEFEF}\textbf{App count}} \\ \hline
	\rowcolor[HTML]{EFEFEF} 
	\multicolumn{2}{|c|}{\cellcolor[HTML]{EFEFEF}\textbf{Location}} \\ \hline
	This application reads location information from all available providers (WiFi, 
	GPS etc\_) & 170 \\ \hline
	\rowcolor[HTML]{EFEFEF} 
	\multicolumn{2}{|c|}{\cellcolor[HTML]{EFEFEF}\textbf{Audio / Video}} \\ \hline
	This application records audio from the 'MIC' source & 100 \\ \hline
	This application captures video from the 'CAMERA' source & 68 \\ \hline
	This application captures video from the 'SURFACE' source & 34 \\ \hline
	This application records audio from the 'CAMCORDER' source & 34 \\ \hline
	\rowcolor[HTML]{EFEFEF} 
	\multicolumn{2}{|c|}{\cellcolor[HTML]{EFEFEF}\textbf{Teleophony}} \\ \hline
	This application makes phone calls & 1682 \\ \hline
	This application sends an SMS message '2' to the '1' phone number & 136 \\ 
	\hline
	This application sends an SMS message 'v9' to the 'v8' phone number & 48 \\ 
	\hline
	This application sends an SMS message 'v10' to the 'v8' phone number & 34 \\ 
	\hline
	This application sends an SMS message 'v8' to the 'v7' phone number & 18 \\ 
	\hline
	This application sends an SMS message '1' to the 'v17' phone number & 34 \\ 
	\hline
	\rowcolor[HTML]{EFEFEF} 
	\multicolumn{2}{|c|}{\cellcolor[HTML]{EFEFEF}\textbf{Teleophony identifiers}} \\ 
	\hline
	This application reads the constant indicating the state of the device SIM card 
	& 1244 \\ \hline
	This application reads the current data connection state & 652 \\ \hline
	This application reads the numeric name (MCC+MNC) of current registered 
	operator & 1580 \\ \hline
	This application reads the ISO country code equivalent of the current registered 
	operator's MCC (Mobile Country Code) & 1188 \\ \hline
	This application reads the MCC+MNC of the provider of the SIM & 1954 \\ \hline
	This application reads the Service Provider Name (SPN) & 684 \\ \hline
	This application reads the device phone type value & 1322 \\ \hline
	This application reads the operator name & 944 \\ \hline
	This application reads the phone number string for line 1, for example, the 
	MSISDN for a GSM phone & 1526 \\ \hline
	This application reads the phone's current state & 930 \\ \hline
	This application reads the unique subscriber ID, for example, the IMSI for a 
	GSM phone & 990 \\ \hline
	This application reads the voice mail number & 238 \\ \hline
	This application reads the ISO country code equivalent for the SIM provider's 
	country code & 1000 \\ \hline
	This application reads the radio technology (network type) currently in use on 
	the device for data transmission & 1008 \\ \hline
	This application reads the unique device ID, i\_e the IMEI for GSM and the 
	MEID or ESN for CDMA phones & 676 \\ \hline
	This application reads the Cell ID value & 242 \\ \hline
	This application reads the SIM's serial number & 454 \\ \hline
	This application reads the alphabetic identifier associated with the voice mail 
	number & 34 \\ \hline
	This application reads the software version number for the device, for example, 
	the IMEI/SV for GSM phones & 68 \\ \hline
	This application reads the type of activity on a data connection & 102 \\ \hline
	This application reads the current location of the device & 344 \\ \hline
	This application reads the Location Area Code value & 242 \\ \hline
	\rowcolor[HTML]{EFEFEF} 
	\multicolumn{2}{|c|}{\cellcolor[HTML]{EFEFEF}\textbf{Personal Information 
	Manager Data leakage}} \\ \hline
	This application accesses data stored in the clipboard & 738 \\ \hline
	This application accesses the contacts list & 272 \\ \hline
	This application accesses the MMS list & 442 \\ \hline
	This application accesses the SMS list & 374 \\ \hline
	This application accesses the SMS/MMS list & 272 \\ \hline
	This application accesses the Bluetooth & 136 \\ \hline
	This application accesses the downloads folder & 204 \\ \hline
	This application accesses the call log & 34 \\ \hline
	This application accesses the calendar & 34 \\ \hline
	This application accesses the synchronisation history & 34 \\ \hline
\end{tabular}
}
	\caption{Androwarn privacy report example for Android10 firmware.}
	\label{tab:AndrowarnPrivacy}
\end{table}

For example, to show what gets detected, we can look at some selected 
categories and see the detected apks. We removed duplicated entries and 
grouped the apps by their package name, and we show this in Table 
\ref{tab:Androwarn_AppsGrouped}. As we can see, there are not as many apks per 
category as it may look in Table \ref{tab:AndrowarnPrivacy} and we have mostly 
com.google.* packages. This is not surprising since our Android10 dataset 
contains mostly official Google stock firmware archives. What is more interesting 
privacy-wise is that we have as well packages from other vendors like Verizon and 
Qualcomm that have access to the camera and location data, according to 
Androwarn.

\begin{table}[H]
	\centering
\resizebox{\textwidth}{!}{
\begin{tabular}{|l|l|l|l|}
	\hline
	\rowcolor[HTML]{EFEFEF} 
	\multicolumn{1}{|c|}{\cellcolor[HTML]{EFEFEF}\textbf{DisplayName}} & 
	\multicolumn{1}{c|}{\cellcolor[HTML]{EFEFEF}\textbf{Filename}} & 
	\multicolumn{1}{c|}{\cellcolor[HTML]{EFEFEF}\textbf{Packagename}} & 
	\multicolumn{1}{c|}{\cellcolor[HTML]{EFEFEF}\textbf{MD5*}} \\ \hline
	\rowcolor[HTML]{EFEFEF} 
	\multicolumn{4}{|c|}{\cellcolor[HTML]{EFEFEF}\textit{This application records 
	audio from the 'MIC' source}} \\ \hline
	My Verizon Services & MyVerizonServices.apk & com.verizon.mips.services & 
	59049cf207598f10c9d9551b99fca31b \\ \hline
	System UI & SystemUIGoogle.apk & com.android.systemui & 
	c3738c06a61bfecb913393d671fde4a8 \\ \hline
	Google Fi & Tycho.apk & com.google.android.apps.tycho & 
	cdbbf2b11dbaeb15bb57374883b2c89f \\ \hline
	Phone & GoogleDialer.apk & com.google.android.dialer & 
	9b89e9b64affdb0b0f3b8f23e100960d \\ \hline
	\rowcolor[HTML]{EFEFEF} 
	\multicolumn{4}{|c|}{\cellcolor[HTML]{EFEFEF}\textit{This application reads the 
	current location of the device}} \\ \hline
	LTE Broadcast Manager & QAS\_DVC\_MSP.apk & com.qti.ltebc & 
	9d2ed99d1d8c158a4dde4f9412c5d29c \\ \hline
	Carrier Services & CarrierServices.apk & com.google.android.ims & 
	48e96d56bb2bf5d4838f01430a4c137e \\ \hline
	My Verizon Services & MyVerizonServices.apk & com.verizon.mips.services & 
	59049cf207598f10c9d9551b99fca31b \\ \hline
	Hidden Menu & SprintHM.apk & com.google.android.hiddenmenu & 
	3bb40d71113c1285d2fcea186d6ca02a \\ \hline
	Google Play services & PrebuiltGmsCoreQt.apk & com.google.android.gms & 
	326f6514b2559a7d19c9971a2da70a1b \\ \hline
	Google App & Velvet.apk & com.google.android.googlequicksearchbox & 
	eb7566588d9cafac77e25cf9596a596e \\ \hline
	Settings & SettingsGoogle.apk & com.android.settings & 
	90ebbff4001869a01235271b3b2e4c66 \\ \hline
	Messenger & PrebuiltBugle.apk & com.google.android.apps.messaging & 
	0acf2716400585bf13af03ce83949101 \\ \hline
	VZW\_Multicast\_MW & QAS\_DVC\_MSP\_VZW.apk & 
	com.qualcomm.ltebc\_vzw & 3756ba0b3f73cdfdc4b16070a19717a8 \\ \hline
	\rowcolor[HTML]{EFEFEF} 
	\multicolumn{4}{|c|}{\cellcolor[HTML]{EFEFEF}\textit{This application reads 
	location information from all available providers (WiFi, GPS etc\_)}} \\ \hline
	Google Play services & PrebuiltGmsCoreQt.apk & com.google.android.gms & 
	326f6514b2559a7d19c9971a2da70a1b \\ \hline
	Google & Velvet.apk & com.google.android.googlequicksearchbox & 
	b9d8009cf051f96a91874b8c820a7106 \\ \hline
	Chrome & Chrome.apk & com.android.chrome & 
	508e65b15292c34f79680c896855f29a \\ \hline
	Maps & Maps.apk & com.google.android.apps.maps & 
	15989c0bd5bc7593c984f38e51b25e41 \\ \hline
	Android System WebView & WebViewGoogle.apk & 
	com.google.android.webview & 87923767a9e46b895413759dd29bf43f \\ \hline
	\rowcolor[HTML]{EFEFEF} 
	\multicolumn{4}{|c|}{\cellcolor[HTML]{EFEFEF}\textit{This application captures 
	video from the 'CAMERA' source}} \\ \hline
	Google Play services & PrebuiltGmsCoreQt.apk & com.google.android.gms & 
	326f6514b2559a7d19c9971a2da70a1b \\ \hline
	Photos & Photos.apk & com.google.android.apps.photos & 
	ec3b52291d88fd0489e600ceda14ecf5 \\ \hline
\end{tabular}
}
	\caption{Androwarn categories grouped by unique apps. *Note: Only one MD5 
	hash shown to save space.}
	\label{tab:Androwarn_AppsGrouped}
\end{table}

We check the three apps' permissions from Verizon and Qualcomm to verify if the 
Androwarn findings are correct. We use AndroGuard to extract the permission and 
get the following results:

\begin{itemize}
	\item MyVerizonServices.apk: It uses 82 permissions in total and defines ten 
	permissions for its services. The app is able with the 
	android.permission.INSTALL\_PACKAGES to install new packages without the 
	user's consent. Moreover, it uses the camera, microphone, location, file, 
	Bluetooth, and NFC permissions. It defines several services, broadcast 
	receivers, and providers that have the exported flag set to true. However, the 
	apps secure most of the exported functions with a permission attribute. 
	
	From looking at the permissions, we can see that MyVerizonServices.apk has 
	permission for the camera and microphone, but Androwarn only detected the 
	microphone access. We can assume that the missing camera message is a 
	false 
	negative by Androwarn. About the MyVerizonServices.apk we can say that it is 
	stored in the \textit{/system/product/priv-app/} folder and that it is a system app 
	with high privileges that seems to have access to all critical privacy services.
	
	\item QAS\_DVC\_MSP.apk: It uses ten permissions in total and requests two 
	permissions from third parties. The app requests the permissions to read and 
	write to the external storage and the signature permission to send embms 
	intents. It requests the following two custom permissions.
	\begin{itemize}
		\item com.qualcomm.permission.USE\_EMBMS\_SERVICE
		\item com.qualcomm.permission.USE\_EMBMS\_SERVICE\_PLATFORM
	\end{itemize}
	 The QAS\_DVC\_MSP.apk itself does not have the location permission, and 
	 we don't know why Androwarn has flagged it to access the device's current 
	 location. The apk itself is stored in \textit{/system/app/QAS\_DVC\_MSP}. 
	 Qualcomm signed it with the same certificate as other Qualcomm apps. For 
	 example, the com.qualcomm.qcrilmsgtunnel (/system/priv-app/qcrilmsgtunnel) 
	 and 
	com.qualcomm.embms(/system/app/embms) packages. We conclude that this 
	app is a system app with the privilege to access the external storage without 
	the user's direct consent but we haven't found any evidence for location access.
	
	\item QAS\_DVC\_MSP\_VZW.apk: The app has ten permissions in total and 
	only the android.-
	permission.READ\_PHONE\_STATE permission is considered 
	dangerous. It allows reading the phone number and the unique device id as well 
	as cellular network information. This permission is likely enough to estimate the 
	location of the phone by using the cellular network information. The app 
	requests these two permissions from other apps: 
	\begin{itemize}
		\item com.qualcomm.permission.USE\_EMBMS\_SERVICE
		\item com.vzw.APNPERMISSION
	\end{itemize}
	The apk itself is stored in /system/app/QAS\_DVC\_MSP\_VZW. It is not 
	signed with the same certificate as the com.qualcomm.embm package but with 
	the same certificate as the qualcomm.com.vzw\_msdc\_api package and some 
	other Qualcomm packages. We conclude, the QAS\_DVC\_MSP\_VZW.apk 
	has access to the phone's state, and we think Androwarn's location message is 
	a correct finding. We do not investigate if the phone state access is justified.
\end{itemize}

The examples above show how we can use an Androwarn report to identify apps 
with access to privacy-related data. Another feature of Androwarn allows us to 
identify if an app accesses native UNIX commands. These can be useful in case 
that we want to search for potential vulnerabilities within an app. Suppose app 
developers don't sanitize a variable input string to a UNIX command shell. In that 
case, it can lead to attacks where attackers may use the input string to conduct 
privileged escalation or confused deputy attacks. Note: Such vulnerabilities can as 
well be found with static taint analysis tools. Table 
\ref{tab:AndrowarnUnixExamples} shows the UNIX 
detection results for Android10 firmware. Compared to the 15'078 apps in our 
Android10 dataset, it seems that not many apps execute UNIX commands directly 
over a shell. If we look at the UNIX commands detected, we can say that most 
commands use a constant string as a command and not a variable input string.

\begin{table}[H]
\centering
\resizebox{\textwidth}{!}{
\begin{tabular}{|l|c|}
	\hline
	\rowcolor[HTML]{EFEFEF} 
	\textbf{Androwarn Category} & \textbf{App count} \\ \hline
	This application executes a UNIX command & 204 \\ \hline
	\begin{tabular}[c]{@{}l@{}}This application executes a UNIX command 
	containing this argument:\\ 
	'Ljava/util/Arrays;-\textgreater{}toString({[}Ljava/lang/Object;)Ljava/lang/String;'\end{tabular}
	 & 136 \\ \hline
	\begin{tabular}[c]{@{}l@{}}This application executes a UNIX command 
	containing this argument: \\ 
	'Ljava/lang/String;-\textgreater{}valueOf(Ljava/lang/Object;)Ljava/lang/String;'\end{tabular}
	 & 102 \\ \hline
	This application executes a UNIX command containing this argument: 'ipconfig 
	/all' & 102 \\ \hline
	This application executes a UNIX command containing this argument: 'cat 
	/proc/meminfo' & 34 \\ \hline
	This application executes a UNIX command containing this argument: 'su -c ls' 
	& 34 \\ \hline
	This application executes a UNIX command containing this argument: 'top -n 1 
	-d 5' & 34 \\ \hline
	This application executes a UNIX command containing this argument: 
	'/system/bin/sync' & 34 \\ \hline
	This application executes a UNIX command containing this argument: '2' & 114 
	\\ \hline
	\begin{tabular}[c]{@{}l@{}}This application executes a UNIX command 
	containing this argument: \\ 
	'Ljava/io/File;-\textgreater{}getAbsolutePath()Ljava/lang/String;'\end{tabular} & 
	34 \\ \hline
	This application executes a UNIX command containing this argument: '1' & 34 \\ 
	\hline
	This application executes a UNIX command containing this argument: 'logcat -d' 
	& 68 \\ \hline
	This application executes a UNIX command containing this argument: 'logcat -t 
	\%d 9' & 34 \\ \hline
	This application executes a UNIX command containing this argument: ' 
	/system/bin/sh -c read  ' & 34 \\ \hline
	This application executes a UNIX command containing this argument: 'ping -c 1 
	-w 100 www.google.com' & 34 \\ \hline
	This application executes a UNIX command containing this argument: 'ping -c 1 
	www.google.com' & 34 \\ \hline
	This application executes a UNIX command containing this argument: 'ping6 -c 
	1 www.google.com' & 34 \\ \hline
	This application executes a UNIX command containing this argument: 
	'/system/bin/getprop fakeTimestamp' & 34 \\ \hline
	This application executes a UNIX command containing this argument: '0' & 80 \\ 
	\hline
	\begin{tabular}[c]{@{}l@{}}This application executes a UNIX command 
	containing this argument: \\ 
	'Ljava/lang/StringBuffer;-\textgreater{}toString()Ljava/lang/String;'\end{tabular} & 
	34 \\ \hline
	This application executes a UNIX command containing this argument: 
	'Ljava/util/List;-\textgreater{}size()I' & 34 \\ \hline
	This application executes a UNIX command containing this argument: '4' & 34 \\ 
	\hline
\end{tabular}
}
	\caption{Androwarn results of UNIX command execution on Android10 firmware.}
	\label{tab:AndrowarnUnixExamples}
\end{table}

To further investigate these, we group the data by unique package names. Table 
\ref{tab:Andowarn_UNIX_grouped} shows the data grouped by package 
names. Some packages in Table \ref{tab:AndrowarnUnixExamples} use variable 
string inputs, and we don't know which commands the apps execute without 
further investigation, or in other words, manual analysis. Another way to use 
Androwarn is to detect the usage of native libraries as shown in Table 
\ref{tab:AndrowarnNativeDetectionExamples}. Detecting libraries can be useful to 
find out if an app use a specific version of native library.

\begin{table}[H]
	\centering
	\resizebox{\textwidth}{!}{
	\begin{tabular}{|c|c|c|c|}
		\hline
		\rowcolor[HTML]{EFEFEF} 
		\textbf{Display Name} & \textbf{Filename} & \textbf{Package Name} & 
		\textbf{MD5*} \\ \hline
		\rowcolor[HTML]{EFEFEF} 
		\multicolumn{4}{|c|}{\cellcolor[HTML]{EFEFEF}\textit{This application 
		executes a UNIX command}} \\ \hline
		My Verizon Services & MyVerizonServices.apk & com.verizon.mips.services 
		& 59049cf207598f10c9d9551b99fca31b \\ \hline
		- & ConnMetrics.apk & com.android.connectivity.metrics & 
		39240d07a0a52d273d2faf3d25e29445 \\ \hline
		Google Play services & PrebuiltGmsCoreQt.apk & com.google.android.gms 
		& 326f6514b2559a7d19c9971a2da70a1b \\ \hline
		Market Feedback Agent & GoogleFeedback.apk & 
		com.google.android.feedback & 4f8b13dd7d67d16bed1356be20d969a7 \\ 
		\hline
		Messages & PrebuiltBugle.apk & com.google.android.apps.messaging & 
		401b2f54be40b118ec47af0d45f0a7db \\ \hline
		Calculator & CalculatorGooglePrebuilt.apk & com.google.android.calculator & 
		7a83b5d080d4962c87d2b0a61303a79c \\ \hline
		\rowcolor[HTML]{EFEFEF} 
		\multicolumn{4}{|c|}{\cellcolor[HTML]{EFEFEF}\textit{'Ljava/util/Arrays;-\textgreater{}toString({[}Ljava/lang/Object;)Ljava/lang/String;'}}
		 \\ \hline
		System Tracing & Traceur.apk & com.android.traceur & 
		131b973a875004aadf150f2d450973cb \\ \hline
		\rowcolor[HTML]{EFEFEF} 
		\multicolumn{4}{|c|}{\cellcolor[HTML]{EFEFEF}\textit{'Ljava/lang/String;-\textgreater{}valueOf(Ljava/lang/Object;)Ljava/lang/String;'}}
		 \\ \hline
		Carrier Services & CarrierServices.apk & com.google.android.ims & 
		040c1a4ea747f53cd0e3a3f6514d33c4 \\ \hline
		Gmail & PrebuiltGmail.apk & com.google.android.gm & 
		96e456cd1418e423c26bfff8b9da9fba \\ \hline
		Messages & PrebuiltBugle.apk & com.google.android.apps.messaging & 
		401b2f54be40b118ec47af0d45f0a7db \\ \hline
		\rowcolor[HTML]{EFEFEF} 
		\multicolumn{4}{|c|}{\cellcolor[HTML]{EFEFEF}\textit{'ipconfig /all'}} \\ \hline
		Carrier Services & CarrierServices.apk & com.google.android.ims & 
		040c1a4ea747f53cd0e3a3f6514d33c4 \\ \hline
		Gmail & PrebuiltGmail.apk & com.google.android.gm & 
		96e456cd1418e423c26bfff8b9da9fba \\ \hline
		Messages & PrebuiltBugle.ap & com.google.android.apps.messaging & 
		401b2f54be40b118ec47af0d45f0a7db \\ \hline
		\rowcolor[HTML]{EFEFEF} 
		\multicolumn{4}{|c|}{\cellcolor[HTML]{EFEFEF}\textit{'cat /proc/meminfo' and 
		'su -c ls' and  'top -n 1 -d 5'}} \\ \hline
		My Verizon Services & MyVerizonServices.apk & com.verizon.mips.services 
		& 59049cf207598f10c9d9551b99fca31b \\ \hline
		\rowcolor[HTML]{EFEFEF} 
		\multicolumn{4}{|c|}{\cellcolor[HTML]{EFEFEF}\textit{'/system/bin/sync'}} \\ 
		\hline
		Google Play Store & Phonesky.apk & com.android.vending & 
		41e574b9e0ab80e5f9849d3950a19036 \\ \hline
		\rowcolor[HTML]{EFEFEF} 
		\multicolumn{4}{|c|}{\cellcolor[HTML]{EFEFEF}\textit{This application 
		executes a UNIX command containing this argument: '2'}} \\ \hline
		Google Play Store & Phonesky.apk & com.android.vending & 
		41e574b9e0ab80e5f9849d3950a19036 \\ \hline
		Phone & GoogleDialer.apk & com.google.android.dialer & 
		3c96d86226ad583848c7bb84b522300b \\ \hline
		Google Play services & PrebuiltGmsCoreQt.apk & com.google.android.gms 
		& 326f6514b2559a7d19c9971a2da70a1b \\ \hline
		- & PrebuiltGmsCoreQt\_MapsDynamite.apk & 
		com.google.android.gms.dynamite\_mapsdynamite & 
		8102cc0587040c740126f0c7d49e82c6 \\ \hline
		\rowcolor[HTML]{EFEFEF} 
		\multicolumn{4}{|c|}{\cellcolor[HTML]{EFEFEF}\textit{'Ljava/io/File;-\textgreater{}getAbsolutePath()Ljava/lang/String;'}}
		 \\ \hline
		Google Play Store & Phonesky.apk & com.android.vending & 
		41e574b9e0ab80e5f9849d3950a19036 \\ \hline
		\rowcolor[HTML]{EFEFEF} 
		\multicolumn{4}{|c|}{\cellcolor[HTML]{EFEFEF}\textit{This application 
		executes a UNIX command containing this argument: '1'}} \\ \hline
		Google Play services & PrebuiltGmsCoreQt.apk & com.google.android.gms 
		& 326f6514b2559a7d19c9971a2da70a1b \\ \hline
		\rowcolor[HTML]{EFEFEF} 
		\multicolumn{4}{|c|}{\cellcolor[HTML]{EFEFEF}\textit{'logcat -d'}} \\ \hline
		Google Play services & PrebuiltGmsCoreQt.apk & com.google.android.gms 
		& 326f6514b2559a7d19c9971a2da70a1b \\ \hline
		Chrome & Chrome.apk & com.android.chrome & 
		f1e48c754b6a3449ced2055f9a9fe73e \\ \hline
		\rowcolor[HTML]{EFEFEF} 
		\multicolumn{4}{|c|}{\cellcolor[HTML]{EFEFEF}\textit{'logcat -t \%d 9'}} \\ \hline
		Google Play services & PrebuiltGmsCoreQt.apk & com.google.android.gms 
		& 326f6514b2559a7d19c9971a2da70a1b \\ \hline
		\rowcolor[HTML]{EFEFEF} 
		\multicolumn{4}{|c|}{\cellcolor[HTML]{EFEFEF}\textit{' /system/bin/sh -c 
		read'}} \\ \hline
		Google & Velvet.apk & com.google.android.googlequicksearchbox & 
		06217435505bca97e1b23ca2cfbb1d12 \\ \hline
		\rowcolor[HTML]{EFEFEF} 
		\multicolumn{4}{|c|}{\cellcolor[HTML]{EFEFEF}\textit{'ping -c 1 -w 100 
		www.google.com' and 'ping -c 1 www.google.com' and  'ping6 -c 1 
		www.google.com'}} \\ \hline
		Settings & SettingsGoogle.apk & com.android.settings & 
		f960cc9309d05aa995ccb3826ccd60a6 \\ \hline
		\rowcolor[HTML]{EFEFEF} 
		\multicolumn{4}{|c|}{\cellcolor[HTML]{EFEFEF}\textit{'/system/bin/getprop 
		fakeTimestamp'}} \\ \hline
		Calendar & CalendarGooglePrebuilt.apk & com.google.android.calendar & 
		1716be0e4471c233994a6c11aec38f74 \\ \hline
		\rowcolor[HTML]{EFEFEF} 
		\multicolumn{4}{|c|}{\cellcolor[HTML]{EFEFEF}\textit{This application 
		executes a UNIX command containing this argument: '0'}} \\ \hline
		Camera & GoogleCamera.apk & com.google.android.GoogleCamera & 
		fc313ea25a8a5c7a9f6d1c05c9137744 \\ \hline
		Google Play Movies & Videos.apk & com.google.android.videos & 
		8c9c65bf6a140583846107e1f2d9c3f4 \\ \hline
		YouTube & YouTube.apk & com.google.android.youtube & 
		3f83d719d05f1c3c1ce9f9e3a58db850 \\ \hline
		\rowcolor[HTML]{EFEFEF} 
		\multicolumn{4}{|c|}{\cellcolor[HTML]{EFEFEF}\textit{'Ljava/lang/StringBuffer;-\textgreater{}toString()Ljava/lang/String;'
		 and 'Ljava/util/List;-\textgreater{}size()I'}} \\ \hline
		Google Play Movies & Videos.apk & com.google.android.videos & 
		8c9c65bf6a140583846107e1f2d9c3f4 \\ \hline
		\rowcolor[HTML]{EFEFEF} 
		\multicolumn{4}{|c|}{\cellcolor[HTML]{EFEFEF}\textit{This application 
		executes a UNIX command containing this argument: '4'}} \\ \hline
		Phone & GoogleDialer.apk & com.google.android.dialer & 
		9b89e9b64affdb0b0f3b8f23e100960d \\ \hline
	\end{tabular}
}
	\caption{Androwarn UNIX execution detection results on Android10 firmware.}
	\label{tab:Andowarn_UNIX_grouped}
\end{table}

\begin{table}[H]
		\centering
	\resizebox{\textwidth}{!}{
\begin{tabular}{|l|c|}
	\hline
	\rowcolor[HTML]{EFEFEF} 
	\textbf{Androwarn message} & \textbf{App count} \\ \hline
	This application loads a native library & 530 \\ \hline
	\begin{tabular}[c]{@{}l@{}}This application loads a native library: \\ 
	'Lcom/qualcomm/ltebc/LTEBCFactory;-\textgreater{}getInstance()Lcom/qualcomm/ltebc/LTEBCFactory;'\end{tabular}
	 & 106 \\ \hline
	This application loads a native library: 'imscamera.jni' & 92 \\ \hline
	This application loads a native library: 'imsmedia.jni' & 92 \\ \hline
	This application loads a native library: 'printspooler.jni' & 136 \\ \hline
	\begin{tabular}[c]{@{}l@{}}This application loads a native library: \\ 
	'Ljava/lang/String;-\textgreater{}valueOf(Ljava/lang/Object;)Ljava/lang/String;'\end{tabular}
	 & 170 \\ \hline
	This application loads a native library: 'tensorflowlite.jni' & 68 \\ \hline
	This application loads a native library: 'curve25519' & 68 \\ \hline
	This application loads a native library: 'flash' & 34 \\ \hline
	\begin{tabular}[c]{@{}l@{}}This application loads a native library: \\ 
	'Ljava/lang/System;-\textgreater{}loadLibrary(Ljava/lang/String;)V'\end{tabular} & 
	28 \\ \hline
\end{tabular}
}
	\caption{Androwarn native library detection examples on Android10 firmware. 
	Note: Only some examples are shown. A complete list of all detected native 
	libraries for Android10 is available in Appendix \ref{Appendix:AndrowarnNative}.}
	\label{tab:AndrowarnNativeDetectionExamples}
\end{table}

So far, we have seen one Androwarn example report on our Android10 firmware. In 
Chapter \ref{Implementation} we will explain how we use FirmareDroid's statistics 
API to generate reports over any arbitrary set of firmware or Android apps. As an 
example, we created reports for Android 7, 8, 9, and 10 and show the results for 
location, audio, and video tracking in Table \ref{tab:AndrowarnV78910}

\begin{table}[H]
		\resizebox{\textwidth}{!}{
	\begin{tabular}{|l|c|c|c|c|}
		\hline
		\rowcolor[HTML]{EFEFEF} 
		\multicolumn{1}{|c|}{\cellcolor[HTML]{EFEFEF}\textbf{Androwarn Category / 
		Android version}} & \textbf{7} & \textbf{8} & \textbf{9} & \textbf{10} \\ \hline
		\rowcolor[HTML]{EFEFEF} 
		\multicolumn{5}{|c|}{\cellcolor[HTML]{EFEFEF}\textit{Location}} \\ \hline
		This application reads location information from all available providers (WiFi, 
		GPS etc.) & 4389 & 1885 & 1498 & 170 \\ \hline
		\rowcolor[HTML]{EFEFEF} 
		\multicolumn{5}{|c|}{\cellcolor[HTML]{EFEFEF}\textit{Microphone / Video}} \\ 
		\hline
		This application records audio from the 'MIC' source & 1252 & 630 & 530 & 
		100 \\ \hline
		This application records audio from the 'CAMCORDER' source & 1372 & 498 
		& 423 & 34 \\ \hline
		This application records audio & 577 & 194 & 250 & 0 \\ \hline
		This application records audio from the 'DEFAULT' source & 237 & 11 & 10 & 
		0 \\ \hline
		This application records audio from the 'VOICE.CALL' source & 76 & 62 & 19 
		& 0 \\ \hline
		This application records audio from the 'VOICE.UPLINK' source & 43 & 59 & 
		35 & 0 \\ \hline
		This application records audio from the 'VOICE.COMMUNICATION' source & 
		1 & 0 & 1 & 0 \\ \hline
		This application records audio from the 'VOICE.DOWNLINK' source & 33 & 
		19 & 1 & 0 \\ \hline
		This application records audio from the 'N/A' source & 21 & 66 & 45 & 0 \\ 
		\hline
		This application records audio from the 'REMOTE.SUBMIX' source & 0 & 49 
		& 0 & 0 \\ \hline
		This application records audio from the 'VOICE.RECOGNITION' source & 33 
		& 2 & 16 & 0 \\ \hline
		\rowcolor[HTML]{EFEFEF} 
		\multicolumn{5}{|c|}{\cellcolor[HTML]{EFEFEF}\textit{Video}} \\ \hline
		This application captures video & 162 & 80 & 27 & 0 \\ \hline
		This application captures video from the 'DEFAULT' source & 27 & 9 & 0 & 0 
		\\ \hline
		This application captures video from the 'N/A' source & 2 & 9 & 14 & 0 \\ 
		\hline
		This application captures video from the 'SURFACE' source & 318 & 305 & 
		390 & 34 \\ \hline
		This application captures video from the 'CAMERA' source & 1601 & 518 & 
		446 & 68 \\ \hline
	\end{tabular}
}
	\caption{Androwarn report results for Android 7, 8, 9 and 10 for the location, 
	audio and video category.}
	\label{tab:AndrowarnV78910}
\end{table}

The numbers in Table \ref{tab:AndrowarnV78910} demonstrate that Androwarn can 
detect apps that use various audio and video input sources. Still, Androwarn does 
not distinguish if the access is legitimate or not. In the case of system apps, at 
least the camera app on every firmware likely has access to the camera and 
microphone. Other system apps should only have access to a microphone and 
camera if the user grants it. However, currently, we cannot verify if these apps can 
access the location, camera, or video without analyzing them manually. Such an 
analysis is outside of the scope of this project. Nevertheless, this report 
demonstrates that we can use FirmwareDroid to find apps with access to privacy 
critical services. 

In summary, we have shown that we can use Androwarn to detect suspicious apps 
and anomalies, but manual analysis is necessary to verify the report results. 
We think security researchers can use the UNIX command detection to find 
vulnerabilities or suspicious commands. However, Androwarn has an unknown 
false positive or negative detection rate, and we don't know how accurate its 
reports are. 

\section{VirusTotal Report Analysis} 
\label{Analysis:VirusTotalQarkAnalysis}

A VirusTotal report contains for every virus scanner a scan result and then 
categorizes the binary as malicious if one of the scanners detects any harmful 
behavior. We are only interested in the apps that were categorized as malicious by 
several virus scanners for our study. Therefore, we defined for our statistics that 
we only flag an app as malicious if it was detected by more than three virus 
scanners as suggested by other researchers \cite{AnAnalysisPrivacyVPN, 
DREBINEffective}. This method should minimize the false positive rate of 
VirusTotal. We scanned 901'181 apps with the VirusTotal API (v.3), and we 
illustrate the results of the scan in Table \ref{tab:VirusTotalResults}.

\begin{table}[H]
		\resizebox{\textwidth}{!}{
\begin{tabular}{|l|c|c|c|c|c|c|c|c|c|c|c|}
	\hline
	\rowcolor[HTML]{EFEFEF} 
	{\backslashbox[40mm]{Category}{Android Version}} & \textbf{Unknown} & 
	\textbf{2} & \textbf{3} & \textbf{4} & \textbf{5} & \textbf{6} & \textbf{7} & 
	\textbf{8} & \textbf{9} & \textbf{10} & \textbf{Total} \\ \hline
	Malicious & \begin{tabular}[c]{@{}c@{}}9 \\ (0.2\%)\end{tabular} & 
	\begin{tabular}[c]{@{}c@{}}22\\ (1.16\%)\end{tabular} & 
	\begin{tabular}[c]{@{}c@{}}0\\ (0\%)\end{tabular} & 
	\begin{tabular}[c]{@{}c@{}}6'621\\ (2.04\%)\end{tabular} & 
	\begin{tabular}[c]{@{}c@{}}1'355\\ (0.95\%)\end{tabular} & 
	\begin{tabular}[c]{@{}c@{}}1'055\\ (0.91\%)\end{tabular} & 
	\begin{tabular}[c]{@{}c@{}}677\\ (0.5\%)\end{tabular} & 
	\begin{tabular}[c]{@{}c@{}}94\\ (0.13\%)\end{tabular} & 
	\begin{tabular}[c]{@{}c@{}}45\\ (0.08\%)\end{tabular} & 
	\begin{tabular}[c]{@{}c@{}}1\\ (0.01\%)\end{tabular} & 
	\cellcolor[HTML]{EFEFEF}\textit{\textbf{\begin{tabular}[c]{@{}c@{}}9'879 \\ 
	(1.1\%)\end{tabular}}} \\ \hline
	Undetected & \begin{tabular}[c]{@{}c@{}}38'087 \\ (99.98\%)\end{tabular} & 
	\begin{tabular}[c]{@{}c@{}}1'876\\ (98.84\%)\end{tabular} & 
	\begin{tabular}[c]{@{}c@{}}151\\ (100\%)\end{tabular} & 
	\begin{tabular}[c]{@{}c@{}}317'174\\ (97.96\%)\end{tabular} & 
	\begin{tabular}[c]{@{}c@{}}141'493\\ (99.05\%)\end{tabular} & 
	\begin{tabular}[c]{@{}c@{}}114'406\\ (99.09\%)\end{tabular} & 
	\begin{tabular}[c]{@{}c@{}}134'254\\ (99.5\%)\end{tabular} & 
	\begin{tabular}[c]{@{}c@{}}72'818\\ (99,87\%)\end{tabular} & 
	\begin{tabular}[c]{@{}c@{}}55'965\\ (99.92\%9\end{tabular} & 
	\begin{tabular}[c]{@{}c@{}}15'078\\ (99.99\%)\end{tabular} & 
	\cellcolor[HTML]{EFEFEF}\textit{\textbf{\begin{tabular}[c]{@{}c@{}}897'296\\ 
	(98.9\%)\end{tabular}}} \\ \hline
	\rowcolor[HTML]{EFEFEF} 
	\textit{\textbf{Total}} & \textit{38'096} & \textit{1'898} & \textit{151} & 
	\textit{323'795} & \textit{142'848} & \textit{115'461} & \textit{134'931} & 
	\textit{72'912} & \textit{56'010} & \textit{15'079} & \textit{\textbf{901'181}} \\ \hline
\end{tabular}
}
	\caption{VirusTotal scanning result of Android apps.}
	\label{tab:VirusTotalResults}
\end{table}

The numbers in Table \ref{tab:VirusTotalResults} show how many app samples we 
have detect as malicious overall. We found with VirusTotal in all Android versions 
except Android 3 malicious apps. Analyzing every malware sample in our corpus 
is outside of this project's scope, but we will give some details on these findings. 
Table \ref{tab:VirustTotal_Malware} shows some more information on the malicious 
samples we detected for Android 8, 9, and 10. We can use the results of other 
scanners like APKiD and AndroGuard to analyze 
these samples or conduct a manual analysis.

\begin{table}[H]
		\resizebox{\textwidth}{!}{
\begin{tabular}{|l|l|l|c|c|}
	\hline
	\rowcolor[HTML]{EFEFEF} 
	\textbf{Filename} & \textbf{Packagename} & \textbf{MD5*} & \textbf{A. 
	Version} & \textbf{Compiler (APKiD)} \\ \hline
	SysStasl.apk or Sys\_stasl.apk & com.transsion.statisticalsales & 
	a24f48fcd9f390800640a9d34b369c6c & 8 & dx, Jack 4.x \\ \hline
	Showcase.apk & com.customermobile.preload.vzw & 
	5a8913dc5df6a14b2673286b04b12539 & 8, 9, 10 & dx, Jack 4.x \\ \hline
	FuriousThiefNew.apk & com.m2fpremium.furiousthief & 
	0bd71ad1afe8df86e289f0c89609898f & 8 & dexlib 2.x \\ \hline
	SpaceMissionGN.apk & com.gameneeti.game.MissionSpaceGN1Pro & 
	419d00d77fcc50e03e2b95ff8d56def0 & 8 & dexlib 2.x \\ \hline
	SCweather\_V4.apk & com.smart.weather & 
	5f6b1cdcb4ea5ca0e973a081a4b0c6c4 & 8 & dexlib 2.x, dx \\ \hline
	com\_cooee\_phenix\_s8.apk & com.cooee.phenix & 
	bd44a85905c7ee2264a08b26ef1c7527 & 8 & dx \\ \hline
	ZDMEUSTOCKPLUS.apk & com.zte.zdm & 
	be81670b117cad27d04de61cb3aed17d & 8 & dx \\ \hline
	PhoneServer.apk & com.google.speed.anisotropic.multistage & 
	3cb9b635fd5b2b3d134bbe55194b4fa0 & 8 & dx \\ \hline
	ID3os.apk & com.aptoide.partners.id3storen & 
	2fd47f2a914da16fa7f2b933a15e7808 & 8, 9 & dx \\ \hline
	DualaidVend.apk & com.excelliance.dualaid.vend & 
	af7a90c7403555b04f53193ae25f1c0d & 8 & dx \\ \hline
	GlobalSearch.apk & com.vivo.globalsearch & 
	fab5f84300876e0920dbf96b0c86aa32 & 8 & dx \\ \hline
	BBKTheme.apk & com.bbk.theme & 3010ce4b192d57fc26e2c32a3374d423 & 
	8, 9 & dx \\ \hline
	iRoaming.apk & com.mobile.iroaming & 5d97ad886275c28534b9ea253be4354e 
	& 8, 9 & dx, Jack 4.x \\ \hline
	ready2Go\_ATT.apk & com.synchronoss.dcs.att.r2g & 
	a10bf6bc3c0c7ca4e37395666a22ec29 & 8 & dx \\ \hline
	SaleTrackerInDemo.apk & com.android.sales & 
	034ff8ce04cb7ab38fe455a4fe1682c3 & 8 & dx, Jack 4.x \\ \hline
	AliensMarsFight\_SDK27\_signed.apk & 
	com.gameneeti.game.aliensmarsfightpro & 
	a652a9e1fb949341ae247e1c985a85c2 & 8 & dexlib 2.x \\ \hline
	MiguVideo.apk & com.cmcc.cmvideo & 2dee8d84291398f354f92629b759fca7 & 
	8 & dx \\ \hline
	MobileMarket.apk & com.aspire.mm & ed73284f6ec22486453fa92987e61b5a & 
	8 & dx \\ \hline
	MobileMusic.apk & cmccwm.mobilemusic & 
	a7a4a4720537395b45d5e14d2c777a1c & 8 & dx \\ \hline
	CricketChampsLeague.apk & com.digi10soft.game.letsplay2020pro & 
	2d1c3e78fed97ef51adfedc54b797017 & 8 & dexlib 2.x \\ \hline
	SamsungExperienceService.apk & com.samsung.android.mobileservice & 
	274a15289edca85044eadd443c362e6c & 8 & dx \\ \hline
	Sinamicroblog.apk & com.sina.weibo & 8a9d1b85e613306e7830ddad92afd69a 
	& 8 & dx, dexmerge \\ \hline
	AppSelect\_ATT.apk & com.dti.att & e04e0fae2f9ef13599ddd4c978b95cd0 & 9 
	& dx \\ \hline
	canid\_stub.apk & com.vzw.ecid & 3f0e107d7b8dbc583fea1352f5e7b7d7 & 9 & 
	dx \\ \hline
	TouchPal.apk & com.emoji.keyboard.touchpal.vivo & 
	a812de7cdd6692380c81a0348ba312a4 & 8 & dx, dexmerge, dexlib 2.x \\ \hline
\end{tabular}
}
	\caption{Detected malware for Android 8, 9, and 10 with APKiD compiler result. 
	*Note: We provide only one MD5 hash to save space.}
	\label{tab:VirustTotal_Malware}
\end{table}

As we can see from Table \ref{tab:VirustTotal_Malware} seven samples use the 
dexlib 2.x compiler. We can assume that attackers repackaged parts of these 
apps. We check whether these malware samples belong to a custom ROM or a 
stock ROM by looking at the build.prop files of the individual firmware where we 
found the malware. As mentioned in Section \ref{Analysis:BuildPropAnalysis} the 
ro.build.tags property contains a string that shows if the developer used 
\textit{release-keys} or other keys for building. In case we find a firmware that 
uses a 
\textit{test-} or \textit{dev-keys} string, we can assume it is a custom ROM or a 
wrongly signed 
stock ROM. Table \ref{tab:MalwareFirmwareBrands} shows the malware samples 
with additional information about the belonging firmware. All firmware with 
a malicious app use the "release-keys" string in the ro.build.tags key. We 
conclude 
that we cannot know if the belonging firmware is a custom or stock ROM because 
using "release-keys" gives us no further information. To our knowledge, the only 
way to verify if these firmware samples are stock ROMs would be to verify their 
system partition signature. However, we cannot verify the signature by ourselves 
since no PKI is in place, and therefore we would need to ask the official OS 
vendor to verify their signatures. In case the vendor would verify that the signature 
is correct, it would mean that the OS vendor has added a packed or tampered 
malware to its firmware. 

\begin{table}[H]
			\resizebox{\textwidth}{!}{
	\begin{tabular}{|l|l|c|c|}
		\hline
		\rowcolor[HTML]{EFEFEF} 
		\textbf{Packagename} & \textbf{MD5} & 
		\textbf{\begin{tabular}[c]{@{}c@{}}Firmware\\ Brand\\ 
		(ro.build.brand)\end{tabular}} & \textbf{\begin{tabular}[c]{@{}c@{}}Firmware \\ 
		build key\\ (ro.build.tags)\end{tabular}} \\ \hline
		com.transsion.statisticalsales & a24f48fcd9f390800640a9d34b369c6c & 
		Tecno & release-keys \\ \hline
		com.customermobile.preload.vzw & 5a8913dc5df6a14b2673286b04b12539 & 
		Google & release-keys \\ \hline
		com.m2fpremium.furiousthief & 0bd71ad1afe8df86e289f0c89609898f & 
		Micromax & release-keys \\ \hline
		com.gameneeti.game.MissionSpaceGN1Pro & 
		419d00d77fcc50e03e2b95ff8d56def0 & Micromax & release-keys \\ \hline
		com.smart.weather & 5f6b1cdcb4ea5ca0e973a081a4b0c6c4 & Imkj & 
		release-keys \\ \hline
		com.cooee.phenix & bd44a85905c7ee2264a08b26ef1c7527 & Imkj & 
		release-keys \\ \hline
		com.zte.zdm & be81670b117cad27d04de61cb3aed17d & ZTE & 
		release-keys \\ \hline
		com.google.speed.anisotropic.multistage & 
		3cb9b635fd5b2b3d134bbe55194b4fa0 & Allview & release-keys \\ \hline
		com.aptoide.partners.id3storen & 2fd47f2a914da16fa7f2b933a15e7808 & 
		Nokia & release-keys \\ \hline
		com.excelliance.dualaid.vend & af7a90c7403555b04f53193ae25f1c0d & 
		Infinix & release-keys \\ \hline
		com.vivo.globalsearch & fab5f84300876e0920dbf96b0c86aa32 & Vivo & 
		release-keys \\ \hline
		com.bbk.theme & 3010ce4b192d57fc26e2c32a3374d423 & Vivo & 
		release-keys \\ \hline
		com.mobile.iroaming & 5d97ad886275c28534b9ea253be4354e & Vivo & 
		release-keys \\ \hline
		com.synchronoss.dcs.att.r2g & a10bf6bc3c0c7ca4e37395666a22ec29 & 
		Samsung & release-keys \\ \hline
		com.android.sales & 034ff8ce04cb7ab38fe455a4fe1682c3 & Lava & 
		release-keys \\ \hline
		com.gameneeti.game.aliensmarsfightpro & 
		a652a9e1fb949341ae247e1c985a85c2 & Symphony & release-keys \\ \hline
		com.cmcc.cmvideo & 2dee8d84291398f354f92629b759fca7 & Samsung & 
		release-keys \\ \hline
		com.aspire.mm & ed73284f6ec22486453fa92987e61b5a & Samsung & 
		release-keys \\ \hline
		cmccwm.mobilemusic & a7a4a4720537395b45d5e14d2c777a1c & Samsung 
		& release-keys \\ \hline
		com.digi10soft.game.letsplay2020pro & 
		2d1c3e78fed97ef51adfedc54b797017 & Micromax & release-keys \\ \hline
		com.samsung.android.mobileservice & 
		274a15289edca85044eadd443c362e6c & Samsung & release-keys \\ \hline
		com.sina.weibo & 8a9d1b85e613306e7830ddad92afd69a & Samsung & 
		release-keys \\ \hline
		com.dti.att & e04e0fae2f9ef13599ddd4c978b95cd0 & Samsung & 
		release-keys \\ \hline
		com.vzw.ecid & 3f0e107d7b8dbc583fea1352f5e7b7d7 & Samsung & 
		release-keys \\ \hline
		com.emoji.keyboard.touchpal.vivo & a812de7cdd6692380c81a0348ba312a4 
		& Vivo & release-keys \\ \hline
	\end{tabular}
}
	\caption{Malware samples with corresponding firmware brand and build key.}
	\label{tab:MalwareFirmwareBrands}
\end{table}

The examples shown in Table \ref{tab:MalwareFirmwareBrands} demonstrate that 
we need better methods to identify if a ROM is an official stock ROM or a custom 
ROM. In any case, using VirusTotal, we have shown that our corpus contains 
several malicious pre-installed apps. In the next section, we will discuss how we 
can use TLSh fuzzy hashing for similarity detection of other malware samples. 

\section{TLSH Fuzzy Hashing} \label{Analysis:FuzzyHashing}

To further explore our dataset, we included TLSH and ssdeep fuzzy hashing 
algorithms into FirmwareDroid. From the literature, we know that researchers have 
used these two algorithms to score the files' similarity. In Section 
\ref{Introduction:FuzzyHashingAndSimiliarityDetection} we mentioned that one of 
the challenges with using fuzzy hashing for malware analysis is how to compare 
the hashes to each other efficiently. One approach is to compare all files to each 
other and then create a matrix of similarity scores. The problem with this approach 
is that the processing time does not scale well or, in other words, is exponential. A 
recent publication \cite{ali2020scalable} overcomes this problem by using a 
multi-stage technique to paralleling the clustering with HAC-T. Their results show 
that it is possible to scale the clustering of TLSH hashes on the cloud. 
Reproducing or integrating their solution into FirmwareDroid is due to the time 
limitation of our project not feasible. It is not the main goal of this thesis to 
discuss TLSH clustering in depth. However, we attempt to implement a custom 
clustering algorithm that could be used on our dataset to find similar binaries for 
malware analysis. 

\begin{table}[H]
	\begin{tabular}{|l|l|}
		\hline
		\rowcolor[HTML]{EFEFEF} 
		\multicolumn{1}{|c|}{\cellcolor[HTML]{EFEFEF}\textbf{\#}} & \textbf{TLSH 
			digest} \\ \hline
		T1 & 
		474110F8EBB3A973188A4383047F4785E73B613CC1E1861668D664C4F213A688379B7C
		\\ \hline
		T2 & 
		CC418AF8E7769873084A4287487F4385E7376A38C2F1961768C615C4F213A284379F3D
		\\ \hline
		T3 & 
		EE418CF8E773993318894296047F8785F73B653CC1E1962664C624D8F213A684379F7D
		\\ \hline
	\end{tabular}
	\caption{TLSH digest examples.}
	\label{tab:tlsh_digest_examples}
\end{table}

\begin{figure}[H]
	\centering
	\includegraphics[width=\linewidth]{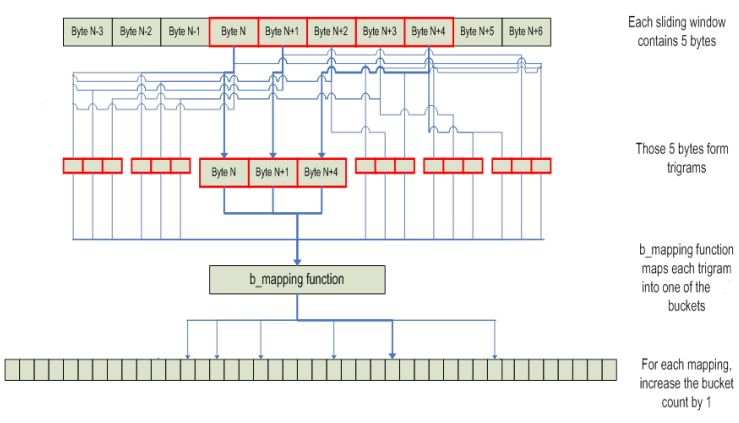}
	\caption{TLSH digest generation process. Image source \cite{DFRWS}}
	\label{fig:17_TLSHMapping.png}
\end{figure}

Interested readers can find the implementation details on TLSH in \cite{TLSH}, and 
we will only explain the parts of the paper necessary to understand our algorithm. 
Table \ref{tab:tlsh_digest_examples} shows some example TLSH digests. The first 
three bytes of every TLSH digest are the header. According to \cite{TLSH} the 
header contains a checksum, the byte length, and the quartile ratios. The rest of 
the TLSH digest is the body containing the buckets counts for similarity. TLSH 
uses a sliding window with a fixed size to create trigrams as illustrated in Figure 
\ref{fig:17_TLSHMapping.png} to create a TSLH digest. TLSH then maps these 
trigrams or triplets to the individual buckets with a mapping function. Jonathan 
Oliver et al. use the Pearson hash \cite{TLSH} as bucket mapping function for 
TLSH. This schema allows TLSH to generate digests with a high probability to 
map similar bytes to the same bucket. We compare two TLSH digests similarities 
by approximating the hamming distance for the header and the body. The resulting 
score measures the distance between two TLSH hash. A score of zero is 
considered an exact match, and the higher the distance between two digests is, 
the more likely it is that they have less in common.

For example, in Table \ref{tab:tlsh_digest_examples} the digests T2 and T3 have 
the same hexadecimal bytes 'F8E77' and 'F213A'. We realize that this means that 
some bytes of both files have a high probability of being similar. Consequently, we 
can say that two TLSH digests with the same hexadecimal digits have a higher 
probability to share similarity than TLSH digest that do not share any hexadecimal 
digits. Moreover, from \cite{TLSH} we know that the fewer differences two TLSH 
digests have, the smaller is the distance between them. Using this information, we 
can implement an algorithm that filters the potential candidates' space by 
comparing 
TLSH digests that share some common bytes. We use the fact that TLSH digests 
with similarity share similar hexadecimal digits in their digest to overcome the 
problem of computing all possible distances in a set of files. 

\begin{figure}[H]
	\centering
	\includegraphics[width=\linewidth]{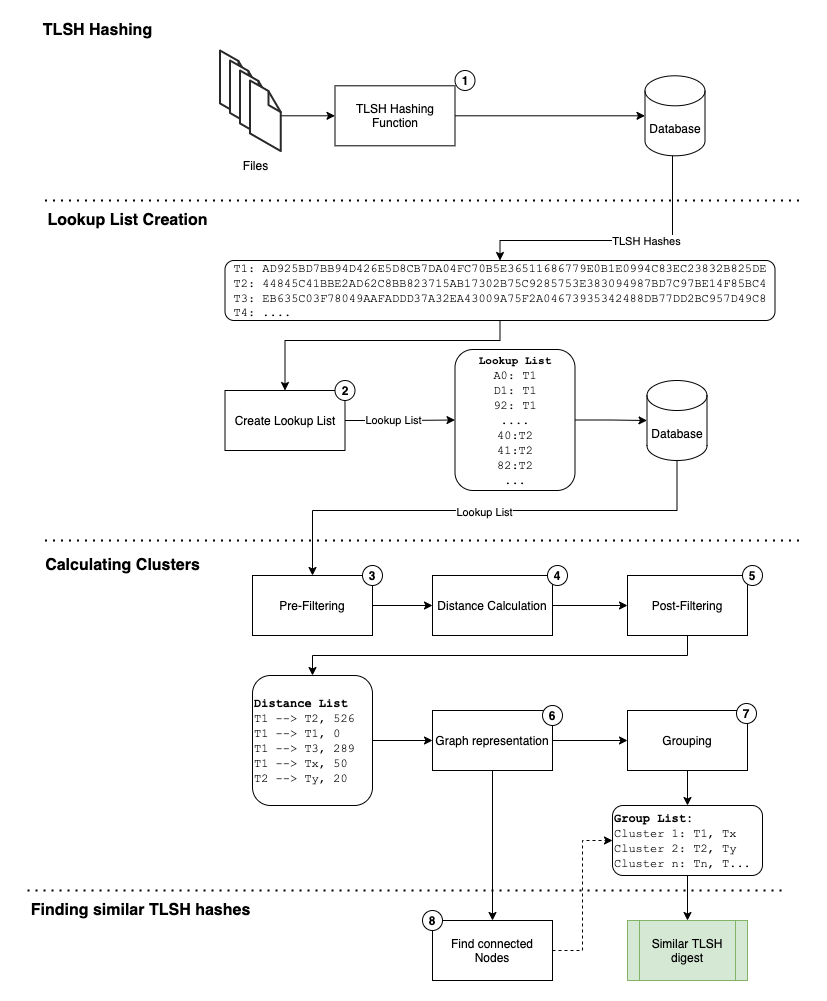}
	\caption{TLSH hashing process in FirmwareDroid.}
	\label{fig:04_tlsh_hash_overview.png}
\end{figure}

In the next subsections, we will go through the complete process from creating 
TLSH hash to filtering them for comparison that we have integrated into 
FirmwareDroid. Figure \ref{fig:04_tlsh_hash_overview.png} illustrate this process.

\textbf{\circled{1} TLSH hashing:} The first step is to hash all files with the TLSH 
hashing function and save the resulting digest in our database. We hashed around 
5.5 million files and stored the resulting digests in our database. To further analyze 
apk files, we create a TLSH digest for the apk and additional TLSH digests for 
containing resources and library files. To do so, we parse the apk file with LIEF 
\cite{LIEF} and hash all containing files. As a result, it allows us to compare 
similar libraries or resources as well, if necessary.

\textbf{\circled{2} Lookup list creation:} As the next step, we create for all TLSH 
hashes in our database a lookup table. This table will later allow us to filter TLSH 
digests that have a small probability of being similar. To create the lookup table, 
we use the algorithm shown in Listing \ref{lst:TLSHLookupTableGen}. We iterate 
once through every TLSH digest in our database to create the lookup table. For 
every digest, we iterate through its hexadecimal digits to store a combination of 
the digit and its index. For example, the T1 hash from Table 
\ref{tab:tlsh_digest_examples} would then generate the following keys 
for the first seven hexadecimal digits: '40', '71', '42', '13', '14', '05', 'F6'. We use all 
these 
keys to fill a python dictionary. We then store the database reference id of the 
complete digest in a python set for each key we generate. It is also possible to 
store the TLSH digest itself instead of a reference for implementations without a 
database. We use python sets to ensure that we do not store duplicate references 
under 
the same key. The result of this algorithm is a dictionary containing sets with 
TLSH digest references. Every set contains the database references for TLSH 
digests with the same digit and index. Consequently, this means if two or more 
TLSH digests generate the same key, their database reference will end up in the 
same set, and we can later counter the number of entries per set to filter the TLSH 
hash.

\begin{lstlisting}[caption={Python code to generate the TLSH lookup table.}, 
label=lst:TLSHLookupTableGen, language=python, basicstyle=\scriptsize] 
TABLE_LENGTH = 70 		# Example Value
BAND_WIDTH = 1			  # Example Value
global_dict = {}


tlsh_hash = tlsh_hash_queue.get() 	# From database
for column_index in range(0, TABLE_LENGTH, BAND_WIDTH):
	row_char = tlsh_hash.tlsh_digest[column_index:column_index + BAND_WIDTH]
	key_label = f"{row_char}{column_index}"
	if key_label not in global_dict:
		global_dict[key_label] = set()
	global_dict[key_label].add(str(tlsh_hash.id))
\end{lstlisting}

\textbf{\circled{3} Pre-Filtering:} After the computation of the lookup table, we can 
start with the clustering process. The goal is to identify similar files without 
computing all the possible combinations for the distance calculation. Therefore, we 
developed a function to find potential candidates as shown in Listing 
\ref{lst:TLSHCandiateMethod}. The function takes as input a TLSH hash and 
returns a list of potential candidate hashes for comparison. To create the 
candidate list, we iterate once through the TLSH digests hexadecimal digits. We 
get for every combination of hexadecimal digit and index the corresponding set of 
TLSH digests from the pre-computed lookup table and merge them to one list. 
This list named "potential\_hashes\_list" in Listing \ref{lst:TLSHCandiateMethod} 
then contains every hash that has at least one hexadecimal digit in common with 
our input TLSH digest. Theoretically, we could remove duplicates from this list, 
and then we would have all candidate hashes that are possible. However, in 
practice, the number of comparisons would still be relatively large and, in some 
cases, infeasible to compute in a rational amount of time. We, therefore, count on 
Line 24 in Listing \ref{lst:TLSHCandiateMethod} the frequency of every TLSH 
database reference to apply a filter. The filter, shown on Line 26 with variable 
name "band\_width\_threshold" defines the minimum number of occurrences a 
candidate hash needs that we consider the hash for comparison. If we set 
the"band\_width\_threshold" variable to zero, we would consider every hash as a 
candidate. Our idea is that when we set an optimal number for the 
"band\_width\_threshold" variable that we filter all the TLSH hashes with a high 
distance and do not need to compute their distance. We will discuss in Section 
\ref{Analysis:TlshEvaluation} how to set an optimal value for the 
"band\_width\_threshold" variable and what limitations this approach has.

\newpage

\begin{lstlisting}[caption={Python code to find candidate hashes.}, 
label=lst:TLSHCandiateMethod, language=python, basicstyle=\scriptsize] 
def get_tlsh_similar_list(tlsh_hash, tlsh_hash_id_list, similiarity_lookup_dict, 
table_length, band_width, band_width_threshold):
   """
	Gets a list of potential candidate hashes by using the number of intersections in 
	the lookup table.
	:param tlsh_hash: class:'TlshHash' - hash to find candidates for.
	:param tlsh_hash_id_list: list(str) - TLSH id's list
	:param similiarity_lookup_dict: dict(str, str) - dict to store candidates.
	:param table_length: int - the length of the TLSH hash.
	:param band_width: int - the width of the lookup table column.
	:param band_width_threshold: int - the minimal number to be considered as 
	potential similar hash.
	:return: list(str) - list of class:'TlshHash' object-ids.
	"""

	similar_hash_id_list = set()
	potential_hashes_list = []
	for column_index in range(0, table_length, band_width):
		row_char = tlsh_hash.tlsh_digest[column_index:column_index + band_width]
		key_label = f"{row_char}{column_index}"
		candidate_list = similiarity_lookup_dict[key_label]
		potential_hashes_list.extend(candidate_list)
	
	counter = Counter(potential_hashes_list)
	for tlsh_candidate_id in potential_hashes_list:
		if counter[tlsh_candidate_id] >= band_width_threshold and tlsh_candidate_id 
			in tlsh_hash_id_list:
				similar_hash_id_list.add(tlsh_candidate_id)
	return similar_hash_id_list
\end{lstlisting}

\textbf{\circled{4} Distance calculation:} In step four, we calculate all the 
candidates' distances with the official TLSH implementation \cite{TLSH}. We then 
apply a variable post-filter (see \circled{5}) to remove all comparison results with a 
variable distance threshold. The post-filtering step is optional, and we use it if we 
want to search for TLSH digest within a specific distance range. As a result of the 
distance calculation, we get a table or matrix with the distance between two 
digests.

\textbf{\circled{6} Graph representation:} As illustrated in Figure 
\ref{fig:04_tlsh_hash_overview.png} 
we store for every digest the distance and TLSH reference. We then visualize this 
data to create undirected weighted graphs. Every edge in one graph represents the 
comparison between two TLSH digests, and the weight is the TLSH distance 
value. In our graph representation every TLSH digest represents one node. A 
connected graph represents all the similar hashes, and we create for every graph a 
group or cluster (see \circled{7}) that we can use for searching. 

\textbf{\circled{8} Search:} If we want to find all files that are similar to one TLSH 
digest, we can query the graph for the TLSH hash and return all nodes connected 
to the hash. Alternatively, we can represent the groups in a hash map for efficient 
query times. Moreover, we can sort the resulting nodes by their weights to get the 
hashes with the smallest distances.

\subsection{Performance of TLSH filtering} \label{Analysis:TlshEvaluation}
\label{Analysis:EvaluationofTLSHClustering}
There are a couple of papers evaluating the performance of fuzzy hashing 
algorithms. Vassil Roussev showed in \cite{EvaluationSimilarityHashes} 
that sdhash outperforms ssdeep in terms of precision. Jonathan Oliver et al. 
describe in their work \cite{TLSH} that under some circumstances, TLSH is 
comparable to ssdeep and sdhash and outperforms them as well. However, in our 
evaluation, we are not comparing the performance of these individual algorithms. 
Instead, we evaluated the performance of our TLSH filtering algorithm. We 
measure the hit-rate to determine how many elements we clustered correctly and 
the missed-rate to determine how many elements we did not consider candidates. 
We test our approach on 5'000 and 10'000 TLSH hashes and compute for all sets 
the ground truth with all distance measurements. We compare our solutions to the 
ground truth and see how it performs. We measure the following values and Tables 
\ref{tab:TLSH_eval_5000} and \ref{tab:TLSH_eval_10000} 
shows the results of our performance measurements.

\begin{itemize}
	\item \textit{Band width threshold:} A variable that defines the minimum 
	threshold to be considered a candidate hash for comparison.
	
	\item T\textit{able creation time:} The time measurement in seconds to create 
	the lookup table and store it in our database.
	
	\item \textit{Cluster creation time: }The time measurement in seconds to create 
	the clusters and store them in the database.
	
	\item \textit{Total time:} The sum of the table creation time and the cluster 
	creation 
	time.
	
	\item \textit{Cluster size average:} The average number of elements in a cluster.
	
	\item \textit{Number of clusters:} The count of clusters.
	
	\item \textit{Cluster count difference:} The difference between the number of 
	clusters of our solution and the number of clusters in the ground truth.
	
	\item \textit{Number of comparisons:} The total amount of TLSH distance 
	comparisons we computed.
	
	\item \textit{Decrease rate:} The difference in percent of the number of 
	comparisons an the ground truth.
	
	\item \textit{Comparison difference:} The difference of the number of 
	comparisons and the ground truth.
	
	\item \textit{Missed rate:} The number of elements that we did not consider as 
	candidates even when valid or in other words the false positive rate.
	
	\item \textit{Hit rate:} The number of elements that we considered correctly as 
	candidates or in other words the true positive rate.
\end{itemize}

\begin{table}[H]
	\centering
	\resizebox{\textwidth}{!}{%
	\begin{tabular}{|c|l|l|l|l|l|l|l|l|l|l|}
		\hline
		\rowcolor[HTML]{EFEFEF} 
		\multicolumn{1}{|l|}{\cellcolor[HTML]{EFEFEF}\textbf{\begin{tabular}[c]{@{}l@{}}Band
		 \\ Width \\ Threshold\end{tabular}}} & \textbf{\begin{tabular}[c]{@{}l@{}}Table 
		\\ creation \\ time (s)\end{tabular}} & \textbf{\begin{tabular}[c]{@{}l@{}}Cluster 
		\\ creation \\ time (s)\end{tabular}} & \textbf{\begin{tabular}[c]{@{}l@{}}Total \\ 
		time (s)\end{tabular}} & \textbf{\begin{tabular}[c]{@{}l@{}}Group \\ size\\ 
		average\end{tabular}} & \textbf{\begin{tabular}[c]{@{}l@{}}Number \\ of\\ 
		clusters\end{tabular}} & \textbf{\begin{tabular}[c]{@{}l@{}}Group \\ 
		difference\end{tabular}} & \textbf{\begin{tabular}[c]{@{}l@{}}Number \\ of \\ 
		comparisons \\ (decrease rate)\end{tabular}} & 
		\textbf{\begin{tabular}[c]{@{}l@{}}Comparisons \\ difference\end{tabular}} & 
		\textbf{\begin{tabular}[c]{@{}l@{}}Missed \\ rate\end{tabular}} & 
		\textbf{\begin{tabular}[c]{@{}l@{}}Hit \\ rate\end{tabular}} \\ \hline
		\cellcolor[HTML]{EFEFEF}\textbf{5} & 502 & 5'724 & 6'226 & 9.4052 & 427 & 
		0 & 15'967'554 (36.13\%\} & 9'032'446 & 0 (0\%) & 4'016 \\ \hline
		\cellcolor[HTML]{EFEFEF}\textbf{10} & 481 & 1'951 & 2'432 & 9.3832 & 428 
		& 1 & 3'004'622 (87.98\%) & 21'995'378 & 0 (0\%) & 4'016 \\ \hline
		\cellcolor[HTML]{EFEFEF}\textbf{13} & 500 & 1'114 & 1'614 & 9.2656 & 433 
		& 6 & 1'260'828 (94.96\%) & 23'739'172 & 4 (0.1\%) & 4'012 \\ \hline
		\cellcolor[HTML]{EFEFEF}\textbf{15} & 602 & 952 & 1'555 & 9.1050 & 438 & 
		11 & 835'422 (96.66\%) & 24'164'578 & 28 (0.67\%) & 3'988 \\ \hline
		\cellcolor[HTML]{EFEFEF}\textbf{18} & 514 & 870 & 1'384 & 8.3879 & 464 & 
		37 & 522'050 (97.91\%) & 24'477'950 & 124 (3.1\%) & 3'892 \\ \hline
		\cellcolor[HTML]{EFEFEF}\textbf{20} & 824 & 812 & 1'636 & 7.6343 & 495 & 
		68 & 396'178 (98.42\%) & 24'603'822 & 237 (5.9\%) & 3'779 \\ \hline
		\cellcolor[HTML]{EFEFEF}\textbf{30} & 722 & 351 & 1'073 & 4.5402 & 659 & 
		232 & 79'996 (99.68\%) & 24'920'004 & 1'024 (25,5\%) & 2'992 \\ \hline
		\cellcolor[HTML]{EFEFEF}\textbf{40} & 671 & 349 & 1'020 & 3.1567 & 804 & 
		377 & 55'940 (99.78\%) & 24'944'060 & 1'478 (36,8\%) & 2'538 \\ \hline
		\cellcolor[HTML]{EFEFEF}\textbf{50} & 692 & 193 & 886 & 2.9413 & 733 & 
		306 & 53'126 (99.79\%) & 24'946'874 & 1'860 (46,3\%) & 2'156 \\ \hline
	\end{tabular}%
	}
	\caption{Performance measurement for 5000 TLSH hashes with variable band 
	width thresholds.}
	\label{tab:TLSH_eval_5000}
\end{table}

\begin{table}[H]
	\centering
	\resizebox{\textwidth}{!}{%
	\begin{tabular}{|
			>{\columncolor[HTML]{EFEFEF}}c |l|l|l|l|l|l|l|l|l|l|}
		\hline
		\multicolumn{1}{|l|}{\cellcolor[HTML]{EFEFEF}\textbf{\begin{tabular}[c]{@{}l@{}}Band
		 \\ Width \\ Threshold\end{tabular}}} & 
		\cellcolor[HTML]{EFEFEF}\textbf{\begin{tabular}[c]{@{}l@{}}Table \\ creation 
		\\ time (s)\end{tabular}} & 
		\cellcolor[HTML]{EFEFEF}\textbf{\begin{tabular}[c]{@{}l@{}}Cluster \\ 
		creation \\ time (s)\end{tabular}} & 
		\cellcolor[HTML]{EFEFEF}\textbf{\begin{tabular}[c]{@{}l@{}}Total \\ time 
		(s)\end{tabular}} & 
		\cellcolor[HTML]{EFEFEF}\textbf{\begin{tabular}[c]{@{}l@{}}Group \\ size\\ 
		average\end{tabular}} & 
		\cellcolor[HTML]{EFEFEF}\textbf{\begin{tabular}[c]{@{}l@{}}Number \\ of\\ 
		clusters\end{tabular}} & 
		\cellcolor[HTML]{EFEFEF}\textbf{\begin{tabular}[c]{@{}l@{}}Group \\ 
		difference\end{tabular}} & 
		\cellcolor[HTML]{EFEFEF}\textbf{\begin{tabular}[c]{@{}l@{}}Number \\ of \\ 
		comparisons\\ (decrease rate)\end{tabular}} & 
		\cellcolor[HTML]{EFEFEF}\textbf{\begin{tabular}[c]{@{}l@{}}Comparisons \\ 
		difference\end{tabular}} & 
		\cellcolor[HTML]{EFEFEF}\textbf{\begin{tabular}[c]{@{}l@{}}Missed \\ 
		rate\end{tabular}} & 
		\cellcolor[HTML]{EFEFEF}\textbf{\begin{tabular}[c]{@{}l@{}}Hit \\ 
		rate\end{tabular}} \\ \hline
		\textbf{1} & 3'475 & 58'162 & 61'637 & 7.3913 & 1'076 & 0 & 99'436'120 
		(0.56\%) & 563'880 & 0 & 7'953 \\ \hline
		\textbf{5} & 3'304 & 46'295 & 49'599 & 7.3913 & 1'076 & 0 & 61'719'110 
		(38.28\%) & 38'280'890 & 0 & 7'953 \\ \hline
		\textbf{10} & 3'298 & 12'219 & 15'517 & 7.3903 & 1'076 & 0 & 8'598'052 
		(91.4\%) & 91'401'948 & 1 (0.01\%) & 7'952 \\ \hline
		\textbf{11} & 3'278 & 9'290 & 12'568 & 7.3935 & 1'075 & 1 & 5'606'898 
		(94.39\%) & 94'393'102 & 5 (0.06\%) & 7'948 \\ \hline
		\textbf{12} & 3'300 & 7'401 & 10'701 & 7.3907 & 1'075 & 1 & 3'848'412 
		(96.15\%) & 96'151'588 & 8 (0.1\%) & 7'945 \\ \hline
		\textbf{13} & 3'275 & 6'207 & 9'483 & 7.3736 & 1'076 & 0 & 2'829'730 
		(97.17\%) & 97'170'270 & 19 (0.24\%) & 7'934 \\ \hline
		\textbf{15} & 3'283 & 4'949 & 8'232 & 7.2631 & 1'087 & 11 & 1'863'104 
		(98.14\%) & 98'136'896 & 58 (0.73\%) & 7'895 \\ \hline
		\textbf{18} & 3'266 & 3'893 & 7'159 & 6.8627 & 1'129 & 53 & 1'220'616 
		(98.78\%) & 98'779'384 & 205 (2.58\%) & 7'748 \\ \hline
		\textbf{20} & 3'291 & 3'281 & 6'572 & 6.4855 & 1'170 & 94 & 924'320 
		(99.08\%) & 99'075'680 & 365 (4.59\%) & 7'588 \\ \hline
		\textbf{30} & 3'255 & 1'194 & 4'448 & 4.5269 & 1'395 & 319 & 142'016 
		(99.86\%) & 99'857'984 & 1'638 (20.6\%) & 6'315 \\ \hline
		\textbf{40} & 3'295 & 933 & 4'228 & 3.8449 & 1'354 & 278 & 83'332 (99.92\%) 
		& 99'916'668 & 2'747 (34.54\%) & 5'206 \\ \hline
		\textbf{50} & 3'302 & 799 & 4'102 & 3.7365 & 1'237 & 161 & 64'442 (99.94\%) 
		& 99'935'558 & 3'331 (41.88\%) & 4'622 \\ \hline
	\end{tabular}%
	}
	\caption{Performance measurement for 10'000 TLSH hashes with variable band 
	width thresholds.}
	\label{tab:TLSH_eval_10000}
\end{table}

In the beginning, we tested our evaluation only on 1'000 files to find fitting band 
width thresholds to test. The numbers showed that an optimal value for the band 
width threshold is likely to lay between 10 and 20; we have increased the number 
of data-points for the range 10 to 20 to find the optimal value for the band width 
threshold. Please note that the calculation of all band width thresholds variable is 
not feasible during this project due to time limitations. Note as well that the 
measurements have different amounts of data-points.

\begin{figure}[H]
	\centering
	\includegraphics[width=0.8\linewidth]{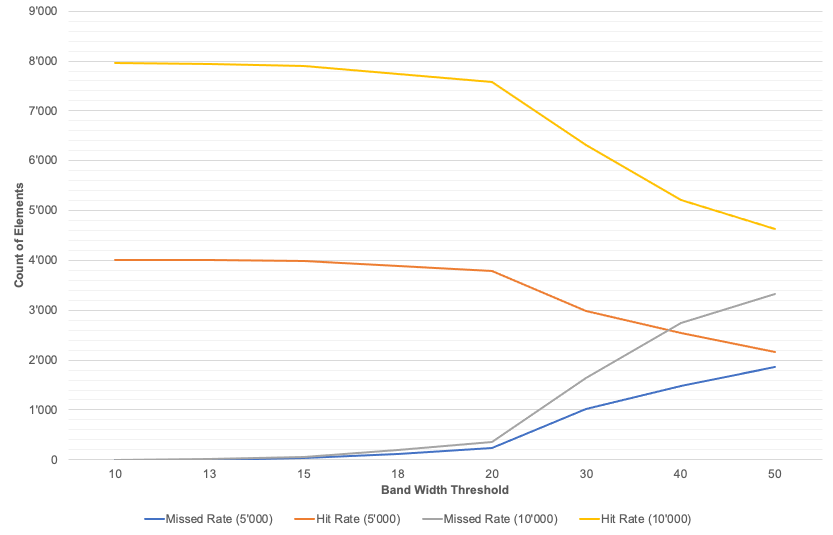}
	\caption{Hit- and missed- rate for 5'000 TLSH hashes with our filtering 
	algorithm.}
	\label{fig:19_TLSH_10000_5000_hit_missed}
\end{figure}

If we look at the missed rate in Table \ref{tab:TLSH_eval_5000} we see that the 
rate increases with selecting a higher band width threshold but has 0\% to 0.1\% 
false positive rate in the range of 0 to 13 for our 5'000 file measurement. We 
illustrated this trend in Figure \ref{fig:19_TLSH_10000_5000_hit_missed}. 
Increasing the band width threshold over 13 increases the missed rate to a point 
where we miss 46,3\% of the elements.

\begin{figure}[H]
	\centering
	\includegraphics[width=0.8\linewidth]{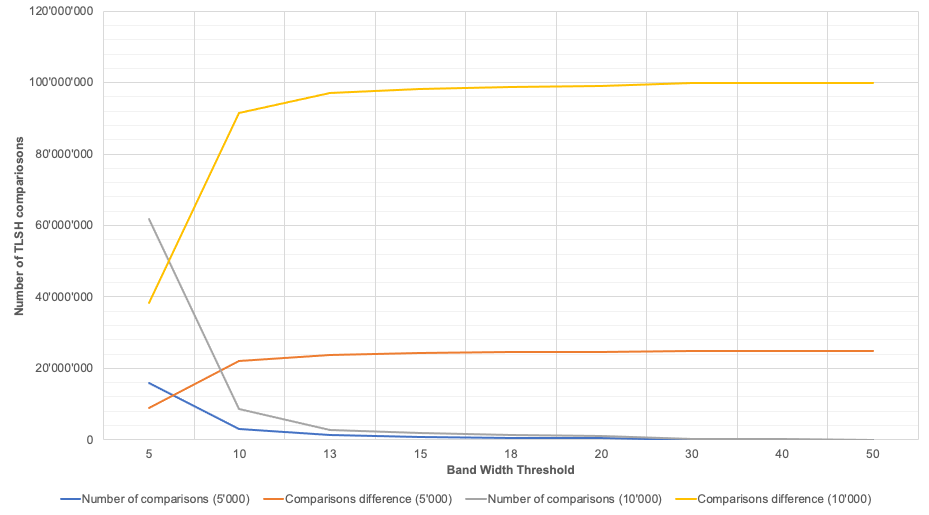}
	\caption{Number of TLSH comparisons }
	\label{fig:18_TLSH_10000_5000_comparisons}
\end{figure}

We illustrated in Figure \ref{fig:18_TLSH_10000_5000_comparisons} the number of 
TLSH comparisons calculated. The curve looks similar to a logarithmic or 
exponential decrease where the curve flattens for higher values. The number of 
comparisons decreases more with smaller band widths thresholds in the range of 1 
to 15. We conclude that this means that increasing the band width threshold to 
some extend does not bring significantly fewer comparisons.

\begin{figure}[H]
	\centering
	\includegraphics[width=0.8\linewidth]{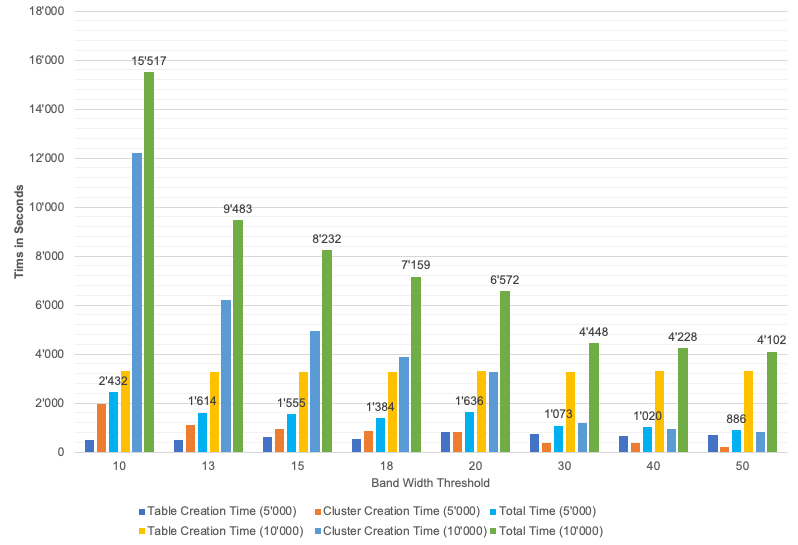}
	\caption{Estimated time necessary to compute the TLSH lookup table and to 
	create clusters for 5000 files.}
	\label{fig:20_TLSH_10000_5000_times}
\end{figure}

We measure our implementation in python to estimate how long the computation 
of the lookup table and clusters would need for a larger dataset. If we look at the 
total time in Table \ref{tab:TLSH_eval_5000}, we see that the computation time 
decreases the most with a threshold range of 1 to 12. The total computation time 
does not significantly decrease if we select higher band width thresholds. Figure 
\ref{fig:20_TLSH_10000_5000_times} illustrates this trend for 5'000 and 10'000 
files. 

We conclude from these numbers that the optimal value for the band width 
threshold lies in the range of 10 to 15. Selecting an optimal value depends on the 
number we want to optimize. If we want to decrease the number of comparisons, 
we can choose a higher band width threshold, resulting in a higher missed rate. 
We think a band width threshold of 14 to 15 is then reasonable with a missed rate 
between 0.1\% to 0.67\% on a small dataset. If we do not tolerate a missed rate at 
all, we select a band width value smaller than 13 for a small dataset. However, If 
we select 13, we have a false positive rate of 0.1\% and can reduce the number of 
comparisons from 25'000'000 down to 1'260'828, which is a decrease of 94.96\% 
as shown in Table \ref{tab:TLSH_eval_5000}. 

When we look at the missed rate for 10'000 files in Table 
\ref{tab:TLSH_eval_10000}, we can see that the missed rate for a band width 
threshold of 13 has more than doubled with nine-teen missed comparisons. 
Compared to the 5'000 files with 0.1\% missed, we have an increase of miss 
elements by 0.14\% for 10'000 files. To further investigate this effect, we have 
repeated the measurements with 10'000 different files and a band with threshold of 
12 and 13 several times. We show the missed rate of these measurements in 
Table \ref{tab:MissedRate_14000}. We can see that the missed rate for three 
measurements with 10'000 files varies from 0.24\% to 0.28\%.  In one 
measurement, we used 10'000 apk files to see if the file type would have a 
significant effect, but the missed rate does not increase significantly by only using 
apk files. It seems that the missed rate does not increase significantly with 
varying 
sets of 10'000 files. 

\begin{table}[H]
	\centering
	\resizebox{\textwidth}{!}{%
		\begin{tabular}{|c|c|c|c|c|c|c|c|}
			\hline
			\rowcolor[HTML]{EFEFEF} 
			\multicolumn{1}{|l|}{\cellcolor[HTML]{EFEFEF}\textbf{\begin{tabular}[c]{@{}l@{}}Band
			 \\ Width \\ Threshold\end{tabular}}} & 
			\multicolumn{1}{l|}{\cellcolor[HTML]{EFEFEF}\textbf{\begin{tabular}[c]{@{}l@{}}Missed
			 rate -\\ 10'000\\ files - 1\#\end{tabular}}} & 
			\multicolumn{1}{l|}{\cellcolor[HTML]{EFEFEF}\textbf{\begin{tabular}[c]{@{}l@{}}Missed
			 rate -\\ 10'000\\ files - 2\#\end{tabular}}} & 
			\multicolumn{1}{l|}{\cellcolor[HTML]{EFEFEF}\textbf{\begin{tabular}[c]{@{}l@{}}Missed
			 rate - \\ 10'000\\ apks - 3\#\end{tabular}}} & 
			\multicolumn{1}{l|}{\cellcolor[HTML]{EFEFEF}\textbf{\begin{tabular}[c]{@{}l@{}}Missed
			 rate -\\ 11'000\\ files - 4\#\end{tabular}}} & 
			\multicolumn{1}{l|}{\cellcolor[HTML]{EFEFEF}\textbf{\begin{tabular}[c]{@{}l@{}}Missed
			 rate -\\ 12'000\\ files - 5\#\end{tabular}}} & 
			\multicolumn{1}{l|}{\cellcolor[HTML]{EFEFEF}\textbf{\begin{tabular}[c]{@{}l@{}}Missed
			 rate -\\ 13'000\\ files - 6\#\end{tabular}}} & 
			\multicolumn{1}{l|}{\cellcolor[HTML]{EFEFEF}\textbf{\begin{tabular}[c]{@{}l@{}}Missed
			 rate -\\ 14'000\\ files - 7\#\end{tabular}}} \\ \hline
			\cellcolor[HTML]{EFEFEF}\textbf{12} & 8 (0.1\%) & 9 (0.11\%) & 9 
			(0.11\%) & 9 (0.09\%) & 10 (0.1\%) & 9 (0.08\%) & 11 (0.09\%) \\ \hline
			\cellcolor[HTML]{EFEFEF}\textbf{13} & 19 (0.24\%) & 22 (0.28\%) & 21 
			(0.26\%) & 21 (0.22\%) & 18 (0.18\%) & 21 (0.19\%) & 28 (0.24\%) \\ \hline
		\end{tabular}%
	}
	\caption{Measurements for the missed rate with increasing number of files.}
	\label{tab:MissedRate_14000}
\end{table}

Moreover, we have increased the number of files up to 14'000 to see if the missed 
rate is growing significantly with more files. However, as we show in Table 
\ref{tab:MissedRate_14000} the missed rate is not significantly increasing in the 
range of 10'000 to 14'000 files. With our measurement we cannot fully explain the 
significant increase of the missed rate in the range from 5'000 to 10'000 files. 
However, it seems to stay for all measurements under 1\% of missed elements 
and we have a reduction of comparisons over 90\% which makes computation 
feasible even for larger datasets.

Another effect we tested is an increase of the band width itself and if it would 
be sufficient to only use the TLSH header in the lookup table. However, tests have 
shown that increasing or using only the header does not bring satisfying results.

\subsection{Complexity of Filtering} \label{Analysis:TlshComplexity}
We assume for all complexity calculations that we use TLSH hashes with a fixed 
size of 70 hexadecimal digits and notate it with $s$. We define $n$ to be the total 
number of TLSH hashes, $z$ to be the number of potential candidate hashes 
before filter, and $x$ to be the number of candidates used for the distance 
calculation.

\textbf{Time complexity:} To create the lookup table, we have to iterate for every 
TLSH digest once through its hexadecimal digits and add then its reference to a 
python set. Adding a hash to the lookup table results in a linear complexity of 
$\mathcal{O}(n*s)$. We then have to calculate for every TLSH a list of candidate 
hashes as shown in Listing \ref{lst:TLSHCandiateMethod}. We iterate for every 
hash once through its hexadecimal digits and create a list of candidate hashes. 
The iteration has the time complexity of $\mathcal{O}(n*s)$. We count the 
frequency of every element in the list with python's counter class. The counter 
class has a linear time complexity of $\mathcal{O}(z)$ for counting the frequencies 
of the elements, where z is the number of potential candidates before filtering. We 
iterate through the list of potential candidate hashes on 
Line 25 in Listing \ref{lst:TLSHCandiateMethod} and filter hashes with the band 
width threshold. This iteration has a linear time complexity of $\mathcal{O}(z)$ 
where z is the number of potential candidate hashes. All operations described 
have linear time complexity, and we conclude that the candidate list creation has 
linear time complexity. When it comes to determining the time complexity of the 
comparisons, we have to iterate through all candidate hashes to generate the list 
of comparisons, which is a linear operation with the complexity of 
$\mathcal{O}(x)$, where x is the number of candidate hashes. In the worst case, x 
is equal to n-1, and all hashes were candidate hashes. In this case, we would 
have a time complexity of $\mathcal{O}(n^2)$ for the comparisons. However, as 
the numbers in Table \ref{tab:TLSH_eval_5000} show 
and Figure \ref{fig:18_TLSH_10000_5000_comparisons} illustrates, it seems that x 
tends 
to follow a exponential decay. We, therefore, assume that on average, we have a 
time complexity of $\mathcal{O}(log(n))$ for x. As a consequence, this would 
mean that our pre-filter with comparisons have, in average time complexity of 
$\mathcal{O}(n*log(n))$.

\textbf{Storage and inserting complexity:} A fact is that our approach trades 
computation time against disk storage. The lookup table has the size of $s*16*n$ 
elements, where 16 is the total number of possible hexadecimal digits. The table 
creation is parallelizable, and updating is possible without recreating the table. To 
insert an element, we have to iterate through each of the new hash's hexadecimal 
digits and add its references to the corresponding lookup table keys. Using python 
dictionaries searching has a complexity of $\mathcal{O}(1)$ for a specific key in 
the lookup table and adding an element to a set in python has as well 
$\mathcal{O}(1)$ according to \cite{PythonTimeComplexity}. We, therefore, need 
only to consider the complexity of iterating through the TLSH hash itself. Inserting 
a hash to the table has a complexity of $\mathcal{O}(n*s)$, where n is the number 
of TLSH hashes to insert and s the length of the TLSH hash digest.

\textbf{Limitations:} Our approach and performance measurement has some 
limitations that we would like to mention. As the numbers show, we trade 
computation time against storage, and therefore this algorithm is only useful when 
we can scale the storage amount. 

The provided measurements were conducted on several thousand files and not on 
millions of files due to resource limitations in calculating the ground truth for 
comparisons. For example, the ground truth computation for 1'000'000 would result 
in 1'000'000$^2/2$ computations, which are infeasible to compute in our 
environment. We used for the measurements different datasets of files with 
varying file types. As shown, the missed rate varies with the number of files and 
the used dataset. With the current measurements, we cannot eliminate the 
possibility that certain datasets or file-types produce worse results or that scaling 
the number of files has a significant effect.

Another limitation we would like to express is that we restrict the comparison to be 
on the same index due to using the lookup table. Consequently, when we have 
TLSH digests that share the same hexadecimal digits but not the same index, we 
may miss them as potential candidates. Another limitation that we did not mention 
so far is that we have more clusters with increasing missed rate. We think this is 
because we do not calculate some distances between the nodes, and therefore, 
we do not have some connections in the graph. Consequently, if we use our graph 
approach for clustering and the band width threshold tolerates false positives, we 
miss some connections between the clusters. We know that this can lead to 
missing elements when using the graph to search for similar TLSH files, and in the 
worst case, the connection between two large clusters is missing, and the search 
is not accurate.

In conclusion, our approach compared to other studies like \cite{ali2020scalable} 
follows a different approach by pre-filtering the hashes in sub-sets instead of 
clustering the complete dataset with Machine Learning techniques like k-means. 
Our approach has shown that we can effectively reduce the search space with a 
low false-positive rate for small (<= 14'000) datasets. However, our performance 
measurement and detailed analysis of the time complexity need more testing to 
come to conclusions. 

	\chapter{FirmwareDroid Implementation} \label{Implementation}

In this chapter, we describe the implementation details of FirmwareDroid. We will 
begin with giving the reader an overview of the systems architecture in 
Section \ref{Implementation:ArchitectureAndTechnologies}. We will then progress 
in Section \ref{Implementation:FirmwareDroidFeatures} with going through the main 
features of FirmwareDroid and discuss in Section 
\ref{Implementation:ExperimentalFeatures} some experimental features.

\section{Architecture and Technologies} 
\label{Implementation:ArchitectureAndTechnologies}

\begin{figure}[H]
	\centering
	\includegraphics[width=1\linewidth]{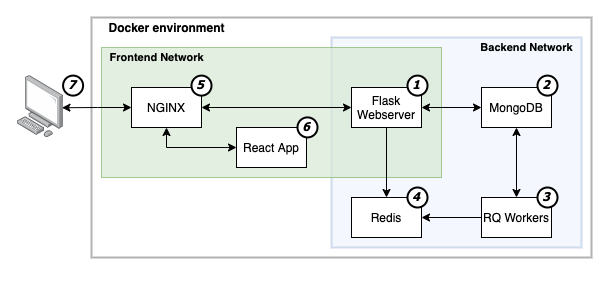}
	\caption{Overview of the docker environment.}
	\label{fig:06_FirmwareDroidArchitecture}
\end{figure}

Figure \ref{fig:06_FirmwareDroidArchitecture} shows an overview of our 
docker environment. We based our main architecture on docker and 
docker-compose. We use several docker containers for our implementation, and if 
necessary, we can extend the environment to contain more containers. 

\begin{enumerate}
	\item Flask: We use as the core of the backend python flask with flask-restx 
	\cite{FlaskRestX}. We implemented a REST API and use Swagger.io 
	\cite{Swagger} to generate an OpenAPI specification. With Swagger, we 
	automate the process of API documentation and make testing less complicated.
	
	We use gunicorn \cite{Gunicorn} as a webserver with the gevent \cite{Gevent} 
	library for multiprocessing. This setup allows the webserver to handle several 
	thousand requests per second.
		
	\item Database: We use MongoDB as the main database, and we store all scan 
	data in one database called FirmwareDroid. We integrated MongoEngine 
	\cite{MongoEngine} as document-object mapper. MongoEngine is the equivalent 
	of an object-related mapper for SQL-based database, and it allows us to map a 
	python class directly to a MongoDB document. All the document classes for 
	MognoDB are defined in the \textit{model} folder of our code repository. 
	
	\item RQ Workers: To modularize the software architecture, we split parts of the 
	software into separate docker containers. To be more precise, every scanner 
	tool has its docker containers. For example, we have a container for 
	AndroGuard and Androwarn. We use the python-RQ \cite{PythonRQ} library for 
	managing a queue system. RQ allows us to manage job queues within a Redis 
	database, and we set up one queue for every scanner. 
	The queue system allows us to spawn various docker containers that all can 
	work on the same queue. For example, we have a queue for every scanner like 
	AndroGuard and Androwarn. In case we want to scan a large number of Android 
	apps with one of the scanners, we can scale the number of docker containers 
	working on the specific queue to increase the performance.
	
	The current implementation does not use Docker Swarm or Kubernetes for 
	orchestration because we wanted to keep the requirements to run 
	FirmwareDroid as minimal as possible. Later versions may require to have a 
	more flexible orchestration for scaling purposes.
		
	\item Memory-Database: Python-RQ uses the in-memory database Redis for 
	storing queues and jobs. We use Redis as a docker container and store our 
	custom jobs and queues in it. The Redis container is configurable over an 
	environment file, and by default, we enforce authentication.
	
	\item Nginx: We use Nginx for caching, routing, and as a reverse proxy. We 
	have integrated a certbot container to integrate HTTPS by default with a 
	self-signed certificate. The certbot container is by default deactivated but can 
	be activated if someone wants to register a domain for a custom FirmwareDroid 
	instance.
	
	\item Frontend: We use React for front-end development and have added a 
	React template app. The react app is deployed with nginx and has an internal 
	routing system. The current version of FirmwareDroid is headless and has no 
	fronted client. We may integrate a front-end in future versions of FirmwareDroid 
	to make the tool more accessible for users.
		
	\item Host-System: The complete FirmwareDroid environment is running within 
	docker. Some docker containers have privileged access to the host system. 
	We run some containers in privileged mode due to the requirement to run some 
	system commands like \textit{mount} that would otherwise not be accessible by 
	a docker container.
\end{enumerate}

All docker services are configurable with environmental files. We store the 
environment files in the \textit{/env} directory of FirmwareDroids root folder. To 
keep the build time during development short, we use multiple dockerfiles that 
include only necessary dependencies for the used service. Splitting the 
dependencies into several dockerfiles has the disadvantage of rebuilding the 
docker images when we add or remove dependencies. Nevertheless, splitting 
allows us to reduce the built time whenever we do not need to add new packages 
and keep a 
separation of docker container dependencies. This separation of dependencies is 
mainly necessary because we use a set of tools with conflicting dependencies and 
want to modularize as much as possible. However, we think splitting allows us to 
customize the docker containers for different environments and scale them if 
necessary.

We cannot always resolve package conflicts because some tools need specific 
software versions to work correctly. Hence, we developed a base dockerfile with 
core dependencies that all worker containers use and then extend the base image 
in a separate dockerfile. We ensure this way that we can run every tool in an 
environment that meets all tool requirements. This approach works fine unless one 
of the tools has a conflict with our base requirements. In this case, we can't 
integrate it into FirmwarDroid with our base Docker image.

\subsection{FirmwareDroid REST API} 
\label{Implementation:FirmwareDroidFeatures}
This section gives an overview of selected FirmwareDroid REST API (v1) 
endpoints in Table \ref{tab:Rest_endpoints}. Note that the
API is still in development, and that some routes are experimental and likely to 
change in future versions. Nevertheless, the core functionality for scanning 
Android apps and importing firmware is working as described in Chapter 
\ref{Fundamentals} and \ref{Analysis}. The REST API's swagger documentation is 
as well available via the route \textit{.../docs}. For example, in our instance, 
swagger 
is found at https://firmwaredroid.cloudlab.zhaw.ch/docs. All API endpoints 
are available under the route \textit{.../api/...}, and admins can monitor all job 
queues under the \textit{.../rq-dashboard/} route. 

\begin{table}[H]
	\resizebox{\textwidth}{!}{
	\begin{tabular}{|l|l|l|}
		\hline
		\rowcolor[HTML]{EFEFEF} 
		\textbf{\begin{tabular}[c]{@{}l@{}}HTTP\\ Method\end{tabular}} & 
		\textbf{Route} & \textbf{Description} \\ \hline
		\rowcolor[HTML]{EFEFEF} 
		\multicolumn{3}{|c|}{\cellcolor[HTML]{EFEFEF}\textit{/v1/androguard/}} \\ \hline
		POST & /\{mode\} & Analyse Android apps with AndroGuard. \\ \hline
		POST & /app\_certificate/download/ & Download a certificate in DER or PEM 
		format. \\ \hline
		POST & /meta\_string\_analysis/ & \begin{tabular}[c]{@{}l@{}}Starts the 
		meta analysis of AndroGuard string \\ analysis.\end{tabular} \\ \hline
		\rowcolor[HTML]{EFEFEF} 
		\multicolumn{3}{|c|}{\cellcolor[HTML]{EFEFEF}\textit{/v1/androwarn/}} \\ \hline
		POST & /\{mode\} & Analysis apps with Androwarn. \\ \hline
		\rowcolor[HTML]{EFEFEF} 
		\multicolumn{3}{|c|}{\cellcolor[HTML]{EFEFEF}\textit{/v1/firmware/}} \\ \hline
		GET & /download/\{firmware\_id\} & Download a firmware as zip archive. \\ 
		\hline
		GET & /mass\_import/ & Starts the import of firmware files from the import 
		folder. \\ \hline
		\rowcolor[HTML]{EFEFEF} 
		\multicolumn{3}{|c|}{\cellcolor[HTML]{EFEFEF}\textit{/v1/qark/}} \\ \hline
		POST & /\{mode\} & \begin{tabular}[c]{@{}l@{}}Analysis apps with Quick 
		Android Review Kit \\ (QARK) and create a report\end{tabular} \\ \hline
		\rowcolor[HTML]{EFEFEF} 
		\multicolumn{3}{|c|}{\cellcolor[HTML]{EFEFEF}\textit{/v1/statistics/}} \\ \hline
		POST & \begin{tabular}[c]{@{}l@{}}/androguard/create\_certificate\_report/\\ 
		\{mode\}/\{report\_name\}\end{tabular} & Create a statistical report for 
		AndroGuard certificate data. \\ \hline
		POST & \begin{tabular}[c]{@{}l@{}}/androguard/create\_report\_plots/\\ 
		\{androguard\_statistics\_report\_id\}\end{tabular} & Create plots for a 
		Androguard statistics report. \\ \hline
		POST & \begin{tabular}[c]{@{}l@{}}/v1/statistics/androguard/\\ 
		create\_statistics\_report/\{mode\}/\{report\_name\}\end{tabular} & Create a 
		statistical report for AndroGuard data \\ \hline
		POST & \begin{tabular}[c]{@{}l@{}}/v1/statistics/androguard/\\ 
		create\_string\_analysis\_report/\{report\_name\}\end{tabular} & Create a 
		statistical report for AndroGuard string meta data. \\ \hline
		POST & \begin{tabular}[c]{@{}l@{}}/v1/statistics/androwarn/\\ 
		create\_statistics\_report/\{mode\}/\\ \{report\_name\}\end{tabular} & Create a 
		statistical report for androwarn data \\ \hline
		POST & \begin{tabular}[c]{@{}l@{}}/v1/statistics/apkid/\\ 
		create\_statistics\_report/\{mode\}/\{report\_name\}\end{tabular} & Create a 
		statistical report for apkid data \\ \hline
		GET & \begin{tabular}[c]{@{}l@{}}/v1/statistics/download/\\ 
		grouped\_by\_version/\{reference\_file\_id\}/\\ 
		\{add\_meta\_data\}\end{tabular} & Download a reference file for a statistics 
		report \\ \hline
		POST & \begin{tabular}[c]{@{}l@{}}/v1/statistics/firmware/\\ 
		create\_statistics\_report/\{mode\}/\{report\_name\}\end{tabular} & Create a 
		statistical report for firmware data \\ \hline
		POST & \begin{tabular}[c]{@{}l@{}}/v1/statistics/qark/\\ 
		create\_statistics\_report/\{mode\}/\{report\_name\}\end{tabular} & Create a 
		statistical report for qark data \\ \hline
		POST & \begin{tabular}[c]{@{}l@{}}/v1/statistics/virustotal/\\ 
		create\_statistics\_report/\{mode\}/\{report\_name\}\end{tabular} & Create a 
		statistical report for VirusTotal data \\ \hline
		\rowcolor[HTML]{EFEFEF} 
		\multicolumn{3}{|c|}{\cellcolor[HTML]{EFEFEF}\textit{/v1/virustotal/}} \\ \hline
		POST & /\{mode\} & Scan a firmware with VirusTotal \\ \hline
		\rowcolor[HTML]{EFEFEF} 
		\multicolumn{3}{|c|}{\cellcolor[HTML]{EFEFEF}\textit{/v1/apkid/}} \\ \hline
		POST & /v1/apkid/\{mode\} & Scan the given apps with APKiD \\ \hline
		\rowcolor[HTML]{EFEFEF} 
		\multicolumn{3}{|c|}{\cellcolor[HTML]{EFEFEF}\textit{/v1/android\_app/}} \\ 
		\hline
		POST & /by\_id/ & Get Android app meta data as json report. \\ \hline
		GET & /download/\{android\_app\_id\} & Download the app with the given id. 
		\\ \hline
		\rowcolor[HTML]{EFEFEF} 
		\multicolumn{3}{|c|}{\cellcolor[HTML]{EFEFEF}\textit{/v1/fuzzy\_hashing/}} \\ 
		\hline
		POST & \begin{tabular}[c]{@{}l@{}}/v1/fuzzy\_hashing/create\_hashes/\\ 
		firmware/\{mode\}\end{tabular} & Creates fuzzy hashes for every file of the 
		given firmware. \\ \hline
		POST & \begin{tabular}[c]{@{}l@{}}/v1/fuzzy\_hashing/\\ 
		download\_cluster\_analysis/\\ 
		\{cluster\_analysis\_id\}/\{fuzzy\_hash\_type\}\end{tabular} & Download a 
		graph file for a cluster analysis. \\ \hline
		POST & \begin{tabular}[c]{@{}l@{}}/v1/fuzzy\_hashing/tlsh/\\ 
		create\_cluster\_analysis/\{regex\_filter\}/\{mode\}/\\ 
		\{distance\_threshold\}/\{compare\_mode\}/\\ 
		\{tlsh\_similiarity\_lookup\_id\}/\{description\}\end{tabular} & Creates a 
		clustering analysis for tlsh digests. \\ \hline
		POST & \begin{tabular}[c]{@{}l@{}}/v1/fuzzy\_hashing/tlsh/\\ 
		create\_similarity\_lookup/\end{tabular} & \begin{tabular}[c]{@{}l@{}}Create a 
		tlsh similarity lookup \\ table for all tlsh hashes.\end{tabular} \\ \hline
		POST & \begin{tabular}[c]{@{}l@{}}/v1/fuzzy\_hashing/tlsh/find\_similar/\\ 
		\{cluster\_analysis\_id\}/\{tlsh\_hash\_id\}\end{tabular} & Find similar tlsh 
		hashes for a given hash. \\ \hline
	\end{tabular}
}
	\caption{List of REST API endpoints with description.}
	\label{tab:Rest_endpoints}
\end{table}

For interested readers, we present a complete list of REST API endpoints in 
Appendix \ref{Appendix:RestAPI}. In the next paragraphs, we will go through 
FirmwareDroid's REST API's main use cases and explain how to use it. 

When we start FirmwareDroid, we create several working directories for 
FirmwarDroid. The working directory path is configurable over FirmareDroid's 
environment variables. By default, we create an import folder under the root folder 
named \textit{00\_file\_storage}. We copy the firmware archives in the 
\textit{import} folder to analyze new firmware samples. After that we start the 
import with a HTTP GET request to \textit{.../api/v1/firmware/mass\_import/}. 
The server will then attempt to import all firmware archives into FirmwareDroid's 
database. FirmwareDroid moves successfully imported Firmware archives to the 
\textit{store} folder. FirmwareDroic moves archives that it could not import to the 
\textit{import\_failed} directory. Note that FirmwareDroid renames all successfully 
imported firmware archives to their md5 hash to prevent parsing errors on later use 
and that FirmwareDroid prevents the import of duplicates. The import script uses 
multi-threading to increase the performance but importing takes time because 
FirmwareDroid has to unpack and hash all files as described in Section 
\ref{Fundamentals:ExtractingPreInstalledApps}.

During the import, FirmwareDroid scans all files of the firmware archives and 
saves references to the database. Moreover, FirmwareDroid copies every found 
apk file to the host system. To be more precise, we store all apk files in 
sub-folders within the \textit{../app\_extract} directory. The md5 hash of the 
firmware names 
the 
sub-folders in the app\_extract folder to prevent naming conflicts. Within the 
firmware folder, we keep the firmware's internal directory hierarchy to avoid naming 
conflicts for the apk files. We extract the apk files to the file-system mainly for two 
reasons. First, it allows analysts to work on the apk files directly with other tools 
without exporting them from the database. Second, it increases the performance 
since most scanners work directly on the file-system, and we then do not need to 
write the apk first to disk before it can be read. 

\textbf{Scanning apps:} The import script will create for every firmware archive and 
apk a document entry in the database. MongoDB gives every document a unique 
id, and we use this id to identify and reference documents within FirmwareDroid. 
When we want to scan an apk with one of the scanners, we use the app 
document's id as a parameter for the REST endpoint. FirmwareDroid defines the 
following routes for scanning apks and takes a JSON object with a list of app ids 
as a parameter.

\begin{itemize}
	\item /v1/androguard/\{mode\}
	\item /v1/qark/\{mode\}
	\item /v1/apkid/\{mode\}
	\item /v1/virustotal/\{mode\}
	\item /v1/androwarn/\{mode\}
\end{itemize}

Additional to the JSON parameter, we have to provide for every route the 
\textit{mode} variable, which is an integer. The mode indicates if we want to scan 
a set of firmware or apk files. The following modes are defined:

\begin{itemize}
	\item Mode 0: Scan only the provided id's in the json.
	\item Mode 1: Scan all apk's in the database.
	\item Mode X: Use x as filter for firmware version. For example, mode 8 would 
	scan all Android 8 apps and mode 9 all Android 9.
\end{itemize}

We ensure that we have only one report for each app in the database and no 
duplicated reports. We configured FirmwareDroid to skip an apk if we already 
scanned the apk with the scanner. FirmwareDroid's current version allows the 
export of the scanning reports as JSON, or admins can access the database 
directly to see the scanning results. 

\textbf{Statistics:} After scanning Android apps, we create statistics of specific 
reports by providing the Android app id to the statistics endpoint. We have 
implemented one or more endpoints for every scanner as shown in Table 
\ref{tab:Rest_endpoints}. The statistic endpoints take the mode parameter for 
filtering the Android version. After creation, we store the reports in the database 
under the "statistics\_reports" collection.

\textbf{TLSH hashing:} During the firmware import, we hash every firmware file 
with \gls{ssdeep} and TLSH. As described in \ref{Analysis:FuzzyHashing} we use 
a 
lookup table for filtering TLSH hashes, and we can create a lookup table with a 
POST request to the \textit{../tlsh/create\_similarity\_lookup/} route. The 
webserver will then index all TLSH hashes in the database. After the lookup table 
creation, we can create a cluster analysis on the file we are interested in with the 
\textit{../tlsh/create\_cluster\_analysis/} route. We added to the route a 
user-settable regex for file-names. Users can use this to cluster-specific file-types 
like apks or elf files instead of all files. As mentioned in 
\ref{Analysis:FuzzyHashing} we create a graph for clustering and the graph can be 
downloaded as .gexf file with the \textit{../tlsh/download\_cluster\_analysis/} route. 
The graph file is compatible with other tools like 
Gephi\footnote{\url{https://gephi.org/}}.

\textbf{Ssdeep hashing:} To cluster ssdeep hashes at scale, we implemented the 
method described by Brian Wallace et. al. in \cite{SSDeepAtScale} which extracts 
from the ssdeep hash base64 chunks of integers. We use the chunks to reduce 
the search space by comparing only the ssdeep digests with matching chunks and 
similar size files. To compare a large dataset of ssdeep digest we store the 
chunks together with the ssdeep digest in our database.

\section{Experimental Features and Ideas} 
\label{Implementation:ExperimentalFeatures}
This section discusses some of the experimental features we integrated or 
attempted to integrate into FirmwareDroid. The development of these features is 
not yet complete and, in some cases, still in progress. Nevertheless, we want to 
give the reader some insights into some ideas we were developing.

\subsection{Diffing Firmware and Android apps} 
\label{Implementation:DiffingFirmware}

The idea was to identify changes between two builds of the same firmware by 
creating a diff of the complete file structure. To our knowledge, there are no 
open-source tools that can compare firmware archives. Therefore we developed an 
own file comparer using the Linux standard tool Diff. Using Diff allows us to identify 
changes in the file system between two firmware archives. The problem is that 
with using Diff, we cannot show which bytes exactly have changed. Instead, we 
can only determine if a file has changed. We use Diff to identify new, 
changed, and removed files, but it has the limitation that we cannot display which 
bytes have changed. For example, other approaches with the rsync algorithm 
\cite{RSync} were checked, but they inherited similar problems. In case we could 
identify the exact bytes that have changed between two firmware archives, this 
may help us determine if some of the changes are unintentional or malicious. For 
example, if an attacker has changed a system setting, we could identify this 
change by diffing two versions of the same firmware.

\subsection{AndroGuard String Enrichment} 
\label{Implementation:StringEnrichment}
Androguard decompiles apks with the DAD decompiler, and it collects 
string values by looking at the \textit{const-string (0x1a)} and 
\textit{const-string/jumbo (0x1b)} smali up-codes. We use the AndroGuards string 
extraction feature to extract strings from our Android apps. Together with 
extracting the string, we save the code references to determine where 
AndroGruard detected the string within the code.

\begin{figure}[H]
	\centering
	\includegraphics[width=\linewidth]{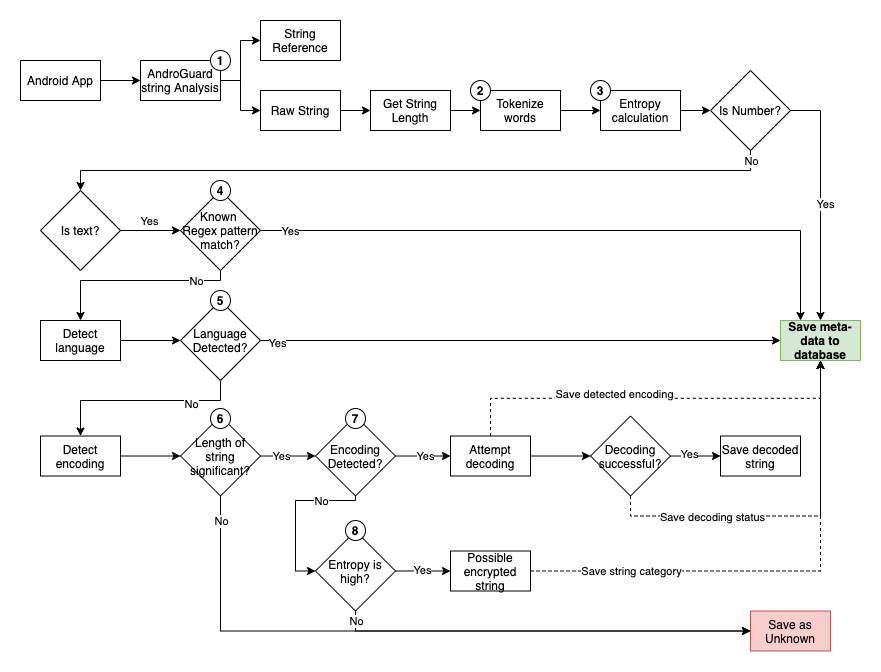}
	\caption{Process of string meta data enrichment.}
	\label{fig:03_StringAnalisys.png}
\end{figure}

Analyzing strings for malware detection is a research topic on its own and outside 
of this project's scope. However, we implemented for FirmwareDroid a module that 
can enrich the raw string data with some metadata. The basic idea of this module 
is to detect some common strings by using a regex classification approach. To be 
more precise, we calculate the entropy, detect languages and encodings, and 
search for known string patterns like SQL statements or URLs. Figure 
\ref{fig:03_StringAnalisys.png} illustrates the implemented metadata 
enrichment process.

\begin{enumerate}
	\item The first step is to scan the app with AndroGuard and save all the strings 
	and references. We calculate the string length and save it as 
	well.
	
	\item The next step is to calculate the number of words in the string. We use 
	python's word regex for this purpose and get an estimated number of words. 
	Note that it is possible to use more sophisticated tokenizers, but an estimate of 
	the number of words is enough for our purposes.
	
	\item We calculate the Shannon, Hartley, and natural entropy for every string 
	and save them to the database. After these calculations, we check if the string 
	is numeric. In case it is numeric, the classification ends.
	 
	\item In case the value is not numeric, we check the string against a list of 
	regex patterns. The current list checks if the string is likely to be an SQL, URL, 
	or path string. Moreover, we check for string patterns that common secret use. 
	For example, we check for "-----BEGIN RSA PRIVATE KEY-----" to test for RSA 
	private keys.
	 
	\item If none of our regex patterns is detected, we progress by attempting to 
	detecting the used language. We use the polyglot library \cite{Polyglot} which 
	can identify over 190 different languages. Polyglot gives for the detected 
	language a confidence score from 0 to 100, and we save both values in our 
	database.
	
	\item If polyglots language detection could not detect a language, we check the 
	length of the string. If the length is under twenty characters, we save the string 
	as an unknown category. Otherwise, we attempt to detect the encoding of the 
	string.
	
	\item We use the chardet library \cite{cchardet} to detect common encodings. In 
	case an encoder is detected, we attempt to decode the string and save the 
	decoded string in the database.
	
	\item If the string length is more than 20 characters, we check if the bit entropy 
	of the string is higher than 7. In this case, we may have detected an encrypted 
	string, and we save this meta-information as well in the database.
\end{enumerate}
We integrated the process above into FirmwareDroid. As mentioned at the 
beginning of this section, string analysis is a research topic on its own, and a 
detailed analysis of the data is outside of this project's scope. However, in total, 
we have collected 2'151'862'783 strings from our app corpus. It should be clear 
that more advanced classification techniques like Machine Learning are necessary 
to handle this data amount.

\section{Dynamic Analysis of pre-installed Apps}
\label{Implementation:DynamicAnalysis}
To our knowledge, no other researcher could conduct a dynamic analysis of 
pre-installed apps on an emulator environment. For example, 
\cite{AnAnalysisofPreinstalled} conducted their dynamic analysis on the Lumen 
mobile traffic dataset and \cite{DroidRay} analyzed iptable rules for their dynamic 
analysis. In this section, we will discuss our plans for integrating dynamic analysis 
of pre-installed apps into FirmwareDroid. We will show which challenges we have 
to overcome to make a dynamic analysis of pre-installed apps feasible. 

DroidBox \cite{droidbox} was used in a recent study \cite{AndroPyTool}. The tool 
offers an Android emulator especially made for Android app analysis that could 
monitor an app's behavior. The tool itself seems no longer to be maintained, and 
we, therefore, decided not to use it for our research. However, we would need 
something similar to analyze pre-installed apps because we do not possess the 
necessary hardware to test all pre-installed apps on real devices. As a 
consequence, we searched for alternative emulator environments that could run 
pre-installed apps for dynamic analysis.

Since we based FirmwaDroid's architecture on docker containers, we searched for 
docker based solutions that we could integrate into FirmwareDroid. Other 
researchers have before attempted to use docker container for Android app 
analysis. For example, Ngoc-Tu Chau and Souhwan Jung discuss in 
\cite{CHAU201838} the challenges to use docker containers for dynamic analysis. 
Hence, we searched for maintained docker containers and found that Google has 
an officially maintained repository \cite{GoogleDockerEmulator} for the Android 
emulator. 

Therefore, we attempted to run the collected pre-installed apps on the official 
Android emulator with a generic Android image to see what problems we would 
need to solve to run the apps. The installation of pre-installed apps on an Android 
emulator is, in many cases, more complicated than with an app from an app 
market. Pre-installed apps have more dependencies on the actual device than 
standard apps. Following, we go through some of the challenges we have to 
overcome to make dynamic analysis possible for pre-installed apps:

\begin{itemize}
	\item \textbf{Emulator environment:} Android stores pre-installed apps of the 
	system partition mainly in the system/app and system/priv-app folders. The 
	system partition is normally read-only, and as a consequence, we cannot install 
	new apps to the app and priv-app folders without remounting the system 
	partition with read-write permissions. In theory it is possible on the standard 
	emulator to remount the system partition with the \textit{adb remount system} 
	\cite{DevAndroidEmuatorCommands} command and the \textit{-writable-system} 
	\cite{DevAndroidEmuatorCommands} parameter. In practice, the tested Android 
	emulator's images had problems rebooting the system after using these 
	commands for remounting.
	
	Running Android apps on an emulator has as well some implications and 
	limitations. Apps that check the runtime environment will realize that they are 
	run within a virtual environment and may behave differently from the hardware. 
	As discussed in Section \ref{Analysis:APKiDAnalysis} there are many variables 
	that Android apps can check to detect the runtime environment. 
	
	\item \textbf{Installation:} We can install an Android app with the \textit{adb 
	install} command. The command will then run the package manager and install 
	the apk in the /data/app/ folder. With pre-installed apps, we can use the 
	\textit{adb push} command to store the apk in the system/app or 
	system/priv-app folder if the system partition is writeable. After copying the apk 
	to the system partition, we have to rewrite the permissions to access the apk. 
	We can do this with the \textit{adb shell chmod} command. However, 
	when trying to automate the installation process on the emulator, we came 
	across some errors:
	
	\begin{itemize}
		\item \textit{INSTALL\_FAILED\_NO\_MATCHING\_ABIS: Failed to extract 
		native libraries}. The installation fails because the package manager could 
		not extract the necessary native libraries. This is the case when the apk has 
		a native library that is incompatible with the emulator's CPU architecture. In 
		some cases, we can solve these incompatibility problems by using another 
		version of the Android emulator.
		
		\item \textit{INSTALL\_FAILED\_UPDATE\_INCOMPATIBLE: Package 
		xx.xxx.xxx signatures do not match previously installed version}. This error 
		occurs when the app we attempt to install already exists on the device and 
		we try to install an older version of the app. We can solve this issue by 
		removing the installed app before the installation of the new one.
		
		\item \textit{INSTALL\_FAILED\_SHARED\_USER\_INCOMPATIBLE 
		xx.xxx.xxx has no signatures that match those in shared user 
		android.uid.system.} The error occurs if we do not use the platform key for 
		signing the app we try to install.
		
		\item \textit{INSTALL\_FAILED\_MISSING\_SHARED\_LIBRARY}: This error 
		occurs when one of the native libraries in system/lib or system/lib64 does 
		not exist or is not accessible.
	\end{itemize}
	
	Another problem we have to face is the installation sequence of apks and the 
	app to app communication. As we have described in Section 
	\ref{Fundamentals:AppSigning} apps can define custom permissions in their 
	AndroidManifest.xml. Therefore some apps are dependent on others in their 
	functionality.
	
	\item \textbf{App permissions:} We have to resign pre-installed apps with the 
	platform key of the emulator image. Otherwise, the system will not give them 
	the system permissions, and we have to install them in the correct directory. 
	Some apps may implement signature checks and will not start if we resign 
	them with a custom platform key.
\end{itemize}
Overall, these problems show that the current Android emulator is not suitable for 
pre-installed app analysis in many cases. What we would need to analyze 
pre-installed apps dynamically is the possibility to emulate a complete firmware. 
However, like other researchers, it is not feasible for us to provide such a solution 
currently.

Moreover, if we manage to install a pre-installed app on an emulator, we need to 
define automated test cases and tools to monitor the runtime environment. We, 
therefore, discuss in the next paragraphs the tools that we could use for 
monitoring.

In many cases \acrfull{dbi} frameworks support multiple platforms, and Android is 
often one of them. For example, Triton \cite{Triton}, DynamoRIO 
\cite{DynamoRIO}, Valgrind \cite{Valgrind} or Frida \cite{FridaSourceCode}. We 
could use these tools to analyze Android apps if we would have the environment 
to run the tools. For FirmwareDroid, we would integrate Frida if possible since it 
seems to be one of the most popular Android frameworks at the time of writing.

\textbf{Frida}: Frida is a \acrshort{dbi} toolkit that allows to hook functions calls at 
runtime and inject code into the process. Its community provides a set of scripts 
to automate the process of dynamic analysis on Android. For example, the 
Medusa framework \cite{Medusa} contains modules for tracing API calls, 
unpacking known packers, or triggering system events. Tools like DexCalibur 
\cite{DexCalibur} or Runtime-Mobile-Security \cite{RMS-Runtime-Mobile-Security} 
use Frida as the core library for their dynamic analysis of Android apps.

Our idea for future FirmwareDroid versions is to integrate the possibility that users 
can dynamically analyze the pre-installed apps on their real phones. To do so, we 
integrated an adb shell library \cite{Adbshell} into FirmwareDroid and are 
experimenting with Frida to automate the process. However, this feature is still 
experimental and only a workaround until we have an environment where we can 
run pre-installed apps on scale in a virtual environment.

	\chapter{Results and Discussion} \label{Results}

In this chapter, we summarize our findings and discuss the limitations of our work. 
We start in Section \ref{Results:Results} with discussing our findings and progress 
in Section \ref{Results:ProjectGoalsReview} with reviewing our project goals. In 
Section \ref{Results:Discussion}, we discuss our work's current state and compare 
it to other studies.

\section{Results} \label{Results:Results}

Our main contribution is FirmwareDroid, a tool that allows us to automate parts of 
the pre-installed Android apps' security testing process. We integrated five 
static-analysis into FirmwareDroid: AndroGuard, APKiD, Androwarn, Qark, and 
VirusTotal. We used FirmwareDroid to scan more than 900'000 Android 
pre-installed apps and explain in Chapter \ref{Analysis} how we use the data of the 
individual scanners to identify privacy and security risk. We summarize the 
scanner results as following:

\begin{itemize}
	\item \textbf{AndroGuard:} In Section \ref{Analysis:AndroGuardAnalysis} we 
	discuss 
	the usage of permission data and show three different examples for Android 8, 
	9, and 10. For our Android 10 data, we calculate how common the normal 
	(43.4\%), dangerous (13.4\%), and signature (41.1\%) base permissions are. 
	Additionally, we calculate the usage of custom permissions and third party 
	permissions. For our Android 9 data, we show the top 30 most used 
	permissions overall and the most used third party permissions. We identify 
	under the top 30 permission eight dangerous, eight signature, 13 normal 
	permissions, and one permission flagged by Google not to be used by 3rd 
	parties. For our Android 8 data, we calculate the distribution of normal, 
	dangerous, and signature permissions and demonstrate that we can identify 
	out-liners with a rather high permission usage. 
	
	In Section \ref{Analysis:Certificates} we show usage statistics of our 
	certificates 
	in our dataset, and we spot a suspicious Samsung certificate for Android 9. We 
	determine the average usage of a certificate per unique package name to be 
	4.45 for Android 8, 4.3 for Android 9, and 2.63 for Android 10 in our dataset. 
	Moreover, we find out that 12.07\% of our Android apps are signed with an 
	insecure AOSP debugging certificate.
	
	\item \textbf{APKiD:} We detected with APKiD in nearly all Android versions, 
	except version three and ten, using the dexlib compiler. Attackers use the 
	dexlib compiler for repackaging in some cases, and we see it as an indicator for 
	suspicious activity. Our scan results show that we have for Android 7 2'395 
	apps, for Android 8'258 apps, and for Android 9'295 apps that use the dexlib 
	complier. Furthermore, we show the usage of obfuscators, packers, and 
	anti-VM techniques for all Android versions. 
	
	\item \textbf{Androwarn:} We demonstrate that we can use the Androwarn 
	reports to identify apps with access to privacy critical services like location, 
	camera, or microphone. We show for the Android versions 7, 8, 9, and 10 the 
	number of detected apps that access the microphone and camera source. 
	Moreover, we demonstrate that we can detect the execution of UNIX 
	commands with Androwarn to find potential vulnerabilities and identify them.
	
	\item \textbf{Quark:} We integrated the scanner into FirmwareDroid, but we 
	could not scan the complete dataset of Android apps with Quark due to the high 
	memory resources Quark needs. However, the tool is functional and can be 
	used to scan smaller datasets.
	
	\item \textbf{VirusTotal:} The scan results show that 1.1\% of the apps in our 
	corpus were flagged as malicious by more than three scanners of VirusTotal. 
	We attempted to identify if the detected malware belongs to stock or custom 
	firmware by checking the firmware's individual build properties for custom ROM 
	indicators. However, we can only assume that some of the found malware 
	belongs to stock firmware and more research is necessary to prove this 
	assumption fully. Moreover, we scan the malware samples with APKiD and 
	detect at least seven samples that use the dexlib 2.x compiler.
\end{itemize}

In total, we have collected 5'934 firmware samples and demonstrate that we can 
automate the extraction process of pre-installed Android apps. In general, 
Android's 
customizations make it difficult for us to develop a generally applicable firmware 
extractor, and we were not able to extract all the data from all firmware samples. 
We find evidence that several terabytes of firmware data are available for free on 
the web, and we can use the data for further research.  

With TLSH and ssdeep, we integrate similarity hashing techniques into 
FirmwareDroid. We develop a filtering and clustering algorithm for TLSH to allow 
similarity analysis on scale. Our fuzzy hashing filtering algorithm describes in 
\ref{Analysis:FuzzyHashing} allows us to reduce the search space for TLSH 
effectively. Our performance evaluation shows good results for datasets with 
10'000 files. In best case, we 
can effectively reduce the search space for comparisons by 97.17\% with a false 
positive rate of 0.24\% and with a band width threshold of 13 on 10'000 files. 
However, 
we conducted our measurements on small datasets, and more time is required to 
test our approach on a larger dataset. Nevertheless, our TLSH filtering approach 
was able to index more than 1.1 million files in our test instance of 
FirmwareDroid.

\section{Project Goals Review} \label{Results:ProjectGoalsReview}
In this section, we review the research questions that we defined at the beginning 
of this document in Section \ref{Introduction:Projectgoals}.

\begin{itemize}
	\item \textit{Can we automate the extraction of Android pre-installed apps from 
	existing android firmware archives?}
	
	In Section \ref{Fundamentals:ExtractingPreInstalledApps} we describe our 
	solution 
	for extracting pre-installed apps from Android firmware. We were able to extract 
	pre-installed apps from 5'934 firmware archives. However, due to 
	Androids firmware build customization, we still have firmware samples that we 
	cannot extract with our current approach.
	
	\item \textit{Can we detect android malware within the standard Android apps 
	like 
	calculator, calendar, browsers, etc.?}
	
	Our VirusTotal scan has not shown any malicious sample for the standard 
	Android apps. We have not found any other evidence for malicious standard 
	apps.
	
	\item \textit{Can we automate the static and dynamic analysis of Android 
	firmware 
	and apps with the existing open-source tools?} 
	
	We showed that the static analysis is possible and FirmwareDroid's architecture 
	allows us to integrate more static analysis tools in the future. However, 
	as described in Section \ref{Implementation:DynamicAnalysis} currently the 
	dynamic analysis of pre-installed apps is not possible due to environmental 
	limitations.
	
	\item \textit{0.1\% of the Android apps in our corpus are malicious.}
	
	As described in Section \ref{Analysis:VirusTotalQarkAnalysis} around 1.1\% of 
	the 
	app in our corpus is tagged malicious by VirusTotal. Therefore, our hypothesis 
	is 
	correct, and we can find malware in pre-installed apps with static analysis tools.
	
	\item \textit{Automate the process of scanning several thousand Android 
		app's with state of the art static and dynamic analysis tools.}
	
	As demonstrated in Chapter \ref{Analysis} with FirmwareDroid, we were able to 
	scan over 900'000 pre-installed apps and automated the process of app 
	extraction 
	and scanning.
	
	\item \textit{Use fuzzy hashing techniques to detect similarities between 
		Android apps.}
	
	In Section \ref{Analysis:FuzzyHashing} we describe how we integrated TLSH 
	into 
	FirmwareDroid. Even if our clustering algorithm isn't the most effective and 
	scaling 
	one, we can use it to cluster and search for similar TLSH hashes.
\end{itemize}

\section{Discussion} \label{Results:Discussion}

As explained in Section \ref{Analysis:BuildPropAnalysis} we have used a wide 
variate of Android firmware versions and have shown that we can extract data 
from numerous vendors. Similar to other studies \cite{firmscope-2020, 
AnAnalysisofPreinstalled} we have developed a tool that can extract data from the 
pre-installed apps for analysis purposes. We integrated state-of-the-art tools into 
FirmwareDroid and scanned 5'931 firmware archives. Compared 
to \cite{firmscope-2020} with 2'017 samples and \cite{DroidRay} with 250 firmware 
samples, we have collected the largest number of Android firmware samples and 
have shown that we can extend our dataset with several terabytes of data from 
various web sources.

Similar to \cite{AnAnalysisofPreinstalled} we were not able to integrate dynamic 
analysis in our project. In Section \ref{Implementation:DynamicAnalysis} we have 
stated the major problems that hindered us in integrating a dynamic analysis into 
FirmwareDroid. We have to conduct more research and engineering before the 
dynamic analysis of Android apps will be feasible in emulated environments.

\subsection{Limitations of our work}

\textbf{App extraction:} Even if it is technically possible with the current 
implementation, we did not extract the pre-installed apps from other partitions than 
the system partition. Images like vendor, oem, or product can contain hundreds of 
pre-installed apps, but we decided not to include them in this study due to time 
and storage limitations. We will discuss further plans of our research in Section 
\ref{Discussion:FutureWork}. As discussed in Section 
\ref{Fundamentals:ExtractingPreInstalledApps} we developed the first version of 
our extractor for Android pre-installed apps and were able to index 6.5 terabytes of 
the firmware. However, we could not develop a tool that can extract all firmware 
archives from every vendor. At the time of writing, we have around 3'200 samples 
in our dataset that we could not import.

\textbf{Dynamic analysis:} As discussed in Section 
\ref{Implementation:DynamicAnalysis} the dynamic analysis of pre-installed apps 
has 
fundamental differences compared to analyzing standard apps. There are several 
challenges to overcome when installing a pre-installed app on another device and 
on scale. We show in Section \ref{Implementation:DynamicAnalysis} that we can 
set up 
several Android emulators in a docker environment but that the dependencies of 
pre-installed apps make testing on an emulator problematic. 

Our goal was to analyze pre-installed apps at scale and without using hardware. 
We tested, therefore, the capabilities of dockerized Android emulators with Google 
and generic Android images. The major problems we had are that the pre-installed 
apps have environment dependencies that we cannot full fill within our virtual 
environment. The docker emulators do not work as intended on our hardware.

\subsection{Android Customization}
The customization of Android Firmware has many advantages when it comes to 
supporting platform-specific architectures or builts. However, a 
clear disadvantage of these customizations is that they make security analysis 
more complicated. There are many customizations and special cases that make it 
often impossible to automate a process that entirely fits all the customizations. An 
example showing these problems is extracting all pre-installed apps from the 
system.img. Vendors do not follow any naming conventions for such files and 
therefore we had to find a solution to detect the correct system.img file for the 
extraction of pre-installed apps. Another 
customization is differing compression-, chunk- or partition 
methods from vendor to vendor. Such customizations make it difficult to automate 
the extraction of pre-installed apps. 

These are just some of many examples that show that the customization of the 
Android firmware has its downside. It can lead to the fact that automated security 
analysis needs to cover these customizations to be effective.

	\chapter{Future Work and Conclusion} \label{FutureWork}

In this chapter, we discuss ideas for enhancing FirmwareDroid and possible 
research directions for future work. We will start in Section 
\ref{Discussion:FutureWork} with giving an overview of possible research topics. 
We then discuss in Section \ref{Discussion:FirmwareDroidEnhancements} 
enhancements to FirmareDroid. At the end of this chapter, we come to our 
conclusions in Section \ref{FutureWork:Conclusion}.

\section{Future Work} \label{Discussion:FutureWork}
In this section, we will discuss possible future research directions. We have 
several topics that we could not discuss in this project and integrate more tools 
and data to FirmwareDroid for research purposes. Following an overview of ideas:

\begin{itemize}
	\item \textbf{Dataset:} As we have explained in Section 
	\ref{Fundamentals:CollectingFirmwareSamples} several websites are offering 
	Android firmware for free. In our study, we were only able to analyze a small 
	subset of the available data on the web. To further investigate the Android 
	firmware eco-system, we could extend our firmware dataset to include 
	device-specific firmware. For example, we could analyze Android firmware for 
	devices like TVs or cars to see any major security and privacy differences.
	
	\item \textbf{Partitions:} Our analysis focused on the system partition of an 
	Android firmware. To further extend our research, we could focus future 
	research on different partitions. For example, we could analyze customizations 
	added to the boot image and the Android kernel.
	
	\item \textbf{Permission analysis:} Our analysis in Chapter \ref{Analysis} 
	shows  example statistics for permission usages. Our research could further 
	progress by automating the analysis of permissions usage with additional data 
	analysis techniques. For example, we could focus on the the custom 
	permission usage of pre-installed apps'. 
	
	\item \textbf{Malware analysis:} As we have shown, scanners like VirusTotal 
	can identify malware samples in our dataset with high confidentiality. During this 
	thesis, we did not perform a detailed analysis of these samples. Future 
	research could analyze the pre-installed malware that we found or use 
	techniques like fuzzy hashing to detect unknown malware.  
	
	\item \textbf{Static Taint Analysis}: In our research, we did not include any 
	static taint analysis tool into FirmwareDroid. During the study, we have tested if 
	it was possible to include FlowDroid \cite{FlowDroid}, AndroGuard's graph 
	representation, and other static taint analysis tools into our environment. 
	However, the main problem we faced is that most of the available static taint 
	analysis tools have high requirements for computation power and do not scale 
	well. We, therefore, decided not to include any of the currently available tools 
	into FirmwareDroid. Nevertheless, in a future release, it would make sense to 
	include static taint analysis tools into FirmwareDroid. 
	
	\item \textbf{Library Detection:} One way to develop Android malware is to 
	create a malicious library that developers then include into their app. To our 
	knowledge, no open-source tools for Android exist that can detect such 
	malicious libraries. In general, identifying libraries can be useful for detecting 
	vulnerabilities and several researchers tried to identify which libraries app 
	developers include in their Android apps. For example, Michael Backes et al. 
	have shown that they were able to identify the use of vulnerable cryptographic 
	libraries in \cite{ReliableThirdPartyLibrary}. The problem with most of these 
	approaches is that they need a reference dataset of libraries and cannot 
	fingerprint unknown libraries. To our knowledge, there are no open-source library 
	detection tools that can be included in FirmwareDroid at the time of writing. The 
	mentioned approaches in Section \ref{Introduction:RelatedWork} were all 
	unmaintained or did not scale on our dataset. 
	
	\item \textbf{Enriching the data:} As shown in Section 
	\ref{Implementation:StringEnrichment} we can enrich some of the collected data 
	with meta-data from other tools and libraries. Such enrichment can help us in 
	analyzing the data when we search for something specific. However, we did not 
	enrich other parts of our dataset. For example, we could classify all the 
	pre-installed apps in categorizing as the Google Play Store does. Such could be 
	useful for filtering the database if we want to conduct studies on a specific 
	class of apps.
	
	\item \textbf{Analysing other files of interest:} From the firmware, we can 
	extract several security-relevant files. For example, CA certificates, host files, 
	cron jobs, or the united.rc. To analyze these files in detail, we need to integrate 
	or build more modules or scanners into FirmwareDroid.
	
	\item \textbf{Firmware Fingerprinting and Rating:} At the time of writing, there 
	are 
	no tools known which allow fingerprinting Android firmware. We think 
	fingerprinting firmware could help identify if the firmware is a custom ROM or an 
	official stock ROM. As we have discussed in Section 
	\ref{Analysis:BuildPropAnalysis} there are some methods to identify if a 
	firmware archive was signed with debugging certificates. However, since 
	Android does not use a PKI, we cannot fully rely on such methods. The 
	challenge in fingerprinting firmware is how to verify if a firmware belongs to a 
	specific vendor without using a PKI. Some vendors like Google release their 
	firmware with checksums, and this allows us to verify their firmware. For other 
	vendors, we cannot rely on an official source to do so. However, it is thinkable 
	to integrate a scoring system for firmware archives that measures a firmware's 
	trustworthiness by different indicators and scores the results as proposed by 
	\cite{uraniborg-scoring-2020}. In some cases where we can detect modified 
	build properties or malicious apps, we could create a blacklist for malicious 
	firmware at least.
\end{itemize}

\subsection{Future of FirmwareDroid} 
\label{Discussion:FirmwareDroidEnhancements}

The current version of FirmwareDroid uses AppVeyor for testing our docker builds. 
However, If we want to progress with this project, we will need to implement more 
unit and integration tests for the core components. The integration of code quality 
tools should be one of the first steps to proceed with this project. We think this will 
help in keeping the source code maintainable and expendable.

Another point is securing the REST API by adding validation and 
sanitation of inputs. The current API version already has some basic validation 
and sanitation methods integrated, but we consider the API as insecure and not 
ready for a release without automated tests. We added basic authentication and 
JWT token-based authentication to the API, and all API endpoints enforce 
authentication, but we do not have a role based security concept for the API so far.

Furthermore, we need to optimize the current database schema for larger 
datasets. Querying on the current implementation can be slow when the query 
contains not index fields like arrays. Moreover, we need to create an index for 
common query fields like the package name and the apk filename. Indexing 
should increase the performance of querying the database and help us in 
managing even larger databases. Another point is to support MongoDB clustering 
to increase the computation power of the database. During development, tests 
have shown that our MongoDB docker container can reject connections when it is 
under heavy load. 
We think this can lead to problems if we try to increase the number of docker 
containers. 

\begin{itemize}
	\item \textbf{Build-prop:} At the moment, the build.prop parser of FirmwareDroid 
	does not index referenced property files. If a build.prop file uses the import 
	statement, it references another build.prop file in the file-system and includes 
	its key-value pairs. Enhancing the parser to follow the referenced files should 
	solve the problem of not getting the version value of some firmware archives.
	
	\item \textbf{DAC and SELinux policies:} Hernandez et al. show in \cite{BicMac} 
	that the static analysis of Android firmware DAC and SELinux policies is 
	feasible and can be used to detect vulnerabilities. The integration of a tool like 
	\cite{BicMac} into FirmwareDroid would be a major step in analyzing one of the 
	core security features of Android firmware.
	
	\item \textbf{Similarity analysis:} Edward Raff et al. describe in 
	\cite{LempelZiv} how the Lempel-Ziv Jaccard distance is an alternative to 
	ssdeep and sdhash. To our knowledge, no publication explains how to use the 
	Lempel-Ziv Jaccard distance at scale for malware clustering. Integrating 
	Lempel-Ziv Jaccard distance into FirmwareDroid could help researchers develop 
	effective cluster algorithms that we could use at scale.
	
	\item \textbf{Detecting malicious certificates:} We show in Section 
	\ref{Analysis:Certificates} that threat actors can create fake certificates, and if 
	poorly created, we can detect such threads in some cases even without a 
	PKI. However, we had to conduct our certificate analysis to spot such a thread 
	manually, and it is desirable to automate the detection process. Creating a 
	collection of certificate fingerprints for genuine certificates from real devices or 
	OS vendors could significantly increase such threads' detection rates.
\end{itemize}

One of our docker architecture's core aspects is that we can integrate more open 
source tools for scanning apps. Therefore, it is a logical next step to integrate and 
test more static analysis tools into FirmwareDroid. Following some tools that we 
are likely to integrate with future versions. 

\begin{itemize}
	\item \textbf{Exodus} \cite{Exodus}: Exodus is a French non-profit platform and 
	static analysis tool for detecting Android \glspl{Tracker}. At the time of writing, 
	Exodus can detect 323 different trackers. We can use the tool as an online 
	service\footnote{https://reports.exodus-privacy.eu.org/en/} or as well as a 
	standalone command-line tool \cite{Exodus}. An Exodus report gives us some 
	insights on which trackers exist within an apk file. We can use this information 
	to model the relationships between app developers and tracker companies.
	
	\item \textbf{Androbug} \cite{AndroBugs}: The AndroBugs framework is a 
	python 2.7 based vulnerability scanner. Its core is based on a modified version 
	of the AndroGuard tool. It extends AndroGuard to scan for common app 
	vulnerabilities like broken WebView configs or exported components without 
	permission checks. The AndroBug framework was designed for scaling mass 
	scans and was used to find several security vulnerabilities \cite{AndroBugs}. 
	The tool itself seems to be no longer maintained. However, refactoring and 
	integrating Androbug into FirmwareDroid could help to find vulnerabilities in older 
	Android apps. 
	
	\item \textbf{Quark-Enigne} \cite{QuarkEngine}: As we have seen in Chapter 
	\ref{Analysis} the current version of FirmwareDroid has some vulnerability 
	scanners included but no tool for malware detection. Quark-Engine is an 
	open-source malware scoring system that claims to be obfuscation resilient.
	
	\item \textbf{SUPER Android Analyzer} \cite{SUPERAndroidAnalyzer}: SUPER 
	is another vulnerability scanner for Android apps. It is based on Rust and has 
	extendible rules set for detecting vulnerabilities. It's a command-line tool, and It 
	can generate JSON and HTML reports of the scanning results.
\end{itemize}

We may include other static analysis tools depending on the data they offer. 
Moreover, if at some point the dynamic analysis of pre-installed apps is feasible 
we will include Frida to automate the analysis process further.

\newpage
\section{Conclusion} \label{FutureWork:Conclusion}
In our work, we have developed a web service capable of extracting and scanning 
Android apps from firmware. Other researchers 
\cite{AnAnalysisofPreinstalled, firmscope-2020}  have implemented 
similar tools. Despite the research on this field to our knowledge, no open-source 
software exists that automated Android firmware analysis. We could not reproduce 
other researchers' results or continue their work. Therefore we implemented 
FirmwareDroid with the ambition to be scaling and capable of reproducing some of 
the results other researchers have shown. FirmwareDroid's docker architecture 
allows it to scale, and we will integrate more tools in future versions.

As discussed in this chapter, we still have many features to be integrated and 
issues to be solved before we can completely automate the process of analyzing 
pre-installed apps. Nevertheless, we think that the current set of features has 
shown its use. As discussed in Chapter \ref{Analysis} we can use FirmwareDroid 
to create statistic reports of any arbitrary set of Android firmware or Android apps. 

Future investigations are necessary to validate the kinds of conclusions that can 
be drawn from this study. However, our analysis shows that pre-installed apps can 
be a risk for the users privacy and that many firmware samples contain 
malicious apps. We showed that Malware detection tools like VirusTotal, together 
with other scanners, can help in identifying malicious pre-installed app. 
Furthermore, we think that enriching FirmwareDroid with TLSH fuzzy hashing and 
clustering of similar binaries is a great feature to identify new malware.

With the release of Android11 new changes to Android's permission model were 
released \cite{DevAndroid11ReleaseNotes}. Including features like one time 
permissions, location monitoring, and API quotas. These features show that 
Android is continuously improving its security and privacy features. Nonetheless, 
we believe that it is well justified to have independent testing software for showing 
where Android's security is still lacking and to monitor vendor made 
customizations. We will therefore progress with improving FirmwareDroid and add 
as well support for Android 11 at some point. Overall we think this study has 
accomplished it's goals even when many questions still remain unanswered.

\section{Acknowledgment}
First of all, I would like to thank my professor Dr. Bernhard Tellenbach for all the 
support. I very much appreciated the excellent technical discussions and the 
feedback. Moreover, I would like to thank the following people or organizations:

\begin{itemize}
	\item Thanks to VirusTotal \cite{VirusTotal} for giving us free access to their 
	web API for over six months. Your support team is awesome!
	
	\item Thanks to all the open-source developers for creating such awesome 
	tools for analyzing Android. 
	
	\item Thanks to all the people taking the time to read and review this document.
\end{itemize}

\textbf{Thanks for all the support of my family, girlfriend, friends, and cats. 
	You are awesome!}

	\begingroup
	\let\cleardoublepage\relax
	\let\clearpage\relax
	\chapter{Directories}
	\bibliography{MasterThesis}

	\listoffigures
	\addtocounter{section}{1}
	\addcontentsline{toc}{section}{\protect\numberline{\thesection} List of Figures}
	\listoftables
	\addtocounter{section}{1}
	\addcontentsline{toc}{section}{\protect\numberline{\thesection} List of Tables}
	\lstlistoflistings
	\addtocounter{section}{1}
	\addcontentsline{toc}{section}{\protect\numberline{\thesection} List of Listings}
	\newpage
	\appendix
	
	\chapter*{Appendix}
	\addcontentsline{toc}{chapter}{Appendix}
	\markboth{Appendix}{}
	\renewcommand{\thesection}{\Alph{section}}
	\captionsetup{labelformat=empty, labelsep=none}
	
	\section{Android Verified Boot code snippets and examples}

\begin{lstlisting}[caption={AVB avb\_slot\_verify from avb\_slot\_verify.c. Code 
source \cite{AndroidVerifiedBoot2Repo}}, basicstyle=\scriptsize, language=C, 
firstline=1250, 
lastline=1492]
AvbSlotVerifyResult avb_slot_verify(AvbOps* ops,
                                    const char* const* requested_partitions,
                                    const char* ab_suffix,
                                    AvbSlotVerifyFlags flags,
                                    AvbHashtreeErrorMode hashtree_error_mode,
                                    AvbSlotVerifyData** out_data) {
  AvbSlotVerifyResult ret;
  AvbSlotVerifyData* slot_data = NULL;
  AvbAlgorithmType algorithm_type = AVB_ALGORITHM_TYPE_NONE;
  bool using_boot_for_vbmeta = false;
  AvbVBMetaImageHeader toplevel_vbmeta;
  bool allow_verification_error =
      (flags & AVB_SLOT_VERIFY_FLAGS_ALLOW_VERIFICATION_ERROR);
  AvbCmdlineSubstList* additional_cmdline_subst = NULL;
  /* Fail early if we're missing the AvbOps needed for slot verification. */
  avb_assert(ops->read_is_device_unlocked != NULL);
  avb_assert(ops->read_from_partition != NULL);
  avb_assert(ops->get_size_of_partition != NULL);
  avb_assert(ops->read_rollback_index != NULL);
  avb_assert(ops->get_unique_guid_for_partition != NULL);
  if (out_data != NULL) {
    *out_data = NULL;
  }
  /* Allowing dm-verity errors defeats the purpose of verified boot so
   * only allow this if set up to allow verification errors
   * (e.g. typically only UNLOCKED mode).
   */
  if (hashtree_error_mode == AVB_HASHTREE_ERROR_MODE_LOGGING &&
      !allow_verification_error) {
    ret = AVB_SLOT_VERIFY_RESULT_ERROR_INVALID_ARGUMENT;
    goto fail;
  }
  /* Make sure passed-in AvbOps support persistent values if
   * asking for libavb to manage verity state.
   */
  if (hashtree_error_mode == AVB_HASHTREE_ERROR_MODE_MANAGED_RESTART_AND_EIO) {
    if (ops->read_persistent_value == NULL ||
        ops->write_persistent_value == NULL) {
      avb_error(
          "Persistent values required for "
          "AVB_HASHTREE_ERROR_MODE_MANAGED_RESTART_AND_EIO "
          "but are not implemented in given AvbOps.\n");
      ret = AVB_SLOT_VERIFY_RESULT_ERROR_INVALID_ARGUMENT;
      goto fail;
    }
  }
  /* Make sure passed-in AvbOps support verifying public keys and getting
   * rollback index location if not using a vbmeta partition.
   */
  if (flags & AVB_SLOT_VERIFY_FLAGS_NO_VBMETA_PARTITION) {
    if (ops->validate_public_key_for_partition == NULL) {
      avb_error(
          "AVB_SLOT_VERIFY_FLAGS_NO_VBMETA_PARTITION was passed but the "
          "validate_public_key_for_partition() operation isn't implemented.\n");
      ret = AVB_SLOT_VERIFY_RESULT_ERROR_INVALID_ARGUMENT;
      goto fail;
    }
  } else {
    avb_assert(ops->validate_vbmeta_public_key != NULL);
  }
  slot_data = avb_calloc(sizeof(AvbSlotVerifyData));
  if (slot_data == NULL) {
    ret = AVB_SLOT_VERIFY_RESULT_ERROR_OOM;
    goto fail;
  }
  slot_data->vbmeta_images =
      avb_calloc(sizeof(AvbVBMetaData) * MAX_NUMBER_OF_VBMETA_IMAGES);
  if (slot_data->vbmeta_images == NULL) {
    ret = AVB_SLOT_VERIFY_RESULT_ERROR_OOM;
    goto fail;
  }
  slot_data->loaded_partitions =
      avb_calloc(sizeof(AvbPartitionData) * MAX_NUMBER_OF_LOADED_PARTITIONS);
  if (slot_data->loaded_partitions == NULL) {
    ret = AVB_SLOT_VERIFY_RESULT_ERROR_OOM;
    goto fail;
  }
  additional_cmdline_subst = avb_new_cmdline_subst_list();
  if (additional_cmdline_subst == NULL) {
    ret = AVB_SLOT_VERIFY_RESULT_ERROR_OOM;
    goto fail;
  }
  if (flags & AVB_SLOT_VERIFY_FLAGS_NO_VBMETA_PARTITION) {
    if (requested_partitions == NULL || requested_partitions[0] == NULL) {
      avb_fatal(
          "Requested partitions cannot be empty when using "
          "AVB_SLOT_VERIFY_FLAGS_NO_VBMETA_PARTITION");
      ret = AVB_SLOT_VERIFY_RESULT_ERROR_INVALID_ARGUMENT;
      goto fail;
    }
    /* No vbmeta partition, go through each of the requested partitions... */
    for (size_t n = 0; requested_partitions[n] != NULL; n++) {
      ret = load_and_verify_vbmeta(ops,
                                   requested_partitions,
                                   ab_suffix,
                                   flags,
                                   allow_verification_error,
                                   0 /* toplevel_vbmeta_flags */,
                                   0 /* rollback_index_location */,
                                   requested_partitions[n],
                                   avb_strlen(requested_partitions[n]),
                                   NULL /* expected_public_key */,
                                   0 /* expected_public_key_length */,
                                   slot_data,
                                   &algorithm_type,
                                   additional_cmdline_subst);
      if (!allow_verification_error && ret != AVB_SLOT_VERIFY_RESULT_OK) {
        goto fail;
      }
    }
  } else {
    /* Usual path, load "vbmeta"... */
    ret = load_and_verify_vbmeta(ops,
                                 requested_partitions,
                                 ab_suffix,
                                 flags,
                                 allow_verification_error,
                                 0 /* toplevel_vbmeta_flags */,
                                 0 /* rollback_index_location */,
                                 "vbmeta",
                                 avb_strlen("vbmeta"),
                                 NULL /* expected_public_key */,
                                 0 /* expected_public_key_length */,
                                 slot_data,
                                 &algorithm_type,
                                 additional_cmdline_subst);
    if (!allow_verification_error && ret != AVB_SLOT_VERIFY_RESULT_OK) {
      goto fail;
    }
  }
  if (!result_should_continue(ret)) {
    goto fail;
  }
  /* If things check out, mangle the kernel command-line as needed. */
  if (!(flags & AVB_SLOT_VERIFY_FLAGS_NO_VBMETA_PARTITION)) {
    if (avb_strcmp(slot_data->vbmeta_images[0].partition_name, "vbmeta") != 0) {
      avb_assert(
          avb_strcmp(slot_data->vbmeta_images[0].partition_name, "boot") == 0);
      using_boot_for_vbmeta = true;
    }
  }
  /* Byteswap top-level vbmeta header since we'll need it below. */
  avb_vbmeta_image_header_to_host_byte_order(
      (const AvbVBMetaImageHeader*)slot_data->vbmeta_images[0].vbmeta_data,
      &toplevel_vbmeta);
  /* Fill in |ab_suffix| field. */
  slot_data->ab_suffix = avb_strdup(ab_suffix);
  if (slot_data->ab_suffix == NULL) {
    ret = AVB_SLOT_VERIFY_RESULT_ERROR_OOM;
    goto fail;
  }
  /* If verification is disabled, we are done ... we specifically
   * don't want to add any androidboot.* options since verification
   * is disabled.
   */
  if (toplevel_vbmeta.flags & AVB_VBMETA_IMAGE_FLAGS_VERIFICATION_DISABLED) {
    /* Since verification is disabled we didn't process any
     * descriptors and thus there's no cmdline... so set root= such
     * that the system partition is mounted.
     */
    avb_assert(slot_data->cmdline == NULL);
    // Devices with dynamic partitions won't have system partition.
    // Instead, it has a large super partition to accommodate *.img files.
    // See b/119551429 for details.
    if (has_system_partition(ops, ab_suffix)) {
      slot_data->cmdline =
          avb_strdup("root=PARTUUID=$(ANDROID_SYSTEM_PARTUUID)");
    } else {
      // The |cmdline| field should be a NUL-terminated string.
      slot_data->cmdline = avb_strdup("");
    }
    if (slot_data->cmdline == NULL) {
      ret = AVB_SLOT_VERIFY_RESULT_ERROR_OOM;
      goto fail;
    }
  } else {
    /* If requested, manage dm-verity mode... */
    AvbHashtreeErrorMode resolved_hashtree_error_mode = hashtree_error_mode;
    if (hashtree_error_mode ==
        AVB_HASHTREE_ERROR_MODE_MANAGED_RESTART_AND_EIO) {
      AvbIOResult io_ret;
      io_ret = avb_manage_hashtree_error_mode(
          ops, flags, slot_data, &resolved_hashtree_error_mode);
      if (io_ret != AVB_IO_RESULT_OK) {
        ret = AVB_SLOT_VERIFY_RESULT_ERROR_IO;
        if (io_ret == AVB_IO_RESULT_ERROR_OOM) {
          ret = AVB_SLOT_VERIFY_RESULT_ERROR_OOM;
        }
        goto fail;
      }
    }
    slot_data->resolved_hashtree_error_mode = resolved_hashtree_error_mode;
    /* Add options... */
    AvbSlotVerifyResult sub_ret;
    sub_ret = avb_append_options(ops,
                                 flags,
                                 slot_data,
                                 &toplevel_vbmeta,
                                 algorithm_type,
                                 hashtree_error_mode,
                                 resolved_hashtree_error_mode);
    if (sub_ret != AVB_SLOT_VERIFY_RESULT_OK) {
      ret = sub_ret;
      goto fail;
    }
  }
  /* Substitute $(ANDROID_SYSTEM_PARTUUID) and friends. */
  if (slot_data->cmdline != NULL && avb_strlen(slot_data->cmdline) != 0) {
    char* new_cmdline;
    new_cmdline = avb_sub_cmdline(ops,
                                  slot_data->cmdline,
                                  ab_suffix,
                                  using_boot_for_vbmeta,
                                  additional_cmdline_subst);
    if (new_cmdline != slot_data->cmdline) {
      if (new_cmdline == NULL) {
        ret = AVB_SLOT_VERIFY_RESULT_ERROR_OOM;
        goto fail;
      }
      avb_free(slot_data->cmdline);
      slot_data->cmdline = new_cmdline;
    }
  }
  if (out_data != NULL) {
    *out_data = slot_data;
  } else {
    avb_slot_verify_data_free(slot_data);
  }
  avb_free_cmdline_subst_list(additional_cmdline_subst);
  additional_cmdline_subst = NULL;
  if (!allow_verification_error) {
    avb_assert(ret == AVB_SLOT_VERIFY_RESULT_OK);
  }
  return ret;
fail:
  if (slot_data != NULL) {
    avb_slot_verify_data_free(slot_data);
  }
  if (additional_cmdline_subst != NULL) {
    avb_free_cmdline_subst_list(additional_cmdline_subst);
  }
  return ret;
}
\end{lstlisting}

\begin{lstlisting}[caption={AVB hash-tree verification method in python. 
Code 
source \cite{AndroidVerifiedBoot2Repo} - filename: avbtool.py}, 
label=lst:AVBTreeVerification, language=python, basicstyle=\scriptsize] 
class AvbHashtreeDescriptor(AvbDescriptor):
  def verify(self, image_dir, image_ext, expected_chain_partitions_map,
		image_containing_descriptor, accept_zeroed_hashtree):
	"""Verifies contents of the descriptor - used in verify_image sub-command.
	
	Arguments:
	image_dir: The directory of the file being verified.
	image_ext: The extension of the file being verified (e.g. '.img').
	expected_chain_partitions_map: A map from partition name to the
	tuple (rollback_index_location, key_blob).
	image_containing_descriptor: The image the descriptor is in.
	accept_zeroed_hashtree: If True, don't fail if hashtree or FEC data is
	zeroed out.
	
	Returns:
	True if the descriptor verifies, False otherwise.
	"""
	if not self.partition_name:
		image_filename = image_containing_descriptor.filename
		image = image_containing_descriptor
	else:
		image_filename = os.path.join(image_dir, self.partition_name + image_ext)
		image = ImageHandler(image_filename, read_only=True)
	# Generate the hashtree and checks that it matches what's in the file.
	digest_size = len(hashlib.new(self.hash_algorithm).digest())
	digest_padding = round_to_pow2(digest_size) - digest_size
	(hash_level_offsets, tree_size) = calc_hash_level_offsets(self.image_size, 
		self.data_block_size, digest_size + digest_padding)
	root_digest, hash_tree = generate_hash_tree(image, self.image_size,
										self.data_block_size,
										self.hash_algorithm, self.salt,
										digest_padding,
										hash_level_offsets,
										tree_size)
	# The root digest must match unless it is not embedded in the descriptor.
	if self.root_digest and root_digest != self.root_digest:
		sys.stderr.write('hashtree of {} does not match descriptor\n'.
			format(image_filename))
		return False
	# ... also check that the on-disk hashtree matches
	image.seek(self.tree_offset)
	hash_tree_ondisk = image.read(self.tree_size)
	is_zeroed = (self.tree_size == 0) or (hash_tree_ondisk[0:8] == b'ZeRoHaSH')
	if is_zeroed and accept_zeroed_hashtree:
		print('{}: skipping verification since hashtree is zeroed and '
			'--accept_zeroed_hashtree was given'
			.format(self.partition_name))
	else:
		if hash_tree != hash_tree_ondisk:
			sys.stderr.write('hashtree of {} contains invalid data\n'.
				format(image_filename))
			return False
		print('{}: Successfully verified {} hashtree of {} for image of {} bytes'
			.format(self.partition_name, self.hash_algorithm, image.filename,
			self.image_size))
	# TODO(zeuthen): we could also verify that the FEC stored in the image is
	# correct but this a) currently requires the 'fec' binary; and b) takes a
	# long time; and c) is not strictly needed for verification purposes as
	# we've already verified the root hash.
	return True
\end{lstlisting}

\begin{lstlisting}[caption={AVB hash-tree generation method in python. Code 
source \cite{AndroidVerifiedBoot2Repo} - filename: avbtool.py}, 
label=lst:AVBTreeGeneration, language=python, basicstyle=\scriptsize] 

def generate_hash_tree(image, image_size, block_size, hash_alg_name, salt,
	digest_padding, hash_level_offsets, tree_size):
	"""Generates a Merkle-tree for a file.
	
	Arguments:
	image: The image, as a file.
	image_size: The size of the image.
	block_size: The block size, e.g. 4096.
	hash_alg_name: The hash algorithm, e.g. 'sha256' or 'sha1'.
	salt: The salt to use.
	digest_padding: The padding for each digest.
	hash_level_offsets: The offsets from calc_hash_level_offsets().
	tree_size: The size of the tree, in number of bytes.
	
	Returns:
	A tuple where the first element is the top-level hash as bytes and the
	second element is the hash-tree as bytes.
	"""
	hash_ret = bytearray(tree_size)
	hash_src_offset = 0
	hash_src_size = image_size
	level_num = 0
	while hash_src_size > block_size:
		level_output_list = []
		remaining = hash_src_size
		while remaining > 0:
			hasher = hashlib.new(hash_alg_name, salt)
			# Only read from the file for the first level - for subsequent
			# levels, access the array we're building.
			if level_num == 0:
				image.seek(hash_src_offset + hash_src_size - remaining)
				data = image.read(min(remaining, block_size))
			else:
				offset = hash_level_offsets[level_num - 1] + hash_src_size - remaining
				data = hash_ret[offset:offset + block_size]
			hasher.update(data)
			
			remaining -= len(data)
			if len(data) < block_size:
				hasher.update(b'\0' * (block_size - len(data)))
				level_output_list.append(hasher.digest())
			if digest_padding > 0:
				level_output_list.append(b'\0' * digest_padding)
		
		level_output = b''.join(level_output_list)
		
		padding_needed = (round_to_multiple(len(level_output), block_size) - 
		len(level_output))
		level_output += b'\0' * padding_needed
		
		# Copy level-output into resulting tree.
		offset = hash_level_offsets[level_num]
		hash_ret[offset:offset + len(level_output)] = level_output
		
		# Continue on to the next level.
		hash_src_size = len(level_output)
		level_num += 1
	
	hasher = hashlib.new(hash_alg_name, salt)
	hasher.update(level_output)
	return hasher.digest(), bytes(hash_ret)
\end{lstlisting}
\newpage
\begin{lstlisting}[caption={AVBTool example output of info\_image.}, 
label=lst:InfoImage, language=sh, basicstyle=\scriptsize] 
Minimum libavb version:   1.0
Header Block:             256 bytes
Authentication Block:     320 bytes
Auxiliary Block:          2048 bytes
Public key (sha1):        ad8569837cc06521720a35357475f87283b59234
Algorithm:                SHA256\_RSA2048
Rollback Index:           1546646400
Flags:                    0
Rollback Index Location:  0
Release String:           'avbtool 1.1.0'
Descriptors:
Chain Partition descriptor:
Partition Name:          system
Rollback Index Location: 1
Public key (sha1):       46b77506c847920e3f00074c9ae005c96b6f416f
Hash descriptor:
Image Size:            30662656 bytes
Hash Algorithm:        sha256
Partition Name:        boot
Salt:                  
5af8a8b131ed8a1f927e2c576fa26b37723ed600f930239a4457ba5fe50c4626
Digest:                
6d474128d079f95b9cc40c556f81bfc3e83697be1878146d8869acf8ccd9602d
Flags:                 0
Hashtree descriptor:
Version of dm-verity:  1
Image Size:            309526528 bytes
Tree Offset:           309526528
Tree Size:             2445312 bytes
Data Block Size:       4096 bytes
Hash Block Size:       4096 bytes
FEC num roots:         2
FEC offset:            311971840
FEC size:              2473984 bytes
Hash Algorithm:        sha1
Partition Name:        product
Salt:                  
5af8a8b131ed8a1f927e2c576fa26b37723ed600f930239a4457ba5fe50c4626
Root Digest:           ded1d335ee0a135137cb95159764f2edd21f9319
Flags:                 0
Hashtree descriptor:
Version of dm-verity:  1
Image Size:            792514560 bytes
Tree Offset:           792514560
Tree Size:             6246400 bytes
Data Block Size:       4096 bytes
Hash Block Size:       4096 bytes
FEC num roots:         2
FEC offset:            798760960
FEC size:              6316032 bytes
Hash Algorithm:        sha1
Partition Name:        vendor
Salt:                  
5af8a8b131ed8a1f927e2c576fa26b37723ed600f930239a4457ba5fe50c4626
Root Digest:           d225f7c4347cb4535254025d21bc7b751e2fa216
Flags:                 0
Hash descriptor:
Image Size:            1745968 bytes
Hash Algorithm:        sha256
Partition Name:        dtbo
Salt:                  
5af8a8b131ed8a1f927e2c576fa26b37723ed600f930239a4457ba5fe50c4626
Digest:                
155003d6e8a6b9b16e7911773d5e148903670885bdd94391d043de331044bcd7
Flags:                 0
\end{lstlisting}

\newpage
\begin{lstlisting}[caption={Example output of the avb meta info\_image command.}, 
label=lst:AVBInfo, language=python, basicstyle=\scriptsize] 
./avbtool info_image --image ./1a.201005.006/vbmeta.img 
Minimum libavb version:   1.0
Header Block:             256 bytes
Authentication Block:     576 bytes
Auxiliary Block:          4096 bytes
Public key (sha1):        e8162d81621cc011ad2dc2cf10880f955500c7e9
Algorithm:                SHA256_RSA4096
Rollback Index:           1601856000
Flags:                    0
Rollback Index Location:  0
Release String:           'avbtool 1.1.0'
Descriptors:
Chain Partition descriptor:
Partition Name:          vbmeta_system
Rollback Index Location: 1
Public key (sha1):       e8162d81621cc011ad2dc2cf10880f955500c7e9
Prop: com.android.build.product.fingerprint -> 
'google/sunfish/sunfish:11/RP1A.201005.006/6828489:user/dev-keys'
Prop: com.android.build.product.os_version -> '11'
Prop: com.android.build.product.security_patch -> '2020-10-05'
Prop: com.android.build.vendor.fingerprint -> 
'google/sunfish/sunfish:11/RP1A.201005.006/6828489:user/release-keys'
Prop: com.android.build.vendor.os_version -> '11'
Prop: com.android.build.vendor.security_patch -> '2020-10-05'
Prop: com.android.build.boot.fingerprint -> 
'google/sunfish/sunfish:11/RP1A.201005.006/6828489:user/release-keys'
Prop: com.android.build.boot.os_version -> '11'
Prop: com.android.build.boot.security_patch -> '2020-10-05'
Prop: com.android.build.dtbo.fingerprint -> 
'google/sunfish/sunfish:11/RP1A.201005.006/6828489:user/release-keys'
Hash descriptor:
Image Size:            33304576 bytes
Hash Algorithm:        sha256
Partition Name:        boot
Salt: ea0769f822eed4e982da4d3d69b16b2bacb1dc1441d6db233eb42fc2ae401b22
Digest:b5a534425ddb667adbb212de571774c672f5458e6a0b2524dc47ef05379435a2
Flags:                 0
Hash descriptor:
Image Size:            1997440 bytes
Hash Algorithm:        sha256
Partition Name:        dtbo
Salt: 867b5747d12fe086925db995e189fe3be8ae3d34bf867284ad5ddac02ba0a15a
Digest:c3dc64ad78723a1020120e8b3e821897972d2a3b03c3ca8d197538660fc5ecbc
Flags:                 0
Hashtree descriptor:
Version of dm-verity:  1
Image Size:            2011947008 bytes
Tree Offset:           2011947008
Tree Size:             15847424 bytes
Data Block Size:       4096 bytes
Hash Block Size:       4096 bytes
FEC num roots:         2
FEC offset:            2027794432
FEC size:              16031744 bytes
Hash Algorithm:        sha1
Partition Name:        product
Salt: 56cf103d96aeb05ebee8cc346dbecac716e765f6e720e8fb7e6890d3cb66d9ec
Root Digest:           ca59a628aea04700db24de76c97502fa7c7b91c7
Flags:                 0
Hashtree descriptor:
Version of dm-verity:  1
Image Size:            541900800 bytes
Tree Offset:           541900800
Tree Size:             4276224 bytes
Data Block Size:       4096 bytes
Hash Block Size:       4096 bytes
FEC num roots:         2
FEC offset:            546177024
FEC size:              4325376 bytes
Hash Algorithm:        sha1
Partition Name:        vendor
Salt: 56cf103d96aeb05ebee8cc346dbecac716e765f6e720e8fb7e6890d3cb66d9ec
Root Digest:           c825b8f0ea955bb543d49be412d7079e8bc92d42
Flags:                 0
\end{lstlisting} \label{Appendix:avbOutput}
	\newpage
	
	\section{AndroGuard Permission Statistics} \label{Appendix:AndroGuardStats}
	\begin{figure}[H]
	\centering
	\includegraphics[width=0.9\linewidth]{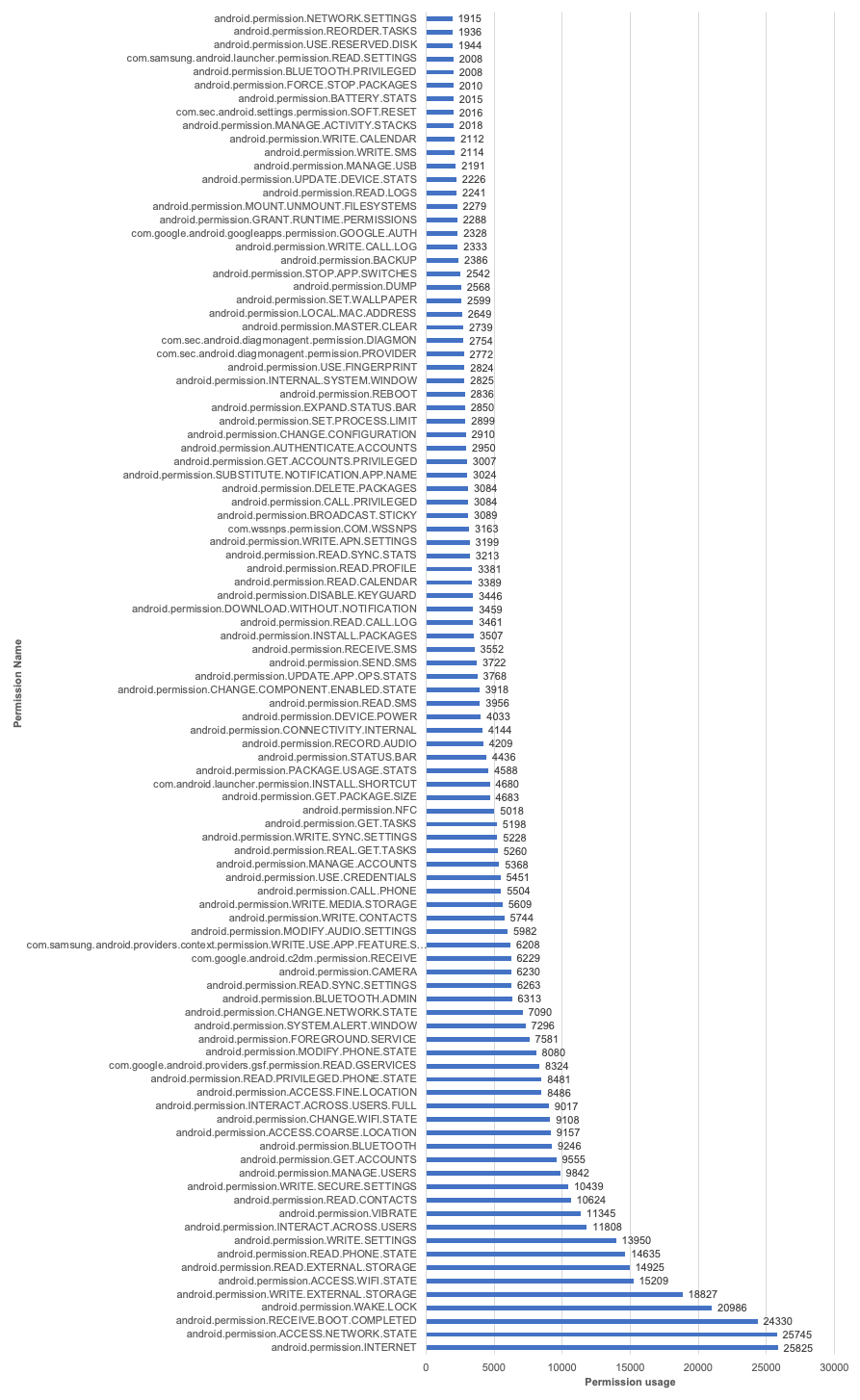}
	\caption{Android 9: Top 100 most used permissions.}
	\label{fig:22_top100_permissions_android9}
\end{figure}

	\newpage
	\section{Androwarn Native Code Loading} \label{Appendix:AndrowarnNative}
	
\fontsize{8}{10}\selectfont
	\begin{longtable}{| p{.94\textwidth} | p{.06\textwidth} |} 
		\hline
		\rowcolor[HTML]{EFEFEF} 
		\textbf{Androwarn message} & \textbf{App count} \\ \hline
		This application loads a native library & 530 \\ \hline
		This application loads a native library: 'networkstackutilsjni' & 136 \\ 
		\hline
		This application loads a native library: 
		
		'Lcom/qualcomm/ltebc/LTEBCFactory;-\textgreater{}getInstance()Lcom/qualcomm/ltebc/LTEBCFactory;'
		 & 106 \\ \hline
		This application loads a native library: 'printspooler.jni' & 136 \\ \hline
		This application loads a native library: 'imscamera.jni' & 92 \\ \hline
		This application loads a native library: 'imsmedia.jni' & 92 \\ \hline
		This application loads a native library: 'jni.pacprocessor' & 136 \\ 
		\hline
		This application loads a native library: 'nfc.nci.jni' & 136 \\ \hline
		This application loads a native library: 'wfds' & 136 \\ \hline
		This application loads a native library: 'bluetooth.jni' & 136 \\ \hline
		This application loads a native library: 
		'Ljava/lang/String;-\textgreater{}valueOf(Ljava/lang/Object;)Ljava/lang/String;'
		 & 170 \\ \hline
		This application loads a native library: 'motricity' & 34 \\ \hline
		This application loads a native library: 
		
		'Lorg/chromium/net/impl/CronetLibraryLoader;-\textgreater{}a(Ljava/lang/Runnable;)V'
		 & 34 \\ \hline
		This application loads a native library: 'brotli' & 34 \\ \hline
		This application loads a native library: 'conscrypt.gmscore.jni' & 68 \\ 
		\hline
		This application loads a native library: 'conscrypt.jni' & 68 \\ \hline
		This application loads a native library: 'game.sdk.device.info.jni' & 34 
		\\ \hline
		This application loads a native library: 'gvr' & 204 \\ \hline
		This application loads a native library: 'tensorflowlite.flex.jni' & 68 \\ 
		\hline
		This application loads a native library: 'tensorflowlite.jni' & 68 \\ \hline
		This application loads a native library: 'AppDataSearch' & 34 \\ \hline
		This application loads a native library: 'curve25519' & 68 \\ \hline
		This application loads a native library: 'inertial-anchor-jni' & 34 \\ \hline
		This application loads a native library: 'vcdiffjni' & 34 \\ \hline
		This application loads a native library: 'wearable-selector' & 34 \\ 
		\hline
		This application loads a native library: 
		'Landroid/os/ConditionVariable;-\textgreater{}open()V' & 66 \\ \hline
		This application loads a native library: 'barhopper' & 124 \\ \hline
		This application loads a native library: 
		'Lorg/chromium/net/impl/k;-\textgreater{}a()Lorg/chromium/net/impl/cm;'
		 & 34 \\ \hline
		This application loads a native library: 'android.dtmf' & 34 \\ \hline
		This application loads a native library: 'arcore.sdk.jni' & 34 \\ \hline
		This application loads a native library: 'framesequence' & 130 \\ \hline
		This application loads a native library: 'google.speech.micro.jni' & 34 
		\\ \hline
		This application loads a native library: 'lens.image.util' & 34 \\ \hline
		This application loads a native library: 'offline.actions.jni' & 34 \\ \hline
		This application loads a native library: 'opuscodec' & 34 \\ \hline
		This application loads a native library: 'sbcdecoder.jni' & 34 \\ \hline
		This application loads a native library: 'sense' & 26 \\ \hline
		This application loads a native library: 'elmyra' & 34 \\ \hline
		This application loads a native library: 'dmengine' & 34 \\ \hline
		This application loads a native library: 'Brella' & 34 \\ \hline
		This application loads a native library: 
		'Ljava/util/Iterator;-\textgreater{}next()Ljava/lang/Object;' & 34 \\ \hline
		This application loads a native library: 'aiai-annotators' & 34 \\ \hline
		This application loads a native library: 'cpuutils' & 34 \\ \hline
		This application loads a native library: 'ocr' & 34 \\ \hline
		This application loads a native library: 'particle-extractor.jni' & 34 \\ 
		\hline
		This application loads a native library: 'textclassifier3.jni.aosp' & 34 \\ 
		\hline
		This application loads a native library: 'textclassifier3.jni.aosp-size' & 
		34 \\ \hline
		This application loads a native library: 'textclassifier3.jni.gmscore' & 
		34 \\ \hline
		This application loads a native library: 'textclassifier3.jni.tclib' & 34 \\ 
		\hline
		This application loads a native library: 'tensorflow.jni' & 68 \\ \hline
		This application loads a native library: 'flash' & 34 \\ \hline
		This application loads a native library: 'sketchology.native' & 34 \\ 
		\hline
		This application loads a native library: 'Lasrp;-\textgreater{}c()Laspa;' 
		& 18 \\ \hline
		This application loads a native library: 
		'Ljava/lang/System;-\textgreater{}loadLibrary(Ljava/lang/String;)V' & 
		28 \\ \hline
		This application loads a native library: 'filterframework.jni' & 68 \\ \hline
		This application loads a native library: 'flacJNI' & 34 \\ \hline
		This application loads a native library: 'native' & 68 \\ \hline
		This application loads a native library: 'oliveoil' & 62 \\ \hline
		This application loads a native library: 'tonemap' & 28 \\ \hline
		This application loads a native library: 'chromium.android.linker' & 68 
		\\ \hline
		This application loads a native library: 
		'webviewchromium.plat.support' & 64 \\ \hline
		This application loads a native library: 'GcamDepsJni' & 22 \\ \hline
		This application loads a native library: 'RSSupport' & 34 \\ \hline
		This application loads a native library: 'RSSupportIO' & 34 \\ \hline
		This application loads a native library: 'SeeDarkJni' & 34 \\ \hline
		This application loads a native library: 'aspect.scorer.jni' & 34 \\ \hline
		This application loads a native library: 'blasV8' & 34 \\ \hline
		This application loads a native library: 'bufferflinger' & 34 \\ \hline
		This application loads a native library: 'cyclops' & 34 \\ \hline
		This application loads a native library: 'entity-clusterer.jni' & 22 \\ \hline
		This application loads a native library: 'halide.hexagon.host' & 34 \\ 
		\hline
		This application loads a native library: 'halide.hexagon.host.pixel1' & 
		30 \\ \hline
		This application loads a native library: 'image' & 22 \\ \hline
		This application loads a native library: 'lightcycle' & 34 \\ \hline
		This application loads a native library: 'link.engine.jni' & 22 \\ \hline
		This application loads a native library: 'refocus' & 30 \\ \hline
		This application loads a native library: 'rsjni' & 34 \\ \hline
		This application loads a native library: 'semanticlift-annotators' & 14 \\ 
		\hline
		This application loads a native library: 'smartcapture.native' & 34 \\ 
		\hline
		This application loads a native library: 'textclassifier.jni' & 22 \\ \hline
		This application loads a native library: 'textclassifier.jni.noicu' & 22 \\ 
		\hline
		This application loads a native library: 'vision.face.jni' & 30 \\ \hline
		This application loads a native library: 'giftranscode' & 34 \\ \hline
		This application loads a native library: 'hobbes.jni' & 34 \\ \hline
		This application loads a native library: 'language.detector.jni' & 34 \\ 
		\hline
		This application loads a native library: 'sensitive.classifier.jni' & 34 \\ 
		\hline
		This application loads a native library: 'gdx' & 34 \\ \hline
		This application loads a native library: 'wallpapers-breel-jni' & 26 \\ 
		\hline
		This application loads a native library: 'WVphoneAPI' & 34 \\ \hline
		This application loads a native library: 'wvmediadrm' & 34 \\ \hline
		This application loads a native library: 'drishti.jni.native' & 34 \\ \hline
		This application loads a native library: 'eko' & 34 \\ \hline
		This application loads a native library: 'gvr.audio' & 34 \\ \hline
		This application loads a native library: 'opusJNI' & 34 \\ \hline
		This application loads a native library: 'vpx' & 34 \\ \hline
		This application loads a native library: 'vpxJNI' & 34 \\ \hline
		This application loads a native library: 'webp.android' & 34 \\ \hline
		This application loads a native library: 'bitmap.parcel' & 34 \\ \hline
		This application loads a native library: 'cello.native' & 34 \\ \hline
		This application loads a native library: 'foxit' & 34 \\ \hline
		This application loads a native library: 'facenet' & 34 \\ \hline
		This application loads a native library: 'borders.scone.leveldb.jni' & 
		12 \\ \hline
		This application loads a native library: 'gpu-retouching-jni' & 12 \\ 
		\hline
		This application loads a native library: 'geswallpapers-jni' & 8 \\ \hline
		This application loads a native library: 
		'Laemk;-\textgreater{}c()Ljava/lang/String;' & 6 \\ \hline
		This application loads a native library: 'crashreporterer' & 6 \\ \hline
		This application loads a native library: 'Lasrr;-\textgreater{}c()Laspc;' 
		& 10 \\ \hline
		\caption{Androwarn native code loading report for Android10 
			firmware.}
		\label{tab:AndrowarnNativeCode}
	\end{longtable}
	\newpage
	
	\section{FirmareDroid Rest API} \label{Appendix:RestAPI}
	{\footnotesize\tabcolsep=3pt
\begin{longtable}[H]{| p{.08\textwidth} | p{.45\textwidth} | p{.47\textwidth} |}
\hline
\rowcolor[HTML]{EFEFEF} 
\textbf{\begin{tabular}[c]{@{}l@{}}HTTP\\ Method\end{tabular}} & \textbf{Route} & 
\textbf{Description} \\ \hline
\rowcolor[HTML]{EFEFEF} 
\multicolumn{3}{|c|}{\cellcolor[HTML]{EFEFEF}\textit{/v1/androguard/}} \\ \hline
POST & /\{mode\} & Analyse Android apps with AndroGuard. \\ \hline
POST & /app\_certificate/download/ & Download a certificate in DER or PEM 
format. \\ \hline
GET & /count/ & \begin{tabular}[c]{@{}l@{}}Gets the number of AndroGuard 
reports in the \\ database.\end{tabular} \\ \hline
GET & /count/app\_certificates & \begin{tabular}[c]{@{}l@{}}Gets the number of 
AndroGuard certificate reports \\ in the database.\end{tabular} \\ \hline
GET & /count/string\_analysis & \begin{tabular}[c]{@{}l@{}}Gets the number of 
AndroGuard string analysis \\ reports in the database.\end{tabular} \\ \hline
POST & /meta\_string\_analysis/ & \begin{tabular}[c]{@{}l@{}}Starts the meta 
analysis of AndroGuard string \\ analysis.\end{tabular} \\ \hline
POST & /set\_packagename/ & \begin{tabular}[c]{@{}l@{}}Copies AndroGuard 
packagenames to the Android \\ app model.\end{tabular} \\ \hline
\rowcolor[HTML]{EFEFEF} 
\multicolumn{3}{|c|}{\cellcolor[HTML]{EFEFEF}\textit{/v1/androwarn/}} \\ \hline
POST & /\{mode\} & Analysis apps with Androwarn. \\ \hline
GET & count/ & Gets the number of Androwarn reports in the database. \\ \hline
GET & /\{androwarn\_id\} & Gets an Androwarn report as json. \\ \hline
\rowcolor[HTML]{EFEFEF} 
\multicolumn{3}{|c|}{\cellcolor[HTML]{EFEFEF}\textit{/v1/auth/}} \\ \hline
POST & /auth/login/ & Get a JWT token for a specific account. \\ \hline
POST & /auth/signup/ & Create an account for JWT based authentication. \\ \hline
\rowcolor[HTML]{EFEFEF} 
\multicolumn{3}{|c|}{\cellcolor[HTML]{EFEFEF}\textit{/v1/comparer/}} \\ \hline
POST & /firmware\_compare/ & \begin{tabular}[c]{@{}l@{}}Compares (diff) the files 
from the first firmware \\ with the second.\end{tabular} \\ \hline
\rowcolor[HTML]{EFEFEF} 
\multicolumn{3}{|c|}{\cellcolor[HTML]{EFEFEF}\textit{/v1/firmware/}} \\ \hline
GET & /by\_md5/\{md5\} & Gets a json report for the firmware. \\ \hline
DELETE & /delete\_all/ & \begin{tabular}[c]{@{}l@{}}Deletes all firmware entries 
from the database \\ and move the firmware archives to the import 
folder.\end{tabular} \\ \hline
POST & /detect\_version/\{mode\} & Attempts to detect the Android version of the 
firmware. \\ \hline
GET & /download/build\_prop\_zip & Download all build.props files as zip file. \\ 
\hline
GET & /download/\{firmware\_id\} & Download a firmware as zip archive. \\ \hline
POST & /get\_app\_id\_list/ & Get a list of Android app id's for a firmware. \\ \hline
GET & /get\_import\_queue\_size/ & Gets the number of files in the import folder. 
\\ \hline
GET & /mass\_import/ & Starts the import of firmware files from the import folder. 
\\ \hline
GET & /\{android\_version\} & \begin{tabular}[c]{@{}l@{}}Get a list of firmware-id's 
with the given version \\ (including subversions).\end{tabular} \\ \hline
\rowcolor[HTML]{EFEFEF} 
\multicolumn{3}{|c|}{\cellcolor[HTML]{EFEFEF}\textit{/v1/jobs/}} \\ \hline
POST & /cancel/\{job\_id\} & Cancel a RQ job. \\ \hline
GET & /get\_result\_file/\{job\_id\} & Get all the currently running job-id's. \\ \hline
POST & /status/\{job\_id\} & Get the status of a background job. \\ \hline
\rowcolor[HTML]{EFEFEF} 
\multicolumn{3}{|c|}{\cellcolor[HTML]{EFEFEF}\textit{/v1/qark/}} \\ \hline
GET & /count/ & Gets the number of Qark reports in the database \\ \hline
POST & /download/json\_reports/\{mode\} & Gets a Qark report as json. \\ \hline
POST & /\{mode\} & \begin{tabular}[c]{@{}l@{}}Analysis apps with Quick Android 
Review Kit \\ (QARK) and create a report\end{tabular} \\ \hline
\rowcolor[HTML]{EFEFEF} 
\multicolumn{3}{|c|}{\cellcolor[HTML]{EFEFEF}\textit{/v1/statistics/}} \\ \hline
POST & \begin{tabular}[c]{@{}l@{}}/androguard/create\_certificate\_report/\\ 
\{mode\}/\{report\_name\}\end{tabular} & Create a statistical report for AndroGuard 
certificate data. \\ \hline
POST & \begin{tabular}[c]{@{}l@{}}/androguard/create\_report\_plots/\\ 
\{androguard\_statistics\_report\_id\}\end{tabular} & Create plots for a Androguard 
statistics report. \\ \hline
POST & \begin{tabular}[c]{@{}l@{}}/v1/statistics/androguard/\\ 
create\_statistics\_report/\\ \{mode\}/\{report\_name\}\end{tabular} & Create a 
statistical report for AndroGuard data \\ \hline
POST & \begin{tabular}[c]{@{}l@{}}/v1/statistics/androguard/\\ 
create\_string\_analysis\_report/\\ \{report\_name\}\end{tabular} & Create a 
statistical report for AndroGuard string meta data. \\ \hline
POST & \begin{tabular}[c]{@{}l@{}}/v1/statistics/androwarn/\\ 
create\_statistics\_report/\\ \{mode\}/\\ \{report\_name\}\end{tabular} & Create a 
statistical report for androwarn data \\ \hline
POST & \begin{tabular}[c]{@{}l@{}}/v1/statistics/apkid/\\ 
create\_statistics\_report/\\ \{mode\}/\{report\_name\}\end{tabular} & Create a 
statistical report for apkid data \\ \hline
GET & \begin{tabular}[c]{@{}l@{}}/v1/statistics/download/\\ grouped\_by\_version/\\ 
filter\_duplicated\_packagename/\\ \{reference\_file\_id\}/\\ 
\{get\_count\}\end{tabular} & \begin{tabular}[c]{@{}l@{}}Download a reference file 
for a statistics report \\ filtered by unqiue packagenames\end{tabular} \\ \hline
GET & \begin{tabular}[c]{@{}l@{}}/v1/statistics/download/\\ grouped\_by\_version/\\ 
\{reference\_file\_id\}/\\ \{add\_meta\_data\}\end{tabular} & Download a reference 
file for a statistics report \\ \hline
GET & \begin{tabular}[c]{@{}l@{}}/v1/statistics/download/images/\\ 
\{image\_file\_id\}\end{tabular} & Download a reference file for a statistics report \\ 
\hline
POST & \begin{tabular}[c]{@{}l@{}}/v1/statistics/download/references/\\ 
\{reference\_file\_id\}\end{tabular} & Download a reference file for a statistics report 
\\ \hline
POST & \begin{tabular}[c]{@{}l@{}}/v1/statistics/firmware/\\ 
create\_statistics\_report/\\ \{mode\}/\{report\_name\}\end{tabular} & Create a 
statistical report for firmware data \\ \hline
POST & \begin{tabular}[c]{@{}l@{}}/v1/statistics/qark/\\ create\_statistics\_report/\\ 
\{mode\}/\{report\_name\}\end{tabular} & Create a statistical report for qark data \\ 
\hline
POST & \begin{tabular}[c]{@{}l@{}}/v1/statistics/virustotal/\\ 
create\_statistics\_report/\\ \{mode\}/\{report\_name\}\end{tabular} & Create a 
statistical report for VirusTotal data \\ \hline
\rowcolor[HTML]{EFEFEF} 
\multicolumn{3}{|c|}{\cellcolor[HTML]{EFEFEF}\textit{/v1/virustotal/}} \\ \hline
POST & /add\_key/ & Add a VirusTotal api key to a user account \\ \hline
GET & /count/ & Gets the number of VirusTotal reports in the database \\ \hline
GET & /\{file\_uuid\_reference\} & /v1/virustotal/\{file\_uuid\_reference\} \\ \hline
POST & /\{mode\} & /v1/virustotal/\{mode\} \\ \hline
\rowcolor[HTML]{EFEFEF} 
\multicolumn{3}{|c|}{\cellcolor[HTML]{EFEFEF}\textit{/v1/apkid/}} \\ \hline
GET & /v1/apkid/count/ & Gets the number of APKiD reports in the database \\ 
\hline
POST & /v1/apkid/\{mode\} & Scan the given apps with APKiD \\ \hline
\rowcolor[HTML]{EFEFEF} 
\multicolumn{3}{|c|}{\cellcolor[HTML]{EFEFEF}\textit{/v1/android\_app/}} \\ \hline
POST & /by\_id/ & Get Android app meta data as json report. \\ \hline
GET & /count/ & Gets the number of Android apps in the database. \\ \hline
GET & /download/\{android\_app\_id\} & Download the app with the given id. \\ 
\hline
\rowcolor[HTML]{EFEFEF} 
\multicolumn{3}{|c|}{\cellcolor[HTML]{EFEFEF}\textit{/v1/firmware\_file/}} \\ \hline
POST & /exporter/export\_file\_by\_id/ & Exports the specific firmware files. \\ 
\hline
POST & \begin{tabular}[c]{@{}l@{}}/exporter/export\_files\_by\_regex/\\ 
\{firmware\_file\_name\_regex\}/\{mode\}\end{tabular} & 
\begin{tabular}[c]{@{}l@{}}Exports the specific firmware files from the given \\ 
firmware list.\end{tabular} \\ \hline
POST & /index\_firmware\_files/\{mode\} & \begin{tabular}[c]{@{}l@{}}Creates an 
index of the files within the firmware \\ images.\end{tabular} \\ \hline
\rowcolor[HTML]{EFEFEF} 
\multicolumn{3}{|c|}{\cellcolor[HTML]{EFEFEF}\textit{/v1/fuzzy\_hashing/}} \\ \hline
POST & \begin{tabular}[c]{@{}l@{}}/v1/fuzzy\_hashing/create\_hashes/\\ 
firmware/\{mode\}\end{tabular} & Creates fuzzy hashes for every file of the given 
firmware. \\ \hline
DELETE & /v1/fuzzy\_hashing/delete\_all/\{mode\} & Deletes all ssDeep/tlsh 
digests from the given firmware. \\ \hline
POST & \begin{tabular}[c]{@{}l@{}}/v1/fuzzy\_hashing/\\ 
download\_cluster\_analysis/\\ 
\{cluster\_analysis\_id\}/\{fuzzy\_hash\_type\}\end{tabular} & Download a graph file 
for a cluster analysis. \\ \hline
POST & /v1/fuzzy\_hashing/ssdeep/compare/ & Compare two ssdeep fuzzy 
hashes. \\ \hline
POST & \begin{tabular}[c]{@{}l@{}}/v1/fuzzy\_hashing/ssdeep/\\ 
create\_cluster\_analysis/\\ \{regex\_filter\}/\{mode\}\end{tabular} & Creates a 
cluster analysis for ssdeep digests. \\ \hline
POST & /v1/fuzzy\_hashing/tlsh/compare/ & Compare two tlsh fuzzy hashes with 
each other. \\ \hline
POST & \begin{tabular}[c]{@{}l@{}}/v1/fuzzy\_hashing/tlsh/\\ 
create\_cluster\_analysis/\\ \{regex\_filter\}/\{mode\}/\\ \{distance\_threshold\}/\\ 
\{compare\_mode\}/\\ \{tlsh\_similiarity\_lookup\_id\}/\\ \{description\}\end{tabular} & 
Creates a clustering analysis for tlsh digests. \\ \hline
POST & \begin{tabular}[c]{@{}l@{}}/v1/fuzzy\_hashing/tlsh/\\ 
create\_similarity\_lookup/\end{tabular} & \begin{tabular}[c]{@{}l@{}}Create a tlsh 
similarity lookup \\ table for all tlsh hashes.\end{tabular} \\ \hline
POST & \begin{tabular}[c]{@{}l@{}}/v1/fuzzy\_hashing/tlsh/find\_similar/\\ 
\{cluster\_analysis\_id\}/\{tlsh\_hash\_id\}\end{tabular} & Find similar tlsh hashes 
for a given hash. \\ \hline
POST & \begin{tabular}[c]{@{}l@{}}/v1/fuzzy\_hashing/tlsh/testing/\\ 
table\_creator\_test/\\ \{mode\}/\{regex\_string\}\\ /\{number\_of\_test\_files\}/\\ 
\{distance\_threshold\}\end{tabular} & \begin{tabular}[c]{@{}l@{}}Creates a static 
number of tlsh lookup tables and \\ clusters with various sizes for evaluation 
purposes.\end{tabular} \\ \hline
\rowcolor[HTML]{EFEFEF} 
\multicolumn{3}{|c|}{\cellcolor[HTML]{EFEFEF}\textit{/v1/cleanup/}} \\ \hline
POST & /androguard\_certificate\_references/ & \begin{tabular}[c]{@{}l@{}}Checks 
if all certificate references are correctly set \\ and correct them if 
necessary.\end{tabular} \\ \hline
GET & /der\_certificates/ & Bugfix: Adds missing certificate file for an Android 
app. \\ \hline
DELETE & /firmware\_files/ & Deletes all dead firmware files references \\ \hline
DELETE & /firmware\_references/ & \begin{tabular}[c]{@{}l@{}}Checks if the 
android app references are still valid \\ and removes invalid references\end{tabular} 
\\ \hline
GET & /firmware\_version/ & Renew all firmware versions \\ \hline
DELETE & /static\_tool\_references & \begin{tabular}[c]{@{}l@{}}Checks if the 
android app references are still valid \\ and removes invalid references\end{tabular} 
\\ \hline
\rowcolor[HTML]{EFEFEF} 
\multicolumn{3}{|c|}{\cellcolor[HTML]{EFEFEF}\textit{/v1/adb/}} \\ \hline
POST & /emulator\_control/run\_monkey/ & Run monkey test for an Android app. 
\\ \hline
\rowcolor[HTML]{EFEFEF} 
\multicolumn{3}{|c|}{\cellcolor[HTML]{EFEFEF}\textit{/v1/frida/}} \\ \hline
POST & /install\_server/ & Install a frida server on a device. \\ \hline
POST & \begin{tabular}[c]{@{}l@{}}/run\_frida\_smoke\_test/\\ \{device\_ip\}/\\ 
\{frida\_port\}\end{tabular} & Test if connection to frida server works. \\ \hline
POST & /upload\_script/ & Upload a frida script to the database. \\ \hline
	\caption{Complete List of FirmareDroid REST API endpoints with description.}
	\label{tab:Rest_endpoints_all}
\end{longtable}
}
	
	\newpage
	\section{Literature Search Results} \label{Appendix:LiteratureSearchResults}
	\begin{longtable}[c]{| p{.8\textwidth} | l | l |}
	\hline
	\rowcolor[HTML]{EFEFEF} 
	\textbf{Title} &
	\textbf{Year} &
	\textbf{Relvant} \\ \hline
	\endhead
	\href{https://ieeexplore.ieee.org/abstract/document/8984540/}{A
	   ransomware detection method using fuzzy hashing for mitigating the risk of   
	occlusion of information systems} &
	2019 &
	TRUE \\ \hline
	\href{https://ieeexplore.ieee.org/abstract/document/8858803/}{Cyberthreat
	   Hunting-Part 1: triaging ransomware using fuzzy hashing, import hashing 
	and   YARA rules} &
	2019 &
	TRUE \\ \hline
	\href{https://ieeexplore.ieee.org/abstract/document/8858825/}{Cyberthreat
	   hunting-part 2: Tracking ransomware threat actors using fuzzy hashing and   
	fuzzy c-means clustering} &
	2019 &
	TRUE \\ \hline
	\href{https://ieeexplore.ieee.org/abstract/document/9177636/}{Fuzzy-Import
	   Hashing: A malware analysis approach} &
	2020 &
	TRUE \\ \hline
	\href{https://ieeexplore.ieee.org/abstract/document/9308189/}{Fuzzy
	   hashing aided enhanced YARA rules for malware triaging} &
	2020 &
	TRUE \\ \hline
	\href{https://ieeexplore.ieee.org/abstract/document/9003148/}{Lockout-Tagout
	   Ransomware: A detection method for ransomware using fuzzy hashing and   
	clustering} &
	2019 &
	TRUE \\ \hline
	\href{https://ieeexplore.ieee.org/abstract/document/9002773/}{Augmented
	   YARA rules fused with fuzzy hashing in ransomware triaging} &
	2019 &
	TRUE \\ \hline
	\href{https://www.usenix.org/conference/cset15/workshop-program/presentation/li}{Experimental
	   study of fuzzy hashing in malware clustering analysis} &
	2015 &
	TRUE \\ \hline
	\href{https://ieeexplore.ieee.org/abstract/document/7575849/}{Detection
	   of malicious portable executables using evidence combinational theory with   
	fuzzy hashing} &
	2016 &
	TRUE \\ \hline
	\href{https://commons.erau.edu/adfsl/2012/wednesday/15/}{A   
	fuzzy hashing approach based on random sequences and hamming distance} &
	2012 &
	TRUE \\ \hline
	\href{https://www.sciencedirect.com/science/article/pii/S1742287613000789}{F2S2:
	   Fast forensic similarity search through indexing piecewise hash signatures} &
	2013 &
	TRUE \\ \hline
	\href{https://core.ac.uk/download/pdf/46911337.pdf}{Security   
	Aspects of Fuzzy Hashing} &
	2011 &
	TRUE \\ \hline
	\href{https://ieeexplore.ieee.org/abstract/document/7847157/}{Forensic
	   malware analysis: The value of fuzzy hashing algorithms in identifying   
	similarities} &
	2016 &
	TRUE \\ \hline
	\href{https://digital.csic.es/handle/10261/135088}{A   new edit 
	distance for fuzzy hashing applications} &
	2015 &
	TRUE \\ \hline
	\href{https://link.springer.com/chapter/10.1007/978-3-319-49445-6\_15}{Malware
	   characterization using windows API call sequences} &
	2016 &
	FALSE \\ \hline
	\href{https://ieeexplore.ieee.org/abstract/document/6754635/}{TLSH--a
	   locality sensitive hash} &
	2013 &
	TRUE \\ \hline
	\href{https://www.koreascience.or.kr/article/JAKO201208138130922.page}{Research
	   on the classification model of similarity malware using fuzzy hash} &
	2012 &
	TRUE \\ \hline
	\href{https://link.springer.com/chapter/10.1007/978-3-642-19931-8\_15}{Web
	   document duplicate detection using fuzzy hashing} &
	2011 &
	FALSE \\ \hline
	\href{https://ieeexplore.ieee.org/abstract/document/6754635/}{TLSH--a
	   locality sensitive hash} &
	2013 &
	TRUE \\ \hline
	\href{https://dl.acm.org/doi/abs/10.1145/3176258.3176306}{Beyond
	   precision and recall: understanding uses (and misuses) of similarity hashes   
	in binary analysis} &
	2018 &
	TRUE \\ \hline
	\href{https://ieeexplore.ieee.org/abstract/document/7847157/}{Forensic
	   malware analysis: The value of fuzzy hashing algorithms in identifying   
	similarities} &
	2016 &
	TRUE \\ \hline
	\href{https://dl.acm.org/doi/abs/10.1145/3077286.3077289}{A   
	comparison of fuzzy hashes: evaluation, guidelines, and future suggestions} &
	2017 &
	TRUE \\ \hline
	\href{https://link.springer.com/chapter/10.1007/978-3-662-45670-5\_19}{Using
	   randomization to attack similarity digests} &
	2014 &
	TRUE \\ \hline
	\href{https://link.springer.com/chapter/10.1007/978-3-030-04503-6\_19}{Dynamic
	   Whitelisting Using Locality Sensitive Hashing} &
	2018 &
	TRUE \\ \hline
	\href{https://link.springer.com/chapter/10.1007/978-3-319-99277-8\_12}{Hashing
	   Incomplete and Unordered Network Streams} &
	2018 &
	FALSE \\ \hline
	\href{https://ieeexplore.ieee.org/abstract/document/7371622/}{Document
	   Copy Detection Using the Improved Fuzzy Hashing} &
	2015 &
	FALSE \\ \hline
	\href{https://ieeexplore.ieee.org/abstract/document/9191381/}{Hac-t
	   and fast search for similarity in security} &
	2020 &
	TRUE \\ \hline
	\href{https://ieeexplore.ieee.org/abstract/document/9280546/}{Scalable
	   Malware Clustering using Multi-Stage Tree Parallelization} &
	2020 &
	TRUE \\ \hline
	\href{https://www.mdpi.com/2076-3417/10/15/5173}{Combined  
	 kNN Classification and hierarchical similarity hash for fast malware   
	detection} &
	2020 &
	TRUE \\ \hline
	\href{https://ieeexplore.ieee.org/abstract/document/9110432/}{Clustering
	   IoT Malware based on Binary Similarity} &
	2020 &
	TRUE \\ \hline
	\href{http://webdiis.unizar.es/$\sim$ricardo/files/TFGs/FuzzyHashingProcesos.pdf}{Evaluación
	   de algoritmos de fuzzy hashing para similitud entre procesos} &
	2017 &
	FALSE \\ \hline
	\href{https://www.mesalab.cn/managercode/achievements
		/\%E8\%AE\%BA\%E6\%96\%87\%EF\%BC\%9ASFH\%20Hashing
		\%20Unordered\%20and\%20Incomplete\%20network\%20traffic\%20on\%20the\%20fly.pdf}{:
		Hashing Unordered and Incomplete Network Streams on the Fly} &
   	2017 &
	FALSE \\ \hline
	\href{https://ieeexplore.ieee.org/abstract/document/8658199/}{An
	   entropy-based solution for identifying android packers} &
	2019 &
	TRUE \\ \hline
	\href{https://mountainscholar.org/handle/11124/173074}{Similarity
	   hashing of malware on IoT devices} &
	2019 &
	TRUE \\ \hline
	\href{https://arxiv.org/abs/1901.02438}{Using   fuzzy bits and 
	neural networks to partially invert few rounds of some   cryptographic hash 
	functions} &
	2019 &
	TRUE \\ \hline
	\href{http://commons.erau.edu/cgi/viewcontent.cgi?article=1379\&context=jdfsl}{Bytewise
	   approximate matching: the good, the bad, and the unknown} &
	2016 &
	TRUE \\ \hline
	\href{https://ieeexplore.ieee.org/abstract/document/6754635/}{TLSH--a
	   locality sensitive hash} &
	2013 &
	TRUE \\ \hline
	\href{https://dl.acm.org/doi/abs/10.1145/3176258.3176306}{Beyond
	   precision and recall: understanding uses (and misuses) of similarity hashes   
	in binary analysis} &
	2018 &
	TRUE \\ \hline
	\href{https://ieeexplore.ieee.org/abstract/document/7847157/}{Forensic
	   malware analysis: The value of fuzzy hashing algorithms in identifying   
	similarities} &
	2016 &
	TRUE \\ \hline
	\href{https://dl.acm.org/doi/abs/10.1145/3077286.3077289}{A   
	comparison of fuzzy hashes: evaluation, guidelines, and future suggestions} &
	2017 &
	TRUE \\ \hline
	\href{https://link.springer.com/chapter/10.1007/978-3-662-45670-5\_19}{Using
	   randomization to attack similarity digests} &
	2014 &
	TRUE \\ \hline
	\href{https://link.springer.com/chapter/10.1007/978-3-030-04503-6\_19}{Dynamic
	   Whitelisting Using Locality Sensitive Hashing} &
	2018 &
	TRUE \\ \hline
	\href{https://link.springer.com/chapter/10.1007/978-3-319-99277-8\_12}{Hashing
	   Incomplete and Unordered Network Streams} &
	2018 &
	FALSE \\ \hline
	\href{https://ieeexplore.ieee.org/abstract/document/7371622/}{Document
	   Copy Detection Using the Improved Fuzzy Hashing} &
	2015 &
	TRUE \\ \hline
	\href{https://ieeexplore.ieee.org/abstract/document/9191381/}{Hac-t
	   and fast search for similarity in security} &
	2020 &
	TRUE \\ \hline
	\href{https://ieeexplore.ieee.org/abstract/document/9280546/}{Scalable
	   Malware Clustering using Multi-Stage Tree Parallelization} &
	2020 &
	TRUE \\ \hline
	\href{https://www.mdpi.com/2076-3417/10/15/5173}{Combined  
	 kNN Classification and hierarchical similarity hash for fast malware   
	detection} &
	2020 &
	TRUE \\ \hline
	\href{https://ieeexplore.ieee.org/abstract/document/9110432/}{Clustering
	   IoT Malware based on Binary Similarity} &
	2020 &
	TRUE \\ \hline
	\href{http://webdiis.unizar.es/$\sim$ricardo/files/TFGs/FuzzyHashingProcesos.pdf}{Evaluación
	   de algoritmos de fuzzy hashing para similitud entre procesos} &
	2017 &
	FALSE \\ \hline
	\multicolumn{2}{|l|}{\href{https://www.mesalab.cn/managercode/achievements/\%E8\%AE\%BA\%E6\%96\%87\%EF\%BC\%9ASFH\%20Hashing\%20Unordered\%20and\%20Incomplete\%20network\%20traffic\%20on\%20the\%20fly.pdf}{SFH:
	   Hashing Unordered and Incomplete Network Streams on the Fly}} &
	FALSE \\ \hline
	\href{https://ieeexplore.ieee.org/abstract/document/8658199/}{An
	   entropy-based solution for identifying android packers} &
	2019 &
	TRUE \\ \hline
	\href{https://mountainscholar.org/handle/11124/173074}{Similarity
	   hashing of malware on IoT devices} &
	2019 &
	TRUE \\ \hline
	\href{https://arxiv.org/abs/1901.02438}{Using   fuzzy bits and 
	neural networks to partially invert few rounds of some   cryptographic hash 
	functions} &
	2019 &
	TRUE \\ \hline
	\href{http://commons.erau.edu/cgi/viewcontent.cgi?article=1379\&context=jdfsl}{Bytewise
	   approximate matching: the good, the bad, and the unknown} &
	2016 &
	TRUE \\ \hline
	\href{https://www.sciencedirect.com/science/article/pii/S1742287617302566}{Lempel-Ziv
	   Jaccard Distance, an effective alternative to ssdeep and sdhash} &
	2018 &
	TRUE \\ \hline
	\href{https://content.sciendo.com/view/journals/slgr/56/1/article-p45.xml}{Application
	   of Normalized Compression Distance and Lempel-Ziv Jaccard Distance in   
	Micro-electrode Signal Stream Classification for the Surgical Treatment of …} &
	2018 &
	FALSE \\ \hline
	\href{https://dl.acm.org/doi/abs/10.1145/3128572.3140446}{Malware
	   classification and class imbalance via stochastic hashed LZJD} &
	2017 &
	TRUE \\ \hline
	\href{https://mountainscholar.org/handle/11124/173074}{Similarity
	   hashing of malware on IoT devices} &
	2019 &
	TRUE \\ \hline
	\href{http://itiis.org/journals/tiis/digital-library/manuscript/file/23099/TIIS\%20Vol\%2013,\%20No\%2012-19.pdf}{Fast
	   k-NN based Malware Analysis in a Massive Malware Environment.} &
	2019 &
	TRUE \\ \hline
	\href{https://ieeexplore.ieee.org/abstract/document/9191381/}{Hac-t
	   and fast search for similarity in security} &
	2020 &
	TRUE \\ \hline
	\href{https://arxiv.org/abs/1910.07283}{Fishdbc:   Flexible, 
	incremental, scalable, hierarchical density-based clustering for   arbitrary data 
	and distance} &
	2019 &
	TRUE \\ \hline
	\href{https://www.mdpi.com/2227-7390/8/4/503}{An   Improved 
	Bytewise Approximate Matching Algorithm Suitable for Files of   Dissimilar 
	Sizes} &
	2020 &
	TRUE \\ \hline
	\href{https://ieeexplore.ieee.org/abstract/document/8514832/}{mrsh-mem:
	   Approximate matching on raw memory dumps} &
	2018 &
	FALSE \\ \hline
	\href{https://link.springer.com/content/pdf/10.1007/978-3-030-39250-5.pdf}{Data
	   Science: New Issues, Challenges and Applications} &
	2020 &
	FALSE \\ \hline
	\href{https://www.sciencedirect.com/science/article/pii/S0167404818303808}{Survey
	   of machine learning techniques for malware analysis} &
	2019 &
	FALSE \\ \hline
	\href{https://www.mdpi.com/1099-4300/21/12/1196}{A   Survey 
	on Using Kolmogorov Complexity in Cybersecurity} &
	2019 &
	FALSE \\ \hline
	\href{https://ieeexplore.ieee.org/abstract/document/8988556/}{A
	   Mechanism for Copyright Protection under Information-Centric Networking} &
	2019 &
	FALSE \\ \hline
	\href{https://www.crysys.hu/publications/files/setit/thesis\_bme\_Bak19bsc.pdf}{Clustering
	   IoT Malware Samples based on Binary Similarity} &
	2019 &
	TRUE \\ \hline
	\href{https://ieeexplore.ieee.org/abstract/document/5766764/}{Understanding
	   plagiarism linguistic patterns, textual features, and detection methods} &
	2011 &
	FALSE \\ \hline
	\href{https://arxiv.org/abs/2006.09271}{A   Survey of Machine 
	Learning Methods and Challenges for Windows Malware   Classification} &
	2020 &
	FALSE \\ \hline
	\href{https://spectrum.library.concordia.ca/978935/}{a   
	framework for automated similarity analysis of malware} &
	2014 &
	TRUE \\ \hline
	\href{https://ieeexplore.ieee.org/abstract/document/8858803/}{Cyberthreat
	   Hunting-Part 1: triaging ransomware using fuzzy hashing, import hashing 
	and   YARA rules} &
	2019 &
	TRUE \\ \hline
	\href{https://ieeexplore.ieee.org/abstract/document/8984540/}{A
	   ransomware detection method using fuzzy hashing for mitigating the risk of   
	occlusion of information systems} &
	2019 &
	TRUE \\ \hline
	\href{https://ieeexplore.ieee.org/abstract/document/9177636/}{Fuzzy-Import
	   Hashing: A malware analysis approach} &
	2020 &
	TRUE \\ \hline
	\href{https://ieeexplore.ieee.org/abstract/document/9002773/}{Augmented
	   YARA rules fused with fuzzy hashing in ransomware triaging} &
	2019 &
	TRUE \\ \hline
	\href{https://www.usenix.org/conference/cset15/workshop-program/presentation/li}{Experimental
	   study of fuzzy hashing in malware clustering analysis} &
	2015 &
	TRUE \\ \hline
	\href{https://commons.erau.edu/adfsl/2012/wednesday/15/}{A   
	fuzzy hashing approach based on random sequences and hamming distance} &
	2012 &
	TRUE \\ \hline
	\href{https://ieeexplore.ieee.org/abstract/document/8858825/}{Cyberthreat
	   hunting-part 2: Tracking ransomware threat actors using fuzzy hashing and   
	fuzzy c-means clustering} &
	2019 &
	TRUE \\ \hline
	\href{https://ieeexplore.ieee.org/abstract/document/9003148/}{Lockout-Tagout
	   Ransomware: A detection method for ransomware using fuzzy hashing and   
	clustering} &
	2019 &
	TRUE \\ \hline
	\href{https://ieeexplore.ieee.org/abstract/document/9308189/}{Fuzzy
	   hashing aided enhanced YARA rules for malware triaging} &
	2020 &
	TRUE \\ \hline
	\href{https://ntnuopen.ntnu.no/ntnu-xmlui/bitstream/handle/11250/144008/KPAstebol.pdf?sequence=1}{mvHash:
	   a new approach for fuzzy hashing} &
	2012 &
	TRUE \\ \hline
	\href{https://ieeexplore.ieee.org/abstract/document/7847157/}{Forensic
	   malware analysis: The value of fuzzy hashing algorithms in identifying   
	similarities} &
	2016 &
	TRUE \\ \hline
	\href{https://ieeexplore.ieee.org/abstract/document/6754635/}{TLSH--a
	   locality sensitive hash} &
	2013 &
	FALSE \\ \hline
	\href{http://www.dasec.h-da.de/wp-content/uploads/2012/08/2012\_08\_Breitinger\_NeSeFo.pdf}{Security
	   and implementation analysis of the similarity digest sdhash} &
	2012 &
	TRUE \\ \hline
	\href{https://www.sciencedirect.com/science/article/pii/S1742287613000789}{F2S2:
	   Fast forensic similarity search through indexing piecewise hash signatures} &
	2013 &
	TRUE \\ \hline
	\href{https://ieeexplore.ieee.org/abstract/document/6320445/}{Properties
	   of a similarity preserving hash function and their realization in sdhash} &
	2012 &
	TRUE \\ \hline
	\href{https://ieeexplore.ieee.org/abstract/document/7575849/}{Detection
	   of malicious portable executables using evidence combinational theory with   
	fuzzy hashing} &
	2016 &
	TRUE \\ \hline
	\href{https://dl.acm.org/doi/abs/10.1145/3077286.3077289}{A   
	comparison of fuzzy hashes: evaluation, guidelines, and future suggestions} &
	2017 &
	TRUE \\ \hline
	\href{https://www.sciencedirect.com/science/article/pii/S1742287617302566}{Lempel-Ziv
	   Jaccard Distance, an effective alternative to ssdeep and sdhash} &
	2018 &
	TRUE \\ \hline
	\href{https://kilthub.cmu.edu/articles/journal\_contribution/Results\_of\_SEI\_Line-Funded\_Exploratory\_New\_Starts\_Projects/6583433/files/12070334.pdf\#page=9}{2
	   Fuzzy Hashing Techniques in Applied Malware Analysis} &
	2012 &
	TRUE \\ \hline
	\href{https://www.usenix.org/conference/cset15/workshop-program/presentation/li}{Experimental
	   study of fuzzy hashing in malware clustering analysis} &
	2015 &
	TRUE \\ \hline
	\href{https://ieeexplore.ieee.org/abstract/document/9002773/}{Augmented
	   YARA rules fused with fuzzy hashing in ransomware triaging} &
	2019 &
	TRUE \\ \hline
	\href{https://ieeexplore.ieee.org/abstract/document/8858803/}{Cyberthreat
	   Hunting-Part 1: triaging ransomware using fuzzy hashing, import hashing 
	and   YARA rules} &
	2019 &
	TRUE \\ \hline
	\href{https://ieeexplore.ieee.org/abstract/document/9177636/}{Fuzzy-Import
	   Hashing: A malware analysis approach} &
	2020 &
	TRUE \\ \hline
	\href{https://ieeexplore.ieee.org/abstract/document/9056506/}{Deep
	   fuzzy hashing network for efficient image retrieval} &
	2020 &
	TRUE \\ \hline
	\href{https://commons.erau.edu/adfsl/2012/wednesday/15/}{A   
	fuzzy hashing approach based on random sequences and hamming distance} &
	2012 &
	TRUE \\ \hline
	\href{https://ieeexplore.ieee.org/abstract/document/9308189/}{Fuzzy
	   hashing aided enhanced YARA rules for malware triaging} &
	2020 &
	TRUE \\ \hline
	\href{https://ieeexplore.ieee.org/abstract/document/8858825/}{Cyberthreat
	   hunting-part 2: Tracking ransomware threat actors using fuzzy hashing and   
	fuzzy c-means clustering} &
	2019 &
	TRUE \\ \hline
	\href{https://ieeexplore.ieee.org/abstract/document/9177856/}{Embedding
	   fuzzy rules with YARA rules for performance optimisation of malware 
	analysis} &
	2020 &
	TRUE \\ \hline
	\href{https://www.irojournals.com/itdw/V2/I3/03.pdf}{A   Novel 
	Information retrieval system for distributed cloud using Hybrid Deep   Fuzzy 
	Hashing Algorithm} &
	2020 &
	TRUE \\ \hline
	\href{https://link.springer.com/chapter/10.1007/978-3-642-35515-8\_12}{Performance
	   issues about context-triggered piecewise hashing} &
	2011 &
	TRUE \\ \hline
	\href{https://dl.acm.org/doi/abs/10.1145/2093698.2093722}{On  
	 fuzzy syndrome hashing with LDPC coding} &
	2011 &
	FALSE \\ \hline
	\href{https://kilthub.cmu.edu/articles/journal\_contribution/Results\_of\_SEI\_Line-Funded\_Exploratory\_New\_Starts\_Projects/6583433/files/12070334.pdf\#page=9}{2
	   Fuzzy Hashing Techniques in Applied Malware Analysis} &
	2012 &
	TRUE \\ \hline
	\href{https://www.igi-global.com/article/scaling-up-software-birthmarks-using-fuzzy-hashing/182539}{Scaling
	   up software birthmarks using fuzzy hashing} &
	2017 &
	FALSE \\ \hline
	\href{https://ieeexplore.ieee.org/abstract/document/7550868/}{A
	   fuzzy hashing technique for large scale software birthmarks} &
	2016 &
	FALSE \\ \hline
	\href{https://ntnuopen.ntnu.no/ntnu-xmlui/bitstream/handle/11250/144008/KPAstebol.pdf?sequence=1}{mvHash:
	   a new approach for fuzzy hashing} &
	2012 &
	TRUE \\ \hline
	\href{https://www.usenix.org/conference/cset15/workshop-program/presentation/li}{Experimental
	   study of fuzzy hashing in malware clustering analysis} &
	2015 &
	TRUE \\ \hline
	\href{https://ieeexplore.ieee.org/abstract/document/9003148/}{Lockout-Tagout
	   Ransomware: A detection method for ransomware using fuzzy hashing and   
	clustering} &
	2019 &
	TRUE \\ \hline
	\href{https://ieeexplore.ieee.org/abstract/document/8858825/}{Cyberthreat
	   hunting-part 2: Tracking ransomware threat actors using fuzzy hashing and   
	fuzzy c-means clustering} &
	2019 &
	TRUE \\ \hline
	\href{https://ieeexplore.ieee.org/abstract/document/8984540/}{A
	   ransomware detection method using fuzzy hashing for mitigating the risk of   
	occlusion of information systems} &
	2019 &
	TRUE \\ \hline
	\href{https://dl.acm.org/doi/abs/10.1145/2684793.2684803}{Clustering
	   spam campaigns with fuzzy hashing} &
	2014 &
	TRUE \\ \hline
	\href{https://ieeexplore.ieee.org/abstract/document/9177636/}{Fuzzy-Import
	   Hashing: A malware analysis approach} &
	2020 &
	TRUE \\ \hline
	\href{https://ieeexplore.ieee.org/abstract/document/8858803/}{Cyberthreat
	   Hunting-Part 1: triaging ransomware using fuzzy hashing, import hashing 
	and   YARA rules} &
	2019 &
	TRUE \\ \hline
	\href{https://ieeexplore.ieee.org/abstract/document/9002773/}{Augmented
	   YARA rules fused with fuzzy hashing in ransomware triaging} &
	2019 &
	TRUE \\ \hline
	\href{https://ieeexplore.ieee.org/abstract/document/7575849/}{Detection
	   of malicious portable executables using evidence combinational theory with   
	fuzzy hashing} &
	2016 &
	TRUE \\ \hline
	\href{https://ieeexplore.ieee.org/abstract/document/9308189/}{Fuzzy
	   hashing aided enhanced YARA rules for malware triaging} &
	2020 &
	TRUE \\ \hline
	\href{https://search.proquest.com/openview/8eb7000746ead07670846cb651554f36/1?pq-origsite=gscholar\&cbl=1976342}{Evaluation
	   on malware classification by combining traffic analysis and fuzzy hashing of   
	malware binary} &
	2015 &
	TRUE \\ \hline
	\href{https://www.irojournals.com/itdw/V2/I3/03.pdf}{A   Novel 
	Information retrieval system for distributed cloud using Hybrid Deep   Fuzzy 
	Hashing Algorithm} &
	2020 &
	TRUE \\ \hline
	\href{https://ieeexplore.ieee.org/abstract/document/6031914/}{Multiple
	   kernel fuzzy clustering} &
	2011 &
	FALSE \\ \hline
	\href{https://ieeexplore.ieee.org/abstract/document/9177856/}{Embedding
	   fuzzy rules with YARA rules for performance optimisation of malware 
	analysis} &
	2020 &
	TRUE \\ \hline
	\href{https://dl.acm.org/doi/abs/10.1145/3041008.3041010}{Emulator
	   vs real phone: Android malware detection using machine learning} &
	2017 &
	TRUE \\ \hline
	\href{https://dl.acm.org/doi/abs/10.1145/2590296.2590325}{Evading
	   android runtime analysis via sandbox detection} &
	2014 &
	TRUE \\ \hline
	\href{https://dl.acm.org/doi/abs/10.1145/2664243.2664250}{Morpheus:
	   automatically generating heuristics to detect android emulators} &
	2014 &
	TRUE \\ \hline
	\href{https://dl.acm.org/doi/abs/10.1145/2799979.2800004}{A   
	robust dynamic analysis system preventing SandBox detection by Android   
	malware} &
	2015 &
	TRUE \\ \hline
	\href{https://ieeexplore.ieee.org/abstract/document/6583806/}{Applying
	   machine learning classifiers to dynamic android malware detection at scale} &
	2013 &
	TRUE \\ \hline
	\href{https://ieeexplore.ieee.org/abstract/document/7389671/}{A
	   study of android malware detection technology evolution} &
	2015 &
	TRUE \\ \hline
	\href{https://dl.acm.org/doi/abs/10.1145/2592791.2592796}{Rage
	   against the virtual machine: hindering dynamic analysis of android malware} &
	2014 &
	TRUE \\ \hline
	\href{https://www.koreascience.or.kr/article/ArticleFullRecord.jsp?cn=JBBHCB\_2015\_v25n5\_1067}{A
	   Study on android emulator detection for mobile game security} &
	2015 &
	TRUE \\ \hline
	\href{https://dl.acm.org/doi/abs/10.1145/2875475.2875487}{Evaluation
	   of android malware detection based on system calls} &
	2016 &
	TRUE \\ \hline
	\href{https://ieeexplore.ieee.org/abstract/document/6982893/}{Malware
	   detection using network traffic analysis in android based mobile devices} &
	2014 &
	TRUE \\ \hline
	\href{https://www.cs.utah.edu/docs/techreports/2015/pdf/UUCS-15-003.pdf}{Android
	   malware detection based on system calls} &
	2015 &
	TRUE \\ \hline
	\href{https://dl.acm.org/doi/abs/10.1145/3386415.3387091}{Out-of-Domain
	   Characteristic Based Hierarchical Emulator Detection for Mobile} &
	2019 &
	TRUE \\ \hline
	\href{https://www.sciencedirect.com/science/article/pii/S2352340918315774}{Data
	   on vulnerability detection in android} &
	2019 &
	TRUE \\ \hline
	\href{https://www.sciencedirect.com/science/article/pii/S1877050916001071}{Anti-Hijack:
	   Runtime detection of malware initiated hijacking in android} &
	2016 &
	TRUE \\ \hline
	\href{https://dl.acm.org/doi/abs/10.1145/2940343.2940348}{CREDROID:
	   Android malware detection by network traffic analysis} &
	2016 &
	TRUE \\ \hline
	\href{https://ieeexplore.ieee.org/abstract/document/7814490/}{Deep4maldroid:
	   A deep learning framework for android malware detection based on linux 
	kernel   system call graphs} &
	2016 &
	TRUE \\ \hline
	\href{https://www.mdpi.com/558048}{AndroShield:   automated 
	android applications vulnerability detection, a hybrid static and   dynamic 
	analysis approach} &
	2019 &
	TRUE \\ \hline
	\href{https://dl.acm.org/doi/abs/10.1145/2480362.2480701}{Mobile-sandbox:
	   having a deeper look into android applications} &
	2013 &
	TRUE \\ \hline
	\href{https://dl.acm.org/doi/abs/10.1145/2808117.2808126}{Android
	   rooting: Methods, detection, and evasion} &
	2015 &
	TRUE \\ \hline
	\href{https://rp.delaat.net/2015-2016/p51/report.pdf}{Jailbreak/root
	   detection evasion study on iOS and Android} &
	2016 &
	TRUE \\ \hline
	\href{https://dl.acm.org/doi/abs/10.1145/2810362.2810364}{All  
	 your root checks are belong to us: The sad state of root detection} &
	2015 &
	TRUE \\ \hline
	\href{https://dl.acm.org/doi/abs/10.1145/2307636.2307663}{Riskranker:
	   scalable and accurate zero-day android malware detection} &
	2012 &
	TRUE \\ \hline
	\href{https://link.springer.com/chapter/10.1007/978-3-319-68542-7\_69}{System
	   log-based android root state detection} &
	2017 &
	TRUE \\ \hline
	\href{https://dl.acm.org/doi/abs/10.1145/2810103.2813714}{Android
	   root and its providers: A double-edged sword} &
	2015 &
	TRUE \\ \hline
	\href{https://ieeexplore.ieee.org/abstract/document/6128277/}{Kernel-based
	   behavior analysis for android malware detection} &
	2011 &
	TRUE \\ \hline
	\href{https://arxiv.org/abs/2012.01812}{On   Root Detection 
	Strategies for Android Devices} &
	2020 &
	TRUE \\ \hline
	\href{https://link.springer.com/chapter/10.1007/978-3-319-04283-1\_6}{Droidapiminer:
	   Mining api-level features for robust malware detection in android} &
	2013 &
	FALSE \\ \hline
	\href{https://link.springer.com/article/10.1007/s10916-018-0966-x}{Root
	   exploit detection and features optimization: mobile device and blockchain   
	based medical data management} &
	2018 &
	FALSE \\ \hline
	\href{https://ieeexplore.ieee.org/abstract/document/8399047/}{User
	   conventional root detection for indoor mall} &
	2018 &
	FALSE \\ \hline
	\href{https://dl.acm.org/doi/abs/10.1145/2557547.2557563}{PREC:
	   practical root exploit containment for android devices} &
	2014 &
	FALSE \\ \hline
	\href{https://www.researchgate.net/profile/Hugo\_Gascon/publication/264785935\_DREBIN\_Effective\_and\_Explainable\_Detection\_of\_Android\_Malware\_in\_Your\_Pocket/links/53efd0020cf26b9b7dcdf395.pdf}{Drebin:
	   Effective and explainable detection of android malware in your pocket.} &
	2014 &
	FALSE \\ \hline
	\href{https://www.usenix.org/conference/usenixsecurity17/technical-sessions/presentation/gasparis}{Detecting
	   android root exploits by learning from root providers} &
	2017 &
	FALSE \\ \hline
	\href{https://ieeexplore.ieee.org/abstract/document/8074847/}{Malware
	   detection in android based on dynamic analysis} &
	2017 &
	TRUE \\ \hline
	\href{https://dl.acm.org/doi/abs/10.1145/2435349.2435377}{Fast,
	   scalable detection of piggybacked mobile applications} &
	2013 &
	FALSE \\ \hline
	\href{https://ieeexplore.ieee.org/abstract/document/6182000/}{Rooting
	   attack detection method on the Android-based smart phone} &
	2011 &
	FALSE \\ \hline
	\href{https://ieeexplore.ieee.org/abstract/document/6735264/}{Machine
	   learning for android malware detection using permission and api calls} &
	2013 &
	FALSE \\ \hline
	\href{https://ieeexplore.ieee.org/abstract/document/7972730/}{Who
	   changed you? Obfuscator identification for Android} &
	2017 &
	TRUE \\ \hline
	\href{https://dl.acm.org/doi/abs/10.1145/3129676.3129708}{An  
	 anti-reverse engineering technique using native code and obfuscator-llvm for   
	android applications} &
	2017 &
	TRUE \\ \hline
	\href{https://arxiv.org/abs/1907.06828}{Automated   
	deobfuscation of Android native binary code} &
	2019 &
	TRUE \\ \hline
	\href{https://link.springer.com/chapter/10.1007/978-3-319-03783-7\_10}{Efficient
	   code obfuscation for Android} &
	2013 &
	TRUE \\ \hline
	\href{https://dl.acm.org/doi/abs/10.1145/2976749.2978333}{Reliable
	   third-party library detection in android and its security applications} &
	2016 &
	TRUE \\ \hline
	\href{https://papers.ssrn.com/sol3/papers.cfm?abstract\_id=3511052}{TargO:
	   A Targeted Obfuscator for Android Applications} &
	2019 &
	TRUE \\ \hline
	\href{https://ieeexplore.ieee.org/abstract/document/8802781/}{Deobfuscating
	   Android native binary code} &
	2019 &
	TRUE \\ \hline
	\href{https://dl.acm.org/doi/abs/10.1145/2843859.2843866}{Enriching
	   reverse engineering through visual exploration of Android binaries} &
	2015 &
	TRUE \\ \hline
	\href{https://ieeexplore.ieee.org/abstract/document/6149548/}{Android:
	   Static analysis using similarity distance} &
	2012 &
	TRUE \\ \hline
	\href{https://net.cs.uni-bonn.de/fileadmin/user\_upload/plohmann/2012-Schulz-Code\_Protection\_in\_Android.pdf}{Code
	   protection in android} &
	2012 &
	FALSE \\ \hline
	\href{https://www.sciencedirect.com/science/article/pii/S0167404816300529}{Control
	   flow obfuscation for android applications} &
	2016 &
	TRUE \\ \hline
	\href{https://link.springer.com/article/10.1007/s11416-016-0282-2}{Testing
	   android malware detectors against code obfuscation: a systematization of   
	knowledge and unified methodology} &
	2017 &
	TRUE \\ \hline
	\href{https://dl.acm.org/doi/abs/10.1145/2751323.2751330}{Exploring
	   reverse engineering symptoms in Android apps} &
	2015 &
	TRUE \\ \hline
	\href{http://media.blackhat.com/bh-ad-11/Desnos/bh-ad-11-DesnosGueguen-Andriod-Reversing\_to\_Decompilation\_WP.pdf}{Android:
	   From reversing to decompilation} &
	2011 &
	FALSE \\ \hline
	\href{https://link.springer.com/chapter/10.1007/978-3-030-01701-9\_10}{Understanding
	   android obfuscation techniques: A large-scale investigation in the wild} &
	2018 &
	TRUE \\ \hline
	\href{https://dl.acm.org/doi/abs/10.1145/2635868.2635869}{Apposcopy:
	   Semantics-based detection of android malware through static analysis} &
	2014 &
	TRUE \\ \hline
	\href{https://link.springer.com/article/10.1007/s10766-018-0609-3}{Prodact:
	   Prefetch-obfuscator to defend against cache timing channels} &
	2019 &
	FALSE \\ \hline
	\href{https://books.google.com/books?hl=en\&lr=\&id=WAaJEN4ZI-QC\&oi=fnd\&pg=PR1\&dq=android+obfuscator\&ots=zGuGGaOTpx\&sig=L9ZK7yWHaoPYijyPys3DNueNMHI}{Decompiling
	   android} &
	2012 &
	FALSE \\ \hline
	\href{https://dl.acm.org/doi/abs/10.1145/2976749.2978422}{Statistical
	   deobfuscation of android applications} &
	2016 &
	TRUE \\ \hline
	\href{https://dl.acm.org/doi/abs/10.1145/3099012.3099020}{Anti-proguard:
	   Towards automated deobfuscation of android apps} &
	2017 &
	TRUE \\ \hline
	\href{https://arxiv.org/abs/1907.06828}{Automated   
	deobfuscation of Android native binary code} &
	2019 &
	TRUE \\ \hline
	\href{https://tests.virusbtn.com/uploads/pdf/magazine/2018/VB2018-Moses-Mordekhay.pdf}{Android
	   app deobfuscation using static-dynamic cooperation} &
	2018 &
	TRUE \\ \hline
	\href{https://www.koreascience.or.kr/article/JAKO201533678768418.page}{A
	   study on deobfuscation method of Android and implementation of automatic   
	analysis tool} &
	2015 &
	TRUE \\ \hline
	\href{https://dl.acm.org/doi/abs/10.1145/3196398.3196433}{Leveraging
	   historical versions of Android apps for efficient and precise taint analysis} &
	2018 &
	FALSE \\ \hline
	\href{https://ieeexplore.ieee.org/abstract/document/8802781/}{Deobfuscating
	   Android native binary code} &
	2019 &
	TRUE \\ \hline
	\href{https://www.usenix.org/conference/usenixsecurity18/presentation/wong}{Tackling
	   runtime-based obfuscation in Android with TIRO} &
	2018 &
	TRUE \\ \hline
	\href{http://jbremer.org/wp-posts/athcon.pdf}{Automated   
	analysis and deobfuscation of android apps \& malware} &
	2013 &
	TRUE \\ \hline
	\href{https://www.sciencedirect.com/science/article/pii/S0167404820300936}{Compile-time
	   code virtualization for android applications} &
	2020 &
	FALSE \\ \hline
	\href{https://link.springer.com/chapter/10.1007/978-3-319-03783-7\_10}{Efficient
	   code obfuscation for Android} &
	2013 &
	FALSE \\ \hline
	\href{https://dl.acm.org/doi/abs/10.1145/3289239.3289243}{Dose:
	   Deobfuscation based on semantic equivalence} &
	2018 &
	TRUE \\ \hline
	\href{https://ieeexplore.ieee.org/abstract/document/7985674/}{Libd:
	   Scalable and precise third-party library detection in android markets} &
	2017 &
	FALSE \\ \hline
	\href{http://manuscriptlink-society-file.s3.amazonaws.com/kism/conference/sma2020/presentation/SMA-2020\_paper\_75.pdf}{Reversing
	   Obfuscated Control Flow Structures in Android Apps using ReDex Optimizer} 
	&
	2020 &
	TRUE \\ \hline
	\href{https://www.sciencedirect.com/science/article/pii/S1574013721000058}{A
	   survey of android application and malware hardening} &
	2021 &
	TRUE \\ \hline
	\href{https://www.ndss-symposium.org/wp-content/uploads/2017/09/harvesting-runtime-values-android-applications-feature-anti-analysis-techniques.pdf}{Harvesting
	   Runtime Values in Android Applications That Feature Anti-Analysis   
	Techniques.} &
	2016 &
	FALSE \\ \hline
	\href{http://isyou.info/inpra/papers/inpra-v4n2-01.pdf}{String   
	deobfuscation scheme based on dynamic code extraction for mobile malwares} 
	&
	2016 &
	TRUE \\ \hline
	\href{https://www.inderscienceonline.com/doi/abs/10.1504/IJAHUC.2019.100730}{Automatic
	   string deobfuscation scheme for mobile applications based on platform-level   
	code extraction} &
	2019 &
	TRUE \\ \hline
	\href{https://dl.acm.org/doi/abs/10.1145/2999572.2999596}{Enabling
	   automatic protocol behavior analysis for android applications} &
	2016 &
	FALSE \\ \hline
	\href{https://link.springer.com/chapter/10.1007/978-3-319-24177-7\_15}{Dexhunter:
	   toward extracting hidden code from packed android applications} &
	2015 &
	TRUE \\ \hline
	\href{https://www.cs.ucr.edu/$\sim$heng/pubs/DroidUnpack\_ndss18.pdf}{Things
	   You May Not Know About Android (Un) Packers: A Systematic Study based 
	on   Whole-System Emulation.} &
	2018 &
	TRUE \\ \hline
	\href{https://link.springer.com/chapter/10.1007/978-3-319-26362-5\_17}{Appspear:
	   Bytecode decrypting and dex reassembling for packed android malware} &
	2015 &
	TRUE \\ \hline
	\href{https://www2.virusbulletin.com/uploads/pdf/conference/vb2014/VB2014-Yu.pdf}{Android
	   packers: facing the challenges, building solutions} &
	2014 &
	TRUE \\ \hline
	\href{https://ieeexplore.ieee.org/abstract/document/7787078/}{Automated
	   detection and classification for packed android applications} &
	2016 &
	TRUE \\ \hline
	\href{https://ieeexplore.ieee.org/abstract/document/7985676/}{Adaptive
	   unpacking of Android apps} &
	2017 &
	TRUE \\ \hline
	\href{https://ieeexplore.ieee.org/abstract/document/9144572/}{DroidPDF:
	   The Obfuscation Resilient Packer Detection Framework for Android Apps} &
	2020 &
	TRUE \\ \hline
	\href{https://ieeexplore.ieee.org/abstract/document/8490853/}{DexX:
	   a double layer unpacking framework for Android} &
	2018 &
	TRUE \\ \hline
	\href{https://www.sciencedirect.com/science/article/pii/S0164121218300311}{AppSpear:
	   automating the hidden-code extraction and reassembling of packed android   
	malware} &
	2018 &
	TRUE \\ \hline
	\href{https://net.cs.uni-bonn.de/fileadmin/user\_upload/plohmann/2012-Schulz-Code\_Protection\_in\_Android.pdf}{Code
	   protection in android} &
	2012 &
	TRUE \\ \hline
	\href{https://ieeexplore.ieee.org/abstract/document/9098088/}{PackerGrind:
	   An Adaptive Unpacking System for Android Apps} &
	2020 &
	TRUE \\ \hline
	\href{https://ieeexplore.ieee.org/abstract/document/8658199/}{An
	   entropy-based solution for identifying android packers} &
	2019 &
	TRUE \\ \hline
	\href{https://www.sciencedirect.com/science/article/pii/S0167404820300936}{Compile-time
	   code virtualization for android applications} &
	2020 &
	FALSE \\ \hline
	\href{https://surface.syr.edu/etd/498/}{Automated   code 
	extraction from packed android applications.} &
	2016 &
	TRUE \\ \hline
	\href{https://ieeexplore.ieee.org/abstract/document/8812043/}{Storydroid:
	   Automated generation of storyboard for Android apps} &
	2019 &
	FALSE \\ \hline
	\href{https://ieeexplore.ieee.org/abstract/document/7961511/}{RepDroid:
	   an automated tool for Android application repackaging detection} &
	2017 &
	TRUE \\ \hline
	\href{https://ieeexplore.ieee.org/abstract/document/7972730/}{Who
	   changed you? Obfuscator identification for Android} &
	2017 &
	TRUE \\ \hline
	\href{https://link.springer.com/chapter/10.1007/978-3-319-18681-8\_24}{Uncovering
	   self code modification in android} &
	2015 &
	TRUE \\ \hline
	\href{https://dl.acm.org/doi/abs/10.1145/2523514.2523539}{AndroSimilar:
	   robust statistical feature signature for Android malware detection} &
	2013 &
	FALSE \\ \hline
	\href{https://link.springer.com/chapter/10.1007/978-3-319-03783-7\_10}{Efficient
	   code obfuscation for Android} &
	2013 &
	FALSE \\ \hline
	\href{https://ieeexplore.ieee.org/abstract/document/7985676/}{Adaptive
	   unpacking of Android apps} &
	2017 &
	TRUE \\ \hline
	\href{https://ieeexplore.ieee.org/abstract/document/8490853/}{DexX:
	   a double layer unpacking framework for Android} &
	2018 &
	TRUE \\ \hline
	\href{https://link.springer.com/chapter/10.1007/978-3-319-24177-7\_15}{Dexhunter:
	   toward extracting hidden code from packed android applications} &
	2015 &
	TRUE \\ \hline
	\href{https://www.cs.ucr.edu/$\sim$heng/pubs/DroidUnpack\_ndss18.pdf}{Things
	   You May Not Know About Android (Un) Packers: A Systematic Study based 
	on   Whole-System Emulation.} &
	2018 &
	TRUE \\ \hline
	\href{https://ieeexplore.ieee.org/abstract/document/9098088/}{PackerGrind:
	   An Adaptive Unpacking System for Android Apps} &
	2020 &
	TRUE \\ \hline
	\href{https://www.sciencedirect.com/science/article/pii/S0164121218300311}{AppSpear:
	   automating the hidden-code extraction and reassembling of packed android   
	malware} &
	2018 &
	TRUE \\ \hline
	\href{https://www.sciencedirect.com/science/article/pii/S0167404820300936}{Compile-time
	   code virtualization for android applications} &
	2020 &
	FALSE \\ \hline
	\href{https://www8.virusbtn.com/uploads/pdf/magazine/2018/VB2018-Stone.pdf}{UNPACKING
	   THE PACKED UNPACKER: REVERSING AN ANDROID ANTI-ANALYSIS 
	NATIVE LIBRARY} &
	2018 &
	TRUE \\ \hline
	\href{https://www.usenix.org/conference/usenixsecurity18/presentation/wong}{Tackling
	   runtime-based obfuscation in Android with TIRO} &
	2018 &
	TRUE \\ \hline
	\href{http://hitcon.org/2015/ENT/PDF/The\%20Terminator\%20to\%20Android\%20Hardening\%20Services\_Yueqian\%20Zhang,\%20Xiapu\%20Luo,\%20Haoyang\%20Yin.pdf}{The
	   terminator to Android hardening services} &
	2015 &
	TRUE \\ \hline
	\href{https://ieeexplore.ieee.org/abstract/document/8293854/}{Android
	   malware familial classification and representative sample selection via   
	frequent subgraph analysis} &
	2018 &
	TRUE \\ \hline
	\href{https://link.springer.com/chapter/10.1007/978-3-319-26362-5\_17}{Appspear:
	   Bytecode decrypting and dex reassembling for packed android malware} &
	2015 &
	TRUE \\ \hline
	\href{https://net.cs.uni-bonn.de/fileadmin/user\_upload/plohmann/2012-Schulz-Code\_Protection\_in\_Android.pdf}{Code
	   protection in android} &
	2012 &
	TRUE \\ \hline
	\href{https://ieeexplore.ieee.org/abstract/document/8812083/}{Graph
	   embedding based familial analysis of android malware using unsupervised   
	learning} &
	2019 &
	FALSE \\ \hline
	\href{https://dl.acm.org/doi/abs/10.1145/2557547.2557558}{Divilar:
	   Diversifying intermediate language for anti-repackaging on android platform} &
	2014 &
	TRUE \\ \hline
	\href{https://link.springer.com/article/10.1007/s42452-019-1124-x}{The
	   Android malware detection systems between hope and reality} &
	2019 &
	TRUE \\ \hline
	\href{https://link.springer.com/chapter/10.1007/978-3-319-93411-2\_4}{Honey,
	   i shrunk your app security: The state of android app hardening} &
	2018 &
	TRUE \\ \hline
	\href{https://ieeexplore.ieee.org/abstract/document/8770262/}{Ctdroid:
	   leveraging a corpus of technical blogs for android malware analysis} &
	2019 &
	FALSE \\ \hline
	\href{https://link.springer.com/chapter/10.1007/978-3-319-24018-3\_12}{How
	   current android malware seeks to evade automated code analysis} &
	2015 &
	TRUE \\ \hline
	\href{https://dl.acm.org/doi/abs/10.1145/2897845.2897914}{revdroid:
	   Code analysis of the side effects after dynamic permission revocation of   
	android apps} &
	2016 &
	TRUE \\ \hline
	\href{https://www.sciencedirect.com/science/article/pii/S0045790617303087}{Machine
	   learning aided Android malware classification} &
	2017 &
	TRUE \\ \hline
	\href{https://arxiv.org/abs/2003.12730}{How   and Why did 
	developers migrate Android Applications from Java to Kotlin? A   study based 
	on code analysis and interviews with developers} &
	2020 &
	FALSE \\ \hline
	\href{https://ieeexplore.ieee.org/abstract/document/8094472/}{DroidFax:
	   A toolkit for systematic characterization of Android applications} &
	2017 &
	TRUE \\ \hline
	\href{https://dl.acm.org/doi/abs/10.1145/2699026.2699105}{Stadyna:
	   Addressing the problem of dynamic code updates in the security analysis of   
	android applications} &
	2015 &
	TRUE \\ \hline
	\href{https://arxiv.org/abs/1410.7749}{Enter   sandbox: Android 
	sandbox comparison} &
	2014 &
	TRUE \\ \hline
	\href{https://link.springer.com/chapter/10.1007/978-3-030-01701-9\_10}{Understanding
	   android obfuscation techniques: A large-scale investigation in the wild} &
	2018 &
	TRUE \\ \hline
	\href{https://www.rasthofer.info/publications/paper/TR2015-Harvester.pdf}{Harvesting
	   runtime data in android applications for identifying malware and enhancing   
	code analysis} &
	2015 &
	TRUE \\ \hline
	\href{https://ieeexplore.ieee.org/abstract/document/7077314/}{Android
	   malware classification using static code analysis and Apriori algorithm   
	improved with particle swarm optimization} &
	2014 &
	TRUE \\ \hline
	\href{https://ieeexplore.ieee.org/abstract/document/6258449/}{Analysis
	   of android applications' permissions} &
	2012 &
	TRUE \\ \hline
	\href{https://net.cs.uni-bonn.de/fileadmin/user\_upload/plohmann/2012-Schulz-Code\_Protection\_in\_Android.pdf}{Code
	   protection in android} &
	2012 &
	TRUE \\ \hline
	\href{https://www.sciencedirect.com/science/article/pii/S0950584912001012}{Static
	   analysis of Android programs} &
	2012 &
	TRUE \\ \hline
	\href{https://ieeexplore.ieee.org/abstract/document/7972730/}{Who
	   changed you? Obfuscator identification for Android} &
	2017 &
	TRUE \\ \hline
	\href{https://link.springer.com/chapter/10.1007/978-3-642-27207-3\_33}{Mobile
	   application compatibility test system design for android fragmentation} &
	2011 &
	FALSE \\ \hline
	\href{https://ieeexplore.ieee.org/abstract/document/7019227/}{Detecting
	   software vulnerabilities in android using static analysis} &
	2014 &
	TRUE \\ \hline
	\href{https://dl.acm.org/doi/abs/10.1145/2382196.2382205}{Why
	   Eve and Mallory love Android: An analysis of Android SSL (in) security} &
	2012 &
	FALSE \\ \hline
	\href{https://ieeexplore.ieee.org/abstract/document/7180133/}{A
	   dataset of open-source android applications} &
	2015 &
	FALSE \\ \hline
	\href{https://dl.acm.org/doi/abs/10.1145/2536853.2536881}{Droidfuzzer:
	   Fuzzing the android apps with intent-filter tag} &
	2013 &
	TRUE \\ \hline
	\href{https://ieeexplore.ieee.org/abstract/document/7985671/}{Making
	   malory behave maliciously: Targeted fuzzing of android execution   
	environments} &
	2017 &
	TRUE \\ \hline
	\href{https://dl.acm.org/doi/abs/10.1145/3407023.3407052}{Template-based
	   Android inter process communication fuzzing} &
	2020 &
	TRUE \\ \hline
	\href{https://www.usenix.org/conference/usenixsecurity20/presentation/liu}{FANS:
	   Fuzzing Android Native System Services via Automated Interface Analysis} 
	&
	2020 &
	TRUE \\ \hline
	\href{http://web.cse.ohio-state.edu/$\sim$zhao.2708/assets/ndss18/NDSS18.pdf}{IoTFuzzer:
	   Discovering Memory Corruptions in IoT Through App-based Fuzzing.} &
	2018 &
	TRUE \\ \hline
	\href{https://dl.acm.org/doi/abs/10.1145/3386901.3388916}{Vulcan:
	   Lessons on reliability of wearables through state-aware fuzzing} &
	2020 &
	FALSE \\ \hline
	\href{https://dl.acm.org/doi/abs/10.1145/2639108.2639131}{Caiipa:
	   Automated large-scale mobile app testing through contextual fuzzing} &
	2014 &
	TRUE \\ \hline
	\href{https://ieeexplore.ieee.org/abstract/document/8576654/}{Automated
	   hybrid analysis of android malware through augmenting fuzzing with forced   
	execution} &
	2018 &
	FALSE \\ \hline
	\href{https://arxiv.org/abs/2008.03585}{Automated Functional Fuzzing of 
	Android Apps} &
	2020 &
	TRUE \\ \hline
	\href{https://ieeexplore.ieee.org/abstract/document/8708183/}{Fuzzing
	   the Android Applications With HTTP/HTTPS Network Data} &
	2019 &
	TRUE \\ \hline
	\href{https://link.springer.com/article/10.1007/s10664-019-09725-6}{Evolutionary
	   Fuzzing of Android OS Vendor System Services} &
	2019 &
	TRUE \\ \hline
	\href{https://www.koreascience.or.kr/article/JAKO201835372300685.page}{A
	   Practical Intent Fuzzing Tool for Robustness of Inter-Component 
	Communication   in Android Apps} &
	2018 &
	TRUE \\ \hline
	\href{https://paper.seebug.org/papers/Security\%20Conf/Blackhat/2015/us-15-Gong-
		Fuzzing-Android-System-Services-By-Binder-Call-To-Escalate-Privilege.pdf}{Fuzzing
	   android system services by binder call to escalate privilege} &
	2015 &
	TRUE \\ \hline
	\href{https://ieeexplore.ieee.org/abstract/document/8371326/}{Fuzzing:
	   State of the art} &
	2018 &
	TRUE \\ \hline
	\href{https://ieeexplore.ieee.org/abstract/document/8854702/}{Efficient
	   Evolutionary Fuzzing for Android Application Installation Process} &
	2019 &
	TRUE \\ \hline
	\href{https://www.koreascience.or.kr/article/JAKO201509057414553.page}{Method
	   of Fuzzing Document Application Based on Android Devices} &
	2015 &
	TRUE \\ \hline
	\href{https://ieeexplore.ieee.org/abstract/document/6228986/}{A
	   whitebox approach for automated security testing of Android applications on   
	the cloud} &
	2012 &
	TRUE \\ \hline
	\href{https://dl.acm.org/doi/abs/10.1145/2590296.2590316}{IntentFuzzer:
	   detecting capability leaks of android applications} &
	2014 &
	TRUE \\ \hline
	\href{https://www.usenix.org/conference/usenixsecurity12/technical-sessions/presentation/yan}{Droidscope:
	   Seamlessly reconstructing the OS and dalvik semantic views for dynamic   
	android malware analysis} &
	2012 &
	TRUE \\ \hline
	\href{https://www.sciencedirect.com/science/article/pii/S0045790617303087}{Machine
	   learning aided Android malware classification} &
	2017 &
	TRUE \\ \hline
	\href{https://dl.acm.org/doi/abs/10.1145/3017427}{The   
	evolution of android malware and android analysis techniques} &
	2017 &
	TRUE \\ \hline
	\href{https://books.google.com/books?hl=en\&lr=\&id=AJXaBAAAQBAJ\&oi=fnd\&pg=PP1\&dq=android+malware+analysis\&ots=JhPi5K11rN\&sig=ipShxoH8rMu6GRQ6hJRY1KRZ6hg}{Android
	   malware and analysis} &
	2014 &
	TRUE \\ \hline
	\href{https://link.springer.com/article/10.1007/s10586-017-0944-y}{Deep
	   neural architectures for large scale android malware analysis} &
	2018 &
	TRUE \\ \hline
	\href{https://ieeexplore.ieee.org/abstract/document/8293854/}{Android
	   malware familial classification and representative sample selection via   
	frequent subgraph analysis} &
	2018 &
	TRUE \\ \hline
	\href{https://ieeexplore.ieee.org/abstract/document/6814679/}{Droid
	   permission miner: Mining prominent permissions for Android malware 
	analysis} &
	2014 &
	TRUE \\ \hline
	\href{https://ieeexplore.ieee.org/abstract/document/6298136/}{Droidmat:
	   Android malware detection through manifest and api calls tracing} &
	2012 &
	TRUE \\ \hline
	\href{https://link.springer.com/article/10.1007/s11036-019-01248-0}{AdDroid:
	   rule-based machine learning framework for android malware analysis} &
	2020 &
	TRUE \\ \hline
	\href{https://link.springer.com/chapter/10.1007/978-3-319-24018-3\_12}{How
	   current android malware seeks to evade automated code analysis} &
	2015 &
	TRUE \\ \hline
	\href{https://ieeexplore.ieee.org/abstract/document/8770262/}{Ctdroid:
	   leveraging a corpus of technical blogs for android malware analysis} &
	2019 &
	FALSE \\ \hline
	\href{https://ieeexplore.ieee.org/abstract/document/6128277/}{Kernel-based
	   behavior analysis for android malware detection} &
	2011 &
	FALSE \\ \hline
	\href{https://ieeexplore.ieee.org/abstract/document/7516385/}{Machine
	   learning classifiers for android malware analysis} &
	2016 &
	TRUE \\ \hline
	\href{https://ieeexplore.ieee.org/abstract/document/7272933/}{Machine
	   learning based hybrid behavior models for Android malware analysis} &
	2015 &
	TRUE \\ \hline
	\href{https://dl.acm.org/doi/abs/10.1145/2799979.2800003}{DRACO:
	   DRoid analyst combo an android malware analysis framework} &
	2015 &
	TRUE \\ \hline
	\href{https://www.artificialstudios.org/alessandro.reina/pubs/eurosec13.pdf}{A
	   system call-centric analysis and stimulation technique to automatically   
	reconstruct android malware behaviors} &
	2013 &
	TRUE \\ \hline
	\href{https://dl.acm.org/doi/abs/10.1145/2635868.2635869}{Apposcopy:
	   Semantics-based detection of android malware through static analysis} &
	2014 &
	TRUE \\ \hline
	\href{https://dl.acm.org/doi/abs/10.1145/2592791.2592796}{Rage
	   against the virtual machine: hindering dynamic analysis of android malware} &
	2014 &
	TRUE \\ \hline
	\href{https://dl.acm.org/doi/abs/10.1145/2924715.2924719}{ScanMe
	   mobile: a cloud-based Android malware analysis service} &
	2016 &
	TRUE \\ \hline
	\href{https://link.springer.com/chapter/10.1007/978-3-319-60876-1\_12}{Deep
	   ground truth analysis of current android malware} &
	2017 &
	TRUE \\ \hline
	\href{https://dl.acm.org/doi/abs/10.1145/2046707.2046779}{Android
	   permissions demystified} &
	2011 &
	FALSE \\ \hline
	\href{https://dl.acm.org/doi/abs/10.1145/2295136.2295141}{Android
	   permissions: a perspective combining risks and benefits} &
	2012 &
	FALSE \\ \hline
	\href{https://dl.acm.org/doi/abs/10.1145/2335356.2335360}{Android
	   permissions: User attention, comprehension, and behavior} &
	2012 &
	FALSE \\ \hline
	\href{https://link.springer.com/chapter/10.1007/978-3-642-34638-5\_6}{A
	   conundrum of permissions: installing applications on an android smartphone} 
	&
	2012 &
	FALSE \\ \hline
	\href{https://dl.acm.org/doi/abs/10.1145/2381934.2381938}{Dr.  
	 android and mr. hide: fine-grained permissions in android applications} &
	2012 &
	FALSE \\ \hline
	\href{https://ieeexplore.ieee.org/abstract/document/6258449/}{Analysis
	   of android applications' permissions} &
	2012 &
	FALSE \\ \hline
	\href{https://link.springer.com/chapter/10.1007/978-3-642-39256-6\_15}{Quantitative
	   security risk assessment of android permissions and applications} &
	2013 &
	FALSE \\ \hline
	\href{https://arxiv.org/abs/1303.0857}{Longitudinal   analysis of 
	android ad library permissions} &
	2013 &
	FALSE \\ \hline
	\href{https://www.usenix.org/conference/usenixsecurity15/technical-sessions/presentation/wijesekera}{Android
	   permissions remystified: A field study on contextual integrity} &
	2015 &
	FALSE \\ \hline
	\href{https://ieeexplore.ieee.org/abstract/document/6693128/}{Flow
	   permissions for android} &
	2013 &
	FALSE \\ \hline
	\href{https://ieeexplore.ieee.org/abstract/document/6529497/}{Studying
	   the effectiveness of android application permissions requests} &
	2013 &
	FALSE \\ \hline
	\href{https://www.usenix.org/legacy/events/webapps11/tech/final\_files/webapps11\_proceedings.pdf\#page=83}{The
	   effectiveness of application permissions} &
	2011 &
	FALSE \\ \hline
	\href{https://ieeexplore.ieee.org/abstract/document/6624000/}{Asking
	   for (and about) permissions used by android apps} &
	2013 &
	FALSE \\ \hline
	\href{https://link.springer.com/chapter/10.1007/978-3-642-30244-2\_4}{Android
	   security permissions–can we trust them?} &
	2011 &
	FALSE \\ \hline
	\href{https://spiral.imperial.ac.uk/handle/10044/1/64236}{Resolving
	   the predicament of android custom permissions} &
	2018 &
	TRUE \\ \hline
	\href{https://books.google.com/books?hl=en\&lr=\&id=DuC64OoJSGQC\&oi=fnd\&pg=PR3\&dq=android+permissions\&ots=CdApbDguTE\&sig=wpdRxBAAgsK0YC1Lrb735uMh8aA}{Application
	   Security for the Android Platform: Processes, Permissions, and Other   
	Safeguards} &
	2011 &
	FALSE \\ \hline
	\href{https://dl.acm.org/doi/abs/10.1145/2420950.2420956}{Permission
	   evolution in the android ecosystem} &
	2012 &
	FALSE \\ \hline
	\href{https://www.igi-global.com/article/a-novel-security-framework-for-managing-android-permissions-using-blockchain-technology/196191}{A
	   novel security framework for managing android permissions using 
	blockchain   technology} &
	2018 &
	FALSE \\ \hline
	\href{https://ieeexplore.ieee.org/abstract/document/7009096/}{Static
	   detection of Android malware by using permissions and API calls} &
	2014 &
	TRUE \\ \hline
	\href{https://dl.acm.org/doi/abs/10.1145/3197231.3197236}{An  
	 investigation into Android run-time permissions from the end users'   
	perspective} &
	2018 &
	TRUE \\ \hline
	\href{https://ieeexplore.ieee.org/abstract/document/6359654/}{Android
	   privacy} &
	2012 &
	FALSE \\ \hline
	\href{https://link.springer.com/chapter/10.1007/978-3-642-30921-2\_17}{AndroidLeaks:
	   automatically detecting potential privacy leaks in android applications on a   
	large scale} &
	2012 &
	FALSE \\ \hline
	\href{https://ieeexplore.ieee.org/abstract/document/7194581/}{Iccta:
	   Detecting inter-component privacy leaks in android apps} &
	2015 &
	TRUE \\ \hline
	\href{https://citeseerx.ist.psu.edu/viewdoc/download?doi=10.1.1.298.7556\&rep=rep1\&type=pdf}{Investigating
	   user privacy in android ad libraries} &
	2012 &
	FALSE \\ \hline
	\href{https://citeseerx.ist.psu.edu/viewdoc/download?doi=10.1.1.706.5640\&rep=rep1\&type=pdf}{ScanDal:
	   Static analyzer for detecting privacy leaks in android applications} &
	2012 &
	FALSE \\ \hline
	\href{https://dl.acm.org/doi/abs/10.1145/2508859.2516676}{Appintent:
	   Analyzing sensitive data transmission in android for privacy leakage   
	detection} &
	2013 &
	TRUE \\ \hline
	\href{https://dl.acm.org/doi/abs/10.1145/2245276.2232009}{A   
	framework for static detection of privacy leaks in android applications} &
	2012 &
	FALSE \\ \hline
	\href{https://link.springer.com/chapter/10.1007/978-3-319-07076-6\_2}{Assessing
	   privacy risks in android: A user-centric approach} &
	2013 &
	FALSE \\ \hline
	\href{https://dl.acm.org/doi/abs/10.1145/2435349.2435380}{Appprofiler:
	   a flexible method of exposing privacy-related behavior in android   
	applications to end users} &
	2013 &
	FALSE \\ \hline
	\href{https://dl.acm.org/doi/abs/10.1145/2766498.2766506}{Securacy:
	   an empirical investigation of Android applications' network usage, privacy   
	and security} &
	2015 &
	TRUE \\ \hline
	\href{https://dl.acm.org/doi/abs/10.1145/2884781.2884855}{Toward
	   a framework for detecting privacy policy violations in android application   
	code} &
	2016 &
	TRUE \\ \hline
	\href{https://www.usenix.org/conference/usenixsecurity13/technical-sessions/presentation/bugiel}{Flexible
	   and fine-grained mandatory access control on android for diverse security 
	and   privacy policies} &
	2013 &
	FALSE \\ \hline
	\href{https://journals.sagepub.com/doi/abs/10.1177/1461444817702397}{Platform
	   privacies: Governance, collaboration, and the different meanings of “privacy”   
	in iOS and Android development} &
	2018 &
	FALSE \\ \hline
	\href{https://dl.acm.org/doi/abs/10.1145/2468356.2468502}{Android
	   and iOS users' differences concerning security and privacy} &
	2013 &
	FALSE \\ \hline
	\href{https://dl.acm.org/doi/abs/10.1145/2766498.2766507}{DroidJust:
	   Automated functionality-aware privacy leakage analysis for Android   
	applications} &
	2015 &
	TRUE \\ \hline
	\href{https://www.usenix.org/conference/soups2016/workshop-program/wpi/presentation/rajivan}{Influence
	   of privacy attitude and privacy cue framing on android app choices} &
	2016 &
	TRUE \\ \hline
	\href{https://dl.acm.org/doi/abs/10.1145/2470654.2466466}{Privacy
	   as part of the app decision-making process} &
	2013 &
	FALSE \\ \hline
	\href{https://www.sciencedirect.com/science/article/pii/S2214212618304356}{Dynamic
	   privacy leakage analysis of Android third-party libraries} &
	2019 &
	TRUE \\ \hline
	\href{https://arxiv.org/abs/1404.7431}{I   know what leaked in 
	your pocket: uncovering privacy leaks on Android Apps   with Static Taint 
	Analysis} &
	2014 &
	TRUE \\ \hline
	\href{https://lib.dr.iastate.edu/cgi/viewcontent.cgi?article=1233\&context=creativecomponents}{An
	   Investigative Study on Android Verified Boot Process} &
	2019 &
	TRUE \\ \hline
	\href{https://books.google.com/books?hl=en\&lr=\&id=y11NBQAAQBAJ\&oi=fnd\&pg=PR5\&dq=android+verified+boot\&ots=nWTwHPvQ4G\&sig=TVJBtqqHX9qpjj0KA2X1Eb0JY7g}{Android
	   security internals: An in-depth guide to Android's security architecture} &
	2014 &
	TRUE \\ \hline
	\href{https://dl.acm.org/doi/abs/10.1145/2808117.2808126}{Android
	   rooting: Methods, detection, and evasion} &
	2015 &
	TRUE \\ \hline
	\href{https://www.usenix.org/conference/woot17/workshop-program/presentation/hay}{fastboot
	   oem vuln: Android bootloader vulnerabilities in vendor customizations} &
	2017 &
	TRUE \\ \hline
	\href{https://arxiv.org/abs/1904.05572}{The   android platform 
	security model} &
	2019 &
	TRUE \\ \hline
	\href{https://odr.chalmers.se/handle/20.500.12380/251869}{Verified
	   boot in embedded systems with hard boot time constraints} &
	2017 &
	TRUE \\ \hline
	\href{https://www.sciencedirect.com/science/article/pii/S0950705118301424}{A
	   novel pattern recognition system for detecting Android malware by analyzing   
	suspicious boot sequences} &
	2018 &
	TRUE \\ \hline
	\href{http://denx.de/wiki/pub/U-Boot/OpenSourceIndia2013/Multi\_image\_booting\_scenarios.pdf}{U-Boot–Multi
	   image booting scenarios} &
	2013 &
	TRUE \\ \hline
	\href{https://dl.acm.org/doi/abs/10.1145/3124680.3124732}{Dm-x:
	   protecting volume-level integrity for cloud volumes and local block devices} &
	2017 &
	TRUE \\ \hline
	\href{https://link.springer.com/chapter/10.1007/978-1-4842-6106-4\_6}{OS
	   Resiliency} &
	2020 &
	FALSE \\ \hline
	\href{https://ieeexplore.ieee.org/abstract/document/6743201/}{Fast
	   boot and fast shutdown of Android on the embedded system} &
	2013 &
	FALSE \\ \hline
	\href{https://www.academia.edu/download/53217693/A\_Research\_On\_Android\_Technology\_With\_New\_Version\_Naugat7.0\_7.1.pdf}{A
	   Research On Android Technology With New Version Naugat (7.0, 7.1)} &
	2017 &
	FALSE \\ \hline
	\href{https://www.linuxsecrets.com/elinux-wiki/images/d/d0/Android\_OTA\_Updates.pdf}{Android*
	   Software Updates} &
	2015 &
	FALSE \\ \hline
	\href{https://link.springer.com/chapter/10.1007/978-3-319-47063-4\_2}{Controlled
	   Android application execution for the IoT infrastructure} &
	2015 &
	FALSE \\ \hline
	\href{https://www.academia.edu/download/56785475/IRJET-V5I3269.pdf}{Recent
	   versions on android and survey on the versions lollipop, marshmallow and   
	nougat} &
	2018 &
	FALSE \\ \hline
	\href{https://ipv4.delaat.net/rp/2013-2014/p40/report.pdf}{Android
	   patching} &
	2014 &
	FALSE \\ \hline
	\href{https://www.sae.org/publications/technical-papers/content/2020-01-1365/}{Android
	   Defense in Depth Strategy in an Automobile Ecosystem} &
	2020 &
	FALSE \\ \hline
	\href{https://dl.acm.org/doi/abs/10.1145/3019612.3019934}{Analysing
	   the vulnerability exploitation in Android with the device-mapper-verity   
	(dm-verity) student research abstract} &
	2017 &
	TRUE \\ \hline
	\href{https://ieeexplore.ieee.org/abstract/document/7962391/}{Euphony:
	   Harmonious unification of cacophonous anti-virus vendor labels for android   
	malware} &
	2017 &
	FALSE \\ \hline
	\href{https://arxiv.org/abs/1904.05572}{The   android platform 
	security model} &
	2019 &
	TRUE \\ \hline
	\href{https://search.proquest.com/openview/ec6c009a0a3f7ef3b2b895605d995aec/1?pq-origsite=gscholar\&cbl=2032677}{Simulating
	   Electronic Circuits on Android Device using Virtual Bread Board} &
	2017 &
	FALSE \\ \hline
	\href{https://link.springer.com/chapter/10.1007/978-981-15-9031-3\_23}{An
	   Android Data Protection Scheme for System-as-Root Architectures} &
	2020 &
	FALSE \\ \hline
	\href{https://link.springer.com/chapter/10.1007/978-3-030-32022-5\_15}{Automation
	   of the municipal inspection process in Ecuador applying mobile-D for android} 
	&
	2019 &
	FALSE \\ \hline
	\href{https://www.researchgate.net/profile/Nuuva\_Yusuf/publication/344642318\_Android\_Application\_Model\_of\_Whats\_Dating\_Violence\_as\_an\_Innovation\_to\_Increase\_the\_Knowledge\_among\_Adolescents/links/5f86aad7a6fdccfd7b5fc918/Android-Application-Model-of-Whats-Dating-Violence-as-an-Innovation-to-Increase-the-Knowledge-among-Adolescents.pdf}{Android
	   Application Model of What's Dating Violence as an Innovation to Increase   
	the Knowledge among Adolescents} &
	2020 &
	FALSE \\ \hline
	\href{https://orbilu.uni.lu/handle/10993/39903}{Creating   better 
	ground truth to further understand Android malware: A large scale   mining 
	approach based on antivirus labels and malicious artifacts} &
	2019 &
	FALSE \\ \hline
	\href{https://books.google.com/books?hl=en\&lr=\&id=wdW1DwAAQBAJ\&oi=fnd\&pg=PA155
		\&dq=android+avb\&ots=6kH9nnPXJ-\&sig=0kPrf-CV3XeruhZrO6J9w-gqW8E}{Automation
	   of the Municipal Inspection Process in Ecuador Applying Mobile-D for 
	Android} &
	2019 &
	FALSE \\ \hline
	\href{https://link.springer.com/chapter/10.1007/978-1-4842-6106-4\_6}{OS
	   Resiliency} &
	2020 &
	FALSE \\ \hline
	\href{https://repositorio.unb.br/handle/10482/39901}{Metodologia
	   para reconstrução de dados em dispositivos móveis com Sistema 
	Operacional Android mediante aplicação de técnicas ISP e Combination 
	firmware} &
	2020 &
	FALSE \\ \hline
	\href{https://www.sciencedirect.com/science/article/pii/S097535831233009X}{Reverse
	   Wenckebach “pseudo-supernormal” conduction or paroxysmal 
	atrioventricular   block} &
	2012 &
	FALSE \\ \hline
	\href{https://riunet.upv.es/bitstream/handle/10251/11535/memoria.pdf?sequence=1}{Android
	   download and search service} &
	2011 &
	FALSE \\ \hline
	\href{http://www.e-journal.janabadra.ac.id/index.php/PSN/article/view/502}{Aplikasi
	   Gamelan Jawa Berbasis Android} &
	2015 &
	FALSE \\ \hline
	\href{https://pubs.asha.org/doi/abs/10.1044/2016\_AJSLP-15-0187}{Integration
	   of motor learning principles into real-time ambulatory voice biofeedback and   
	example implementation via a clinical case study with vocal fold nodules} &
	2017 &
	FALSE \\ \hline
	\href{https://link.springer.com/article/10.1134/S2079057016030085}{Association
	   of aging-related obesity and metabolic syndrome in men} &
	2016 &
	FALSE \\ \hline
	\href{https://ieeexplore.ieee.org/abstract/document/7066494/}{Androi-based
	   C\&D (connected \& downloadable) IVI (in-vehicle infotainment)   platform} &
	2015 &
	FALSE \\ \hline
	\href{https://dl.acm.org/doi/abs/10.1145/3019612.3019934}{Analysing
	   the vulnerability exploitation in Android with the device-mapper-verity   
	(dm-verity) student research abstract} &
	2017 &
	TRUE \\ \hline
	\href{https://dl.acm.org/doi/abs/10.1145/2818000.2818036}{Baredroid:
	   Large-scale analysis of android apps on real devices} &
	2015 &
	FALSE \\ \hline
	\href{https://dl.acm.org/doi/abs/10.1145/3124680.3124732}{Dm-x:
	   protecting volume-level integrity for cloud volumes and local block devices} &
	2017 &
	TRUE \\ \hline
	\href{https://www.linuxsecrets.com/elinux-wiki/images/d/d0/Android\_OTA\_Updates.pdf}{Android*
	   Software Updates} &
	2015 &
	FALSE \\ \hline
	\href{https://link.springer.com/chapter/10.1007/978-3-319-47063-4\_2}{Controlled
	   Android application execution for the IoT infrastructure} &
	2015 &
	FALSE \\ \hline
	\href{https://homepages.staff.os3.nl/$\sim$delaat/rp/2015-2016/p30/report.pdf}{Subverting
	   Android 6.0 fingerprint authentication} &
	2016 &
	FALSE \\ \hline
	\href{https://dl.acm.org/doi/abs/10.1145/2994459.2994470}{Secure
	   containers in Android: the Samsung KNOX case study} &
	2016 &
	FALSE \\ \hline
	\href{https://www.usenix.org/conference/woot17/workshop-program/presentation/hay}{fastboot
	   oem vuln: Android bootloader vulnerabilities in vendor customizations} &
	2017 &
	FALSE \\ \hline
	\href{https://ipv4.delaat.net/rp/2013-2014/p40/report.pdf}{Android
	   patching} &
	2014 &
	FALSE \\ \hline
	\href{https://link.springer.com/chapter/10.1007/978-1-4842-6106-4\_6}{OS
	   Resiliency} &
	2020 &
	FALSE \\ \hline
	\href{https://www.sae.org/publications/technical-papers/content/2020-01-1365/}{Android
	   Defense in Depth Strategy in an Automobile Ecosystem} &
	2020 &
	FALSE \\ \hline
	\href{http://ictactjournals.in/paper/IJCT\_Paper\_1\_863\_to\_867.pdf}{Analysis
	   of android vulnerabilities and modern exploitation techniques} &
	2014 &
	FALSE \\ \hline
	\href{https://books.google.com/books?hl=en\&lr=\&id=y11NBQAAQBAJ\&oi=fnd\&pg=PR5\&dq=android+\%22dm+verity\%22\&ots=nWTwHPvR4G\&sig=\_-d7Cdyyir3JNN5l6Q1zQQ3LhxA}{Android
	   security internals: An in-depth guide to Android's security architecture} &
	2014 &
	FALSE \\ \hline
	\href{https://work.delaat.net/rp/2016-2017/p45/report.pdf}{Android
	   7 File Based Encryption and the Attacks Against It} &
	2017 &
	FALSE \\ \hline
	\href{https://www.usenix.org/conference/linuxfastsummit18/presentation/tso}{File
	   System-level Integrity Protection} &
	2018 &
	TRUE \\ \hline
	\href{https://trepo.tuni.fi/bitstream/handle/123456789/24231/Parviainen.pdf?sequence=3}{Android
	   Camera Tuning Application} &
	2016 &
	FALSE \\ \hline
	\href{https://ieeexplore.ieee.org/abstract/document/9153399/}{Hybrid
	   firmware analysis for known mobile and iot security vulnerabilities} &
	2020 &
	TRUE \\ \hline
	\href{https://www.usenix.org/conference/usenixsecurity20/presentation/hernandez}{BigMAC:
	   Fine-Grained Policy Analysis of Android Firmware} &
	2020 &
	TRUE \\ \hline
	\href{https://ieeexplore.ieee.org/abstract/document/8947099/}{Detect
	   malware in android firmware based on distributed network environment} &
	2019 &
	TRUE \\ \hline
	\href{https://ieeexplore.ieee.org/abstract/document/8787829/}{Toward
	   Automated Firmware Analysis in the IoT Era} &
	2019 &
	TRUE \\ \hline
	\href{https://www.sciencedirect.com/science/article/pii/S014036641731294X}{The
	   Nexmon firmware analysis and modification framework: Empowering 
	researchers   to enhance Wi-Fi devices} &
	2018 &
	FALSE \\ \hline
	\href{https://ieeexplore.ieee.org/abstract/document/9152633/}{An
	   analysis of pre-installed android software} &
	2020 &
	TRUE \\ \hline
	\href{https://www.usenix.org/conference/usenixsecurity18/presentation/tian}{Attention
	   spanned: Comprehensive vulnerability analysis of AT commands within the   
	android ecosystem} &
	2018 &
	FALSE \\ \hline
	\href{https://link.springer.com/content/pdf/10.1007/s11277-016-3618-9.pdf}{Vulnerabilities
	   of android OS-based telematics system} &
	2017 &
	TRUE \\ \hline
	\href{https://hernan.de/u/papers/Toward\_Automated\_Firmware\_Analysis\_in\_the\_IoT\_Era.pdf}{Toward
	   Automated Firmware Analysis in the IoT Era} &
	2019 &
	TRUE \\ \hline
	\href{https://ieeexplore.ieee.org/ielaam/8013/8821441/8787829-aam.pdf}{Towards
	   Automated Firmware Analysis in the IoT Era} &
	2019 &
	TRUE \\ \hline
	\href{https://books.google.com/books?hl=en\&lr=\&id=i-yWIVd4z7MC\&oi=fnd\&pg=PP1\&dq=android+firmware+analysis\&ots=L\_t-6GyyOm\&sig=dBwut0XRYe4TCyNZFgGJGyo8gAE}{Android
	   forensics: investigation, analysis and mobile security for Google Android} &
	2011 &
	FALSE \\ \hline
	\href{https://ieeexplore.ieee.org/abstract/document/8823941/}{Proxray:
	   Protocol model learning and guided firmware analysis} &
	2019 &
	FALSE \\ \hline
	\href{https://www.usenix.org/conference/usenixsecurity20/presentation/elsabagh}{FIRMSCOPE:
	   Automatic Uncovering of Privilege-Escalation Vulnerabilities in Pre-Installed   
	Apps in Android Firmware} &
	2020 &
	TRUE \\ \hline
	\href{https://www.mdpi.com/2079-9292/8/1/91}{Assessment   
	of dual frequency GNSS observations from a Xiaomi Mi 8 Android smartphone 
	and   positioning performance analysis} &
	2019 &
	FALSE \\ \hline
	\href{http://citeseerx.ist.psu.edu/viewdoc/download?doi=10.1.1.295.7508\&rep=rep1\&type=pdf}{Android
	   anti-virus analysis} &
	2012 &
	FALSE \\ \hline
	\href{https://dl.acm.org/doi/abs/10.1145/3017427}{The   
	evolution of android malware and android analysis techniques} &
	2017 &
	FALSE \\ \hline
	\href{https://dl.acm.org/doi/abs/10.1145/2590296.2590313}{DroidRay:
	   a security evaluation system for customized android firmwares} &
	2014 &
	TRUE \\ \hline
	\href{https://ieeexplore.ieee.org/abstract/document/7446031/}{Andrubis--1,000,000
	   apps later: A view on current Android malware behaviors} &
	2014 &
	FALSE \\ \hline
	\href{https://www.sciencedirect.com/science/article/pii/S0950584917302987}{Static
	   analysis of android apps: A systematic literature review} &
	2017 &
	FALSE \\ \hline
	\href{https://ieeexplore.ieee.org/abstract/document/9152633/}{An
	   analysis of pre-installed android software} &
	2020 &
	TRUE \\ \hline
	\href{https://dl.acm.org/doi/abs/10.1145/2976749.2978370}{Scalable
	   graph-based bug search for firmware images} &
	2016 &
	TRUE \\ \hline
	\href{https://www.usenix.org/conference/usenixsecurity20/presentation/elsabagh}{FIRMSCOPE:
	   Automatic Uncovering of Privilege-Escalation Vulnerabilities in Pre-Installed   
	Apps in Android Firmware} &
	2020 &
	TRUE \\ \hline
	\href{https://ieeexplore.ieee.org/abstract/document/8947099/}{Detect
	   malware in android firmware based on distributed network environment} &
	2019 &
	FALSE \\ \hline
	\href{https://ieeexplore.ieee.org/abstract/document/9153399/}{Hybrid
	   firmware analysis for known mobile and iot security vulnerabilities} &
	2020 &
	TRUE \\ \hline
	\href{https://ieeexplore.ieee.org/abstract/document/7446031/}{Andrubis--1,000,000
	   apps later: A view on current Android malware behaviors} &
	2014 &
	FALSE \\ \hline
	\href{https://ieeexplore.ieee.org/abstract/document/7962474/}{Sensitive
	   data leakage detection in pre-installed applications of custom Android   
	firmware} &
	2017 &
	FALSE \\ \hline
	\href{https://www.usenix.org/conference/usenixsecurity18/presentation/tian}{Attention
	   spanned: Comprehensive vulnerability analysis of AT commands within the   
	android ecosystem} &
	2018 &
	FALSE \\ \hline
	\href{https://link.springer.com/chapter/10.1007/978-3-319-58469-0\_16}{Towards
	   automated classification of firmware images and identification of embedded   
	devices} &
	2017 &
	TRUE \\ \hline
	\href{https://link.springer.com/chapter/10.1007/978-3-319-93411-2\_6}{Binarm:
	   Scalable and efficient detection of vulnerabilities in firmware images of   
	intelligent electronic devices} &
	2018 &
	FALSE \\ \hline
	\href{https://www.sciencedirect.com/science/article/pii/S0167739X19308490}{Static
	   malware detection and attribution in android byte-code through an end-to-end   
	deep system} &
	2020 &
	FALSE \\ \hline
	\href{https://www.usenix.org/conference/usenixsecurity14/technical-sessions/presentation/costin}{A
	   large-scale analysis of the security of embedded firmwares} &
	2014 &
	FALSE \\ \hline
	\href{https://ieeexplore.ieee.org/abstract/document/8418369/}{The
	   university of sussex-huawei locomotion and transportation dataset for   
	multimodal analytics with mobile devices} &
	2018 &
	FALSE \\ \hline
	\href{https://ieeexplore.ieee.org/abstract/document/9229869/}{Abusing
	   Android Runtime for Application Obfuscation} &
	2020 &
	FALSE \\ \hline
	\href{https://link.springer.com/chapter/10.1007/978-3-319-98446-9\_35}{Evaluation
	   of tree based machine learning classifiers for android malware detection} &
	2018 &
	FALSE \\ \hline
	\href{https://ieeexplore.ieee.org/abstract/document/6234407/}{Dissecting
	   android malware: Characterization and evolution} &
	2012 &
	FALSE \\ \hline
	\href{https://dl.acm.org/doi/abs/10.1145/3301326.3301390}{Evaluating
	   machine learning models for android malware detection: a comparison study} 
	&
	2018 &
	FALSE \\ \hline
	\href{https://ieeexplore.ieee.org/abstract/document/8433203/}{Efficient
	   signature generation for classifying cross-architecture IoT malware} &
	2018 &
	FALSE \\ \hline
	\href{https://dl.acm.org/doi/abs/10.1145/3330482.3330522}{Proposing
	   Automatic Dataset Generation System to Support Android Sensitive Data 
	Leakage   Detection Systems} &
	2019 &
	FALSE \\ \hline
	\href{https://dl.acm.org/doi/abs/10.1145/2590296.2590312}{Efficient,
	   context-aware privacy leakage confinement for android applications without   
	firmware modding} &
	2014 &
	FALSE \\ \hline
	\href{https://ieeexplore.ieee.org/abstract/document/9152633/}{An
	   analysis of pre-installed android software} &
	2020 &
	TRUE \\ \hline
	\href{https://dl.acm.org/doi/abs/10.1145/2307636.2307663}{Riskranker:
	   scalable and accurate zero-day android malware detection} &
	2012 &
	FALSE \\ \hline
	\href{https://www.usenix.org/conference/usenixsecurity20/presentation/clements}{HALucinator:
	   Firmware re-hosting through abstraction layer emulation} &
	2020 &
	FALSE \\ \hline
	\href{https://ieeexplore.ieee.org/abstract/document/6234407/}{Dissecting
	   android malware: Characterization and evolution} &
	2012 &
	FALSE \\ \hline
	\href{https://www.sciencedirect.com/science/article/pii/S0956566318304755}{Cost-effective
	   smartphone-based reconfigurable electrochemical instrument for alcohol   
	determination in whole blood samples} &
	2018 &
	FALSE \\ \hline
	\href{http://www.csd.uoc.gr/$\sim$hy558/papers/mal\_apps.pdf}{Hey,
	   you, get off of my market: detecting malicious apps in official and   alternative 
	android markets.} &
	2012 &
	FALSE \\ \hline
	\href{https://dl.acm.org/doi/abs/10.1145/2590296.2590313}{DroidRay:
	   a security evaluation system for customized android firmwares} &
	2014 &
	FALSE \\ \hline
	\href{https://ieeexplore.ieee.org/abstract/document/7801876/}{Monet:
	   a user-oriented behavior-based malware variants detection system for 
	android} &
	2016 &
	FALSE \\ \hline
	\href{https://dl.acm.org/doi/abs/10.1145/2660267.2660359}{Semantics-aware
	   android malware classification using weighted contextual api dependency   
	graphs} &
	2014 &
	FALSE \\ \hline
	\href{https://dl.acm.org/doi/abs/10.1145/3291061}{A   close 
	look at a daily dataset of malware samples} &
	2019 &
	FALSE \\ \hline
	\href{https://ieeexplore.ieee.org/abstract/document/7962474/}{Sensitive
	   data leakage detection in pre-installed applications of custom Android   
	firmware} &
	2017 &
	FALSE \\ \hline
	\href{https://dl.acm.org/doi/abs/10.1145/3301326.3301390}{Evaluating
	   machine learning models for android malware detection: a comparison study} 
	&
	2018 &
	FALSE \\ \hline
	\href{https://dl.acm.org/doi/abs/10.1145/3381991.3395608}{Parallel
	   Space Traveling: A Security Analysis of App-Level Virtualization in Android} &
	2020 &
	FALSE \\ \hline
	\href{https://dl.acm.org/doi/abs/10.1145/2976749.2978370}{Scalable
	   graph-based bug search for firmware images} &
	2016 &
	FALSE \\ \hline
	\href{https://link.springer.com/chapter/10.1007/978-3-319-98446-9\_35}{Evaluation
	   of tree based machine learning classifiers for android malware detection} &
	2018 &
	FALSE \\ \hline
	\href{https://ieeexplore.ieee.org/abstract/document/7446031/}{Andrubis--1,000,000
	   apps later: A view on current Android malware behaviors} &
	2014 &
	FALSE \\ \hline
	\href{https://link.springer.com/chapter/10.1007/978-3-030-15462-2\_12}{Breaking
	   all the things—A systematic survey of firmware extraction techniques for 
	IoT   devices} &
	2018 &
	FALSE \\ \hline
	\href{https://www.usenix.org/conference/usenixsecurity20/presentation/hernandez}{BigMAC:
	   Fine-Grained Policy Analysis of Android Firmware} &
	2020 &
	TRUE \\ \hline
	\href{https://dl.acm.org/doi/abs/10.1145/3063955.3063981}{Android
	   digital forensics: data, extraction and analysis} &
	2017 &
	FALSE \\ \hline
	\href{https://www.usenix.org/conference/usenixsecurity18/presentation/tian}{Attention
	   spanned: Comprehensive vulnerability analysis of AT commands within the   
	android ecosystem} &
	2018 &
	FALSE \\ \hline
	\href{https://repositum.tuwien.at/handle/20.500.12708/2673}{Human
	   assisted information extraction through Bluetooth low energy on android} &
	2019 &
	FALSE \\ \hline
	\href{https://ieeexplore.ieee.org/abstract/document/8931782/}{A
	   firmware code gene extraction technology for IoT terminal} &
	2019 &
	FALSE \\ \hline
	\href{https://www.mdpi.com/2076-3417/10/12/4231}{Methodology
	   for Forensics Data Reconstruction on Mobile Devices with Android 
	Operating   System Applying In-System Programming and Combination 
	Firmware} &
	2020 &
	FALSE \\ \hline
	\href{https://dl.acm.org/doi/abs/10.1145/3349623.3355477}{Free
	   your CSI: A channel state information extraction platform for modern Wi-Fi   
	chipsets} &
	2019 &
	FALSE \\ \hline
	\href{https://search.proquest.com/openview/b0b2e9cb242bda4074a221dc912cfb22/1?pq-origsite=gscholar\&cbl=1686341}{Automatic
	   feature extraction, categorization and detection of malicious code in Android   
	applications} &
	2014 &
	FALSE \\ \hline
	\href{https://www.academia.edu/download/16807224/Thesis.pdf}{Android
	   forensic capability and evaluation of extraction tools} &
	2012 &
	FALSE \\ \hline
	\href{https://docs.lib.purdue.edu/open\_access\_theses/615/}{WeChat
	   forensic artifacts: Android phone extraction and analysis} &
	2015 &
	FALSE \\ \hline
	\href{https://ieeexplore.ieee.org/abstract/document/8787829/}{Toward
	   Automated Firmware Analysis in the IoT Era} &
	2019 &
	FALSE \\ \hline
	\href{https://iopscience.iop.org/article/10.1088/1757-899X/771/1/012024/meta}{Logical
	   Acquisition in the Forensic Investigation Process of Android Smartphones   
	based on Agent using Open Source Software} &
	2020 &
	FALSE \\ \hline
	\href{https://www.sciencedirect.com/science/article/pii/S014036641731294X}{The
	   Nexmon firmware analysis and modification framework: Empowering 
	researchers   to enhance Wi-Fi devices} &
	2018 &
	FALSE \\ \hline
	\href{https://ieeexplore.ieee.org/abstract/document/9153399/}{Hybrid
	   firmware analysis for known mobile and iot security vulnerabilities} &
	2020 &
	FALSE \\ \hline
	\href{https://books.google.com/books?hl=en\&lr=\&id=2qo6AwAAQBAJ\&oi=fnd\&pg=PR25\&dq=android+firmware+extraction\&ots=0nuNRfi7kZ\&sig=TfPFqY-AIqect\_P-2ZdE5Qhrw\_A}{Android
	   hacker's handbook} &
	2014 &
	FALSE \\ \hline
	\href{https://ieeexplore.ieee.org/abstract/document/7813792/}{Android
	   phone forensic: Tools and techniques} &
	2016 &
	FALSE \\ \hline
	\href{https://www.uit.edu.mm/storage/2020/09/No.-4.pdf}{ADetect:
	   Hybrid Analysis Feature Extraction for Android malware Detection} &
	2018 &
	FALSE \\ \hline
	\href{https://papers.ssrn.com/sol3/papers.cfm?abstract\_id=3356336}{Android
	   Forensics: Tools and Techniques for Manual Data Extraction} &
	2019 &
	FALSE \\ \hline
	\href{https://dl.acm.org/doi/abs/10.1145/3395351.3399360}{BaseSAFE:
	   baseband sanitized fuzzing through emulation} &
	2020 &
	TRUE \\ \hline
	\href{https://www.usenix.org/conference/usenixsecurity20/presentation/ruge}{Frankenstein:
	   Advanced Wireless Fuzzing to Exploit New Bluetooth Escalation Targets} &
	2020 &
	TRUE \\ \hline
	\href{https://dl.acm.org/doi/abs/10.1145/3133956.3134069}{Difuze:
	   Interface aware fuzzing for kernel drivers} &
	2017 &
	TRUE \\ \hline
	\href{https://www.cspensky.info/pdfs/ndss2019\_04A-1\_Song\_paper.pdf}{Periscope:
	   An effective probing and fuzzing framework for the hardware-os boundary} &
	2019 &
	TRUE \\ \hline
	\href{https://www.ndss-symposium.org/wp-content/uploads/2020/02/24018-paper.pdf}{HFL:
	   Hybrid fuzzing on the linux kernel} &
	2020 &
	TRUE \\ \hline
	\href{https://www.usenix.org/conference/usenixsecurity20/presentation/hernandez}{BigMAC:
	   Fine-Grained Policy Analysis of Android Firmware} &
	2020 &
	TRUE \\ \hline
	\href{https://www.usenix.org/conference/usenixsecurity20/presentation/song}{Agamotto:
	   Accelerating Kernel Driver Fuzzing with Lightweight Virtual Machine   
	Checkpoints} &
	2020 &
	TRUE \\ \hline
	\href{https://ieeexplore.ieee.org/abstract/document/9152767/}{Ex-vivo
	   dynamic analysis framework for Android device drivers} &
	2020 &
	TRUE \\ \hline
	\href{https://onlinelibrary.wiley.com/doi/abs/10.1002/cpe.5756}{Fw‐fuzz:
	   A code coverage‐guided fuzzing framework for network protocols on 
	firmware} &
	2020 &
	TRUE \\ \hline
	\href{https://ieeexplore.ieee.org/abstract/document/8371326/}{Fuzzing:
	   State of the art} &
	2018 &
	TRUE \\ \hline
	\href{https://ieeexplore.ieee.org/abstract/document/8368433/}{Vulnerability
	   detection in iot firmware: A survey} &
	2017 &
	TRUE \\ \hline
	\href{https://www.usenix.org/conference/usenixsecurity20/presentation/elsabagh}{FIRMSCOPE:
	   Automatic Uncovering of Privilege-Escalation Vulnerabilities in Pre-Installed   
	Apps in Android Firmware} &
	2020 &
	TRUE \\ \hline
	\href{https://www.usenix.org/conference/woot20/presentation/fioraldi}{AFL++:
	   Combining incremental steps of fuzzing research} &
	2020 &
	TRUE \\ \hline
	\href{https://ieeexplore.ieee.org/abstract/document/8708183/}{Fuzzing
	   the Android Applications With HTTP/HTTPS Network Data} &
	2019 &
	TRUE \\ \hline
	\href{https://ieeexplore.ieee.org/abstract/document/8923779/}{Kernel
	   vulnerability analysis: a survey} &
	2019 &
	TRUE \\ \hline
	\href{https://www.usenix.org/conference/usenixsecurity18/presentation/talebi}{Charm:
	   Facilitating dynamic analysis of device drivers of mobile systems} &
	2018 &
	TRUE \\ \hline
	\href{https://www.sciencedirect.com/science/article/pii/S1742287617303948}{New
	   flash memory acquisition methods based on firmware update protocols for 
	LG   Android smartphones} &
	2018 &
	TRUE \\ \hline
	\href{https://dl.acm.org/doi/abs/10.1145/3319535.3363212}{Slake:
	   facilitating slab manipulation for exploiting vulnerabilities in the Linux kernel} 
	&
	2019 &
	TRUE \\ \hline
	\href{https://www.usenix.org/conference/usenixsecurity20/presentation/hernandez}{BigMAC:
	   Fine-Grained Policy Analysis of Android Firmware} &
	2020 &
	TRUE \\ \hline
	\href{https://link.springer.com/chapter/10.1007/978-3-030-15462-2\_12}{Breaking
	   all the things—A systematic survey of firmware extraction techniques for 
	IoT   devices} &
	2018 &
	TRUE \\ \hline
	\href{http://proceeding.vap.ac.vn/index.php/proceedingvap/article/view/2017.00011}{DETECT
	   SECURITY THREAT IN ANDROID CUSTOM FIRMWARE BY ANALYZING 
	APPLICATIONS   FRAMEWORK AND DEFAULT SETTINGS} &
	2019 &
	TRUE \\ \hline
	\href{https://ieeexplore.ieee.org/abstract/document/8947099/}{Detect
	   malware in android firmware based on distributed network environment} &
	2019 &
	TRUE \\ \hline
	\href{https://ieeexplore.ieee.org/abstract/document/7881801/}{Welcome
	   to Binder: A kernel level attack model for the Binder in Android operating   
	system} &
	2016 &
	TRUE \\ \hline
	\href{https://www.sciencedirect.com/science/article/pii/S1742287617301317}{Live
	   acquisition of main memory data from Android smartphones and 
	smartwatches} &
	2017 &
	FALSE \\ \hline
	\href{https://dl.acm.org/doi/abs/10.1145/3063955.3063981}{Android
	   digital forensics: data, extraction and analysis} &
	2017 &
	FALSE \\ \hline
	\href{https://link.springer.com/chapter/10.1007/978-3-319-46279-0\_19}{A
	   tool for volatile memory acquisition from Android devices} &
	2016 &
	FALSE \\ \hline
	\href{https://books.google.com/books?hl=en\&lr=\&id=2qo6AwAAQBAJ\&oi=fnd\&pg=PR25\&dq=android+firmware+kernel+extraction\&ots=0nuNRfi8dZ\&sig=ogyMcRCJtQfOIPdNwJRZOlThU4M}{Android
	   hacker's handbook} &
	2014 &
	FALSE \\ \hline
	\href{https://ieeexplore.ieee.org/abstract/document/8585583/}{A
	   Novel approach for bootkit detection in Android Platform} &
	2018 &
	TRUE \\ \hline
	\href{https://www.usenix.org/conference/usenixsecurity18/presentation/tian}{Attention
	   spanned: Comprehensive vulnerability analysis of AT commands within the   
	android ecosystem} &
	2018 &
	TRUE \\ \hline
	\href{https://dl.acm.org/doi/abs/10.1145/2590296.2590313}{DroidRay:
	   a security evaluation system for customized android firmwares} &
	2014 &
	TRUE \\ \hline
	\href{https://dl.acm.org/doi/abs/10.1145/3349623.3355477}{Free
	   your CSI: A channel state information extraction platform for modern Wi-Fi   
	chipsets} &
	2019 &
	FALSE \\ \hline
	\href{https://search.proquest.com/openview/ac7e946ecfb8249c1ee9695183004121/1?pq-origsite=gscholar\&cbl=1976342}{Mobile
	   root exploit detection based on system events extracted from android   
	platform} &
	2013 &
	FALSE \\ \hline
	\href{https://ijict.itrc.ac.ir/article-1-37-en.html}{An   
	Attack-Defense Model for the Binder on the Android Kernel Level} &
	2017 &
	TRUE \\ \hline
	\href{https://ieeexplore.ieee.org/abstract/document/8931782/}{A
	   firmware code gene extraction technology for IoT terminal} &
	2019 &
	TRUE \\ \hline
	\href{https://dl.acm.org/doi/abs/10.1145/3133956.3134069}{Difuze:
	   Interface aware fuzzing for kernel drivers} &
	2017 &
	TRUE \\ \hline
	\href{https://www.mdpi.com/2076-3417/10/12/4231}{Methodology
	   for Forensics Data Reconstruction on Mobile Devices with Android 
	Operating   System Applying In-System Programming and Combination 
	Firmware} &
	2020 &
	FALSE \\ \hline
	\href{https://www.usenix.org/conference/usenixsecurity20/presentation/hernandez}{BigMAC:
	   Fine-Grained Policy Analysis of Android Firmware} &
	2020 &
	TRUE \\ \hline
	\href{https://www.usenix.org/conference/usenixsecurity20/presentation/elsabagh}{FIRMSCOPE:
	   Automatic Uncovering of Privilege-Escalation Vulnerabilities in Pre-Installed   
	Apps in Android Firmware} &
	2020 &
	TRUE \\ \hline
	\href{https://ieeexplore.ieee.org/abstract/document/7446031/}{Andrubis--1,000,000
	   apps later: A view on current Android malware behaviors} &
	2014 &
	FALSE \\ \hline
	\href{https://dl.acm.org/doi/abs/10.1145/3191737}{Anatomy   
	of a vulnerable fitness tracking system: Dissecting the fitbit cloud, app,   and 
	firmware} &
	2018 &
	FALSE \\ \hline
	\href{https://dl.acm.org/doi/abs/10.1145/2307636.2307663}{Riskranker:
	   scalable and accurate zero-day android malware detection} &
	2012 &
	FALSE \\ \hline
	\href{https://link.springer.com/chapter/10.1007/978-3-030-15462-2\_12}{Breaking
	   all the things—A systematic survey of firmware extraction techniques for 
	IoT   devices} &
	2018 &
	FALSE \\ \hline
	\href{https://ieeexplore.ieee.org/abstract/document/8947099/}{Detect
	   malware in android firmware based on distributed network environment} &
	2019 &
	FALSE \\ \hline
	\href{https://dl.acm.org/doi/abs/10.1145/2590296.2590313}{DroidRay:
	   a security evaluation system for customized android firmwares} &
	2014 &
	FALSE \\ \hline
	\href{https://ieeexplore.ieee.org/abstract/document/7962474/}{Sensitive
	   data leakage detection in pre-installed applications of custom Android   
	firmware} &
	2017 &
	FALSE \\ \hline
	\href{http://www.csd.uoc.gr/$\sim$hy558/papers/mal\_apps.pdf}{Hey,
	   you, get off of my market: detecting malicious apps in official and   alternative 
	android markets.} &
	2012 &
	FALSE \\ \hline
	\href{https://dl.acm.org/doi/abs/10.1145/2133601.2133640}{Detecting
	   repackaged smartphone applications in third-party android marketplaces} &
	2012 &
	FALSE \\ \hline
	\href{https://ieeexplore.ieee.org/abstract/document/9152633/}{An
	   analysis of pre-installed android software} &
	2020 &
	TRUE \\ \hline
	\href{https://ieeexplore.ieee.org/abstract/document/7849904/}{ADROIT:
	   Android malware detection using meta-information} &
	2016 &
	FALSE \\ \hline
	\href{https://dl.acm.org/doi/abs/10.1145/2766498.2766508}{Droideagle:
	   seamless detection of visually similar android apps} &
	2015 &
	FALSE \\ \hline
	\href{https://dl.acm.org/doi/abs/10.1145/3063955.3063981}{Android
	   digital forensics: data, extraction and analysis} &
	2017 &
	FALSE \\ \hline
	\href{https://dl.acm.org/doi/abs/10.1145/2660267.2660359}{Semantics-aware
	   android malware classification using weighted contextual api dependency   
	graphs} &
	2014 &
	FALSE \\ \hline
	\href{https://ieeexplore.ieee.org/abstract/document/8787829/}{Toward
	   Automated Firmware Analysis in the IoT Era} &
	2019 &
	FALSE \\ \hline
	\href{https://www.sciencedirect.com/science/article/pii/S0950584916300386}{Effective
	   detection of android malware based on the usage of data flow APIs and 
	machine   learning} &
	2016 &
	FALSE \\ \hline
	\href{http://proceeding.vap.ac.vn/index.php/proceedingvap/article/view/2017.00011}{DETECT
	   SECURITY THREAT IN ANDROID CUSTOM FIRMWARE BY ANALYZING 
	APPLICATIONS   FRAMEWORK AND DEFAULT SETTINGS} &
	2019 &
	FALSE \\ \hline
	\href{https://ieeexplore.ieee.org/abstract/document/9152633/}{An
	   analysis of pre-installed android software} &
	2020 &
	TRUE \\ \hline
	\href{https://www.usenix.org/conference/usenixsecurity20/presentation/elsabagh}{FIRMSCOPE:
	   Automatic Uncovering of Privilege-Escalation Vulnerabilities in Pre-Installed   
	Apps in Android Firmware} &
	2020 &
	TRUE \\ \hline
	\href{https://www.sciencedirect.com/science/article/pii/S1084804520300527}{Pleasure
	   or pain? An evaluation of the costs and utilities of bloatware applications   in 
	android smartphones} &
	2020 &
	TRUE \\ \hline
	\href{https://ieeexplore.ieee.org/abstract/document/7551989/}{Jred:
	   Program customization and bloatware mitigation based on static analysis} &
	2016 &
	FALSE \\ \hline
	\href{https://dl.acm.org/doi/abs/10.1145/2897845.2897904}{Checking
	   intent-based communication in android with intent space analysis} &
	2016 &
	FALSE \\ \hline
	\href{https://books.google.com/books?hl=en\&lr=\&id=cfOBDwAAQBAJ\&oi=fnd\&pg=PP1\&dq=android+\%22pre+installed\%22+apps+bloatware+analysis\&ots=4UlpN2dpIW\&sig=rCZMNtfbA9lkOQg41tI3pzdx58g}{Learning
	   Android Forensics: Analyze Android devices with the latest forensic tools 
	and   techniques} &
	2018 &
	FALSE \\ \hline
	\href{https://link.springer.com/chapter/10.1007/978-3-319-72389-1\_15}{Smartphone
	   bloatware: an overlooked privacy problem} &
	2017 &
	TRUE \\ \hline
	\href{http://jictra.com.pk/index.php/jictra/article/view/7}{There   
	is no such thing as free Lunch: An Investigation of Bloatware Effects on   Smart 
	Devices} &
	2018 &
	TRUE \\ \hline
	\href{https://arxiv.org/abs/2010.10088}{How   Did That Get In 
	My Phone? Unwanted App Distribution on Android Devices} &
	2020 &
	TRUE \\ \hline
	\href{https://ieeexplore.ieee.org/abstract/document/8023140/}{Ghost
	   installer in the shadow: Security analysis of app installation on android} &
	2017 &
	FALSE \\ \hline
	\href{https://dl.acm.org/doi/abs/10.1145/2815675.2815686}{Characterizing
	   smartphone usage patterns from millions of android users} &
	2015 &
	FALSE \\ \hline
	\href{https://books.google.com/books?hl=en\&lr=\&id=bABCDwAAQBAJ\&oi=fnd\&pg=PA169\&dq=android+\%22pre+installed\%22+apps+bloatware+analysis\&ots=50amIOx2Z2\&sig=QK\_-w9wNTbpwsXXdX6W6dbt90Q8}{Privacy
	   Problem} &
	2017 &
	FALSE \\ \hline
	\href{https://ieeexplore.ieee.org/abstract/document/6488439/}{Do
	   android users write about electric sheep? examining consumer reviews in   
	google play} &
	2013 &
	FALSE \\ \hline
	\href{https://dl.acm.org/doi/abs/10.1145/2851613.2851713}{Insights
	   into rooted and non-rooted android mobile devices with behavior analytics} &
	2016 &
	FALSE \\ \hline
	\href{https://www.usenix.org/conference/atc15/technical-session/presentation/martins}{Selectively
	   taming background android apps to improve battery lifetime} &
	2015 &
	FALSE \\ \hline
	\href{https://aaltodoc.aalto.fi/handle/123456789/27914}{Intrusion
	   Detection System for Android: Linux kernel system calls analysis} &
	2017 &
	FALSE \\ \hline
	\href{http://www.teis-workshop.org/papers/2016/TEIS\_2016\_6\_Geng.pdf}{Bloatware
	   and Jailbreaking: How Consumer-Initiated Modification Interacts with 
	Product   Pricing} &
	2016 &
	FALSE \\ \hline
	\href{http://sedici.unlp.edu.ar/handle/10915/73288}{Mobile   
	application development approaches: a comparative analysis on the use of   
	storage space} &
	2018 &
	FALSE \\ \hline
	\href{https://ieeexplore.ieee.org/abstract/document/6838907/}{Rootguard:
	   Protecting rooted android phones} &
	2014 &
	FALSE \\ \hline
	\href{https://ieeexplore.ieee.org/abstract/document/9152633/}{An
	   analysis of pre-installed android software} &
	2020 &
	TRUE \\ \hline
	\href{https://www.usenix.org/conference/usenixsecurity20/presentation/elsabagh}{FIRMSCOPE:
	   Automatic Uncovering of Privilege-Escalation Vulnerabilities in Pre-Installed   
	Apps in Android Firmware} &
	2020 &
	TRUE \\ \hline
	\href{https://ieeexplore.ieee.org/abstract/document/7107450/}{Towards
	   mutation analysis of android apps} &
	2015 &
	FALSE \\ \hline
	\href{https://dl.acm.org/doi/abs/10.1145/2420950.2420956}{Permission
	   evolution in the android ecosystem} &
	2012 &
	FALSE \\ \hline
	\href{https://dl.acm.org/doi/abs/10.1145/2508859.2516689}{Vetting
	   undesirable behaviors in android apps with permission use analysis} &
	2013 &
	FALSE \\ \hline
	\href{https://www.academia.edu/download/52584050/5513797455.pdf}{System
	   calls analysis of malwares on android} &
	2013 &
	FALSE \\ \hline
	\href{https://dl.acm.org/doi/abs/10.1145/2590296.2590313}{DroidRay:
	   a security evaluation system for customized android firmwares} &
	2014 &
	FALSE \\ \hline
	\href{https://arxiv.org/abs/1304.4915}{Forensic   analysis of 
	instant messenger applications on android devices} &
	2013 &
	FALSE \\ \hline
	\href{https://re.public.polimi.it/handle/11311/1047427}{Going   
	native: Using a large-scale analysis of android apps to create a practical   
	native-code sandboxing policy} &
	2016 &
	FALSE \\ \hline
	\href{https://dl.acm.org/doi/abs/10.1145/2046707.2046779}{Android
	   permissions demystified} &
	2011 &
	FALSE \\ \hline
	\href{https://dl.acm.org/doi/abs/10.1145/2818000.2818036}{Baredroid:
	   Large-scale analysis of android apps on real devices} &
	2015 &
	FALSE \\ \hline
	\href{https://dl.packetstormsecurity.net/papers/general/NDSS12\_WOODPECKER.pdf}{Systematic
	   detection of capability leaks in stock android smartphones.} &
	2012 &
	FALSE \\ \hline
	\href{https://www.usenix.org/conference/usenixsecurity16/technical-sessions/presentation/aafer}{Harvesting
	   inconsistent security configurations in custom android roms via differential   
	analysis} &
	2016 &
	FALSE \\ \hline
	\href{https://dl.acm.org/doi/abs/10.1145/3143361.3143400}{Studying
	   TLS usage in Android apps} &
	2017 &
	FALSE \\ \hline
	\href{https://www.researchgate.net/profile/Franklin\_Tchakounte/publication/271199472\_Permission-based\_Malware\_Detection\_Mechanisms\_on\_Android\_Analysis\_and\_Perspectives/links/54c11b5c0cf2d03405c4de97.pdf}{Permission-based
	   malware detection mechanisms on android: Analysis and perspectives} &
	2014 &
	FALSE \\ \hline
	\href{https://www.researchgate.net/profile/Hans\_Hoefken/publication/260320851\_Forensic\_Analysis\_of\_Geodata\_in\_Android\_Smartphones/links/0046353148e8fccc09000000/Forensic-Analysis-of-Geodata-in-Android-Smartphones.pdf}{Forensic
	   analysis of geodata in android smartphones} &
	2011 &
	FALSE \\ \hline
	\href{https://www.sciencedirect.com/science/article/pii/S174228761500002X}{APK
	   Auditor: Permission-based Android malware detection system} &
	2015 &
	TRUE \\ \hline
	\href{https://www.sciencedirect.com/science/article/pii/S0950584916300684}{Mutation
	   operators for testing Android apps} &
	2017 &
	TRUE \\ \hline
	\href{https://dl.acm.org/doi/abs/10.1145/2494603.2480308}{Insights
	   into layout patterns of mobile user interfaces by an automatic analysis of   
	android apps} &
	2013 &
	TRUE \\ \hline
	\href{https://dl.acm.org/doi/abs/10.1145/2590296.2590325}{Evading
	   android runtime analysis via sandbox detection} &
	2014 &
	TRUE \\ \hline
	\href{https://ieeexplore.ieee.org/abstract/document/9152633/}{An
	   analysis of pre-installed android software} &
	2020 &
	TRUE \\ \hline
	\href{https://dl.acm.org/doi/abs/10.1145/2746266.2746271}{Android
	   malware static analysis techniques} &
	2015 &
	TRUE \\ \hline
	\href{https://re.public.polimi.it/handle/11311/1047427}{Going   
	native: Using a large-scale analysis of android apps to create a practical   
	native-code sandboxing policy} &
	2016 &
	TRUE \\ \hline
	\href{https://www.usenix.org/conference/usenixsecurity20/presentation/elsabagh}{FIRMSCOPE:
	   Automatic Uncovering of Privilege-Escalation Vulnerabilities in Pre-Installed   
	Apps in Android Firmware} &
	2020 &
	TRUE \\ \hline
	\href{https://dl.acm.org/doi/abs/10.1145/2590296.2590313}{DroidRay:
	   a security evaluation system for customized android firmwares} &
	2014 &
	TRUE \\ \hline
	\href{https://dl.acm.org/doi/abs/10.1145/2420950.2420956}{Permission
	   evolution in the android ecosystem} &
	2012 &
	FALSE \\ \hline
	\href{https://dl.acm.org/doi/abs/10.1145/2046707.2046779}{Android
	   permissions demystified} &
	2011 &
	FALSE \\ \hline
	\href{https://ieeexplore.ieee.org/abstract/document/7557676/}{Merging
	   permission and api features for android malware detection} &
	2016 &
	FALSE \\ \hline
	\href{https://ieeexplore.ieee.org/abstract/document/7961980/}{Open
	   doors for bob and mallory: Open port usage in android apps and security   
	implications} &
	2017 &
	FALSE \\ \hline
	\href{https://dl.acm.org/doi/abs/10.1145/1999995.2000018}{Analyzing
	   inter-application communication in Android} &
	2011 &
	FALSE \\ \hline
	\href{https://www.academia.edu/download/52584050/5513797455.pdf}{System
	   calls analysis of malwares on android} &
	2013 &
	TRUE \\ \hline
	\href{https://dl.acm.org/doi/abs/10.1145/2508859.2516689}{Vetting
	   undesirable behaviors in android apps with permission use analysis} &
	2013 &
	FALSE \\ \hline
	\href{https://ieeexplore.ieee.org/abstract/document/6280226/}{A
	   case study in open source software security and privacy: Android adware} &
	2012 &
	TRUE \\ \hline
	\href{https://link.springer.com/chapter/10.1007/978-3-030-34175-6\_15}{Android
	   multitasking mechanism: Formal semantics and static analysis of apps} &
	2019 &
	TRUE \\ \hline
	\href{https://dl.acm.org/doi/abs/10.1145/3243734.3243843}{Invetter:
	   Locating insecure input validations in android services} &
	2018 &
	TRUE \\ \hline
	\href{https://ieeexplore.ieee.org/abstract/document/7107450/}{Towards
	   mutation analysis of android apps} &
	2015 &
	TRUE \\ \hline
	\href{https://dl.acm.org/doi/abs/10.1145/3143361.3143400}{Studying
	   TLS usage in Android apps} &
	2017 &
	TRUE \\ \hline
	\href{https://dl.packetstormsecurity.net/papers/general/NDSS12\_WOODPECKER.pdf}{Systematic
	   detection of capability leaks in stock android smartphones.} &
	2012 &
	TRUE \\ \hline
	\href{https://ieeexplore.ieee.org/abstract/document/6531746/}{A
	   new android malware detection approach using bayesian classification} &
	2013 &
	FALSE \\ \hline
	\href{https://ieeexplore.ieee.org/abstract/document/9152633/}{An
	   analysis of pre-installed android software} &
	2020 &
	TRUE \\ \hline
	\href{https://www.academia.edu/download/52584050/5513797455.pdf}{System
	   calls analysis of malwares on android} &
	2013 &
	TRUE \\ \hline
	\href{https://dl.acm.org/doi/abs/10.1145/2590296.2590313}{DroidRay:
	   a security evaluation system for customized android firmwares} &
	2014 &
	TRUE \\ \hline
	\href{https://www.researchgate.net/profile/Victor\_Van\_Der\_Veen/publication/270576401\_Dynamic\_Analysis\_of\_Android\_Malware/links/54ae68470cf2828b29fcf321/Dynamic-Analysis-of-Android-Malware.pdf}{Dynamic
	   analysis of android malware} &
	2013 &
	TRUE \\ \hline
	\href{https://dl.acm.org/doi/abs/10.1145/2897073.2897716}{Droidmate:
	   a robust and extensible test generator for android} &
	2016 &
	TRUE \\ \hline
	\href{https://dl.acm.org/doi/abs/10.1145/2590296.2590325}{Evading
	   android runtime analysis via sandbox detection} &
	2014 &
	TRUE \\ \hline
	\href{https://dl.acm.org/doi/abs/10.1145/2508859.2516689}{Vetting
	   undesirable behaviors in android apps with permission use analysis} &
	2013 &
	TRUE \\ \hline
	\href{https://www.usenix.org/conference/usenixsecurity20/presentation/elsabagh}{FIRMSCOPE:
	   Automatic Uncovering of Privilege-Escalation Vulnerabilities in Pre-Installed   
	Apps in Android Firmware} &
	2020 &
	TRUE \\ \hline
	\href{https://ieeexplore.ieee.org/abstract/document/8255546/}{Samadroid:
	   a novel 3-level hybrid malware detection model for android operating system} 
	&
	2018 &
	FALSE \\ \hline
	\href{http://rasthofer.info/publications/paper/NDSS2014-SuSi.pdf}{A
	   machine-learning approach for classifying and categorizing android sources   
	and sinks.} &
	2014 &
	FALSE \\ \hline
	\href{https://re.public.polimi.it/handle/11311/1047427}{Going   
	native: Using a large-scale analysis of android apps to create a practical   
	native-code sandboxing policy} &
	2016 &
	TRUE \\ \hline
	\href{https://dl.acm.org/doi/abs/10.1145/1999995.2000018}{Analyzing
	   inter-application communication in Android} &
	2011 &
	TRUE \\ \hline
	\href{https://ieeexplore.ieee.org/abstract/document/6876208/}{Permission
	   use analysis for vetting undesirable behaviors in android apps} &
	2014 &
	TRUE \\ \hline
	\href{https://dl.acm.org/doi/abs/10.1145/2046707.2046779}{Android
	   permissions demystified} &
	2011 &
	TRUE \\ \hline
	\href{https://ieeexplore.ieee.org/abstract/document/8038362/}{Analyzing
	   android application in real-time at kernel level} &
	2017 &
	TRUE \\ \hline
	\href{https://dl.acm.org/doi/abs/10.1145/2818000.2818036}{Baredroid:
	   Large-scale analysis of android apps on real devices} &
	2015 &
	TRUE \\ \hline
	\href{https://cs.gmu.edu/media/techreports/GMU-CS-TR-2015-5.pdf}{Automated
	   dynamic enforcement of synthesized security policies in android} &
	2015 &
	TRUE \\ \hline
	\href{https://link.springer.com/chapter/10.1007/978-3-319-13257-0\_20}{Analyzing
	   android browser apps for file://vulnerabilities} &
	2014 &
	TRUE \\ \hline
	\href{https://ieeexplore.ieee.org/abstract/document/6956578/}{The
	   peril of fragmentation: Security hazards in android device driver   
	customizations} &
	2014 &
	TRUE \\ \hline
	\href{https://dl.acm.org/doi/abs/10.1145/2590296.2590313}{DroidRay:
	   a security evaluation system for customized android firmwares} &
	2014 &
	TRUE \\ \hline
	\href{https://ieeexplore.ieee.org/abstract/document/7557676/}{Merging
	   permission and api features for android malware detection} &
	2016 &
	TRUE \\ \hline
	\href{https://link.springer.com/chapter/10.1007/978-3-319-04283-1\_6}{Droidapiminer:
	   Mining api-level features for robust malware detection in android} &
	2013 &
	TRUE \\ \hline
	\href{https://www.sciencedirect.com/science/article/pii/S0167404817300640}{PIndroid:
	   A novel Android malware detection system using ensemble learning 
	methods} &
	2017 &
	TRUE \\ \hline
	\href{https://dl.acm.org/doi/abs/10.1145/2420950.2420956}{Permission
	   evolution in the android ecosystem} &
	2012 &
	TRUE \\ \hline
	\href{https://www.researchgate.net/profile/Franklin\_Tchakounte/publication/271199472\_Permission-based\_Malware\_Detection\_Mechanisms\_on\_Android\_Analysis\_and\_Perspectives/links/54c11b5c0cf2d03405c4de97.pdf}{Permission-based
	   malware detection mechanisms on android: Analysis and perspectives} &
	2014 &
	TRUE \\ \hline
	\href{https://ieeexplore.ieee.org/abstract/document/6531746/}{A
	   new android malware detection approach using bayesian classification} &
	2013 &
	TRUE \\ \hline
	\href{https://www.sciencedirect.com/science/article/pii/S174228761500002X}{APK
	   Auditor: Permission-based Android malware detection system} &
	2015 &
	TRUE \\ \hline
	\href{https://dl.acm.org/doi/abs/10.1145/2746266.2746271}{Android
	   malware static analysis techniques} &
	2015 &
	TRUE \\ \hline
	\href{https://ieeexplore.ieee.org/abstract/document/6313688/}{Malicious
	   android applications in the enterprise: What do they do and how do we fix   
	it?} &
	2012 &
	TRUE \\ \hline
	\href{https://ieeexplore.ieee.org/abstract/document/6550526/}{Android
	   SMS malware: Vulnerability and mitigation} &
	2013 &
	TRUE \\ \hline
	\href{https://ieeexplore.ieee.org/abstract/document/9152633/}{An
	   analysis of pre-installed android software} &
	2020 &
	TRUE \\ \hline
	\href{https://arxiv.org/abs/1904.00736}{Using   deep neural 
	network for Android malware detection} &
	2019 &
	TRUE \\ \hline
	\href{http://www.llonely.com/goto/https://public.gdatasoftware.com/Presse/Publikationen/Malware\_Reports/G\_DATA\_MobileMWR\_Q2\_2015\_EN.pdf}{Mobile
	   malware report} &
	2015 &
	TRUE \\ \hline
	\href{https://ieeexplore.ieee.org/abstract/document/7389671/}{A
	   study of android malware detection technology evolution} &
	2015 &
	TRUE \\ \hline
	\href{https://ieeexplore.ieee.org/abstract/document/6956577/}{Upgrading
	   your android, elevating my malware: Privilege escalation through mobile os   
	updating} &
	2014 &
	TRUE \\ \hline
	\href{https://www.academia.edu/download/52584050/5513797455.pdf}{System
	   calls analysis of malwares on android} &
	2013 &
	TRUE \\ \hline
	\href{http://www.itiis.org/digital-library/manuscript/file/20858/TIISVol9No8-12.pdf}{Intelligent
	   Approach for Android Malware Detection.} &
	2015 &
	TRUE \\ \hline
	\href{https://www.researchgate.net/profile/Victor\_Van\_Der\_Veen/publication/270576401\_Dynamic\_Analysis\_of\_Android\_Malware/links/54ae68470cf2828b29fcf321/Dynamic-Analysis-of-Android-Malware.pdf}{Dynamic
	   analysis of android malware} &
	2013 &
	TRUE \\ \hline
	\href{https://ieeexplore.ieee.org/abstract/document/7985674/}{Libd:
	   Scalable and precise third-party library detection in android markets} &
	2017 &
	TRUE \\ \hline
	\href{https://dl.acm.org/doi/abs/10.1145/2976749.2978333}{Reliable
	   third-party library detection in android and its security applications} &
	2016 &
	TRUE \\ \hline
	\href{https://dl.acm.org/doi/abs/10.1145/2889160.2889178}{Libradar:
	   Fast and accurate detection of third-party libraries in android apps} &
	2016 &
	TRUE \\ \hline
	\href{https://ieeexplore.ieee.org/abstract/document/6827639/}{Addetect:
	   Automated detection of android ad libraries using semantic analysis} &
	2014 &
	TRUE \\ \hline
	\href{https://ieeexplore.ieee.org/abstract/document/8543426/}{Orlis:
	   Obfuscation-resilient library detection for Android} &
	2018 &
	TRUE \\ \hline
	\href{https://www.sciencedirect.com/science/article/pii/S0167404818311301}{Securing
	   android applications via edge assistant third-party library detection} &
	2019 &
	TRUE \\ \hline
	\href{https://ieeexplore.ieee.org/abstract/document/8330204/}{Detecting
	   third-party libraries in android applications with high precision and recall} &
	2018 &
	TRUE \\ \hline
	\href{https://ieeexplore.ieee.org/abstract/document/8478000/}{Large-scale
	   third-party library detection in android markets} &
	2018 &
	TRUE \\ \hline
	\href{https://ieeexplore.ieee.org/abstract/document/9286020/}{Automated
	   Third-Party Library Detection for Android Applications: Are We There Yet?} &
	2020 &
	TRUE \\ \hline
	\href{https://ieeexplore.ieee.org/abstract/document/6985631/}{Andarwin:
	   Scalable detection of android application clones based on semantics} &
	2014 &
	TRUE \\ \hline
	\href{https://dl.acm.org/doi/abs/10.1145/3293882.3330563}{Libid:
	   reliable identification of obfuscated third-party android libraries} &
	2019 &
	TRUE \\ \hline
	\href{https://link.springer.com/article/10.1007/s11277-018-5456-4}{Identify
	   and inspect libraries in android applications} &
	2018 &
	TRUE \\ \hline
	\href{https://ieeexplore.ieee.org/abstract/document/6665098/}{Automatic
	   detection of inter-application permission leaks in android applications} &
	2013 &
	TRUE \\ \hline
	\href{https://www.usenix.org/conference/enigma2018/presentation/derr}{The
	   impact of third-party code on android app security} &
	2018 &
	TRUE \\ \hline
	\href{https://www.sciencedirect.com/science/article/pii/S0164121219301001}{Revisiting
	   the impact of common libraries for android-related investigations} &
	2019 &
	TRUE \\ \hline
	\href{https://ieeexplore.ieee.org/abstract/document/7972730/}{Who
	   changed you? Obfuscator identification for Android} &
	2017 &
	TRUE \\ \hline
	\href{https://link.springer.com/chapter/10.1007/978-3-642-40203-6\_11}{Andarwin:
	   Scalable detection of semantically similar android applications} &
	2013 &
	FALSE \\ \hline
	\href{https://dl.acm.org/doi/abs/10.1145/3133956.3134059}{Keep me updated: 
	An empirical study of third-party library updatability on android } 
	&
	2017 &
	TRUE \\ \hline
	\href{https://dl.acm.org/doi/abs/10.1145/2771783.2771800}{Dynamic
	   detection of inter-application communication vulnerabilities in Android} &
	2015 &
	FALSE \\ \hline
	\href{https://ieeexplore.ieee.org/abstract/document/6566868/}{Networkprofiler:
	   Towards automatic fingerprinting of android apps} &
	2013 &
	TRUE \\ \hline
	\href{https://link.springer.com/article/10.1007/s11704-017-6493-y}{Fingerprinting
	   Android malware families} &
	2019 &
	FALSE \\ \hline
	\href{https://www.sciencedirect.com/science/article/pii/S0167404820301437}{Scalable
	   and robust unsupervised android malware fingerprinting using 
	community-based   network partitioning} &
	2020 &
	TRUE \\ \hline
	\href{https://dl.acm.org/doi/abs/10.1145/2513228.2513331}{Design
	   of a mobile inspector for detecting illegal Android applications using   
	fingerprinting} &
	2013 &
	FALSE \\ \hline
	\href{https://link.springer.com/chapter/10.1007/978-3-319-98989-1\_4}{Investigating
	   Fingerprinters and Fingerprinting-Alike Behaviour of Android Applications} &
	2018 &
	FALSE \\ \hline
	\href{https://spectrum.library.concordia.ca/986733/}{Resilient   
	and Scalable Android Malware Fingerprinting and Detection} &
	2020 &
	FALSE \\ \hline
	\href{http://rua.ua.es/dspace/handle/10045/76082}{Android   
	library for recognition of activities of daily living: implementation   
	considerations, challenges, and solutions} &
	2018 &
	FALSE \\ \hline
	\href{https://ieeexplore.ieee.org/abstract/document/8399710/}{WiFi
	   RSSI preprocessing library for Android} &
	2018 &
	FALSE \\ \hline
	\href{https://ieeexplore.ieee.org/abstract/document/8835276/}{Sensorid:
	   Sensor calibration fingerprinting for smartphones} &
	2019 &
	FALSE \\ \hline
	\href{https://www.mdpi.com/1424-8220/21/2/432}{Development 
	  of a Smartphone-Based University Library Navigation and Information 
	Service   Employing Wi-Fi Location Fingerprinting} &
	2021 &
	FALSE \\ \hline
	\href{https://ieeexplore.ieee.org/abstract/document/8543426/}{Orlis:
	   Obfuscation-resilient library detection for Android} &
	2018 &
	TRUE \\ \hline
	\href{https://ieeexplore.ieee.org/abstract/document/8456014/}{Website
	   Fingerprinting by Power Estimation Based Side-Channel Attacks on Android 
	7} &
	2018 &
	FALSE \\ \hline
	\href{https://arxiv.org/abs/1408.1416}{Mobile   device 
	identification via sensor fingerprinting} &
	2014 &
	FALSE \\ \hline
	\href{https://link.springer.com/chapter/10.1007/978-3-319-58469-0\_23}{Binsign:
	   fingerprinting binary functions to support automated analysis of code   
	executables} &
	2017 &
	FALSE \\ \hline
	\href{http://modul.repo.mercubuana-yogya.ac.id/modul/files/openjournal/OpenJournnalOfIndustry/canvas.pdf}{Pixel
	   perfect: Fingerprinting canvas in HTML5} &
	2012 &
	FALSE \\ \hline
	\href{https://ieeexplore.ieee.org/abstract/document/8057232/}{POWERFUL:
	   Mobile app fingerprinting via power analysis} &
	2017 &
	FALSE \\ \hline
	\href{https://www.usenix.org/conference/usenixsecurity18/presentation/bonett}{Discovering
	   flaws in security-focused static analysis tools for android using systematic   
	mutation} &
	2018 &
	TRUE \\ \hline
	\href{https://ieeexplore.ieee.org/abstract/document/7467346/}{HornDroid:
	   Practical and sound static analysis of Android applications by SMT solving} &
	2016 &
	TRUE \\ \hline
	\href{https://dl.acm.org/doi/abs/10.1145/2996358}{droid:   
	Assessment and evaluation of android application analysis tools} &
	2016 &
	TRUE \\ \hline
	\href{https://ieeexplore.ieee.org/abstract/document/7180133/}{A
	   dataset of open-source android applications} &
	2015 &
	TRUE \\ \hline
	\href{https://www.sciencedirect.com/science/article/pii/S0950584912001012}{Static
	   analysis of Android programs} &
	2012 &
	TRUE \\ \hline
	\href{https://ieeexplore.ieee.org/abstract/document/6149548/}{Android:
	   Static analysis using similarity distance} &
	2012 &
	TRUE \\ \hline
	\href{https://dl.acm.org/doi/abs/10.1145/3243734.3243835}{Jn-saf:
	   Precise and efficient ndk/jni-aware inter-language static analysis framework   
	for security vetting of android applications with native code} &
	2018 &
	TRUE \\ \hline
	\href{https://ieeexplore.ieee.org/abstract/document/7582763/}{HybriDroid:
	   static analysis framework for Android hybrid applications} &
	2016 &
	TRUE \\ \hline
	\href{https://dl.acm.org/doi/abs/10.1145/3236024.3236029}{Do  
	 android taint analysis tools keep their promises?} &
	2018 &
	TRUE \\ \hline
	\href{https://ieeexplore.ieee.org/abstract/document/8817113/}{Where
	   are we looking? Understanding android static analysis techniques} &
	2019 &
	TRUE \\ \hline
	\href{https://dl.acm.org/doi/abs/10.1145/2259051.2259056}{Dexpler:
	   converting android dalvik bytecode to jimple for static analysis with soot} &
	2012 &
	TRUE \\ \hline
	\href{https://www.sciencedirect.com/science/article/pii/S0950584917302987}{Static
	   analysis of android apps: A systematic literature review} &
	2017 &
	TRUE \\ \hline
	\href{https://dl.acm.org/doi/abs/10.1145/3293882.3339001}{Androlic:
	   an extensible flow, context, object, field, and path-sensitive static   analysis 
	framework for Android} &
	2019 &
	TRUE \\ \hline
	\href{https://dl.acm.org/doi/abs/10.1145/2970276.2970277}{Reflection-aware
	   static analysis of android apps} &
	2016 &
	TRUE \\ \hline
	\href{https://ieeexplore.ieee.org/abstract/document/8445034/}{Software
	   quality through the eyes of the end-user and static analysis tools: A study   
	on android oss applications} &
	2018 &
	TRUE \\ \hline
	\href{https://ieeexplore.ieee.org/abstract/document/7446031/}{Andrubis--1,000,000
	   apps later: A view on current Android malware behaviors} &
	2014 &
	FALSE \\ \hline
	\href{https://ieeexplore.ieee.org/abstract/document/8539081/}{RedDroid:
	   Android application redundancy customization based on static analysis} &
	2018 &
	TRUE \\ \hline
	\href{https://ieeexplore.ieee.org/abstract/document/8229907/}{An
	   app based on static analysis for Android ransomware} &
	2017 &
	TRUE \\ \hline
	\href{https://ieeexplore.ieee.org/abstract/document/8094430/}{Evaluating
	   state-of-the-art free and open source static analysis tools against buffer   
	errors in android apps} &
	2017 &
	TRUE \\ \hline
	\href{https://dl.acm.org/doi/abs/10.1145/2837614.2837661}{Combining
	   static analysis with probabilistic models to enable market-scale android   
	inter-component analysis} &
	2016 &
	TRUE \\ \hline
	\href{https://ieeexplore.ieee.org/abstract/document/6149548/}{Android:
	   Static analysis using similarity distance} &
	2012 &
	TRUE \\ \hline
	\href{https://ieeexplore.ieee.org/abstract/document/6112328/}{Using
	   static analysis for automatic assessment and mitigation of unwanted and   
	malicious activities within Android applications} &
	2011 &
	TRUE \\ \hline
	\href{https://ieeexplore.ieee.org/abstract/document/7019227/}{Detecting
	   software vulnerabilities in android using static analysis} &
	2014 &
	TRUE \\ \hline
	\href{https://www.sciencedirect.com/science/article/pii/S0950584917302987}{Static
	   analysis of android apps: A systematic literature review} &
	2017 &
	TRUE \\ \hline
	\href{http://isyou.info/jisis/vol6/no4/jisis-2016-vol6-no4-03.pdf}{Android
	   Application Protection against Static Reverse Engineering based on   
	Multidexing.} &
	2016 &
	TRUE \\ \hline
	\href{https://onlinelibrary.wiley.com/doi/abs/10.1002/spe.2643}{VAnDroid:
	   A framework for vulnerability analysis of Android applications using a   
	model‐driven reverse engineering technique} &
	2019 &
	TRUE \\ \hline
	\href{https://dl.acm.org/doi/abs/10.1145/2751323.2751330}{Exploring
	   reverse engineering symptoms in Android apps} &
	2015 &
	TRUE \\ \hline
	\href{https://link.springer.com/article/10.1631/FITEE.1601491}{Discovering
	   optimal features using static analysis and a genetic search based method 
	for   Android malware detection} &
	2018 &
	TRUE \\ \hline
	\href{https://www.mdpi.com/558048}{AndroShield:   automated 
	android applications vulnerability detection, a hybrid static and   dynamic 
	analysis approach} &
	2019 &
	TRUE \\ \hline
	\href{https://www.itc.ktu.lt/index.php/ITC/article/view/21457}{What
	   static analysis can utmost offer for Android malware detection} &
	2019 &
	TRUE \\ \hline
	\href{https://ieeexplore.ieee.org/abstract/document/8539081/}{RedDroid:
	   Android application redundancy customization based on static analysis} &
	2018 &
	TRUE \\ \hline
	\href{https://www.researchgate.net/profile/Mark-Stamp/publication/314521542\_Static\_and\_Dynamic\_Analysis\_of\_Android\_Malware/links/5bdee8ad299bf1124fb9cab6/Static-and-Dynamic-Analysis-of-Android-Malware.pdf}{Static
	   and dynamic analysis of android malware.} &
	2017 &
	TRUE \\ \hline
	\href{https://arxiv.org/abs/1410.7751}{Andlantis:   Large-scale 
	Android dynamic analysis} &
	2014 &
	TRUE \\ \hline
	\href{https://link.springer.com/article/10.1007/s10515-018-0237-6}{Static
	   window transition graphs for Android} &
	2018 &
	TRUE \\ \hline
	\href{https://link.springer.com/chapter/10.1007/978-3-030-58793-2\_30}{Reverse
	   Engineering of Android Applications: REiMPAcT} &
	2020 &
	TRUE \\ \hline
	\href{https://dl.acm.org/doi/abs/10.1145/2259051.2259056}{Dexpler:
	   converting android dalvik bytecode to jimple for static analysis with soot} &
	2012 &
	TRUE \\ \hline
	\href{https://ieeexplore.ieee.org/abstract/document/8716380/}{Android
	   applications scanning: The guide} &
	2019 &
	TRUE \\ \hline
	\href{https://journal.utem.edu.my/index.php/jtec/article/view/2870}{A
	   hybrid approach for reverse engineering GUI model from android apps for   
	automated testing} &
	2017 &
	TRUE \\ \hline
	\href{https://ieeexplore.ieee.org/abstract/document/6079840/}{An
	   Android security case study with bauhaus} &
	2011 &
	TRUE \\ \hline
	\href{https://link.springer.com/chapter/10.1007/978-3-319-68690-5\_12}{Detecting
	   energy bugs in android apps using static analysis} &
	2017 &
	TRUE \\ \hline
	\href{https://ieeexplore.ieee.org/abstract/document/6813664/}{Static
	   analysis for extracting permission checks of a large scale framework: The   
	challenges and solutions for analyzing android} &
	2014 &
	TRUE \\ \hline
	\href{https://dl.acm.org/doi/abs/10.1145/2382196.2382222}{Pscout:
	   analyzing the android permission specification} &
	2012 &
	TRUE \\ \hline
	\href{https://ieeexplore.ieee.org/abstract/document/6814679/}{Droid
	   permission miner: Mining prominent permissions for Android malware 
	analysis} &
	2014 &
	TRUE \\ \hline
	\href{https://www.researchgate.net/profile/Mark-Stamp/publication/314521542\_Static\_and\_Dynamic\_Analysis\_of\_Android\_Malware/links/5bdee8ad299bf1124fb9cab6/Static-and-Dynamic-Analysis-of-Android-Malware.pdf}{Static
	   and dynamic analysis of android malware.} &
	2017 &
	TRUE \\ \hline
	\href{https://ieeexplore.ieee.org/abstract/document/6494934/}{Automatically
	   securing permission-based software by reducing the attack surface: An   
	application to android} &
	2012 &
	TRUE \\ \hline
	\href{https://ieeexplore.ieee.org/abstract/document/7417621/}{Reevaluating
	   android permission gaps with static and dynamic analysis} &
	2015 &
	TRUE \\ \hline
	\href{https://dl.acm.org/doi/abs/10.1145/2791405.2791558}{Two
	   phase static analysis technique for android malware detection} &
	2015 &
	FALSE \\ \hline
	\href{https://ieeexplore.ieee.org/abstract/document/7792435/}{ANASTASIA:
	   ANdroid mAlware detection using STatic analySIs of Applications} &
	2016 &
	FALSE \\ \hline
	\href{https://dl.acm.org/doi/abs/10.1145/2484313.2484359}{DroidAlarm:
	   an all-sided static analysis tool for Android privilege-escalation malware} &
	2013 &
	FALSE \\ \hline
	\href{https://www.sciencedirect.com/science/article/pii/S0925231217312870}{DroidDet:
	   effective and robust detection of android malware using static analysis along   
	with rotation forest model} &
	2018 &
	FALSE \\ \hline
	\href{https://ieeexplore.ieee.org/abstract/document/6735264/}{Machine
	   learning for android malware detection using permission and api calls} &
	2013 &
	FALSE \\ \hline
	\href{https://www.sciencedirect.com/science/article/pii/S0164121220300169}{An
	   Android application risk evaluation framework based on minimum permission 
	set   identification} &
	2020 &
	TRUE \\ \hline
	\href{https://ieeexplore.ieee.org/abstract/document/7203053/}{Analysis
	   of android inter-app security vulnerabilities using covert} &
	2015 &
	FALSE \\ \hline
	\href{https://scholarworks.sjsu.edu/etd\_projects/488/}{Static   
	and dynamic analysis for android malware detection} &
	2016 &
	FALSE \\ \hline
	\href{https://ieeexplore.ieee.org/abstract/document/7009096/}{Static
	   detection of Android malware by using permissions and API calls} &
	2014 &
	FALSE \\ \hline
	\href{https://link.springer.com/chapter/10.1007/978-981-10-7200-0\_13}{Static
	   and dynamic analysis for android malware detection} &
	2018 &
	FALSE \\ \hline
	\href{https://link.springer.com/article/10.1007/s12652-020-01957-5}{A
	   novel permission ranking system for android malware detection—the 
	permission   grader} &
	2020 &
	TRUE \\ \hline
	\href{https://aip.scitation.org/doi/abs/10.1063/1.4992953}{An   
	Android malware detection system based on machine learning} &
	2017 &
	FALSE \\ \hline
	\href{https://digital-library.theiet.org/content/journals/10.1049/iet-ifs.2014.0099}{High
	   accuracy android malware detection using ensemble learning} &
	2015 &
	FALSE \\ \hline
	\href{https://ieeexplore.ieee.org/abstract/document/7537161/}{Detection
	   of android malware by static analysis on permissions and sensitive 
	functions} &
	2016 &
	TRUE \\ \hline
	\href{https://link.springer.com/chapter/10.1007/978-94-007-2792-2\_50}{A
	   study on static analysis model of mobile application for privacy protection} &
	2012 &
	FALSE \\ \hline
	\href{https://iopscience.iop.org/article/10.1088/1742-6596/1288/1/012053/meta}{Vulnerability
	   Parser: A Static Vulnerability Analysis System for Android Applications} &
	2019 &
	TRUE \\ \hline
	\href{https://ieeexplore.ieee.org/abstract/document/6298136/}{Droidmat:
	   Android malware detection through manifest and api calls tracing} &
	2012 &
	FALSE \\ \hline
	\href{https://dl.acm.org/doi/abs/10.1145/2185448.2185466}{Droidchecker:
	   analyzing android applications for capability leak} &
	2012 &
	FALSE \\ \hline
	\href{https://ieeexplore.ieee.org/abstract/document/8697714/}{S3ntinel:
	   An Extensible Static Analysis Framework for Android Applications} &
	2018 &
	TRUE \\ \hline
	\href{https://dl.acm.org/doi/abs/10.1145/2837614.2837661}{Combining
	   static analysis with probabilistic models to enable market-scale android   
	inter-component analysis} &
	2016 &
	FALSE \\ \hline
	\href{https://dl.acm.org/doi/abs/10.1145/2480362.2480701}{Mobile-sandbox:
	   having a deeper look into android applications} &
	2013 &
	FALSE \\ \hline
	\href{https://dl.acm.org/doi/abs/10.1145/3243734.3243835}{Jn-saf:
	   Precise and efficient ndk/jni-aware inter-language static analysis framework   
	for security vetting of android applications with native code} &
	2018 &
	FALSE \\ \hline
	\href{https://ieeexplore.ieee.org/abstract/document/7307791/}{Android
	   malware detection based on static analysis of characteristic tree} &
	2015 &
	FALSE \\ \hline
	\href{https://dl.acm.org/doi/abs/10.1145/2897845.2897927}{R-droid:
	   Leveraging android app analysis with static slice optimization} &
	2016 &
	FALSE \\ \hline
	\href{https://ieeexplore.ieee.org/abstract/document/9283770/}{ASAF:
	   Android Static Analysis Framework} &
	2020 &
	FALSE \\ \hline
	\href{https://dl.acm.org/doi/abs/10.1145/2484313.2484359}{DroidAlarm:
	   an all-sided static analysis tool for Android privilege-escalation malware} &
	2013 &
	FALSE \\ \hline
	\href{https://www.usenix.org/conference/usenixsecurity13/technical-sessions/presentation/octeau}{Effective
	   inter-component communication mapping in android: An essential step 
	towards   holistic security analysis} &
	2013 &
	FALSE \\ \hline
	\href{https://people.csail.mit.edu/rinard/paper/ndss15.droidsafe.pdf}{Information
	   flow analysis of android applications in droidsafe.} &
	2015 &
	FALSE \\ \hline
	\href{https://ieeexplore.ieee.org/abstract/document/7446031/}{Andrubis--1,000,000
	   apps later: A view on current Android malware behaviors} &
	2014 &
	FALSE \\ \hline
	\href{https://dl.acm.org/doi/abs/10.1145/2382196.2382222}{Pscout:
	   analyzing the android permission specification} &
	2012 &
	FALSE \\ \hline
	\href{https://dl.acm.org/doi/abs/10.1145/2666356.2594299}{Flowdroid:
	   Precise context, flow, field, object-sensitive and lifecycle-aware taint   
	analysis for android apps} &
	2014 &
	FALSE \\ \hline
	\href{https://digital-library.theiet.org/content/journals/10.1049/iet-ifs.2013.0095}{Analysis
	   of Bayesian classification-based approaches for Android malware detection} 
	&
	2014 &
	FALSE \\ \hline
	\href{https://ieeexplore.ieee.org/abstract/document/7163067/}{Effective
	   real-time android application auditing} &
	2015 &
	FALSE \\ \hline
	\href{https://www.usenix.org/conference/usenixsecurity20/presentation/hernandez}{BigMAC:
	   Fine-Grained Policy Analysis of Android Firmware} &
	2020 &
	TRUE \\ \hline
	\href{https://dl.acm.org/doi/abs/10.1145/2590296.2590312}{Efficient,
	   context-aware privacy leakage confinement for android applications without   
	firmware modding} &
	2014 &
	FALSE \\ \hline
	\href{https://www.researchgate.net/profile/Victor\_Van\_Der\_Veen/publication/270576401\_Dynamic\_Analysis\_of\_Android\_Malware/links/54ae68470cf2828b29fcf321/Dynamic-Analysis-of-Android-Malware.pdf}{Dynamic
	   analysis of android malware} &
	2013 &
	FALSE \\ \hline
	\href{https://www.usenix.org/conference/usenixsecurity20/presentation/harrison}{PARTEMU:
	   Enabling Dynamic Analysis of Real-World TrustZone Software Using 
	Emulation} &
	2020 &
	TRUE \\ \hline
	\href{https://dl.acm.org/doi/abs/10.1145/3359789.3359833}{Opening
	   Pandora's box through ATFuzzer: dynamic analysis of AT interface for 
	Android   smartphones} &
	2019 &
	FALSE \\ \hline
	\href{https://ieeexplore.ieee.org/abstract/document/7226692/}{Code
	   coverage measurement for Android dynamic analysis tools} &
	2015 &
	TRUE \\ \hline
	\href{https://www.usenix.org/conference/usenixsecurity20/presentation/elsabagh}{FIRMSCOPE:
	   Automatic Uncovering of Privilege-Escalation Vulnerabilities in Pre-Installed   
	Apps in Android Firmware} &
	2020 &
	TRUE \\ \hline
	\href{https://dl.acm.org/doi/abs/10.1145/2590296.2590313}{DroidRay:
	   a security evaluation system for customized android firmwares} &
	2014 &
	TRUE \\ \hline
	\href{https://www.usenix.org/conference/usenixsecurity18/presentation/talebi}{Charm:
	   Facilitating dynamic analysis of device drivers of mobile systems} &
	2018 &
	TRUE \\ \hline
	\href{https://www.sciencedirect.com/science/article/pii/S1742287617303948}{New
	   flash memory acquisition methods based on firmware update protocols for 
	LG   Android smartphones} &
	2018 &
	FALSE \\ \hline
	\href{https://ieeexplore.ieee.org/abstract/document/9153399/}{Hybrid
	   firmware analysis for known mobile and iot security vulnerabilities} &
	2020 &
	TRUE \\ \hline
	\href{https://www.researchgate.net/profile/Jonas\_Zaddach/publication/269197057\_Avatar\_A\_Framework\_to\_Support\_Dynamic\_Security\_Analysis\_of\_Embedded\_Systems'\_Firmwares/links/5e0b2725299bf10bc3852355/Avatar-A-Framework-to-Support-Dynamic-Security-Analysis-of-Embedded-Systems-Firmwares.pdf}{AVATAR:
	   A Framework to Support Dynamic Security Analysis of Embedded Systems'   
	Firmwares.} &
	2014 &
	TRUE \\ \hline
	\href{https://www.usenix.org/conference/usenixsecurity20/presentation/clements}{HALucinator:
	   Firmware re-hosting through abstraction layer emulation} &
	2020 &
	TRUE \\ \hline
	\href{https://ieeexplore.ieee.org/abstract/document/8368433/}{Vulnerability
	   detection in iot firmware: A survey} &
	2017 &
	TRUE \\ \hline
	\href{https://dl.acm.org/doi/abs/10.1145/2939918.2939926}{Evading
	   android runtime analysis through detecting programmed interactions} &
	2016 &
	FALSE \\ \hline
	\href{https://ieeexplore.ieee.org/abstract/document/9152633/}{An
	   analysis of pre-installed android software} &
	2020 &
	TRUE \\ \hline
	\href{https://ieeexplore.ieee.org/abstract/document/7446031/}{Andrubis
	   1,000,000 apps later: A view on current Android malware behaviors} &
	2014 &
	FALSE \\ \hline
	\href{https://ieeexplore.ieee.org/abstract/document/8787829/}{Toward
	   Automated Firmware Analysis in the IoT Era} &
	2019 &
	FALSE \\ \hline
	\href{https://content.iospress.com/articles/journal-of-intelligent-and-fuzzy-systems/ifs169424}{Detecting
	   Android malware using long short-term memory (LSTM)} &
	2018 &
	FALSE \\ \hline
	\href{https://ieeexplore.ieee.org/abstract/document/8278529/}{A
	   collaborative framework for android malware detection using DNS \& 
	dynamic   analysis} &
	2017 &
	FALSE \\ \hline
	\href{https://ieeexplore.ieee.org/abstract/document/9152633/}{An
	   analysis of pre-installed android software} &
	2020 &
	TRUE \\ \hline
	\href{https://dl.acm.org/doi/abs/10.1145/2590296.2590313}{DroidRay:
	   a security evaluation system for customized android firmwares} &
	2014 &
	TRUE \\ \hline
	\href{https://www.academia.edu/download/52584050/5513797455.pdf}{System
	   calls analysis of malwares on android} &
	2013 &
	FALSE \\ \hline
	\href{https://www.researchgate.net/profile/Victor\_Van\_Der\_Veen/publication/270576401\_Dynamic\_Analysis\_of\_Android\_Malware/links/54ae68470cf2828b29fcf321/Dynamic-Analysis-of-Android-Malware.pdf}{Dynamic
	   analysis of android malware} &
	2013 &
	FALSE \\ \hline
	\href{https://dl.acm.org/doi/abs/10.1145/2897073.2897716}{Droidmate:
	   a robust and extensible test generator for android} &
	2016 &
	FALSE \\ \hline
	\href{https://ieeexplore.ieee.org/abstract/document/8255546/}{Samadroid:
	   a novel 3-level hybrid malware detection model for android operating system} 
	&
	2018 &
	FALSE \\ \hline
	\href{https://www.usenix.org/conference/usenixsecurity20/presentation/elsabagh}{FIRMSCOPE:
	   Automatic Uncovering of Privilege-Escalation Vulnerabilities in Pre-Installed   
	Apps in Android Firmware} &
	2020 &
	TRUE \\ \hline
	\href{http://rasthofer.info/publications/paper/NDSS2014-SuSi.pdf}{A
	   machine-learning approach for classifying and categorizing android sources   
	and sinks.} &
	2014 &
	FALSE \\ \hline
	\href{https://dl.acm.org/doi/abs/10.1145/2508859.2516689}{Vetting
	   undesirable behaviors in android apps with permission use analysis} &
	2013 &
	FALSE \\ \hline
	\href{https://dl.acm.org/doi/abs/10.1145/3278532.3278558}{Beyond
	   google play: A large-scale comparative study of chinese android app 
	markets} &
	2018 &
	TRUE \\ \hline
	\href{https://cs.gmu.edu/media/techreports/GMU-CS-TR-2015-5.pdf}{Automated
	   dynamic enforcement of synthesized security policies in android} &
	2015 &
	FALSE \\ \hline
	\href{https://ieeexplore.ieee.org/abstract/document/6956578/}{The
	   peril of fragmentation: Security hazards in android device driver   
	customizations} &
	2014 &
	FALSE \\ \hline
	\href{https://re.public.polimi.it/handle/11311/1047427}{Going   
	native: Using a large-scale analysis of android apps to create a practical   
	native-code sandboxing policy} &
	2016 &
	FALSE \\ \hline
	\href{https://link.springer.com/chapter/10.1007/978-3-319-04283-1\_6}{Droidapiminer:
	   Mining api-level features for robust malware detection in android} &
	2013 &
	FALSE \\ \hline
	\href{https://dl.acm.org/doi/abs/10.1145/2590296.2590325}{Evading
	   android runtime analysis via sandbox detection} &
	2014 &
	FALSE \\ \hline
	\href{https://link.springer.com/chapter/10.1007/978-3-319-13257-0\_20}{Analyzing
	   android browser apps for file://vulnerabilities} &
	2014 &
	FALSE \\ \hline
	\href{https://ieeexplore.ieee.org/abstract/document/6280226/}{A
	   case study in open source software security and privacy: Android adware} &
	2012 &
	FALSE \\ \hline
	\href{https://dl.acm.org/doi/abs/10.1145/3300061.3345452}{Device
	   administrator use and abuse in Android: Detection and characterization} &
	2019 &
	FALSE \\ \hline
	\href{https://dl.packetstormsecurity.net/papers/general/NDSS12\_WOODPECKER.pdf}{Systematic
	   detection of capability leaks in stock android smartphones.} &
	2012 &
	FALSE \\ \hline
	\href{https://dl.acm.org/doi/abs/10.1145/2664243.2664245}{Design
	   and implementation of an android host-based intrusion prevention system} &
	2014 &
	FALSE \\ \hline
	\href{https://ieeexplore.ieee.org/abstract/document/6982893/}{Malware
	   detection using network traffic analysis in android based mobile devices} &
	2014 &
	TRUE \\ \hline
	\href{https://arxiv.org/abs/1410.7751}{Andlantis:   Large-scale 
	Android dynamic analysis} &
	2014 &
	TRUE \\ \hline
	\href{https://dl.acm.org/doi/abs/10.1145/2940343.2940348}{CREDROID:
	   Android malware detection by network traffic analysis} &
	2016 &
	TRUE \\ \hline
	\href{https://ieeexplore.ieee.org/abstract/document/6657309/}{Ananas-a
	   framework for analyzing android applications} &
	2013 &
	TRUE \\ \hline
	\href{https://dl.acm.org/doi/abs/10.1145/2480362.2480701}{Mobile-sandbox:
	   having a deeper look into android applications} &
	2013 &
	TRUE \\ \hline
	\href{https://link.springer.com/chapter/10.1007/978-3-319-72550-5\_46}{Android
	   malware detection based on network traffic using decision tree algorithm} &
	2018 &
	TRUE \\ \hline
	\href{https://www.sciencedirect.com/science/article/pii/S016412122030087X}{Identifying
	   vulnerabilities of SSL/TLS certificate verification in Android apps with   static 
	and dynamic analysis} &
	2020 &
	TRUE \\ \hline
	\href{https://ieeexplore.ieee.org/abstract/document/9153400/}{LIBSPECTOR:
	   Context-Aware Large-Scale Network Traffic Analysis of Android Applications} 
	&
	2020 &
	TRUE \\ \hline
	\href{https://ieeexplore.ieee.org/abstract/document/7013120/}{Comparative
	   study of k-means and mini batch k-means clustering algorithms in android   
	malware detection using network traffic analysis} &
	2014 &
	TRUE \\ \hline
	\href{https://dl.acm.org/doi/abs/10.1145/2592791.2592796}{Rage
	   against the virtual machine: hindering dynamic analysis of android malware} &
	2014 &
	TRUE \\ \hline
	\href{https://ieeexplore.ieee.org/abstract/document/7043530/}{Malware
	   detection in Android by network traffic analysis} &
	2015 &
	TRUE \\ \hline
	\href{https://ieeexplore.ieee.org/abstract/document/7035780/}{Manilyzer:
	   automated android malware detection through manifest analysis} &
	2014 &
	TRUE \\ \hline
	\href{https://www.mdpi.com/558048}{AndroShield:   automated 
	android applications vulnerability detection, a hybrid static and   dynamic 
	analysis approach} &
	2019 &
	TRUE \\ \hline
	\href{https://www.sciencedirect.com/science/article/pii/S2214212619303837}{Visualizing
	   the outcome of dynamic analysis of Android malware with VizMal} &
	2020 &
	TRUE \\ \hline
	\href{https://www.ndss-symposium.org/wp-content/uploads/2017/09/intellidroid-targeted-input-generator-dynamic-analysis-android-malware.pdf}{IntelliDroid:
	   A Targeted Input Generator for the Dynamic Analysis of Android Malware.} &
	2016 &
	TRUE \\ \hline
	\href{https://ieeexplore.ieee.org/abstract/document/8455983/}{NTPDroid:
	   a hybrid android malware detector using network traffic and system   
	permissions} &
	2018 &
	TRUE \\ \hline
	\href{https://www.researchgate.net/profile/Mark-Stamp/publication/314521542\_Static\_and\_Dynamic\_Analysis\_of\_Android\_Malware/links/5bdee8ad299bf1124fb9cab6/Static-and-Dynamic-Analysis-of-Android-Malware.pdf}{Static
	   and dynamic analysis of android malware.} &
	2017 &
	TRUE \\ \hline
	\href{https://www.sciencedirect.com/science/article/pii/S1742287617303183}{An
	   in-depth analysis of Android malware using hybrid techniques} &
	2018 &
	TRUE \\ \hline
	\href{https://ieeexplore.ieee.org/abstract/document/8374030/}{A
	   novel dynamic Android malware detection system with ensemble learning} &
	2018 &
	TRUE \\ \hline
	\href{https://link.springer.com/article/10.1007/s10207-014-0250-0}{Mobile-Sandbox:
	   combining static and dynamic analysis with machine-learning techniques} &
	2015 &
	TRUE \\ \hline
	\href{https://search.proquest.com/openview/7dedd0447f7a9400cb60eba599536da6/1?pq-origsite=gscholar\&cbl=18750\&diss=y}{DroidDocker:
	   Dynamic analysis of android applications on Docker} &
	2016 &
	TRUE \\ \hline
	\href{https://www.worldscientific.com/doi/abs/10.1142/9789813273238\_0066}{A
	   new tool for static and dynamic Android malware analysis} &
	2018 &
	TRUE \\ \hline
	\href{http://128.171.57.22/handle/10125/64555}{Detecting   
	repackaged android applications using perceptual hashing} &
	2020 &
	FALSE \\ \hline
	\href{https://ieeexplore.ieee.org/abstract/document/7214136/}{Dynalize:
	   Dynamic analysis of mobile apps in a platform-as-a-service cloud} &
	2015 &
	TRUE \\ \hline
	\href{https://www.sciencedirect.com/science/article/pii/S1877050916302873}{Scredent:
	   Scalable real-time anomalies detection and notification of targeted malware   
	in mobile devices} &
	2016 &
	FALSE \\ \hline
	\href{https://www.sciencedirect.com/science/article/pii/S1566253518306778}{Android
	   malware detection through hybrid features fusion and ensemble classifiers:   
	The AndroPyTool framework and the OmniDroid dataset} &
	2019 &
	FALSE \\ \hline
	\href{https://ieeexplore.ieee.org/abstract/document/8622324/}{R2-d2:
	   Color-inspired convolutional neural network (cnn)-based android malware   
	detections} &
	2018 &
	FALSE \\ \hline
	\href{https://content.iospress.com/articles/journal-of-intelligent-and-fuzzy-systems/ifs169953}{Grefenstette
	   Bias based genetic algorithm for multi-site offloading using docker container   
	in edge computing} &
	2019 &
	FALSE \\ \hline
	\href{https://ieeexplore.ieee.org/abstract/document/8679493/}{Analysis
	   and detection of malware in Android applications using machine learning} &
	2019 &
	FALSE \\ \hline
	\href{http://www.iaeng.org/IJCS/issues\_v45/issue\_4/IJCS\_45\_4\_03.pdf}{Android
	   malware classification based on mobile security framework} &
	2018 &
	FALSE \\ \hline
	\href{https://link.springer.com/chapter/10.1007/978-3-030-58951-6\_13}{Understanding
	   the Security Risks of Docker Hub} &
	2020 &
	FALSE \\ \hline
	\href{https://ieeexplore.ieee.org/abstract/document/9284058/}{Time-travel
	   testing of Android apps} &
	2020 &
	FALSE \\ \hline
	\href{https://arxiv.org/abs/2005.10762}{Concurrency-related   
	flaky test detection in android apps} &
	2020 &
	FALSE \\ \hline
	\href{https://dl.acm.org/doi/abs/10.1145/3329786}{Dynamic   
	malware analysis in the modern era. A state of the art survey} &
	2019 &
	FALSE \\ \hline
	\href{https://ieeexplore.ieee.org/abstract/document/8667832/}{An
	   empirical case study on the temporary file smell in dockerfiles} &
	2019 &
	FALSE \\ \hline
	\href{https://ieeexplore.ieee.org/abstract/document/8026871/}{Performance
	   tuning and modeling for big data applications in docker containers} &
	2017 &
	FALSE \\ \hline
	\href{http://scholar.sun.ac.za/handle/10019.1/102802}{Verifying
	   Android Applications Using Java PathFinder} &
	2017 &
	FALSE \\ \hline
	\href{https://ieeexplore.ieee.org/abstract/document/8890109/}{FastBuild:
	   Accelerating Docker image building for efficient development and 
	deployment   of container} &
	2019 &
	FALSE \\ \hline
	\href{https://arxiv.org/abs/1410.7751}{Andlantis:   Large-scale 
	Android dynamic analysis} &
	2014 &
	FALSE \\ \hline
	\href{https://dl.acm.org/doi/abs/10.1145/3041008.3041010}{Emulator
	   vs real phone: Android malware detection using machine learning} &
	2017 &
	TRUE \\ \hline
	\href{https://dl.acm.org/doi/abs/10.1145/2592791.2592796}{Rage
	   against the virtual machine: hindering dynamic analysis of android malware} &
	2014 &
	FALSE \\ \hline
	\href{https://dl.acm.org/doi/abs/10.1145/2799979.2800004}{A   
	robust dynamic analysis system preventing SandBox detection by Android   
	malware} &
	2015 &
	FALSE \\ \hline
	\href{https://ieeexplore.ieee.org/abstract/document/6657309/}{Ananas-a
	   framework for analyzing android applications} &
	2013 &
	FALSE \\ \hline
	\href{https://www.mdpi.com/558048}{AndroShield:   automated 
	android applications vulnerability detection, a hybrid static and   dynamic 
	analysis approach} &
	2019 &
	FALSE \\ \hline
	\href{https://cecs.wright.edu/$\sim$pmateti/Courses/4440/Lectures/Detection/summercon12-bouncer.pdf}{Dissecting
	   the android bouncer} &
	2012 &
	FALSE \\ \hline
	\href{https://www.researchgate.net/profile/Mark-Stamp/publication/314521542\_Static\_and\_Dynamic\_Analysis\_of\_Android\_Malware/links/5bdee8ad299bf1124fb9cab6/Static-and-Dynamic-Analysis-of-Android-Malware.pdf}{Static
	   and dynamic analysis of android malware.} &
	2017 &
	FALSE \\ \hline
	\href{https://ieeexplore.ieee.org/abstract/document/7502337/}{DynaLog:
	   An automated dynamic analysis framework for characterizing android   
	applications} &
	2016 &
	FALSE \\ \hline
	\href{http://ceur-ws.org/Vol-1575/paper\_10.pdf}{ARTDroid:   A 
	Virtual-Method Hooking Framework on Android ART Runtime.} &
	2016 &
	FALSE \\ \hline
	\href{https://ieeexplore.ieee.org/abstract/document/9152767/}{Ex-vivo
	   dynamic analysis framework for Android device drivers} &
	2020 &
	FALSE \\ \hline
	\href{https://ieeexplore.ieee.org/abstract/document/7098023/}{Condroid:
	   Targeted dynamic analysis of android applications} &
	2015 &
	FALSE \\ \hline
	\href{https://dl.acm.org/doi/abs/10.1145/2480362.2480701}{Mobile-sandbox:
	   having a deeper look into android applications} &
	2013 &
	TRUE \\ \hline
	\href{https://arxiv.org/abs/1609.04718}{Glassbox:   dynamic 
	analysis platform for malware android applications on real devices} &
	2016 &
	TRUE \\ \hline
	\href{https://ieeexplore.ieee.org/abstract/document/8326293/}{DDefender:
	   Android application threat detection using static and dynamic analysis} &
	2018 &
	FALSE \\ \hline
	\href{https://ieeexplore.ieee.org/abstract/document/6906344/}{DroidTrace:
	   A ptrace based Android dynamic analysis system with forward execution   
	capability} &
	2014 &
	FALSE \\ \hline
	\href{https://dl.acm.org/doi/abs/10.1145/3134600.3134601}{Droid-AntiRM:
	   Taming control flow anti-analysis to support automated dynamic analysis of   
	android malware} &
	2017 &
	FALSE \\ \hline
	\href{https://books.google.com/books?hl=en\&lr=\&id=72nnDwAAQBAJ\&oi=fnd\&pg=PA73\&dq=android+dynamic+analysis+emulator\&ots=42pKkKSpv8\&sig=Xl8zujvsUB4AXehU6G48hDZ5DfM}{Effectiveness
	   of state-of-the-art dynamic analysis techniques in identifying diverse   Android 
	malware and future enhancements} &
	2020 &
	FALSE \\ \hline
	\href{https://link.springer.com/chapter/10.1007/978-3-030-61638-0\_26}{DaVinci:
	   Android App Analysis Beyond Frida via Dynamic System Call 
	Instrumentation} &
	2020 &
	TRUE \\ \hline
	\href{http://ceur-ws.org/Vol-1575/paper\_10.pdf}{ARTDroid:   A 
	Virtual-Method Hooking Framework on Android ART Runtime.} &
	2016 &
	TRUE \\ \hline
	\href{https://dl.acm.org/doi/abs/10.1145/3400286.3418266}{EDAroid:
	   An Efficient Dynamic Analysis Tool for Android Applications} &
	2020 &
	TRUE \\ \hline
	\href{https://ieeexplore.ieee.org/abstract/document/8029446/}{T2Droid:
	   A TrustZone-based dynamic analyser for Android applications} &
	2017 &
	TRUE \\ \hline
	\href{https://ieeexplore.ieee.org/abstract/document/8585560/}{Toward
	   developing a systematic approach to generate benchmark android malware   
	datasets and classification} &
	2018 &
	TRUE \\ \hline
	\href{https://ieeexplore.ieee.org/abstract/document/9206660/}{On
	   the (un) reliability of privacy policies in android apps} &
	2020 &
	TRUE \\ \hline
	\href{https://link.springer.com/chapter/10.1007/978-3-319-93411-2\_4}{Honey,
	   i shrunk your app security: The state of android app hardening} &
	2018 &
	FALSE \\ \hline
	\href{https://www.mdpi.com/2073-8994/12/7/1128}{Two   
	Anatomists Are Better than One—Dual-Level Android Malware Detection} &
	2020 &
	FALSE \\ \hline
	\href{https://ieeexplore.ieee.org/abstract/document/9286110/}{An
	   Automated Assessment of Android Clipboards} &
	2020 &
	TRUE \\ \hline
	\href{https://i.blackhat.com/us-18/Thu-August-9/us-18-Schranz-ARTist-A-Novel-Instrumentation-Framework-for-Reversing-and-Analyzing-Android-Apps-and-the-Middleware.pdf}{ARTist-A
	   novel ins trumentation framework for reversing and analyzing android apps 
	and   the middleware} &
	2019 &
	TRUE \\ \hline
	\href{https://publications.cispa.saarland/id/eprint/3340}{Bringing
	   Balance to the Force: Dynamic Analysis of the Android Application 
	Framework} &
	2021 &
	TRUE \\ \hline
	\href{https://link.springer.com/chapter/10.1007/978-3-030-44038-1\_120}{Android
	   Hook Detection Based on Machine Learning and Dynamic Analysis} &
	2020 &
	TRUE \\ \hline
	\href{https://essay.utwente.nl/79200/}{Remote   runtime 
	detection of tampering and of dynamic analysis attempts for Android   apps} &
	2019 &
	TRUE \\ \hline
	\href{https://www.diva-portal.org/smash/record.jsf?pid=diva2:1109568}{Evaluating
	   Dynamic AnalysisMethods for Android Applications} &
	2017 &
	TRUE \\ \hline
	\href{https://www.usenix.org/conference/raid2019/presentation/nisi}{Exploring
	   Syscall-Based Semantics Reconstruction of Android Applications} &
	2019 &
	TRUE \\ \hline
	\href{https://link.springer.com/chapter/10.1007/978-981-15-9609-4\_7}{A
	   Hidden File Extraction Scheme Defeating Malware Using Android Dynamic   
	Loading} &
	2019 &
	TRUE \\ \hline
	\href{https://dl.acm.org/doi/abs/10.1145/2857705.2857739}{Aspectdroid:
	   Android app analysis system} &
	2016 &
	TRUE \\ \hline
	\href{https://ieeexplore.ieee.org/abstract/document/7502337/}{DynaLog:
	   An automated dynamic analysis framework for characterizing android   
	applications} &
	2016 &
	TRUE \\ \hline
	\href{https://link.springer.com/article/10.1007/s10586-016-0703-5}{Droidward:
	   an effective dynamic analysis method for vetting android applications} &
	2018 &
	TRUE \\ \hline
	\href{https://ieeexplore.ieee.org/abstract/document/7417621/}{Reevaluating
	   android permission gaps with static and dynamic analysis} &
	2015 &
	TRUE \\ \hline
	\href{https://www.ndss-symposium.org/wp-content/uploads/2017/09/intellidroid-targeted-input-generator-dynamic-analysis-android-malware.pdf}{IntelliDroid:
	   A Targeted Input Generator for the Dynamic Analysis of Android Malware.} &
	2016 &
	TRUE \\ \hline
	\href{https://dl.acm.org/doi/abs/10.1145/3212480.3212504}{Towards
	   dynamically monitoring Android applications on non-rooted devices in the   
	wild} &
	2018 &
	FALSE \\ \hline
	\href{https://www.mdpi.com/558048}{AndroShield:   automated 
	android applications vulnerability detection, a hybrid static and   dynamic 
	analysis approach} &
	2019 &
	TRUE \\ \hline
	\href{https://ieeexplore.ieee.org/abstract/document/7226692/}{Code
	   coverage measurement for Android dynamic analysis tools} &
	2015 &
	FALSE \\ \hline
	\href{https://ieeexplore.ieee.org/abstract/document/6906344/}{DroidTrace:
	   A ptrace based Android dynamic analysis system with forward execution   
	capability} &
	2014 &
	TRUE \\ \hline
	\href{https://ieeexplore.ieee.org/abstract/document/7898406/}{Security
	   assessment of code obfuscation based on dynamic monitoring in android 
	things} &
	2017 &
	TRUE \\ \hline
	\href{https://dl.acm.org/doi/abs/10.1145/2663761.2664223}{DroidDolphin:
	   a dynamic Android malware detection framework using big data and 
	machine   learning} &
	2014 &
	TRUE \\ \hline
	\href{https://link.springer.com/chapter/10.1007/978-981-15-1675-7\_2}{The
	   state of Android security} &
	2020 &
	TRUE \\ \hline
	\href{https://books.google.com/books?hl=en\&lr=\&id=72nnDwAAQBAJ\&oi=fnd\&pg=PA73\&dq=android+dynamic+analysis+api+monitoring\&ots=42pKkKSqp8\&sig=w0EpaqnSTh9vXWgcjOYUCnLmkxY}{Effectiveness
	   of state-of-the-art dynamic analysis techniques in identifying diverse   Android 
	malware and future enhancements} &
	2020 &
	TRUE \\ \hline
	\href{https://dl.acm.org/doi/abs/10.1145/2592791.2592796}{Rage
	   against the virtual machine: hindering dynamic analysis of android malware} &
	2014 &
	TRUE \\ \hline
	\href{https://www.sciencedirect.com/science/article/pii/S1877050916001599}{Analysis
	   of malicious behavior of android apps} &
	2016 &
	FALSE \\ \hline
	\href{https://www.sciencedirect.com/science/article/pii/S0167404817301207}{DroidInjector:
	   A process injection-based dynamic tracking system for runtime behaviors of   
	Android applications} &
	2017 &
	TRUE \\ \hline
	\href{http://ijns.jalaxy.com.tw/contents/ijns-v18-n1/ijns-2016-v18-n1-p182-192.pdf}{Malicious
	   behavior analysis for android applications} &
	2016 &
	FALSE \\ \hline
	\href{https://ieeexplore.ieee.org/abstract/document/8374030/}{A
	   novel dynamic Android malware detection system with ensemble learning} &
	2018 &
	FALSE \\ \hline
	\href{https://books.google.com/books?hl=en\&lr=\&id=rBMqCgAAQBAJ\&oi=fnd\&pg=PA93\&dq=android+dynamic+analysis+api+monitoring\&ots=g2qvPymWkq\&sig=udNLczIfTd7rrN0JUfjt6v6Vl9I}{DroidMonitor:
	   a high-level programming model for dynamic API monitoring on Android} &
	2015 &
	TRUE \\ \hline
	\href{https://dl.acm.org/doi/abs/10.1145/2480362.2480701}{Mobile-sandbox:
	   having a deeper look into android applications} &
	2013 &
	FALSE \\ \hline
	\href{https://ieeexplore.ieee.org/abstract/document/6906344/}{DroidTrace:
	   A ptrace based Android dynamic analysis system with forward execution   
	capability} &
	2014 &
	TRUE \\ \hline
	\href{https://link.springer.com/article/10.1007/s10586-016-0703-5}{Droidward:
	   an effective dynamic analysis method for vetting android applications} &
	2018 &
	TRUE \\ \hline
	\href{https://ieeexplore.ieee.org/abstract/document/8706349/}{Real
	   time android ransomware detection by analyzed android applications} &
	2019 &
	TRUE \\ \hline
	\href{https://www.ndss-symposium.org/wp-content/uploads/2017/09/intellidroid-targeted-input-generator-dynamic-analysis-android-malware.pdf}{IntelliDroid:
	   A Targeted Input Generator for the Dynamic Analysis of Android Malware.} &
	2016 &
	TRUE \\ \hline
	\href{https://dl.acm.org/doi/abs/10.1145/3212480.3212504}{Towards
	   dynamically monitoring Android applications on non-rooted devices in the   
	wild} &
	2018 &
	FALSE \\ \hline
	\href{https://dl.acm.org/doi/abs/10.1145/2857705.2857739}{Aspectdroid:
	   Android app analysis system} &
	2016 &
	TRUE \\ \hline
	\href{https://ieeexplore.ieee.org/abstract/document/8374030/}{A
	   novel dynamic Android malware detection system with ensemble learning} &
	2018 &
	FALSE \\ \hline
	\href{https://ieeexplore.ieee.org/abstract/document/8255546/}{Samadroid:
	   a novel 3-level hybrid malware detection model for android operating system} 
	&
	2018 &
	FALSE \\ \hline
	\href{https://www.sciencedirect.com/science/article/pii/S074373151630140X}{A
	   hybrid approach of mobile malware detection in Android} &
	2017 &
	FALSE \\ \hline
	\href{https://ieeexplore.ieee.org/abstract/document/8074847/}{Malware
	   detection in android based on dynamic analysis} &
	2017 &
	FALSE \\ \hline
	\href{https://citeseerx.ist.psu.edu/viewdoc/download?doi=10.1.1.1073.3909\&rep=rep1\&type=pdf}{Android
	   based mobile application development and its security} &
	2012 &
	FALSE \\ \hline
	\href{https://www.sciencedirect.com/science/article/pii/S2214212619303837}{Visualizing
	   the outcome of dynamic analysis of Android malware with VizMal} &
	2020 &
	TRUE \\ \hline
	\href{http://ceur-ws.org/Vol-1575/paper\_10.pdf}{ARTDroid:   A 
	Virtual-Method Hooking Framework on Android ART Runtime.} &
	2016 &
	FALSE \\ \hline
	\href{https://dl.acm.org/doi/abs/10.1145/2480362.2480701}{Mobile-sandbox:
	   having a deeper look into android applications} &
	2013 &
	FALSE \\ \hline
	\href{https://books.google.com/books?hl=en\&lr=\&id=72nnDwAAQBAJ\&oi=fnd\&pg=PA73\&dq=android+dynamic+analysis+file+access+monitoring\&ots=42pKkKSqr5\&sig=nrqD9lZkPLoM6xuvqojzkmqQ1O0}{Effectiveness
	   of state-of-the-art dynamic analysis techniques in identifying diverse   Android 
	malware and future enhancements} &
	2020 &
	FALSE \\ \hline
	\href{https://dl.acm.org/doi/abs/10.1145/2619239.2631434}{Droid-sec:
	   deep learning in android malware detection} &
	2014 &
	FALSE \\ \hline
	\href{https://www.mdpi.com/558048}{AndroShield:   automated 
	android applications vulnerability detection, a hybrid static and   dynamic 
	analysis approach} &
	2019 &
	FALSE \\ \hline
	\href{https://dl.acm.org/doi/abs/10.1145/2046614.2046619}{Crowdroid:
	   behavior-based malware detection system for android} &
	2011 &
	FALSE \\ \hline
	\href{https://ieeexplore.ieee.org/abstract/document/7417621/}{Reevaluating
	   android permission gaps with static and dynamic analysis} &
	2015 &
	FALSE \\ \hline
	\href{https://dl.acm.org/doi/abs/10.1145/2614628.2614633}{Android
	   taint flow analysis for app sets} &
	2014 &
	TRUE \\ \hline
	\href{https://ieeexplore.ieee.org/abstract/document/6394931/}{Leakminer:
	   Detect information leakage on android with static taint analysis} &
	2012 &
	TRUE \\ \hline
	\href{https://dl.acm.org/doi/abs/10.1145/2666356.2594299}{Flowdroid:
	   Precise context, flow, field, object-sensitive and lifecycle-aware taint   
	analysis for android apps} &
	2014 &
	TRUE \\ \hline
	\href{https://orbilu.uni.lu/bitstream/10993/25162/1/Highly\%20Precise\%20Taint\%20Analysis\%20for\%20Android\%20Application.pdf}{Highly
	   precise taint analysis for android applications} &
	2013 &
	TRUE \\ \hline
	\href{https://dl.acm.org/doi/abs/10.1145/3236024.3236029}{Do  
	 android taint analysis tools keep their promises?} &
	2018 &
	TRUE \\ \hline
	\href{https://dl.acm.org/doi/abs/10.1145/2771783.2771803}{Scalable
	   and precise taint analysis for android} &
	2015 &
	TRUE \\ \hline
	\href{https://arxiv.org/abs/1404.7431}{I   know what leaked in 
	your pocket: uncovering privacy leaks on Android Apps   with Static Taint 
	Analysis} &
	2014 &
	TRUE \\ \hline
	\href{https://ieeexplore.ieee.org/abstract/document/8952502/}{A
	   qualitative analysis of Android taint-analysis results} &
	2019 &
	TRUE \\ \hline
	\href{https://dl.acm.org/doi/abs/10.1145/2635868.2635869}{Apposcopy:
	   Semantics-based detection of android malware through static analysis} &
	2014 &
	TRUE \\ \hline
	\href{https://ieeexplore.ieee.org/abstract/document/7582763/}{HybriDroid:
	   static analysis framework for Android hybrid applications} &
	2016 &
	TRUE \\ \hline
	\href{https://www.sciencedirect.com/science/article/pii/S0950584917302987}{Static
	   analysis of android apps: A systematic literature review} &
	2017 &
	TRUE \\ \hline
	\href{https://ieeexplore.ieee.org/abstract/document/8010886/}{TaintMan:
	   An ART-compatible dynamic taint analysis framework on unmodified and   
	non-rooted Android devices} &
	2017 &
	TRUE \\ \hline
	\href{https://ieeexplore.ieee.org/abstract/document/6461017/}{TrustDroid:
	   Preventing the use of SmartPhones for information leaking in corporate   
	networks through the used of static analysis taint tracking} &
	2012 &
	TRUE \\ \hline
	\href{https://ieeexplore.ieee.org/abstract/document/9284080/}{Heaps'n
	   leaks: how heap snapshots improve Android taint analysis} &
	2020 &
	TRUE \\ \hline
	\href{https://ieeexplore.ieee.org/abstract/document/7011272/}{Appcaulk:
	   Data leak prevention by injecting targeted taint tracking into android apps} &
	2014 &
	FALSE \\ \hline
	\href{https://ieeexplore.ieee.org/abstract/document/8410576/}{BridgeTaint:
	   a bi-directional dynamic taint tracking method for JavaScript bridges in   
	android hybrid applications} &
	2018 &
	TRUE \\ \hline
	\href{https://ieeexplore.ieee.org/abstract/document/7223198/}{On
	   the effectiveness of dynamic taint analysis for protecting against private   
	information leaks on Android-based devices} &
	2013 &
	TRUE \\ \hline
	\href{https://dl.acm.org/doi/abs/10.1145/3196398.3196433}{Leveraging
	   historical versions of Android apps for efficient and precise taint analysis} &
	2018 &
	TRUE \\ \hline
	\href{https://ieeexplore.ieee.org/abstract/document/6828062/}{Overview
	   of information flow tracking techniques based on taint analysis for android} &
	2014 &
	TRUE \\ \hline
	\href{https://dl.acm.org/doi/abs/10.1145/3243218.3243222}{Poking
	   the bear: Lessons learned from probing three android malware datasets} &
	2018 &
	FALSE \\ \hline
	\href{https://link.springer.com/chapter/10.1007/978-3-319-58469-0\_23}{Binsign:
	   fingerprinting binary functions to support automated analysis of code   
	executables} &
	2017 &
	FALSE \\ \hline
	\href{https://www.usenix.org/conference/usenixsecurity12/technical-sessions/presentation/yan}{Droidscope:
	   Seamlessly reconstructing the OS and dalvik semantic views for dynamic   
	android malware analysis} &
	2012 &
	FALSE \\ \hline
	\href{https://www.sciencedirect.com/science/article/pii/S016740481630164X}{DroidNative:
	   Automating and optimizing detection of Android native code malware 
	variants} &
	2017 &
	FALSE \\ \hline
	\href{https://arxiv.org/abs/1903.10560}{Don't   Pick the Cherry: 
	An Evaluation Methodology for Android Malware Detection   Methods} &
	2019 &
	FALSE \\ \hline
	\href{https://dl.acm.org/doi/abs/10.1145/3319535.3354240}{Automatic
	   fingerprinting of vulnerable ble iot devices with static uuids from mobile   
	apps} &
	2019 &
	FALSE \\ \hline
	\href{https://ieeexplore.ieee.org/abstract/document/8514207/}{Unmasking
	   android obfuscation tools using spatial analysis} &
	2018 &
	TRUE \\ \hline
	\href{https://dl.acm.org/doi/abs/10.1145/3176258.3176322}{Authorship
	   attribution of android apps} &
	2018 &
	TRUE \\ \hline
	\href{https://core.ac.uk/download/pdf/286999186.pdf}{Obfuscation
	   fingerprinting in Android binaries} &
	2015 &
	TRUE \\ \hline
	\href{https://ieeexplore.ieee.org/abstract/document/6999911/}{Android
	   security: a survey of issues, malware penetration, and defenses} &
	2014 &
	FALSE \\ \hline
	\href{https://www.usenix.org/conference/usenixsecurity19/presentation/pham}{Hidemyapp:
	   Hiding the presence of sensitive apps on android} &
	2019 &
	FALSE \\ \hline
	\href{https://www.sciencedirect.com/science/article/pii/S0167404817302377}{Toward
	   a more dependable hybrid analysis of android malware using aspect-oriented   
	programming} &
	2018 &
	FALSE \\ \hline
	\href{https://link.springer.com/chapter/10.1007/978-3-030-34238-8\_3}{Compiler
	   Provenance Attribution} &
	2020 &
	FALSE \\ \hline
	\href{https://ieeexplore.ieee.org/abstract/document/8684307/}{EveDroid:
	   Event-aware Android malware detection against model degrading for IoT   
	devices} &
	2019 &
	FALSE \\ \hline
	\href{https://link.springer.com/chapter/10.1007/978-3-319-59608-2\_3}{Exploiting
	   android system services through bypassing service helpers} &
	2016 &
	FALSE \\ \hline
	\href{https://www.sciencedirect.com/science/article/pii/S0167739X17329965}{Big
	   Data fingerprinting information analytics for sustainability} &
	2018 &
	FALSE \\ \hline
	\caption{}
	\label{tab:my-table}\\
\end{longtable}

	\endgroup
\end{document}